\documentclass[structabstract]{aa}
\pdfoutput=1
\usepackage{pictex}
\usepackage{longtable,lscape}
\usepackage{latexsym}
\usepackage{graphics}
\usepackage{afterpage}
\usepackage{graphicx}
\usepackage{float}
\usepackage{txfonts}
\begin{document}

\title
{
Unusual quasars from the Sloan Digital Sky Survey selected 
by means of Kohonen self-organising maps\thanks{The full catalogue (Table 3)
is only available in electronic form at the CDS via anonymous
ftp to cdsarc.u-strasbg.fr (130.79.128.5)
or via http://cdsweb.u-strasbg.fr/cgi-bin/qcat?J/A+A/
}
}

\author{H. Meusinger \inst{1}
	 \and 
	P. Schalldach \inst{1}
         \and 
	R.-D. Scholz \inst{2}
      	 \and 
	A. in der Au \inst{1}
         \and
	M. Newholm \inst{1}  
	       \and 
	A. de Hoon \inst{2} 
	       \and
	B. Kaminsky \inst{3} 
}

\institute{Th\"uringer Landessternwarte Tautenburg, Sternwarte 5, D--07778
         Tautenburg, Germany
         \and 
	 Leibniz-Institut f\"ur Astrophysik Potsdam (AIP),
	 An der Sternwarte 16,
	 D--14482 Potsdam, Germany
         \and
	 Universit\"at Leipzig,
	 Fakult\"at f\"ur Physik und Geowissenschaften,
	 Linn\'estr. 5
	 D--04103 Leipzig, Germany		 
}
\date{Received 23 September 2011; accepted 03 February 2012}


\abstract 
{
Large spectroscopic surveys have discovered very peculiar and hitherto
unknown types of active galactic nuclei (AGN). Such rare objects may hold
clues to the accretion history of the supermassive black holes at
the centres of galaxies.   
}
{
We aim to create a sizeable sample of unusual quasars from the unprecedented
spectroscopic database of the Sloan Digital Sky Survey (SDSS).
}
{
We exploit the spectral archive of the SDSS Data Release 7 to select unusual
quasar spectra. The selection method is based on a combination of the power of 
self-organising maps and the visual inspection of a huge number of spectra.
Self-organising maps were applied to nearly $10^5$ spectra classified as quasars
at redshifts from $z=0.6$\ to 4.3 by the SDSS pipeline. Particular attention was paid to
minimise possible contamination by rare peculiar stellar spectral types. All
selected quasar spectra were individually studied to determine the object
type and the redshift.
}
{
We present a catalogue of 1005 quasars with unusual spectra. These spectra
are dominated by either broad absorption lines (BALs; 42\%),
unusual red continua (27\%), weak emission lines (18\%), or conspicuously 
strong optical and/or UV iron emission (11\%).  
This large sample provides a useful resource for both studying properties and
relations of/between different types of unusual quasars and selecting particularly
interesting objects, even though the compilation is not aimed at completeness
in a quantifiable sense. The spectra are grouped
into six types for which composite spectra are constructed and mean properties are 
computed. Remarkably, all these types turn out to be on average more luminous than 
comparison samples of normal quasars after a statistical correction is made
for intrinsic reddening ($E(B-V) \approx 0$ to 0.4 for SMC-like extinction).
Both the unusual BAL quasars and the strong iron emitters have significantly 
lower radio luminosities than normal quasars. We also confirm that strong BALs 
avoid the most radio-luminous quasars. For 32 particularly interesting objects,  
individual spectra are presented. Among these objects are quasars with many
narrow BAL troughs and one quasar where the continuum is strongly suppressed by
overlapping BAL troughs across nearly the whole SDSS spectrum. Finally, we
create a sample of quasars similar to the two ``mysterious'' objects discovered
by Hall et al. (2002) and briefly discuss the quasar properties and possible
explanations of their highly peculiar spectra.
}
{}



\titlerunning{Kohonen-selected unusual SDSS quasars}
\authorrunning{H. Meusinger et al.}

\maketitle

%
\section{Introduction}
%

From the very beginning of quasar astronomy 
(Schmidt \cite{Schmidt63}; 
Greenstein \& Matthews \cite{Greenstein63}; 
Sandage \cite{Sandage65}), 
optical spectra have been one of the most important sources of information about
this class of fascinating objects. 
Quasars are the most luminous types of active galactic nuclei (AGNs). 
The average ultraviolet (UV) and optical spectral energy distribution (SED)
of AGNs is offen found to be widely consistent
with the predictions of the standard model of a geometrically thin and
optically thick accretion disc around a massive or supermassive black hole
(Shakura \& Sunyaev \cite{Shakura73};
Shields \cite{Shields78}; 
Laor \cite{Laor90}; 
Blaes et al. \cite{Blaes01}). 
The continuum radiation field from the accretion disc provides the photons
for the ionisation of the surrounding gas and powers broad and narrow emission
lines. The unified AGN model 
(Antonucci \cite{Antonucci93}; 
Urry \& Padovani \cite{Urry95})
allows us to understand some differences in the spectra of various AGN types
as being caused by different orientations of the sightline towards the central engine.
For the vast majority of known quasars, the UV-optical SED is well represented
by global composite spectra produced from 
large samples (e.g., Francis et al. \cite{Francis91}; Zheng et al. \cite{Zheng97}; 
Vanden Berk et al. \cite{VandenBerk01}), where discretion should be exercised
however when using the composite of one sample as a template for other samples
(Vanden Berk et al. \cite{VandenBerk01}; 
Richards et al. \cite{Richards03}). 
       
It is to be expected that various predicted spectral features, such as 
the spectral slope or the strength of broad emission lines, can change 
from quasar to quasar owing to
variations in physical parameters
(e.g., Koratkar \& Blaes \cite{Koratkar99}; 
Laor \& Davis \cite{Laor11}), and there is empirical evidence that 
spectral properties can also be modified by variability
(Kinney et al. \cite{Kinney90};
Wilhite et al. \cite{Wilhite05}; 
Hall et al. \cite{Hall11};
Meusinger et al. \cite{Meusinger11}).  
Further modifications of the spectrum can be caused by 
absorption from material related to the quasar itself, 
to the host galaxy, or from unrelated intervening matter along the 
line of sight. In addition, some spectral features can be diluted by 
unrelated emission components 
(Johnston et al. \cite{Johnston03}; 
Kishimoto et al. \cite{Kishimoto08})
or magnified by gravitational lensing 
(Abajas et al. \cite{Abajas07};
Yonehara et al. \cite{Yonehara08}; 
Blackburne et al. \cite{Blackburne11}).

On the other hand, it is not to be expected that quasars are fully 
described by a static model, because  
``quasars are not {\it things} so much as {\it processes}''
(Richards et al. \cite{Richards11}). 
Over approximately the past decade, spectroscopic surveys have
revealed a variety of quasar spectra that dramatically differ from the
standard SED, and confirmed the existence of populations
of exotic and hitherto unknown quasar types
(Fan et al. \cite{Fan99};
White et al. \cite{White00};
Menou et al. \cite{Menou01};
Hines \cite{Hines01};
Hall et al. \cite{Hall02}, \cite{Hall04};
Brunner et al. \cite{Brunner03};
Meusinger et al. \cite{Meusinger05};
Plotkin et al. \cite{Plotkin08};
Diamond-Stanic et al. \cite{Diamond09};
Urrutia et al. \cite{Urrutia09}). 
The discovery and investigation of these rare peculiar objects is important
because they may represent links to special evolutionary stages and provide 
an excellent laboratory to study feedback mechanisms between star formation and 
accretion activity. There are strong indications from
different lines of thought for a tight relationship between the activity and 
growth of massive black holes and the evolution of their host galaxies 
(e.g., Sanders et al. \cite{Sanders88};
Magorrian et al. \cite{Magorrian98};
Kauffmann \& H\"ahnelt \cite{Kauffmann00};
Taniguchi \cite{Taniguchi03};
H\"aring \& Rix \cite{Haering04};
Hopkins et al. \cite{Hopkins06};
Kauffmann \& Heckman \cite{Kauffmann09};
Meng et al. \cite{Meng10};
Shabala et al. \cite{Shabala11}).
Much attention has been focused in particular on identifying the
youthful quasar population. 
     
There are several tentative arguments supporting the view that young quasars 
are associated with gaseous outflows and absorption by dust and gas.     
Many AGNs show blueshifted absorption from outflowing matter 
(Ganguly \& Brotherton \cite{Ganguly08}). Very broad absorption lines (BALs) 
corresponding to a wide variety of velocities are observed in luminous quasars.
These BALs are naturally explained as footprints of powerful, sub-relativistic (up to $\sim 0.2c$)
outflows (Weymann et al. \cite{Weymann91}), which are perhaps related to
radiation-driven winds from the accretion disc (Murray \& Chiang \cite{Murray98}).
Broad absorption lines are thus the most obvious
manifestation of matter accelerated by the central engine of AGNs.
It has been frequently argued that the BAL fraction can be explained as being caused by 
an orientation effect within the unification model where otherwise normal quasars 
are seen at sightlines close to the equatorial plane of the AGN 
(Weymann et al. \cite{Weymann91}). Typical BAL quasars are often subdivided into those
showing broad absorption lines
from only high-ionisation species such as \ion{C}{iv} and \ion{N}{v} (HiBALs)
and those showing in addition also absorption from low-ionisation species such
as \ion{Mg}{ii} and \ion{Al}{iii} (LoBALs). A rare subclass of the LoBAL quasars 
are the FeLoBALs with absorption from metastable excited states of \ion{Fe}{II}\, 
and \ion{Fe}{III}\, (Hazard et al. \cite{Hazard87}; Becker et al. \cite{Becker00}). 
The BAL quasars are thought to account for between 10\% and $\sim 40$\% of all quasars 
(Hewett \& Foltz \cite{Hewett03}; 
Reichard et al. \cite{Reichard03}; 
Carballo et al. \cite{Carballo06};
Trump et al. \cite{Trump06}; 
Dai et al. \cite{Dai08};
Gibson et al. \cite{Gibson09};
Allen et al. \cite{Allen11}). 
We note that the exact BAL fraction is not trivial
to determine because of differential selection effects between BAL quasars 
and non-BAL quasars. In addition, different definitions of BAL quasars have been
used in different studies.  According to the criteria applied by 
Trump et al. (\cite{Trump06}), BAL quasars comprise 26\% of quasars, LoBALs 
about 1.3\%, and FeLoBALs about 0.3\% (these being the raw fractions without correction for the
additional reddening of BAL quasars compared to non-BALs).
  
Outflows are believed to be fundamental to understanding both the overall picture of AGNs and
the AGN feedback. Despite their importance, BAL quasars remain poorly understood. 
In particular, there is currently no concensus 
about the role of evolutionary effects in determining the properties of BAL quasars.
The properties of HiBAL quasars appear to be consistent
with the pure orientation model
(e.g., 
Surdej \& Hutsemekers \cite{Surdej87};
Gallagher et al. \cite{Gallagher07};
Doi et al. \cite{Doi09}). 
On the other hand, LoBAL quasars  may not be explained by orientation alone 
(e.g.,  Ghosh  \& Punsly \cite{Ghosh07};
Montenegro-Montes et al. \cite{Montenegro09};
Zhang et al. \cite{Zhang10};
but see Gallagher et al. \cite{Gallagher07}
and Hall \& Chajet \cite{Hall11b}).
Their properties probably depend on both orientation and evolution effects
(Richards et al. \cite{Richards11}; Allen et al. \cite{Allen11}).
Based on the shape of the continuum in the UV and X-rays and on the 
BAL fraction in the infrared, it has been suggested that FeLoBAL quasars
in particular probe a much more obscured quasar population than non-BAL
quasars and may represent a young 
evolutionary stage when a thick shroud of gas and dust is being expelled 
from the central region of the AGN 
(Voit et al. \cite{Voit93}; 
Becker et al. \cite{Becker97}; 
Canalizo \& Stockton \cite{Canalizo01};
L\'ipari et al. \cite{Lipari09};
Dai et al. \cite{Dai10};
Farrah et al. \cite{Farrah10}).

It was suggested by early studies that AGNs spend a substantial fraction
of their lifetimes in a dust-enshrouded
environment. This picture was based on the scenario of merger-driven
AGN activity 
(e.g., Sanders et al. \cite{Sanders88};
Canalizo \& Stockton \cite{Canalizo01};
Komossa et al. \cite{Komossa03};
Hopkins et al. \cite{Hopkins05},\cite{Hopkins06};
Guyon et al. \cite{Guyon06};
Bennert et al. \cite{Bennert08}). 
Major mergers of gas-rich galaxies are very efficient in driving gas 
and dust to the central regions thus feeding both circumnuclear starbursts
and accretion onto massive black holes
(e.g., Hernquist \cite{Hernquist89};
Springel et al. \cite{Springel05}). 
The natural consequences are not only a connection between nuclear accretion 
activity and starbursts but also the prediction of a substantial fraction of 
very dusty quasars. Studies in the radio, infrared, and hard X-ray domain 
(e.g. Webster et al. \cite{Webster95}; 
Cutri et al. \cite{Cutri01};
White et al. \cite{White03}; 
Glikman et al. \cite{Glikman07};
Leipski et al. \cite{Leipski08};
Polletta et al. \cite{Polletta08})
have revealed a large population (perhaps $>$ 50\%) of dust-reddened quasars
that optical colour surveys tend to miss and that are not fully explained by
obscuration in the framework of the (static) unified model
(Mart\'inez-Sansigre et al. \cite{Martinez06}; 
Georgakakis et al. \cite{Georgakakis09}; 
Urrutia et al. \cite{Urrutia09}). 
Though it is much more 
difficult to find dust-obscured quasars in the optical, it is also not unrealistic 
to expect moderately reddened quasars in optical surveys.
   
Another category of unusual quasar spectra that has been discovered by large
spectroscopic surveys, typically display the unobscured UV quasar continuum
but extremely weak or undectable
emission lines. The first such quasars were detected more than one decade ago
(McDowell et al. \cite{McDowell95}; Fan et al. \cite{Fan99}). 
About 80 luminous sources of this type with $2.2 \le z \le 5.9$
have since been identified by the Sloan Digital Sky Survey (Shemmer et al. \cite{Shemmer10} 
and references therein).  
Various suggestions have been made to explain the absence
of broad lines in unobscured quasars (e.g., Shemmer et al. \cite{Shemmer10}; 
Laor \& Davis \cite{Laor11}; and references therein) but this AGN type remains 
puzz\-ling. 
A subset of quasars with weak UV emission lines have been identified that
are known as PHL 1811 analogs 
(Leighly et al. \cite{Leighly01}; 
Leighly et al. \cite{Leighly07}; 
Wu et al. \cite{Wu11},\cite{Wu12}).
Leighly et al. (\cite{Leighly07}) suggested that the weak UV emission lines
of these objects are due to an unusual soft SED deficient
in ionising photons, perhaps due to exceptionally high Eddington ratios
(see e.g., Shemmer et al. \cite{Shemmer09} and references therein). 
On the basis of the multi-epoch photometric data of the quasars in the SDSS
stripe 82, Meusinger et al. (\cite{Meusinger11}) found that weak-line 
quasars are less variable than normal. This seems to be consistent with the 
suggestion (Hryniewicz et al. \cite{Hryniewicz10}) that these quasars are in 
an early evolutionary stage.
However, we note that the sample of weak-line quasars with reliable
variability data is small and needs to be checked carefully for possible 
stellar contamination.

Unusually strong \ion{Fe}{ii} emission is observed for a small fraction of quasars. 
The \ion{Fe}{ii} emission extends from the near-UV to the optical region and 
is often one of the most prominent spectral features. 
The strength of the optical
\ion{Fe}{ii} emission is one of the parameters that control a set of correlations 
between various emission line properties, 
known as the Eigenvector 1 (E1, Boroson \& Green \cite{Boroson92a}).
Eigenvector 1 is thought
to serve as a surrogate H-R diagram for AGNs (Sulentic et al. \cite{Sulentic07}). 
The principal driver of E1 seems to be the accretion rate, which may be related 
to the evolutionary state. Strong iron emission is correlated to the occurrence
of BAL troughs (Boroson \& Meyers \cite{Boroson92b}; Zhang et al. \cite{Zhang10}).

We finally note that double-peaked broad emission lines are another
type of rare feature in quasar spectra
(Halpern et al. \cite{Halpern96};
Strateva et al. \cite{Strateva03};
Luo et al. \cite{Luo09};
Chornock et al. \cite{Chornock10}).  
Strateva et al. (\cite{Strateva03}) presented a sample
of 116 double-peaked Balmer line AGNs (corresponding to $\sim 4$\% of low-$z$ AGNs). 
The spectral signature of this subclass is displaced red and blue peaks of the
emission lines similar to double-peaked emission lines found in the
spectra of cataclysmic variables. Double-peaked emission lines are believed to originate 
from rotational motion in a relativistic accretion disc. High rotational velocities in
combination with a highly inclined disc and strong reddening can alter the
AGN spectrum significantly  (Hall et al. \cite{Hall02}). To date, the double-peak
effect has only been observed in low-ionisation lines, while the UV lines, which
dominate the spectra of high-$z$ quasars, are usually single-peaked.  For some double-peaked 
AGNs, a subparsec supermassive black hole binary has been suggested as explanation (Gaskell 
\cite{Gaskell83}; Tang et al. \cite{Tang09}). The existence of close massive 
black-hole binaries appears as a natural consequence of galaxy mergers.

About one decade ago, only a few peculiar quasars had been discovered, 
mainly from follow-up of radio sources from the FIRST survey 
(Becker et al. \cite{Becker95},\cite{Becker00}; 
White et al. \cite{White00}). 
The Sloan Digital Sky Survey (SDSS; York et al. \cite {York06}) 
demonstrated in its early phase that these rare objects are
not unique, but are members of populations of very peculiar quasars
that were occasionally found to have spectral properties never seen before
(Hall et al. \cite{Hall02}).
Meanwhile, the SDSS has dramatically increased the number of spectroscopically 
confirmed quasars. With its efficient quasar selection technique and 
huge number of high-quality spectra, the SDSS is uniquely qualified
to significantly increase the samples of unusual quasars. 

The present paper aims to select spectroscopic outliers among
the quasars from
the SDSS Seventh Data Release (DR7; Abazajian \cite{Abazajian09}).
The database is shortly described in Sect.\,\ref{sec:SDSS}.
The selection itself is based on an artificial neural network algorithm 
that uses unsupervised learning (Sect.\,\ref{sec:kohonen}).
The strategy of unusual quasar selection, the rejection of contaminating sources, 
and the classification of the selected objects are described in 
Sect.\,\ref{sec:selection}. The final catalogue and sample-averaged properties, 
including composite spectra, are discussed in Sect.\,\ref{sec:groups},
followed by the presentation and discussion of 32 particularly 
strange spectra. 
Summary and conclusions are given in Sect.\,\ref{sec:summary}.
Standard cosmological parameters 
$H_0=71$ km s$^{-1}$ Mpc$^{-1}, \Omega_{\rm m}=0.27$, and $\Omega_{\Lambda}=0.73$
are used throughout the paper.

%
\section{Database: Sloan Digital Sky Survey}\label{sec:SDSS}
%

The Sloan Digital Sky Survey (SDSS; York et al. \cite{York06}) 
mapped one quarter of the sky and performed a redshift survey of galaxies, 
quasars, and stars. The total spectroscopic survey area covers 
9380 square degrees of the high Galactic latitude sky.
The spectra were taken with the 2.5~m SDSS telescope at Apache Point Observatory
equipped with a pair of double fibre-fed spectrographs.
As the spectra are collected, the data is passed 
on and filtered through the spectroscopic pipeline. The automated system
produces wavelength and flux-calibrated spectra, measures emission and
absorption lines, classifies spectra, and determines the redshifts, which 
are, among other data, written into the headers of the fits files containing
the spectra. The spectra are made publicly available via the SDSS Data 
Archive Server (DAS). The present study uses quasar spectra from the SDSS 
Seventh Data Release (DR7; Abazajan et al. \cite{Abazajian09}) 
available from  the DAS. The DR7, marking the completion of the original goals 
of the SDSS and the 
end of the phase known as SDSS-II, contains 1\,440\,961 spectra in total 
(after removing skies and duplicates), including 
121\,363 quasar spectra. 
The spectra cover the wavelength range 
3800-9200\AA\ with a resolution of $\sim 2000$ and sampling of $\sim 2.4$ pixels per 
resolution element. 

The majority of quasars were selected for spectroscopic follow-up according to
their deviation away from the ``stellar locus'' in the multidimensional SDSS
colour space. The SDSS filter system
allows us to discover
 quasars over the redshift range from 
$\sim 0$ to 7. In addition to the multicolour selection, a small fraction 
of unresolved objects were selected as primary quasar candidates because their 
SDSS positions closely match either the positions of radio sources from the VLA FIRST
survey (Becker et al. \cite{Becker95}) or of {\it ROSAT} X-ray sources (Voges et al. 
\cite{Voges00}).
A fully detailed description of the SDSS method of selecting quasars is
given by Richards et al. (\cite{Richards02}).

A catalogue of 105\,783 quasars from the SDSS DR7 was compiled and published by
Schneider et al. (\cite{Schneider10}) as the Fifth Edition of the SDSS Quasar
Catalogue (hereafter QCDR7). This catalogue illustrates the unprecedented impact that
SDSS has made for the quasar database. Nevertheless, since the present study is aimed 
at the search for very unusual SDSS quasar spectra, our quasar selection is based 
on the spectra available via the Data Archive Sever of the SDSS DR7
instead of using the QCDR7.

%
\section{Kohonen self-organising maps}\label{sec:kohonen}
%

\begin{figure*}[htbp]
\begin{tabbing}
\includegraphics[bb=185 306 432 545,scale=0.356,clip]{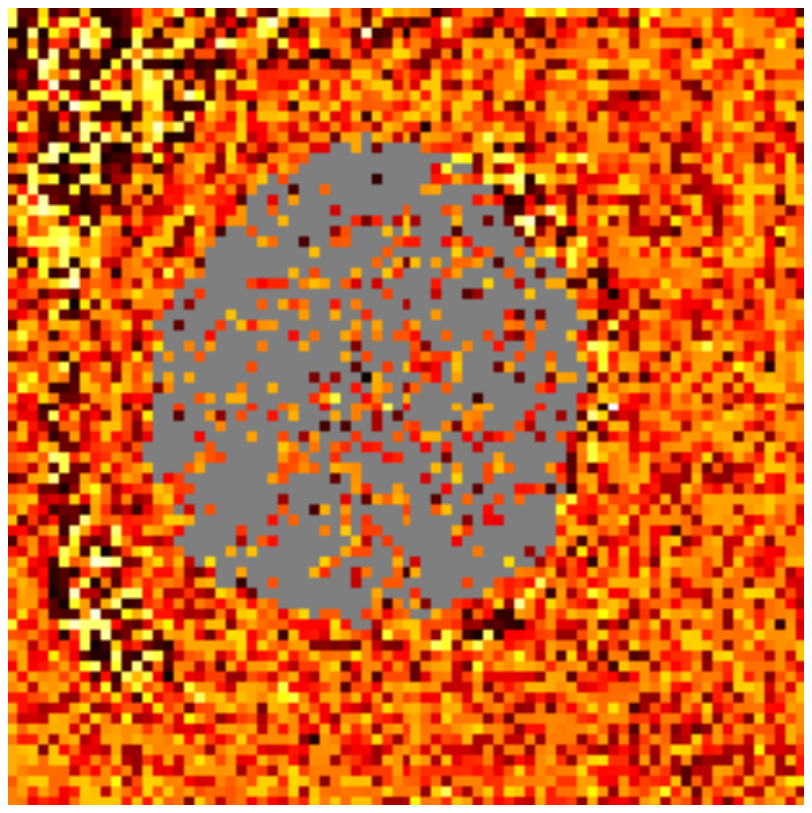}\hfill \=
\includegraphics[bb=185 306 430 545,scale=0.356,clip]{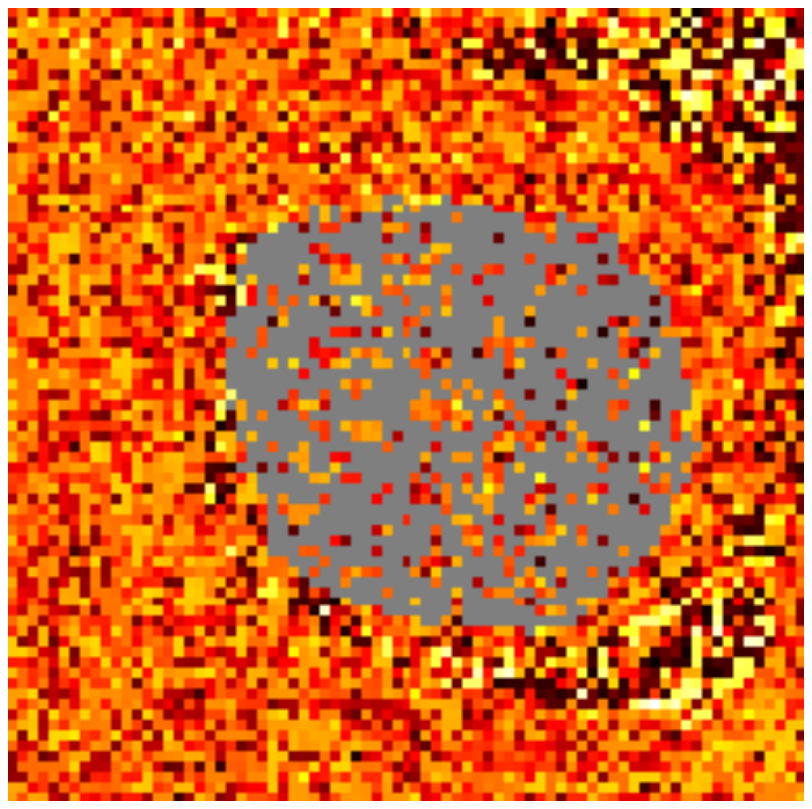}\hfill \=
\includegraphics[bb=185 306 430 545,scale=0.356,clip]{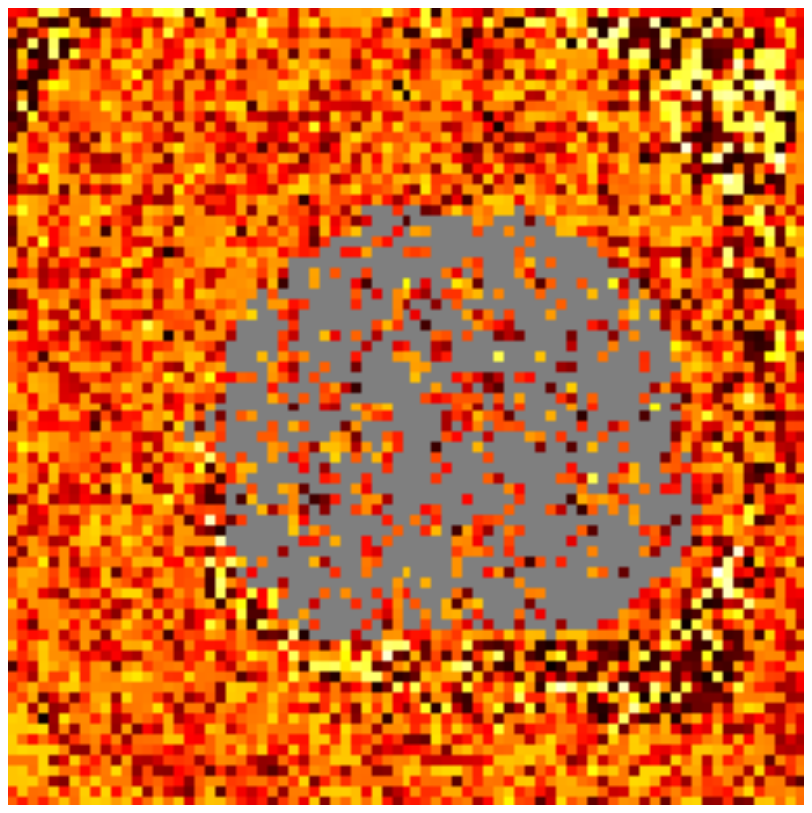}\hfill \=
\includegraphics[bb=185 306 430 545,scale=0.356,clip]{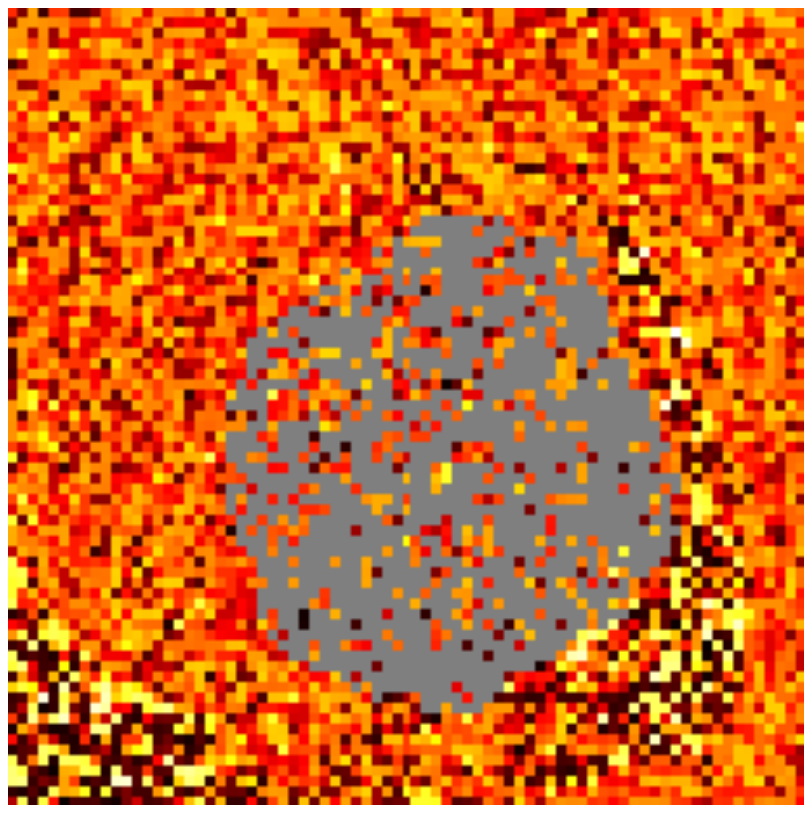}\hfill \=
\includegraphics[bb=185 306 430 545,scale=0.356,clip]{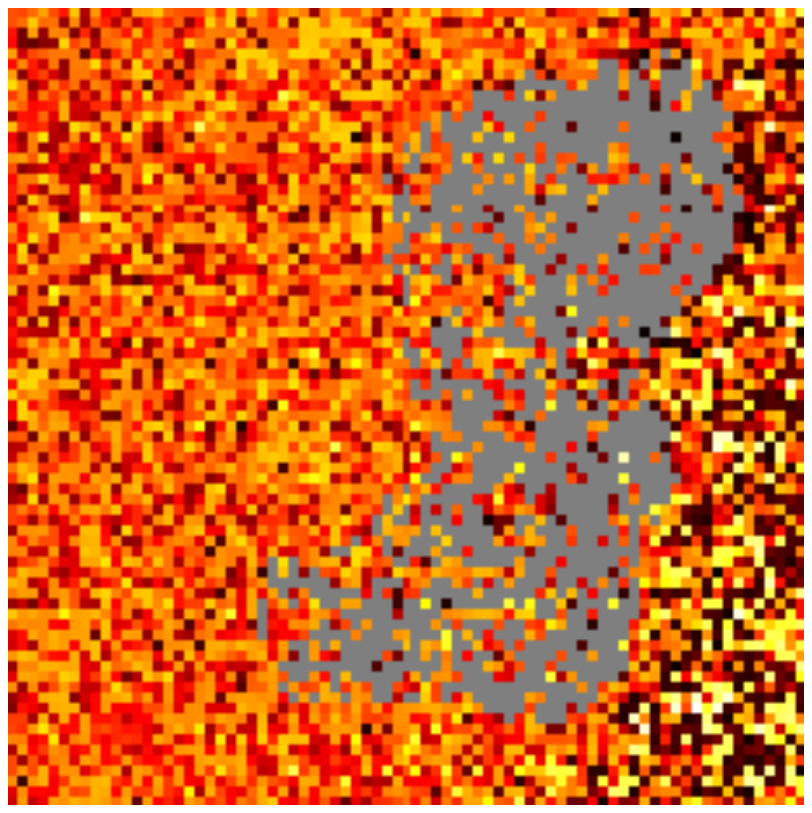}\hfill \=
\includegraphics[bb=185 306 430 545,scale=0.356,clip]{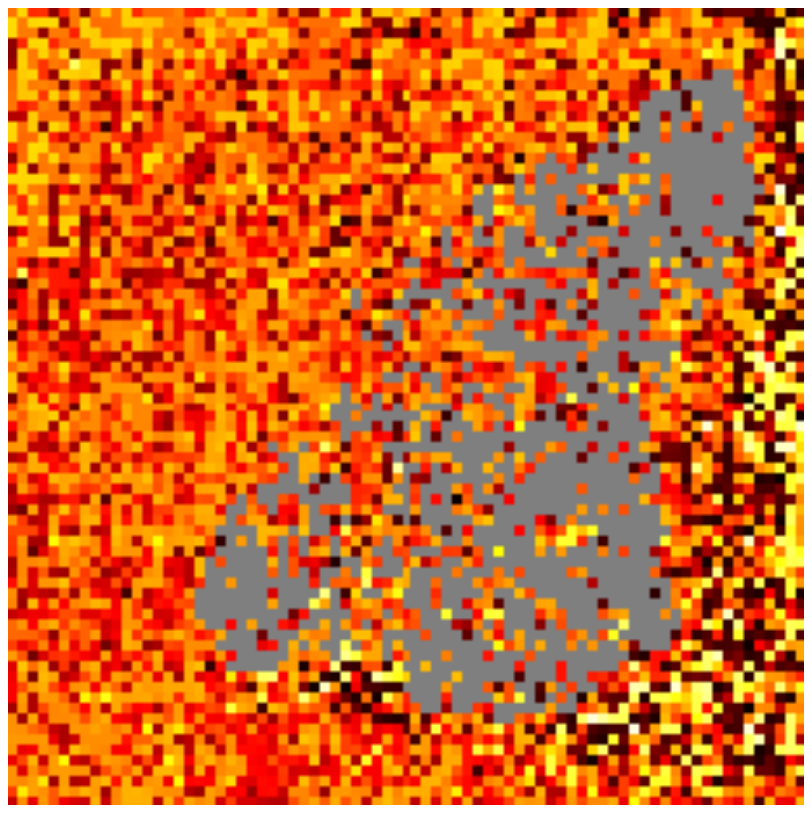}\hfill \\
\includegraphics[bb=185 306 432 545,scale=0.356,clip]{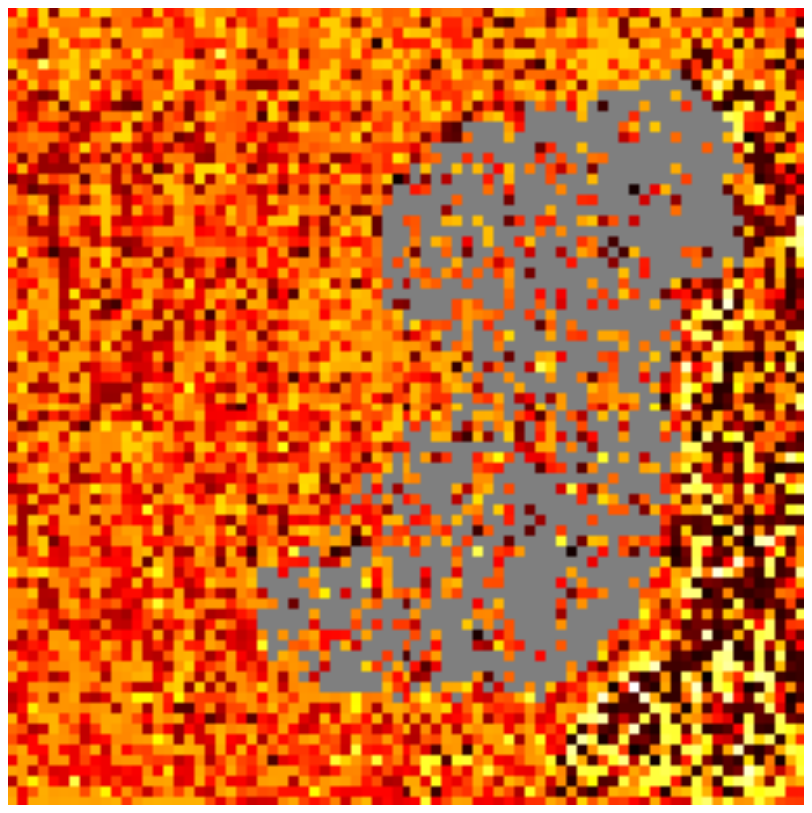}\hfill \=
\includegraphics[bb=185 306 430 545,scale=0.356,clip]{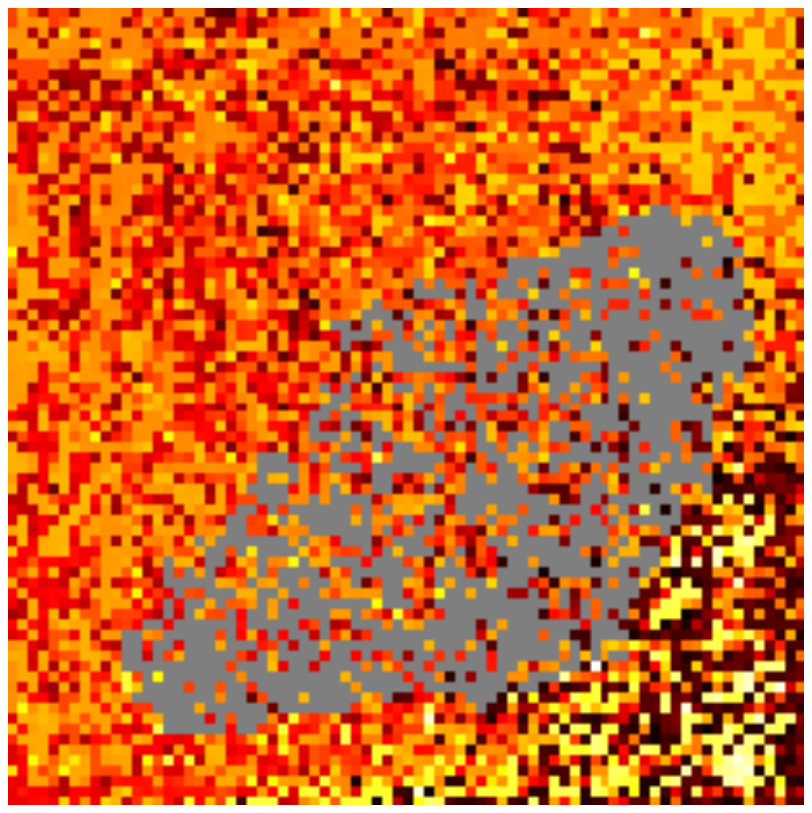}\hfill \=
\includegraphics[bb=185 306 430 545,scale=0.356,clip]{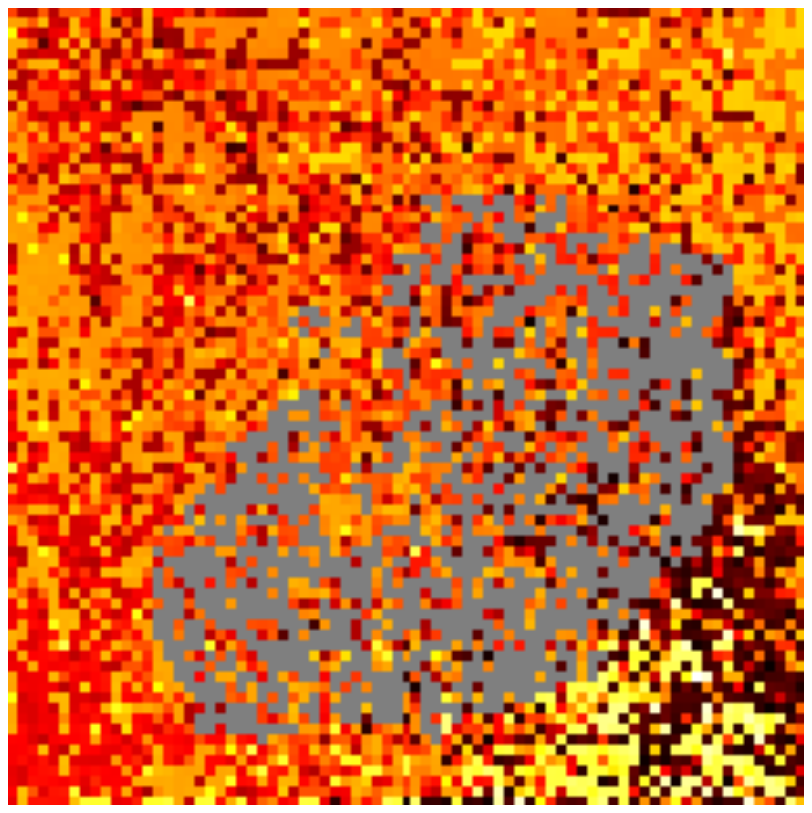}\hfill \=
\includegraphics[bb=185 306 430 545,scale=0.356,clip]{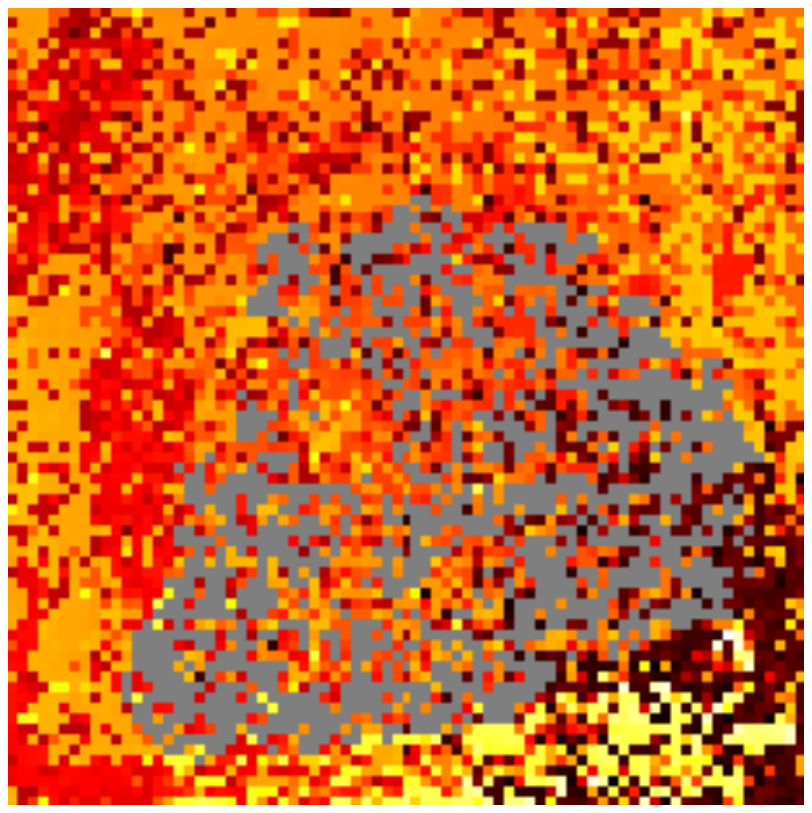}\hfill \=
\includegraphics[bb=185 306 430 545,scale=0.356,clip]{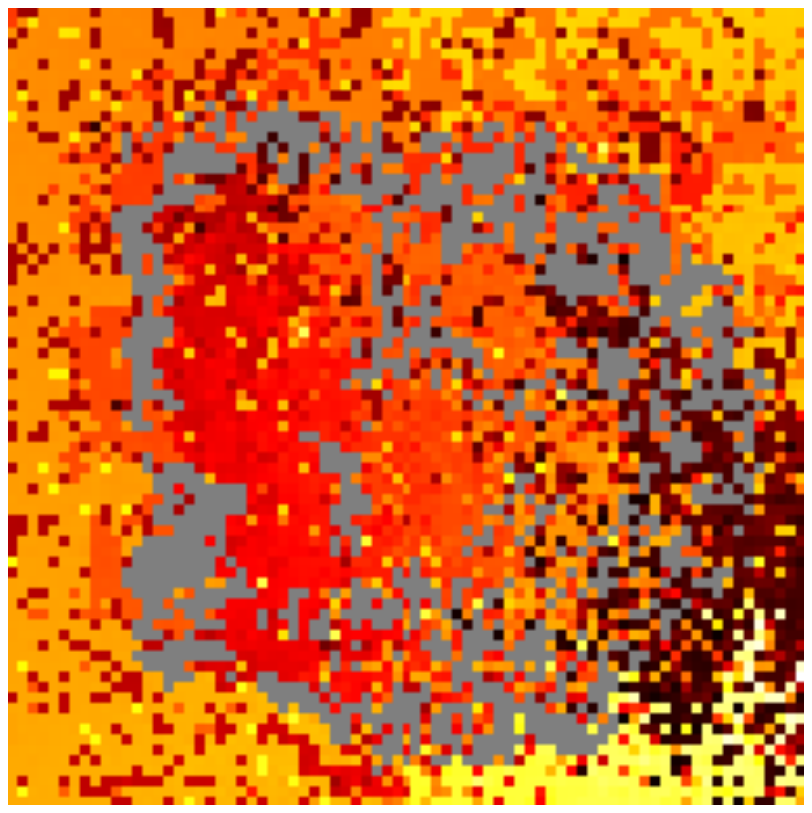}\hfill \=
\includegraphics[bb=185 306 430 545,scale=0.356,clip]{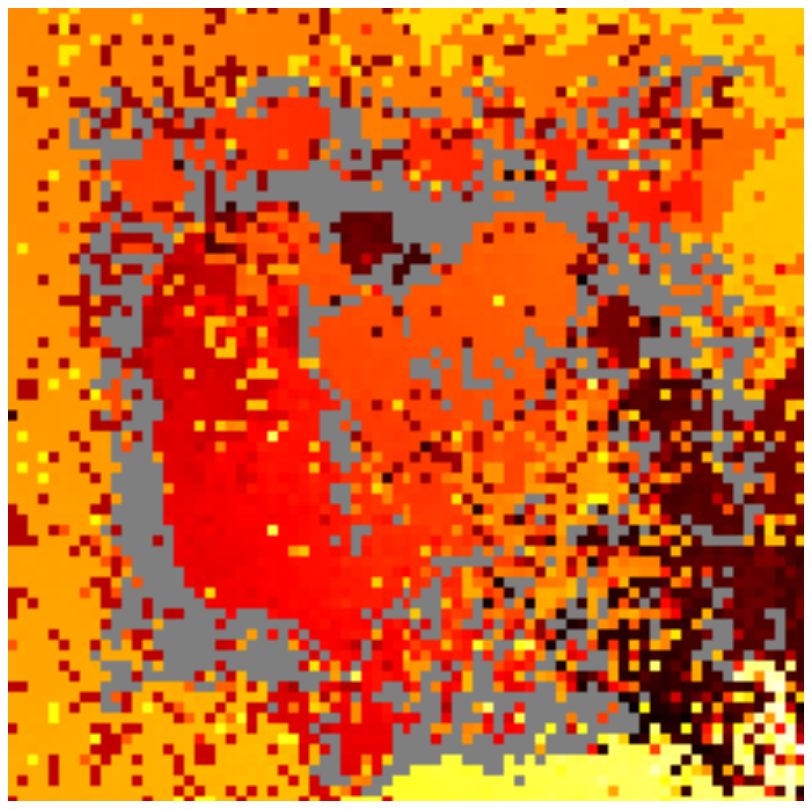}\hfill \\
\includegraphics[bb=185 306 432 545,scale=0.356,clip]{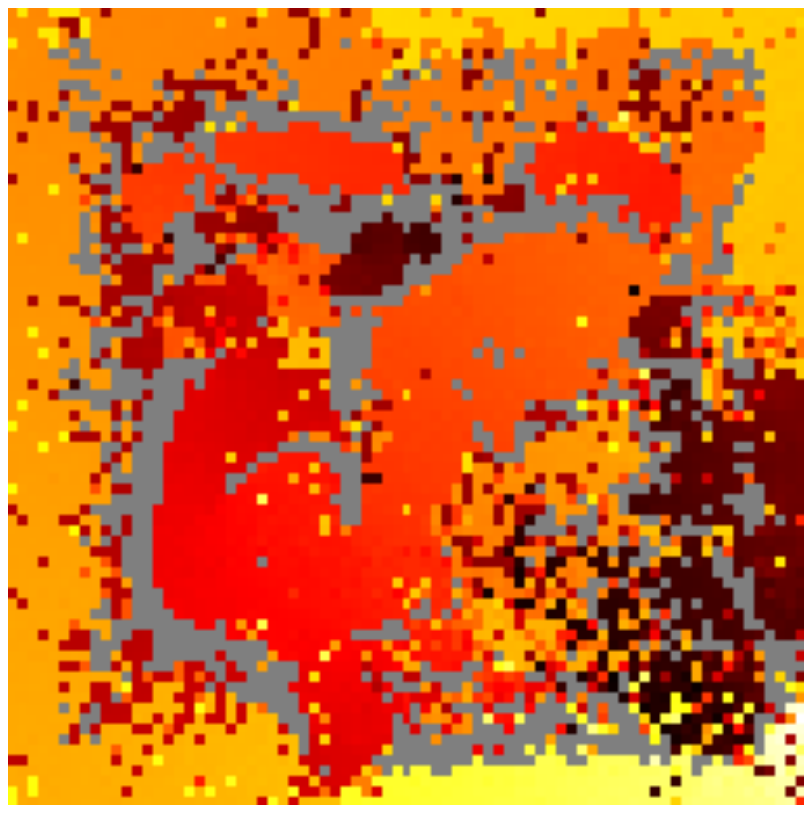}\hfill \=
\includegraphics[bb=185 306 430 545,scale=0.356,clip]{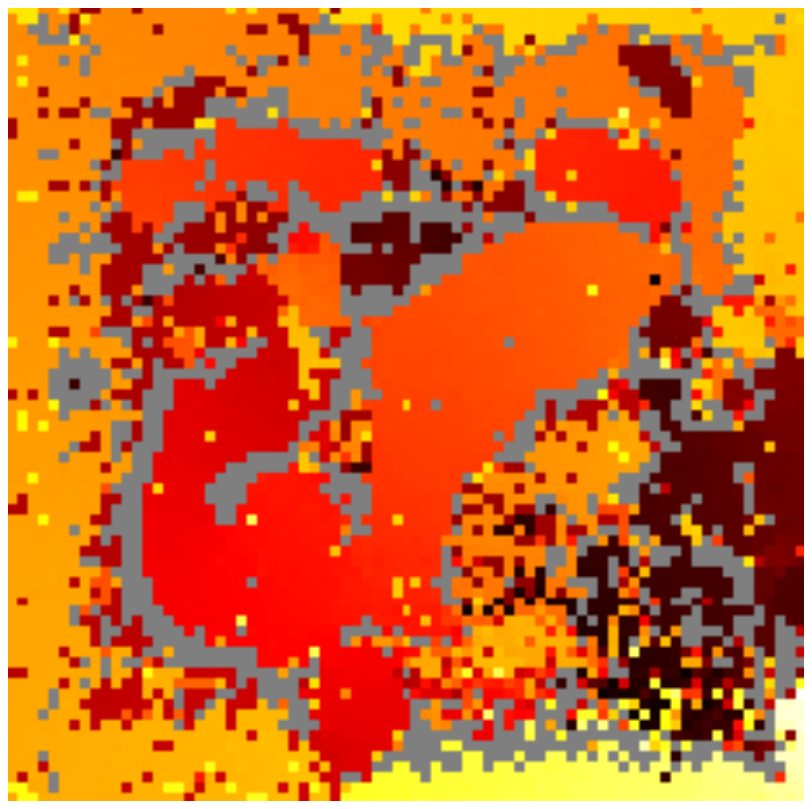}\hfill \=
\includegraphics[bb=185 306 430 545,scale=0.356,clip]{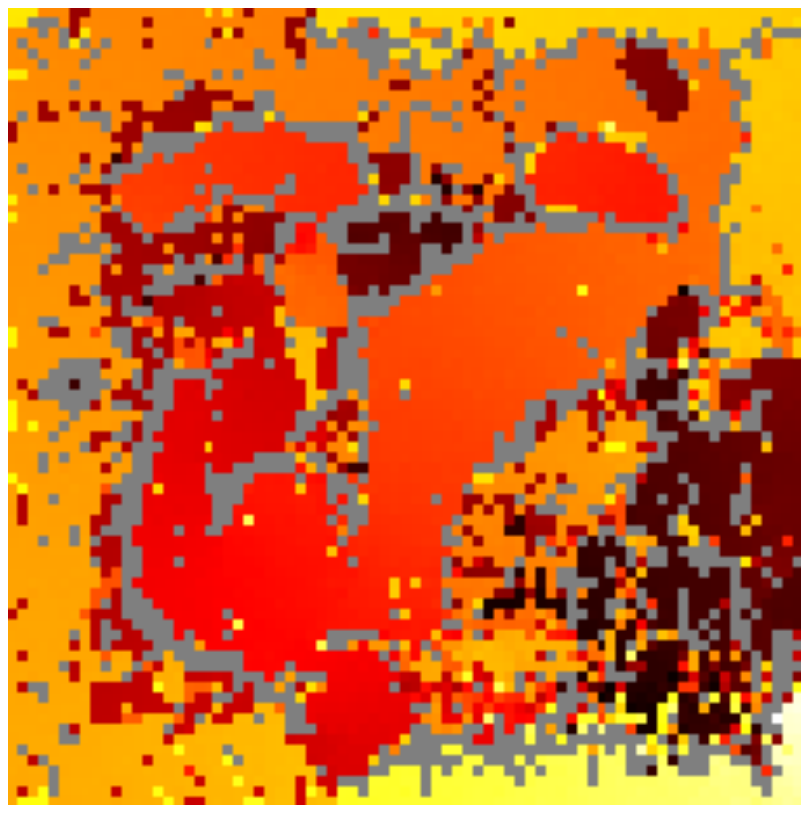}\hfill \=
\includegraphics[bb=185 306 430 545,scale=0.356,clip]{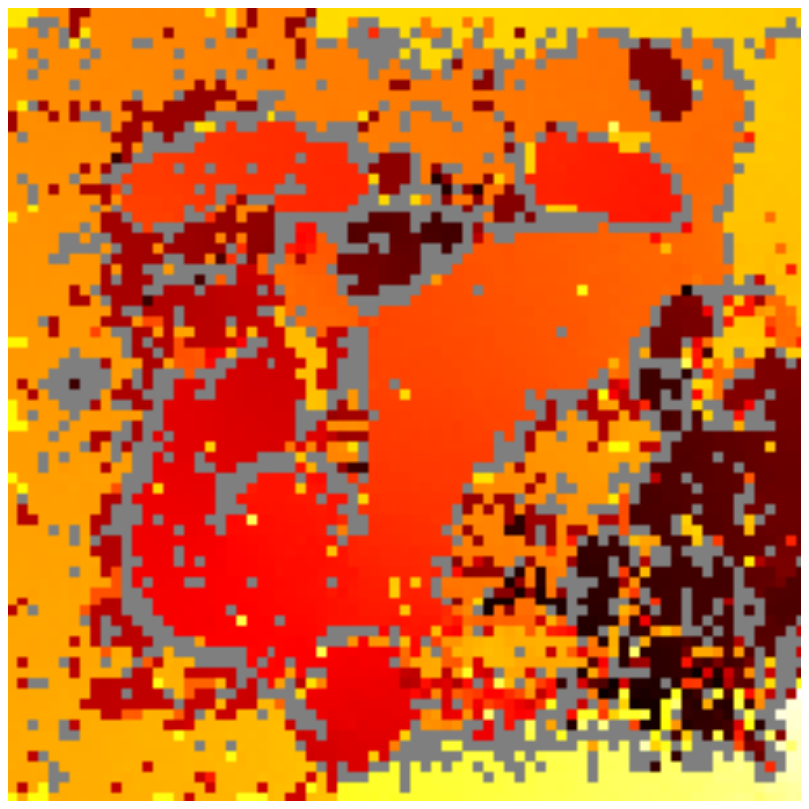}\hfill \=
\includegraphics[bb=185 306 430 545,scale=0.356,clip]{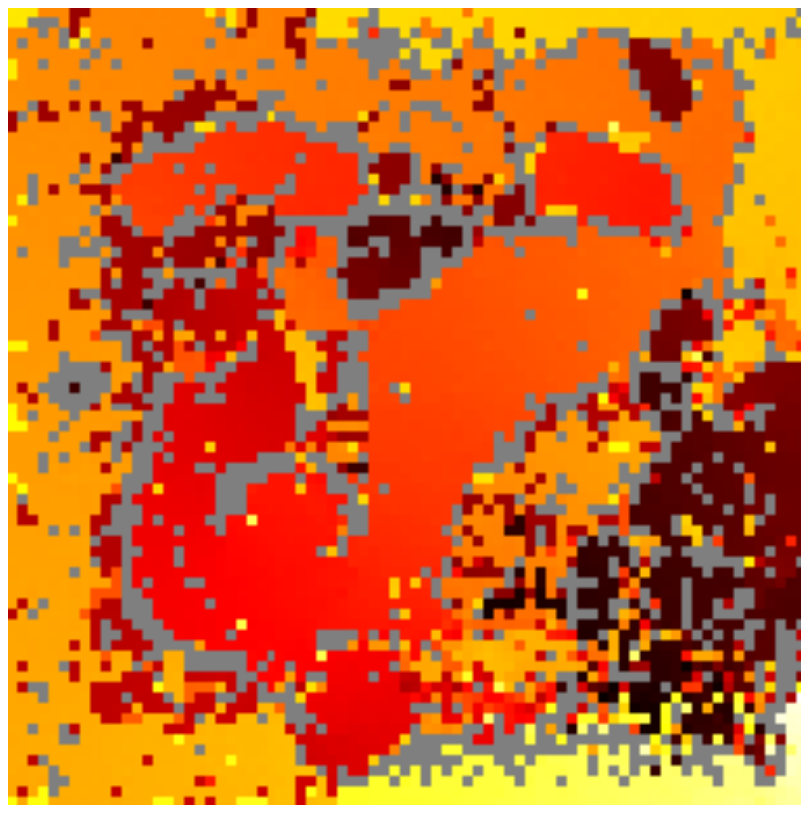}\hfill \=
\includegraphics[bb=185 306 430 545,scale=0.356,clip]{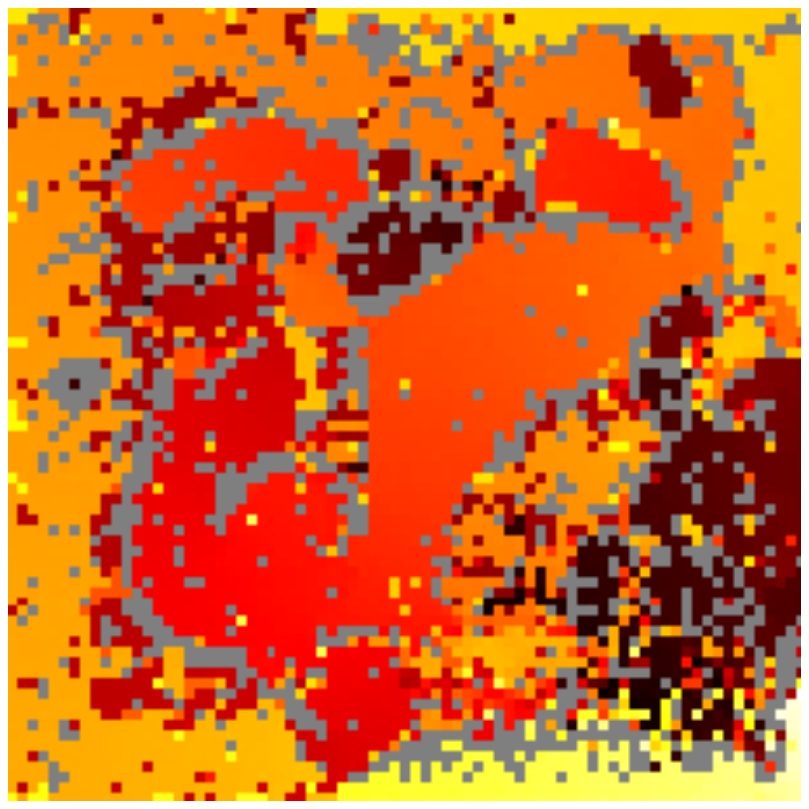}\hfill \\
\end{tabbing}
\caption{Temporal evolution (left to right then top to bottom) of the Kohonen SOM
of $5\,10^3$ SDSS quasar spectra. The quasar redshifts are highlighted by means of 
colour coding. The gradient from dark (black) to bright (light yellow) represents the range from 
low to high redshift.
}
\label{fig:z-map}
\end{figure*}

\begin{figure*}[htbp]
\begin{tabbing}
\includegraphics[bb=185 306 432 545,scale=0.356,clip]{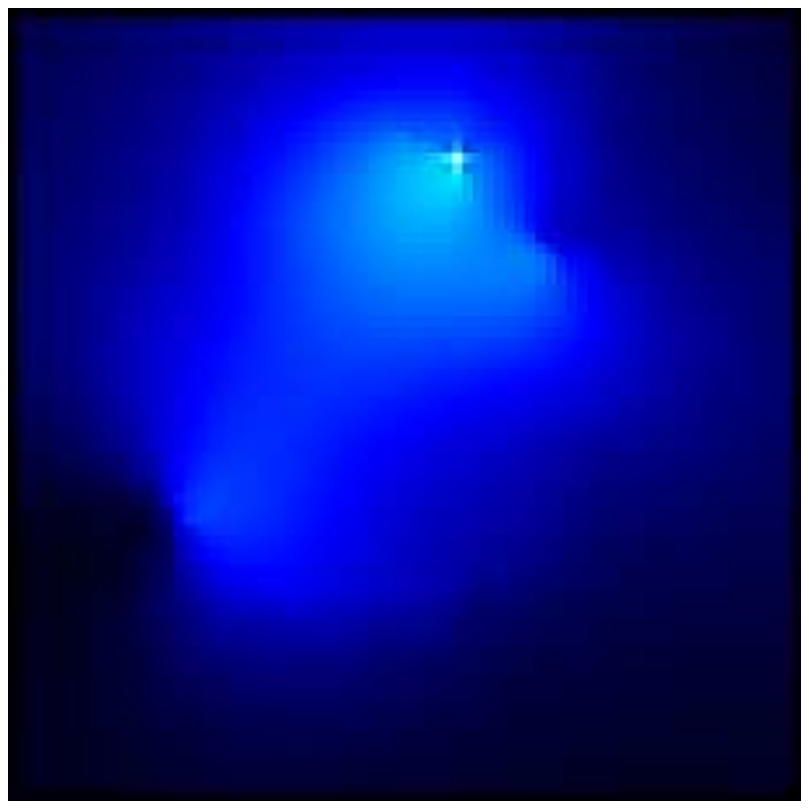}\hfill \=
\includegraphics[bb=185 306 430 545,scale=0.356,clip]{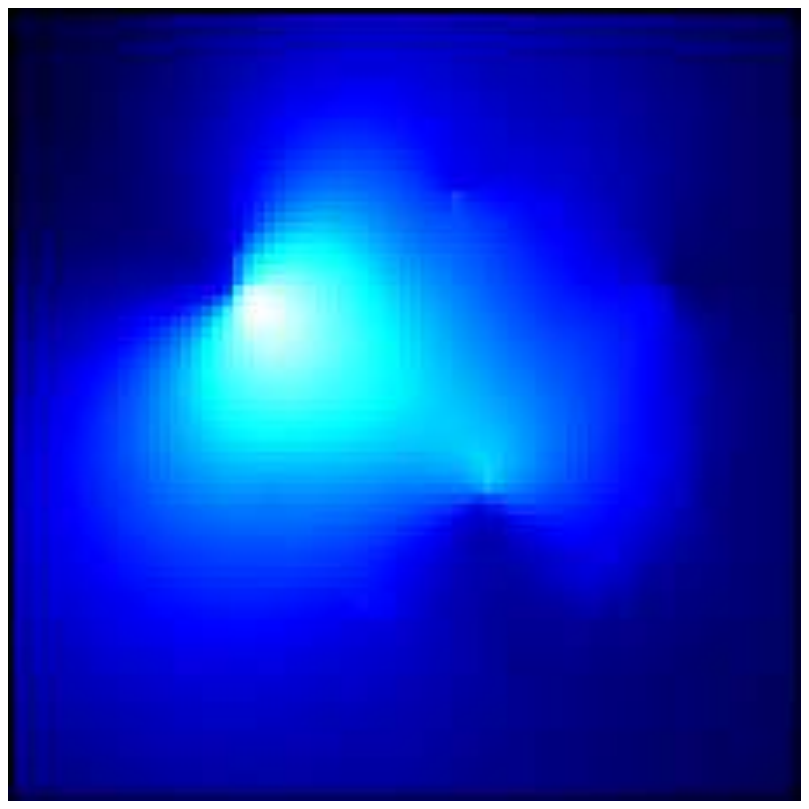}\hfill \=
\includegraphics[bb=185 306 430 545,scale=0.356,clip]{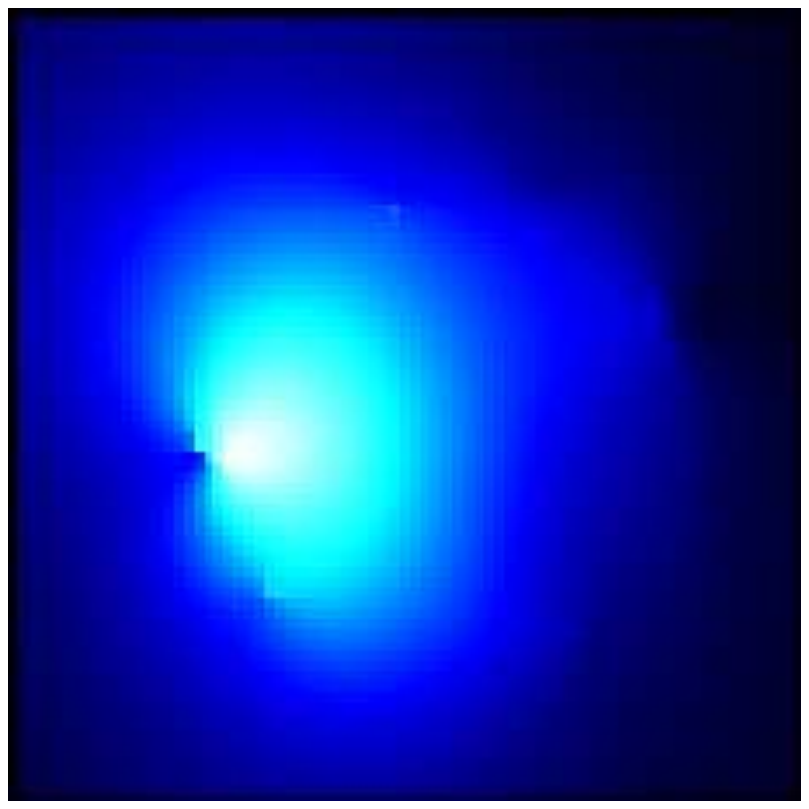}\hfill \=
\includegraphics[bb=185 306 430 545,scale=0.356,clip]{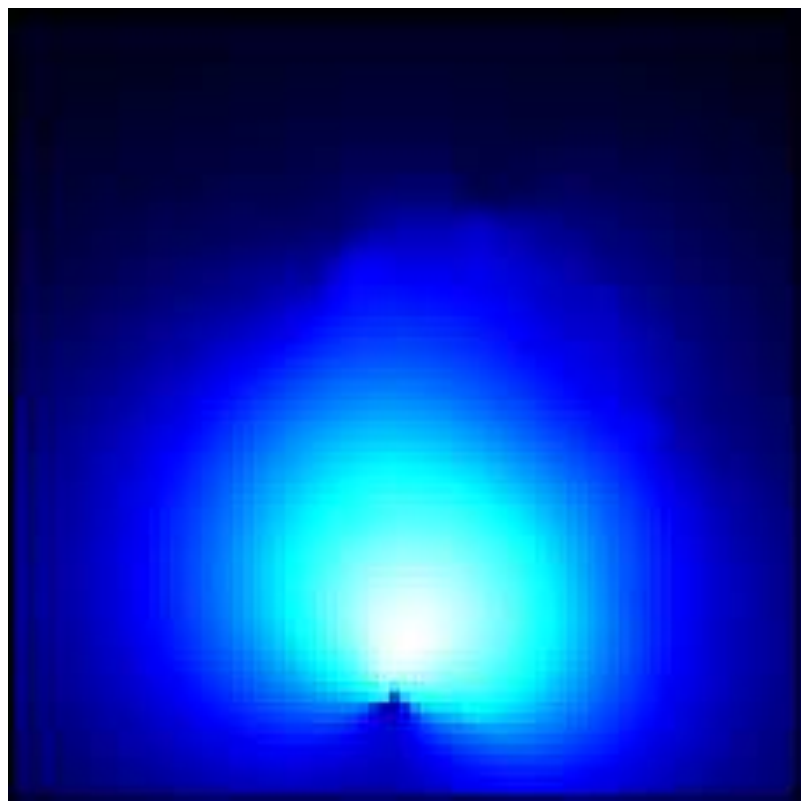}\hfill \=
\includegraphics[bb=185 306 430 545,scale=0.356,clip]{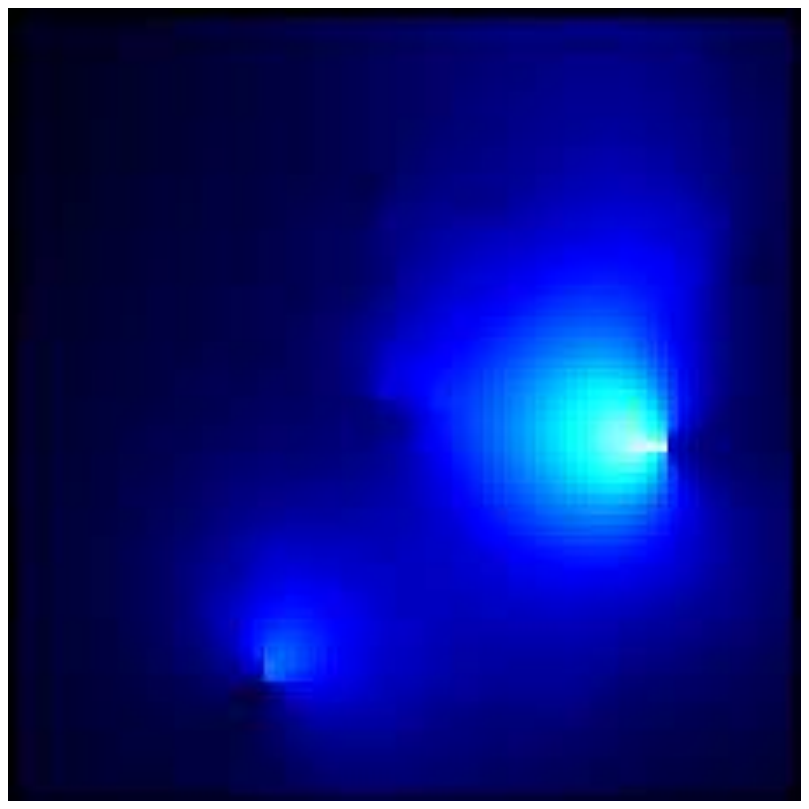}\hfill \=
\includegraphics[bb=185 306 430 545,scale=0.356,clip]{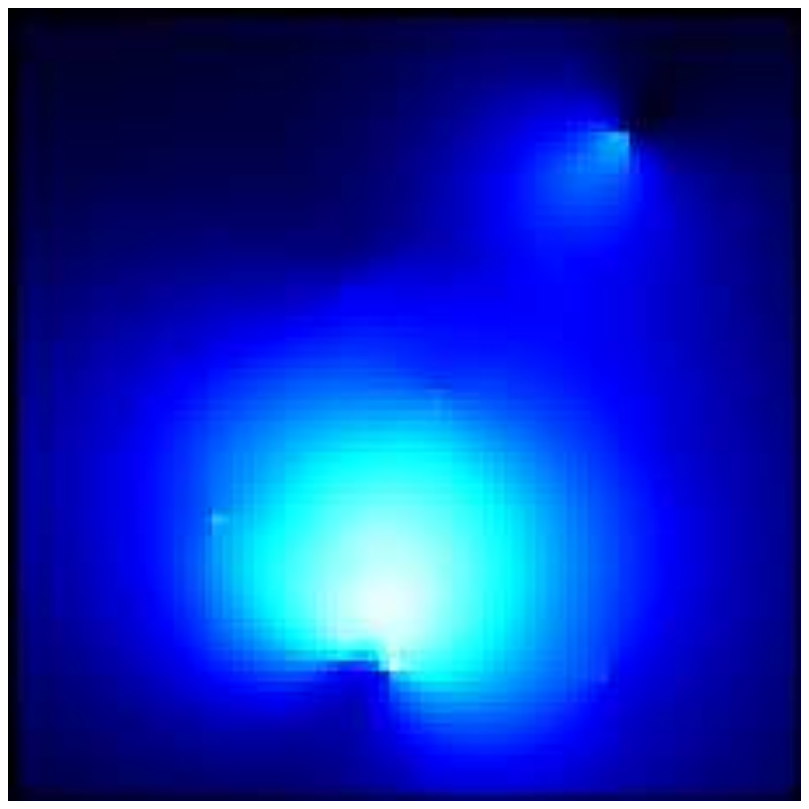}\hfill \\
\includegraphics[bb=185 306 432 545,scale=0.356,clip]{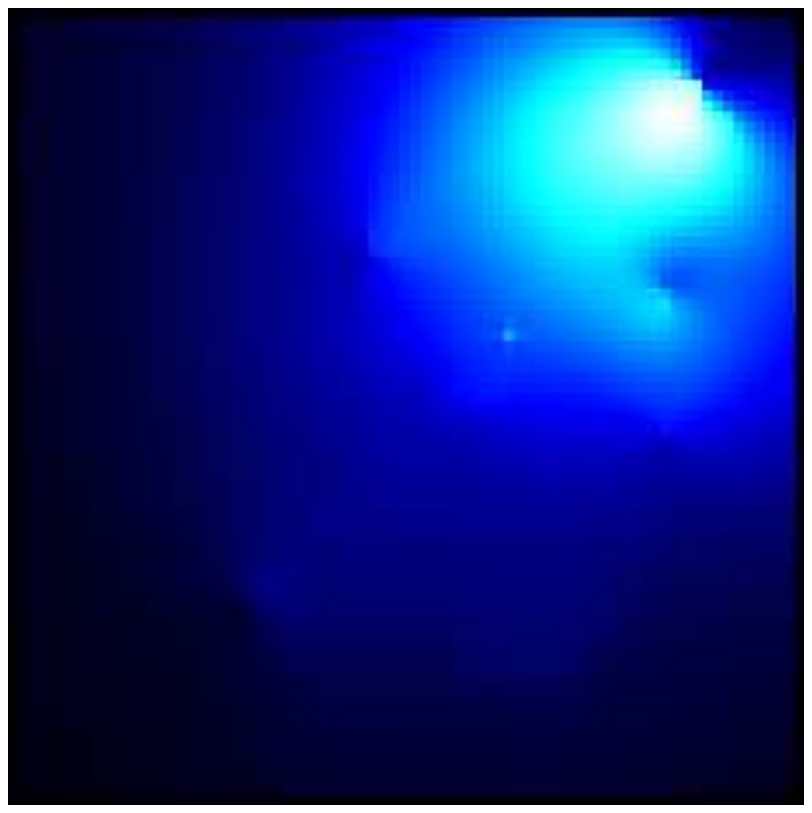}\hfill \=
\includegraphics[bb=185 306 430 545,scale=0.356,clip]{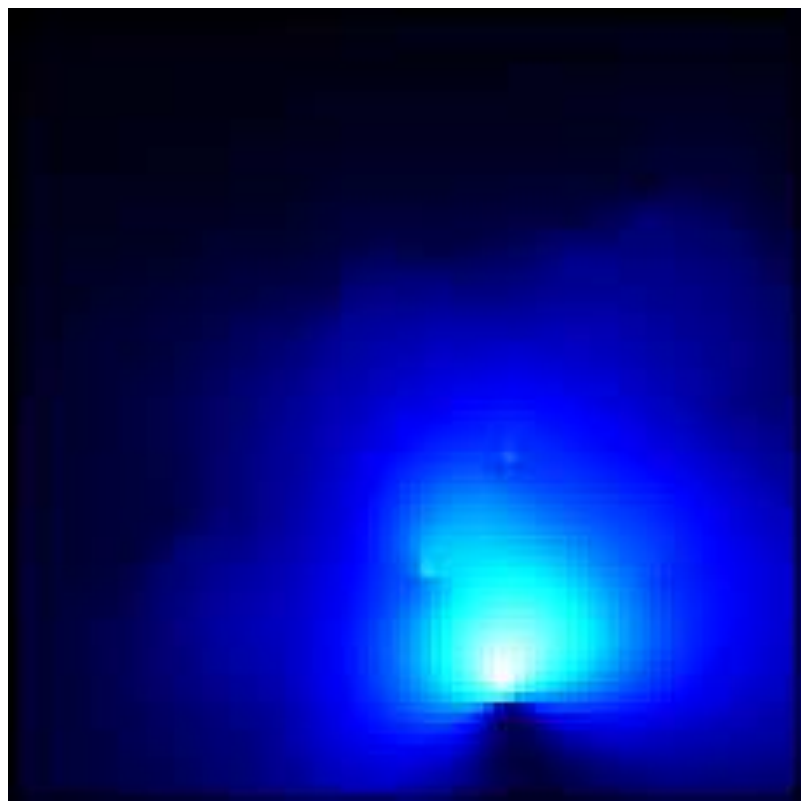}\hfill \=
\includegraphics[bb=185 306 430 545,scale=0.356,clip]{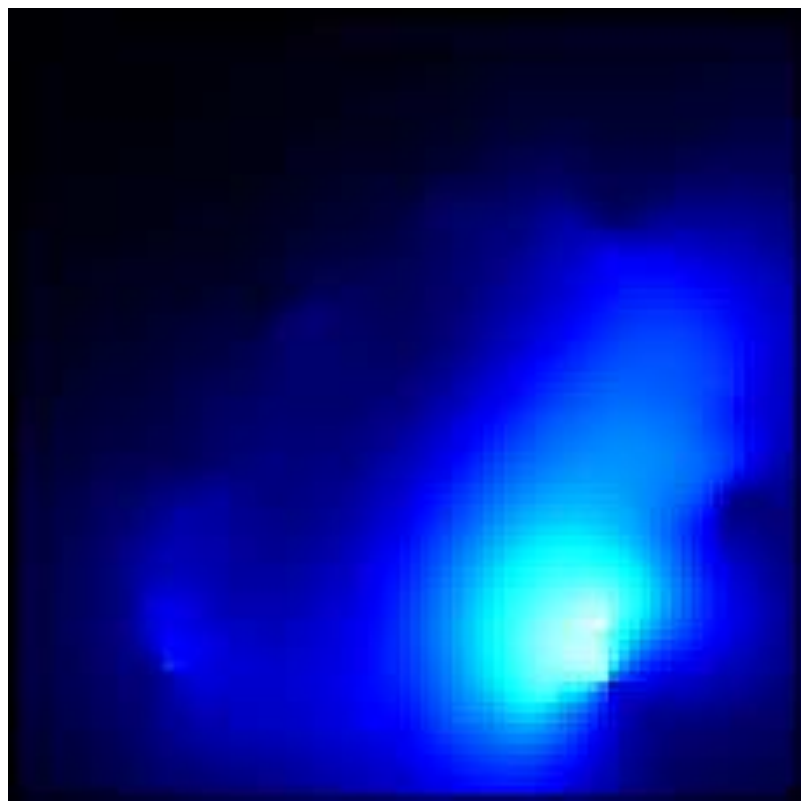}\hfill \=
\includegraphics[bb=185 306 430 545,scale=0.356,clip]{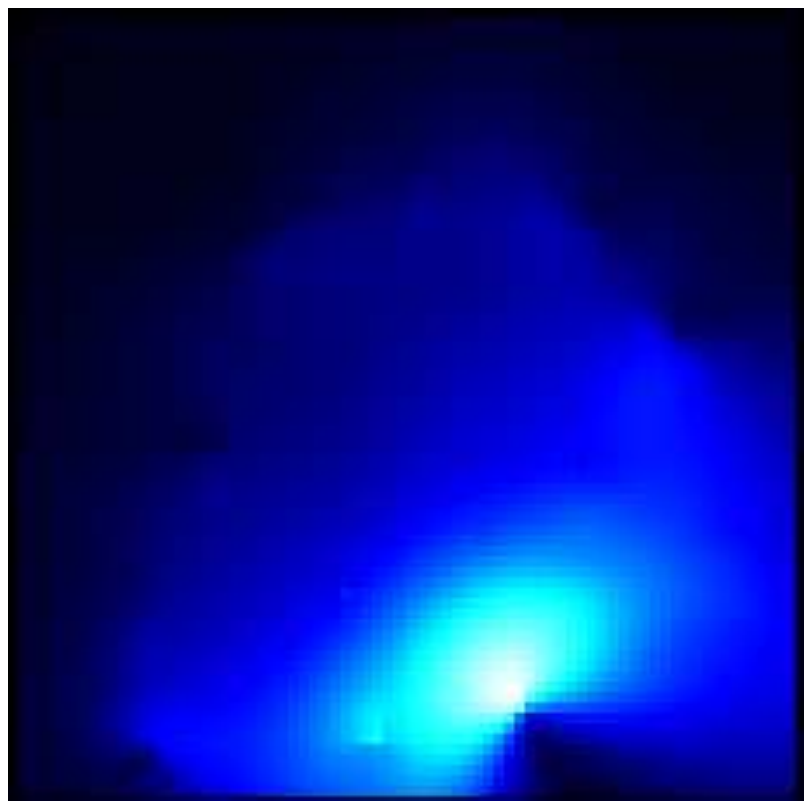}\hfill \=
\includegraphics[bb=185 306 430 545,scale=0.356,clip]{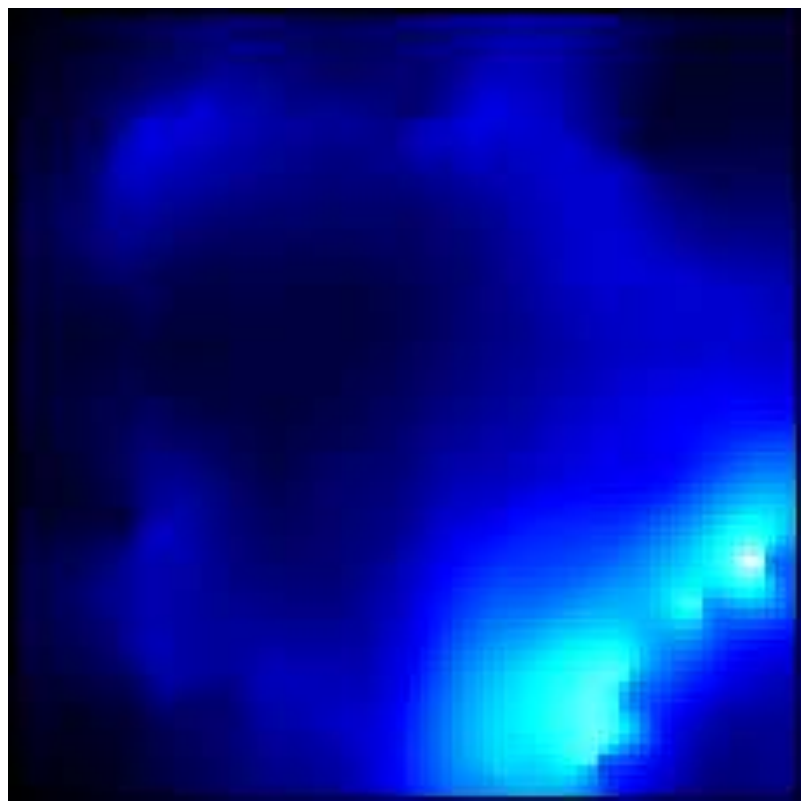}\hfill \=
\includegraphics[bb=185 306 430 545,scale=0.356,clip]{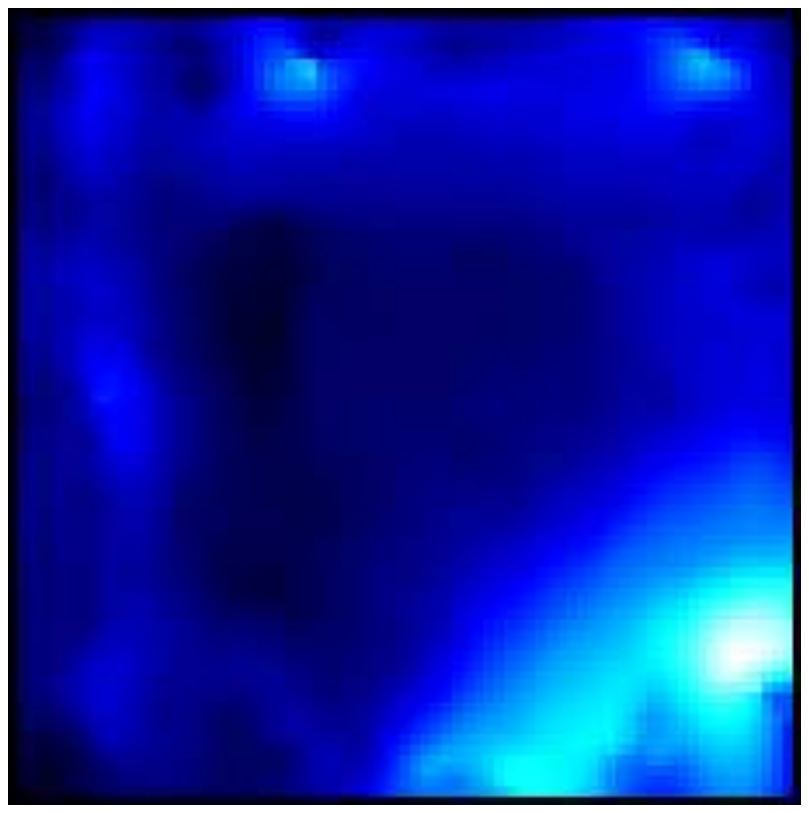}\hfill \\
\includegraphics[bb=185 306 432 545,scale=0.356,clip]{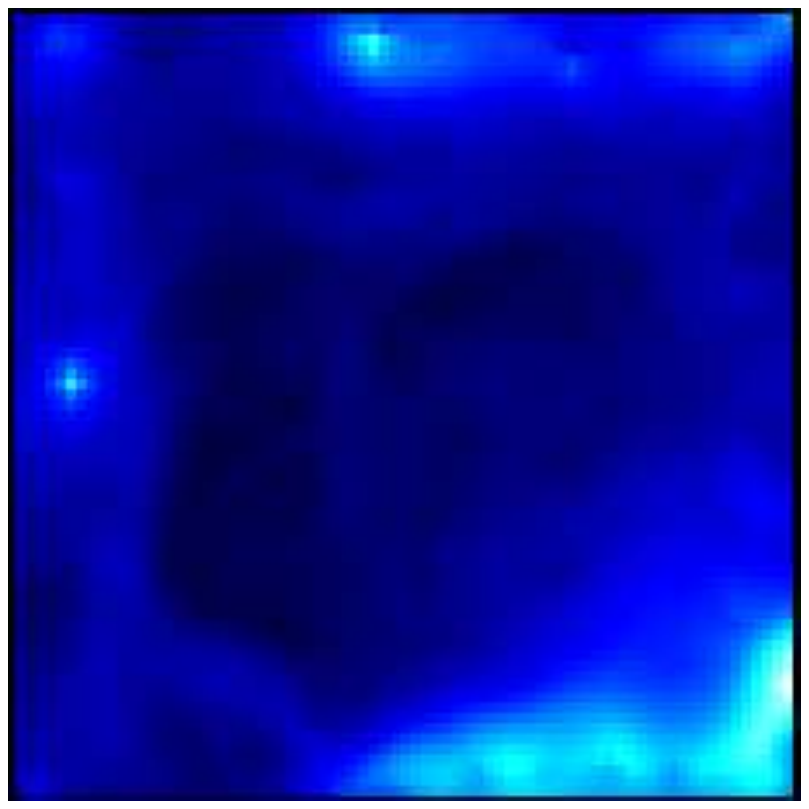}\hfill \=
\includegraphics[bb=185 306 430 545,scale=0.356,clip]{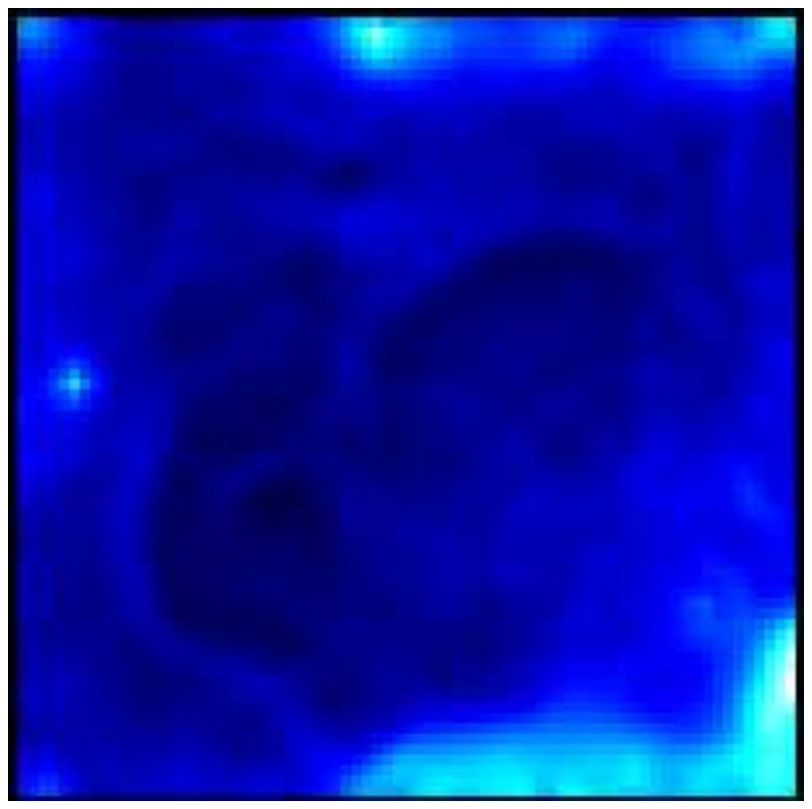}\hfill \=
\includegraphics[bb=185 306 430 545,scale=0.356,clip]{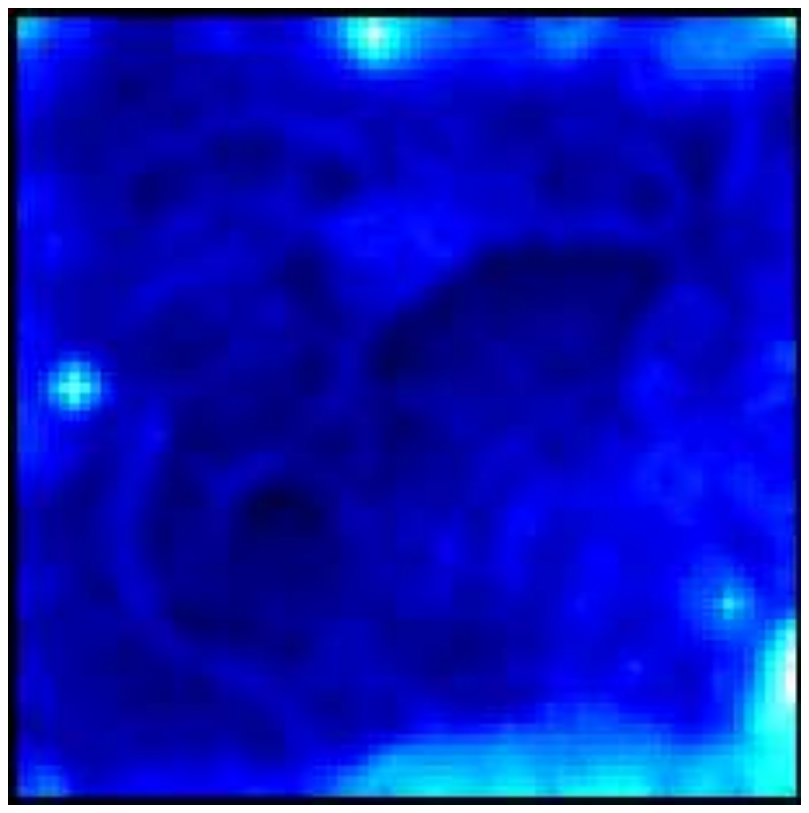}\hfill \=
\includegraphics[bb=185 306 430 545,scale=0.356,clip]{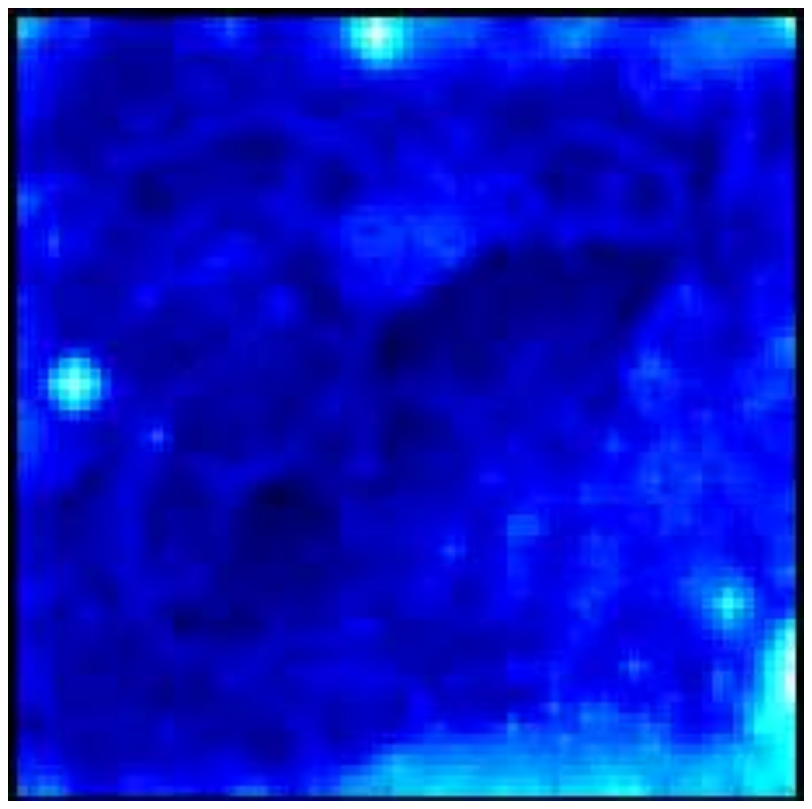}\hfill \=
\includegraphics[bb=185 306 430 545,scale=0.356,clip]{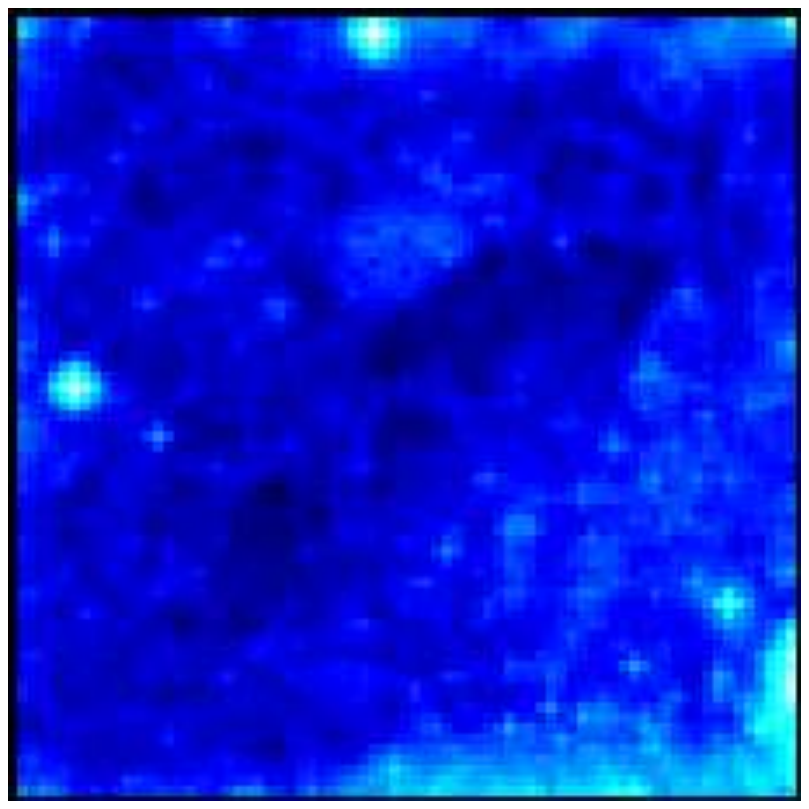}\hfill \=
\includegraphics[bb=185 306 430 545,scale=0.356,clip]{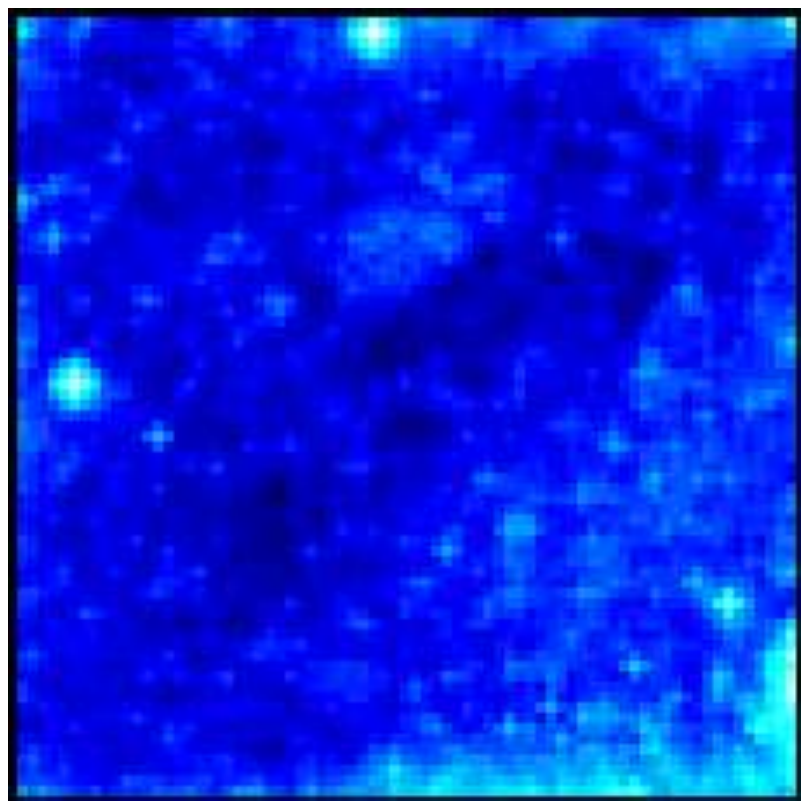}\hfill \\
\end{tabbing}
\caption{As Fig.\,\ref{fig:z-map}, but showing the U-matrices.  
}
\label{fig:u-map}
\end{figure*}

The self-organising map (SOM) algorithm (Kohonen \cite{Kohonen82, Kohonen01})
is an example of unsupervised learning of an artificial neural network.
It is an effective tool to project higher-dimensional input data onto a
two-dimensional (2D) topological map.
In this way, the algorithm has the capability of sorting input data onto a
plane according to similarity and thus ideally prepares the data for visual 
inspection. To achieve this goal, each input element is assigned to one 
neuron in the network. All neurons of the network are organised onto a
2D grid. In each learning step, the input element is reassigned to
some neuron, depending on matching criteria described in Sect.\,\ref{sec:som-algorithm}. 
Through this iterative stochastic process, the neural network is capable of learning, 
since it is designed to adapt to the input space. Therefore, input
elements are not placed randomly, but increasingly closer to similar
input elements, thereby clustering the input data.

The characteristic evolution of the Kohonen map of quasar spectra is illustrated
in Figs.\,\ref{fig:z-map} and \ref{fig:u-map}. About $5\,10^3$ quasar spectra from
the Fourth Data Release (DR4; Adelman-McCarthy et al. \cite{Adelman06}) 
were arranged in 200 iteration
steps\footnote{ 
The only reason for the use of DR4, instead of DR7, is that it
can be analysed using substantially less computing time. As this map was produced
for illustrative purposes only, we therefore chose to use the smaller database
from the DR4.
}. 
The 18 panels show the 2D arrangement of the spectra
at different iteration steps
({\it top}: 1, 2, 3, 4, 6, 8,
{\it middle}: 10, 20, 30, 40, 60, 80, 
{\it bottom}: 100, 120, 140, 160, 180, 200).
In Fig.\,\ref{fig:z-map}, the redshifts $z$ from the spectroscopic pipeline of the SDSS 
are highlighted by means of colour coding. The gradient from dark to bright 
(black to yellow in the colour image) represents the range from low to high redshift.
About 20\% of the pixels are not associated with spectra and are marked in grey (see below). 
It is clearly seen that different redshifts are separated in the final Kohonen map.

A usual way of representing a SOM is the unified distance matrix
(U-matrix; Ultsch \& Siemon \cite{Ultsch90}) shown in Fig.\,\ref{fig:u-map} for the
same iteration steps as in Fig.\,\ref{fig:z-map}.
The U-matrix visualises the differences between the spectra and their
neighbours. Light colours indicate high degrees of variation between 
adjacent spectra. 
Strong differences between the neighbours on the map are measured, in  particular,
for rare peculiar spectra. Fig.\,\ref{fig:u-map} illustrates that these 
outliers tend to settle at the edges and corners of the final SOM.
This is a useful property of the Kohonen maps that allows the efficient selection
of unusual spectra.

For the present study, the quasar spectra from the SDSS DR7 were clustered according 
to their relative differences. At a given redshift, these differences are 
dominated by the shape of the continuum and the presence of 
strong/broad absorption and emission features. Ideally, the spectra will
be sorted according to their spectral subtypes, thus differentiating
between quasar subclasses. 

In this Section we describe the preparation of 
the spectral dataset and  the SOM algorithm in a mathematically hand-waving manner. 
A more detailed description of the method and the visualisation of results from
the Kohonen mapping of large samples of spectra will be given elsewhere
(in der Au et al., in preparation).

\subsection{Data set}\label{sec:preprocessing}
 
An SQL query asking for quasars (\textsc{\texttt{spec\_cln}} = 3 or 4, 
i.e., quasar or high-$z$ quasar) with redshifts $z>0.5$ in the SDSS DR7
returned $103\,955$ spectra. 
We are aware of the possibility that some unusual quasar spectra were 
classified by the SDSS spectroscopic pipeline as unknown objects 
(spectral class = {\sc unknown}, \textsc{\texttt{spec\_cln}} = 0). 
However, the subsample of the {\sc unknowns} is quite heterogeneous in nature 
and contains a high fraction of spectra with low signal-to-noise ratios 
(S/Ns). In addition, there are, of course, no reliable redshifts given for these
objects. As a consequence, analysing the SOMs of this subsample requires
a different approach than for the quasars. The present paper is concerned
with the SOMs of the quasar sample (but see Sect.\,\ref{subsect:myst}
for an exception). The computation and systematic analysis of the SOMs 
for the {\sc unknowns} is in preparation and will be the subject of 
a separate investigation.  

As a first step, the overall size of the data set was reduced. We extracted
the spectra from the \textsc{\texttt{fits}} files, together with the selected
header keywords \textsc{\texttt{z, spec\_cln, mjd, fiberid}}, and
\textsc{\texttt{plateid}}.  
As our search for unusual quasars was designed to identify spectral peculiarities
covering a wide spectral range, such as strong BALs, overlapping absorption
troughs, and strong reddening, we reduced the resolution of the spectra used
for the Kohonen method by a factor of four, which resulted in $\sim975$ pixels
covering the wavelength range of $3800-9200\AA$ in the observer frame. 
The consequence was a significant gain of computation time. In the subsequent
process of evaluating the selected spectra we used, however, the original SDSS 
spectra. All spectra were normalised to the integrated flux density.

The Kohonen method applied to the whole sample of quasar 
spectra is expected to separate the spectra according to their redshifts $z$ \
(Fig.\,\ref{fig:z-map}). However, we repeat that it was our aim to search for
unusual spectra. It is therefore advisable to apply the method to the quasars in
narrow $z$ intervals.
The bin size $\Delta z$ should be small enough to ensure that the differences
between the spectra, as seen by the SOM, caused by their different redshifts are
smaller than the differences due to any spectral peculiarities. On the other hand,
the intervals must be wide
enough to cover a large enough number of spectra. After some trials, we found that
$\Delta z \sim 0.1$ is a good choice. Hence, we produced SOMs for the quasar
spectra in redshift bins with a 0.1 redshift step size. The mean number of spectra
per bin amounts to  2\,680, ranging from $165$ to $6\,847$.

\subsection{SOM algorithm}\label{sec:som-algorithm}

The set of input data (i.e. spectra) is defined as vectors
$\vec{x}(j)=\left[\xi_1(j),..,\xi_n(j)\right]\in\Re^{n}$, where $n=975$ is
the number of pixels in each spectrum and $j$ denotes the index in the sequence of
source spectra $j = 0 \ldots j_{\rm max}$ (where $j_{\rm max}$ strongly varies 
with $z$). The neural network consists of $i\in\left\{1\ldots N\right\}$ neurons,
represented by weight vectors 
$\vec{m}_i = \left[\mu_{i, 1},\ldots,\mu_{i,n}\right]\in\Re^{n}$,
that are organised on a two-dimensional grid. We use a flat grid with
closed boundaries (in contrast to cylindrical or toroidal boundary conditions).
The weight vectors are modified in each iteration step and thus vary with
the discrete time coordinate
$t=0,1,2,\ldots,t_{\rm max}$, i.e., $\vec{m}_i = \vec{m}_i(t)$.
Each weight vector $\vec{m}_i$ can be considered as an artificial
spectrum. The entire set of weight vectors approximates the distribution 
of input spectra.

The SOM algorithm is essentially based on two processes that are responsible for
the self-organising properties of the neural network: (1) determining the
best-matching unit (BMU) for a given randomly chosen input element and (2) the 
successive adaptation of the weight vectors in the neighbourhood of the BMU
towards the given input element.

For a given spectrum $\vec{x}(j)$, the Euclidean distance 
$\left\|\vec{x}(j)-\vec{m}_i\right\|$
to each neuron $\vec{m}_i$ is computed and the BMU is identified with
the winning neuron $\vec{m}_{\rm \, c}$, i.e. the neuron for which the 
distance is minimised, i.e.,
\begin{equation}
 \left\|\vec{x}(j)-\vec{m}_{\rm \, c}\right\| 
    = \mbox{min}_i \left\{\left\|\vec{x}(j)-\vec{m}_i\right\|\right\}.
\end{equation}
In the next iteration step ($t \rightarrow t+1$), an adaptation
of all neurons is performed, according to their difference from a spectrum $\vec{x}_j$ and the
neighbourhood function $h_{{\rm c},i}(\tau)$
\begin{equation}
\label{eq:adaption}
\vec{m}_i(t+1)=\vec{m}_i(t)+h_{{\rm c},i}(\tau)\cdot\big[\vec{x}(j)-\vec{m}_i(t)\big],
\end{equation}
where $\tau = t/t_{\rm max}$.
The neighbourhood function 
\begin{equation}
h_{{\rm c},i}(\tau)=\alpha(\tau)\cdot 
\exp{\Big(-\frac{\left\|\vec{r}_{\rm c}-\vec{r}_i\right\|}{2\sigma^2(\tau)} \Big) }
\label{eq:hci}
\end{equation}
acts as a smoothing kernel over the network and converges to zero with an increasing number
of learning steps. The vectors $\vec{r}_{\rm c} \in \Re^2$ and  $\vec{r}_i \in \Re^2$
are the location vector of the BMU and and the location vector of the weight
vector $\vec{m}_i$, respectively.
Compared to the frequently used Gaussian smoothing kernel, Eq.\,(\ref{eq:hci}) uses 
a slightly modified version that has broader wings and a sharper peak.
We found from various trials that Eq.\,(\ref{eq:hci}) yields more reliable 
clustering results.

The process of finding the BMU and adapting the weight vectors is performed for 
each input spectrum in random order within each learning step. 
The neighbourhood function is modified over time by the learning rate
$\alpha(\tau)$ and the radius function $\sigma(t)$ in such a way that
{\it (a)} the map develops large-scale structures in the early phase, while 
{\it (b)} finer adjustments occur in later steps. Both functions are assumed to
monotonically decrease with time $\tau\in\left\{0 \ldots 1\right\}$ and are
parametrised in the simple way
\begin{equation}
\alpha(\tau)=\alpha_{0}\left( \frac{\alpha_{1}}{\alpha_{0}} \right)^{\tau},
\ \ \ \sigma(\tau)=\sigma_{0} \left( \frac{\sigma_{1}}{\sigma_{0}} \right)^{\tau},
\end{equation}
where $\alpha_{0}, \alpha_{1}, \sigma_{0}, \sigma_{1}$
are the learning parameters of the Kohonen network
with $\alpha_{0} \geq \alpha_{1}$ and $\sigma_{0} \geq \sigma_{1}$.
The values used here are listed in Tab.\,\ref{tab:NWParams}.

\begin{table}[h]
\caption{Network parameters used for the clustering.}
\centering
\begin{tabular}{l r}                
\hline\hline                                   
Parameter                  & Values \\
\hline
Number of neurons $N$ & $196 \ldots 8281$ \\
Number of iteration steps $t_{\rm max}$ & 100 \\
Learn radius  $\sigma_{0}$ & 1.0 \\
Learn radius  $\sigma_{1}$ & 0.125 \\
Learning rate $\alpha_{0}$ & 0.25 \\
Learning rate $\alpha_{1}$ & 0.01 \\
\hline       
\end{tabular}
\label{tab:NWParams}
\end{table}

The number of neurons in the network must at least correspond to 
the number of source spectra but can be larger, i.e., $N \ge j_{\rm max}$. 
Our experience has shown that better results 
can be achieved when a certain fraction of positions within the map is not 
occupied with source spectra. These unoccupied neurons open up space to 
form cluster boundaries between distinct spectral types, which 
settle into clearly separated areas of the map. Outlier spectra have 
enough space to roam the neural landscape. We found that a ratio
of $N/j_{\rm max}\approx1.2$ represents a good trade-off.
Hence, the grid size varies between $14\times 14$ and $91 \times 91$ neurons
for the highest-$z$ bin and the $1.5<z\le1.6$ bin, respectively.  

In the beginning of the iteration process at $t=0$, each weight vector 
$\vec{m}_i(0)$ is initialised with a random input spectrum
$\vec{x}(j)$. The initialisation with purely random weight vectors would
require much more learning steps to obtain comparable results.
In each iteration step, we start by computing the Euclidean distances for
the first input spectrum to all weight vectors within the network. The weight
vector with the shortest distance is identified with the BMU for this 
particular input spectrum. Hence, the input spectrum is moved to its 
network location for this particular learning step. The BMU and all other
weight vectors are then updated according to the neighbourhood function. This 
process is repeated for all remaining input spectra and over multiple learning
steps until convergence is reached, i.e. until the BMUs and the associated input 
spectra essentially remain at the same positions in successive iteration 
steps.
In our case of relatively small maps, the re-distribution of the spectra
reaches sufficient convergence after about 100 iteration steps.\footnote{There
are two reasons for such a small number of iterations in a neural network.
First, our interest is focused on broadband spectral
features and the spectral resolution is correspondingly low. Secondly, the 
final spectra maps are inspected visually. For an automated classification and 
higher resolution spectra, $t_{\rm max}$ must be considerably increased.} 

\begin{figure*}[htbp]
\begin{tabbing}
\includegraphics[bb=45 00 495 780,scale=0.178,angle=270,clip]{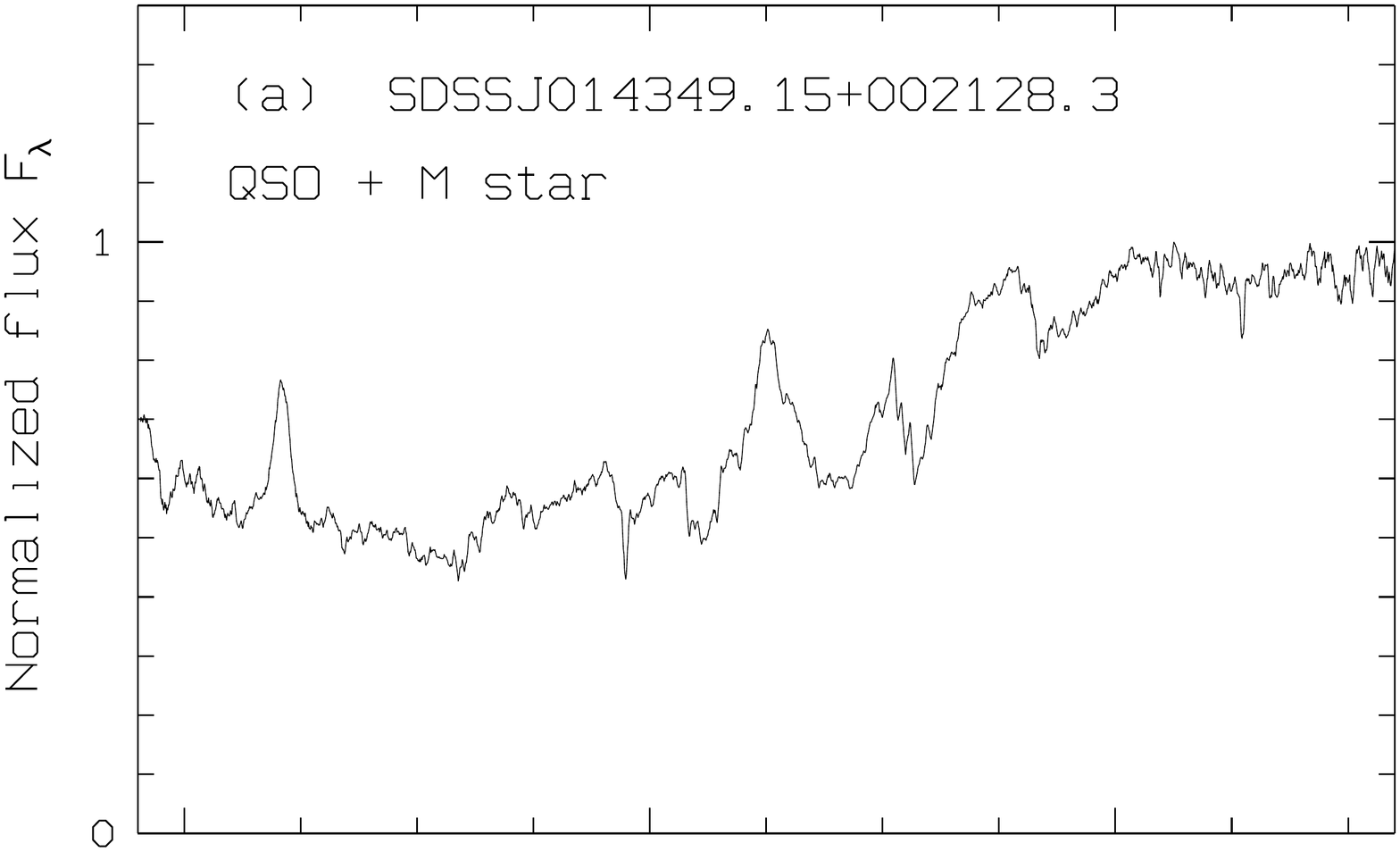} \=
\includegraphics[bb=45 80 495 780,scale=0.178,angle=270,clip]{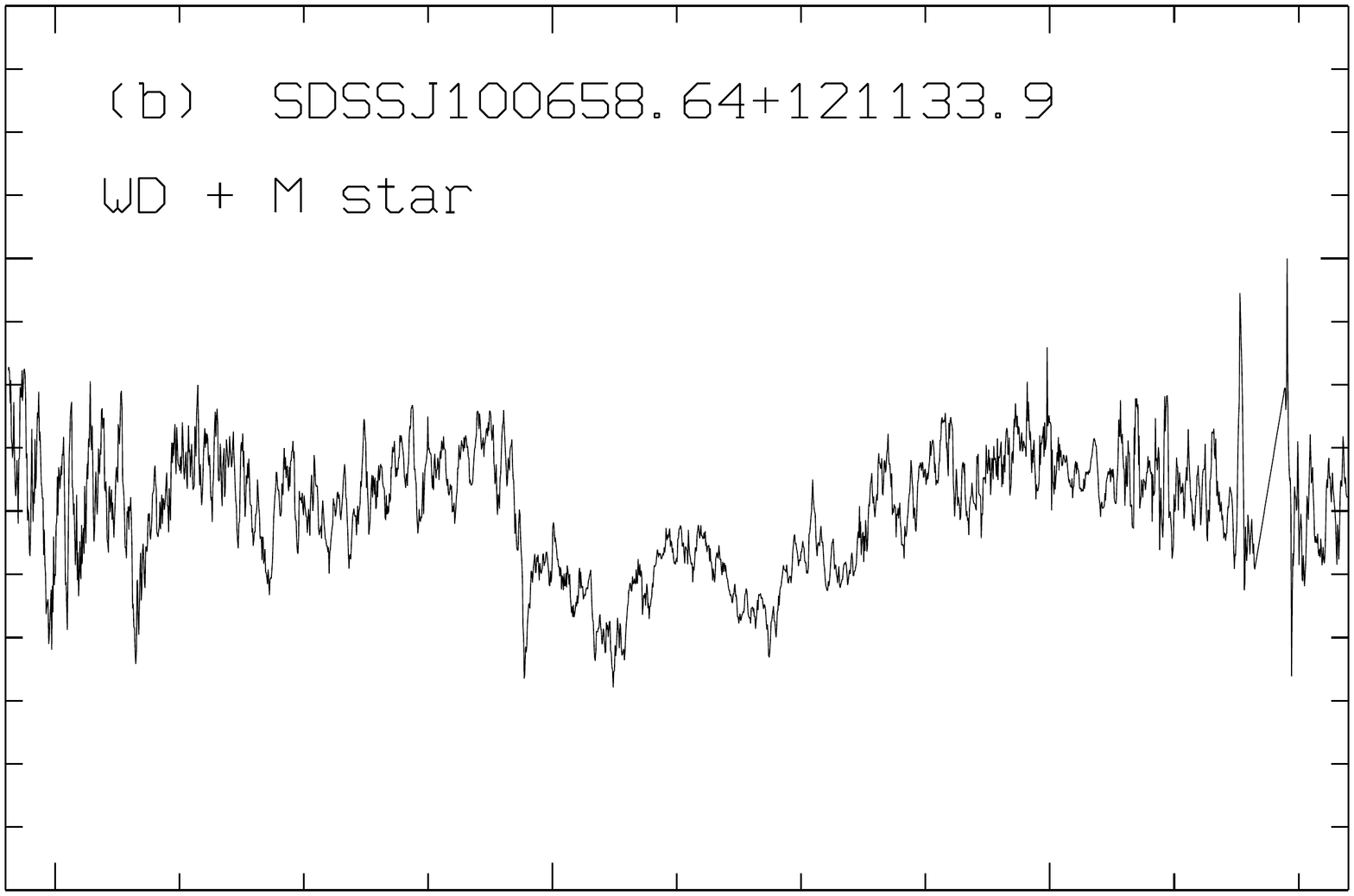} \=
\includegraphics[bb=45 80 495 780,scale=0.178,angle=270,clip]{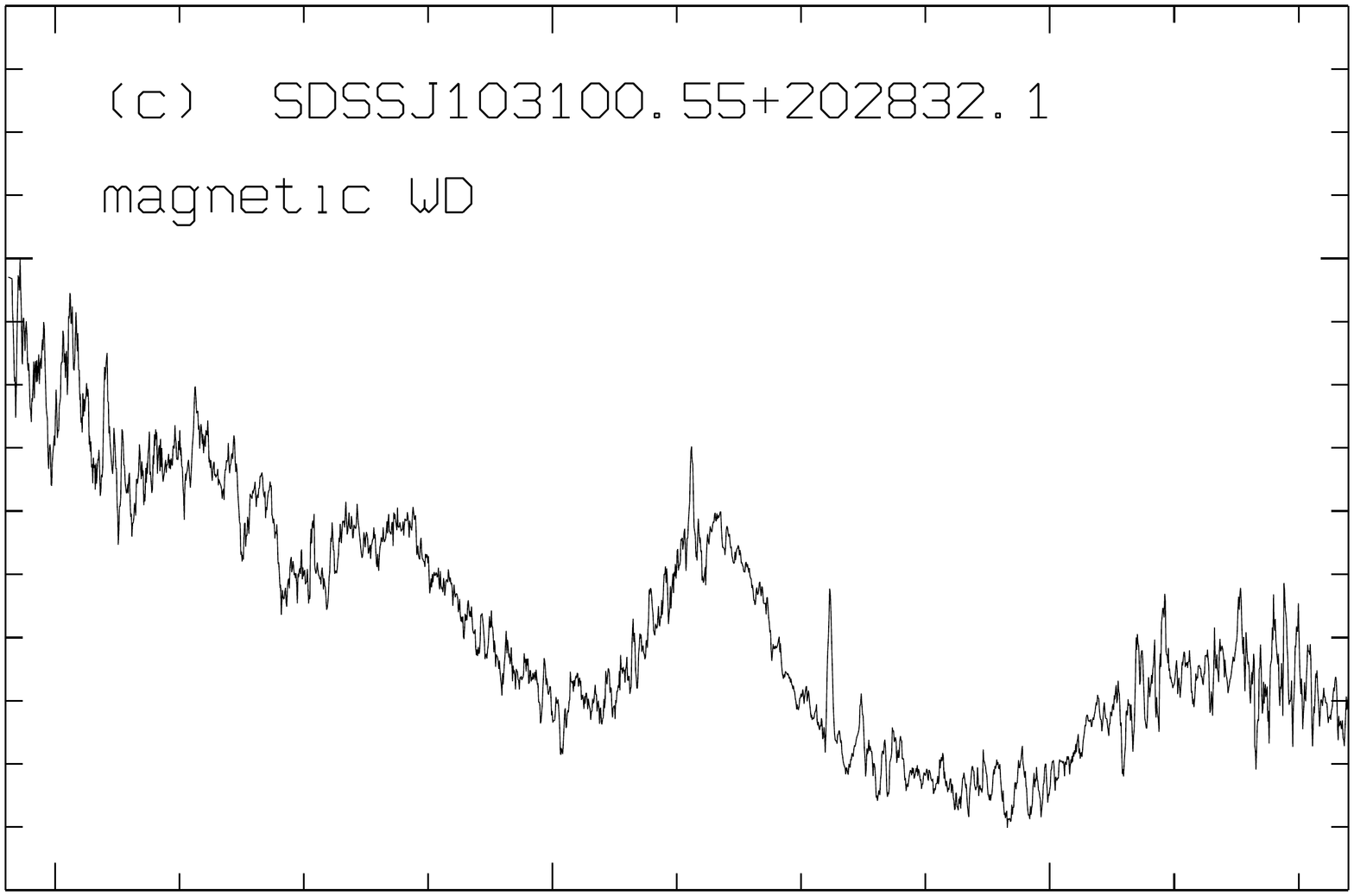} \=
\includegraphics[bb=45 80 495 780,scale=0.178,angle=270,clip]{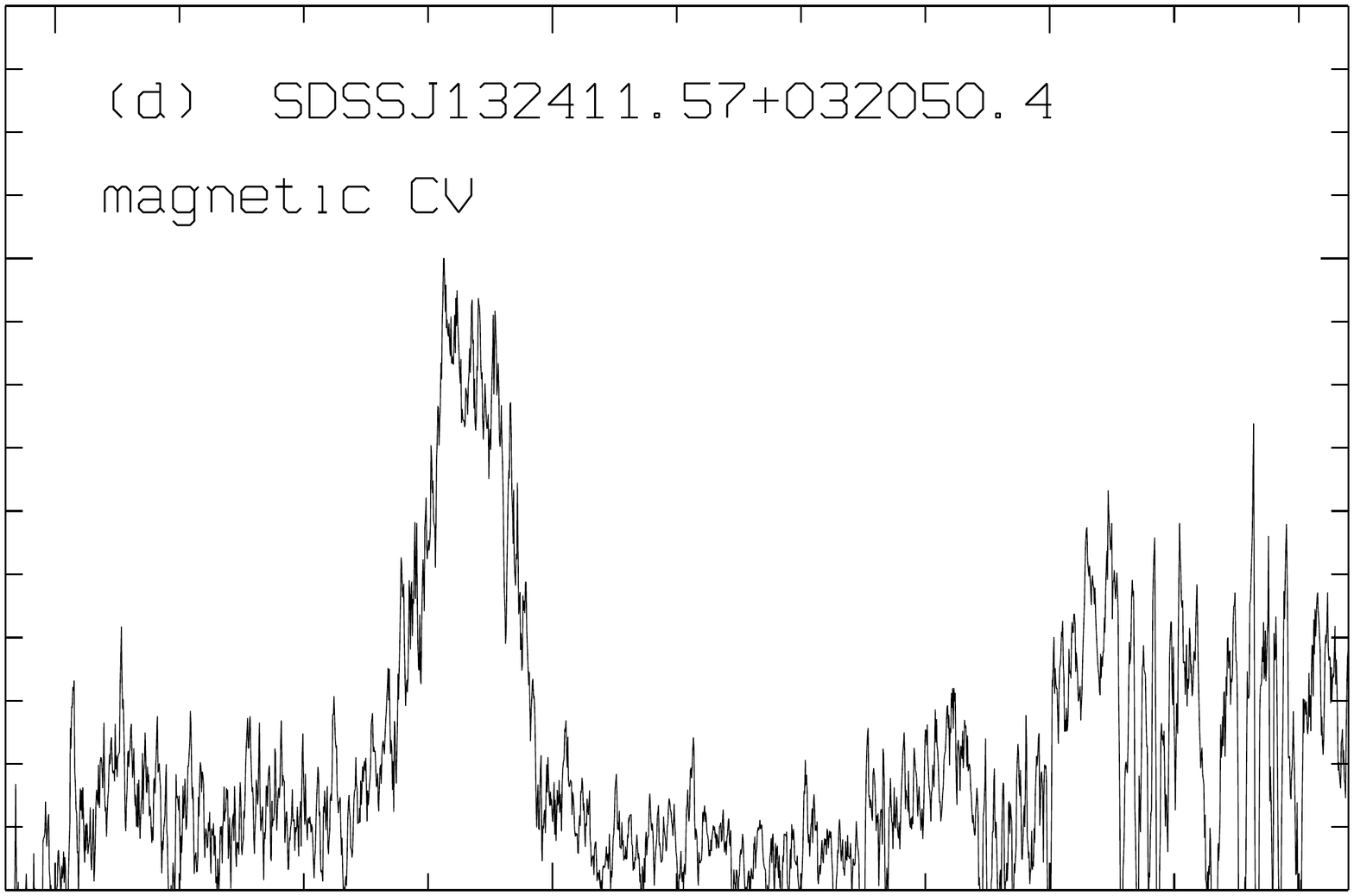} \\
\includegraphics[bb=45 00 580 780,scale=0.178,angle=270,clip]{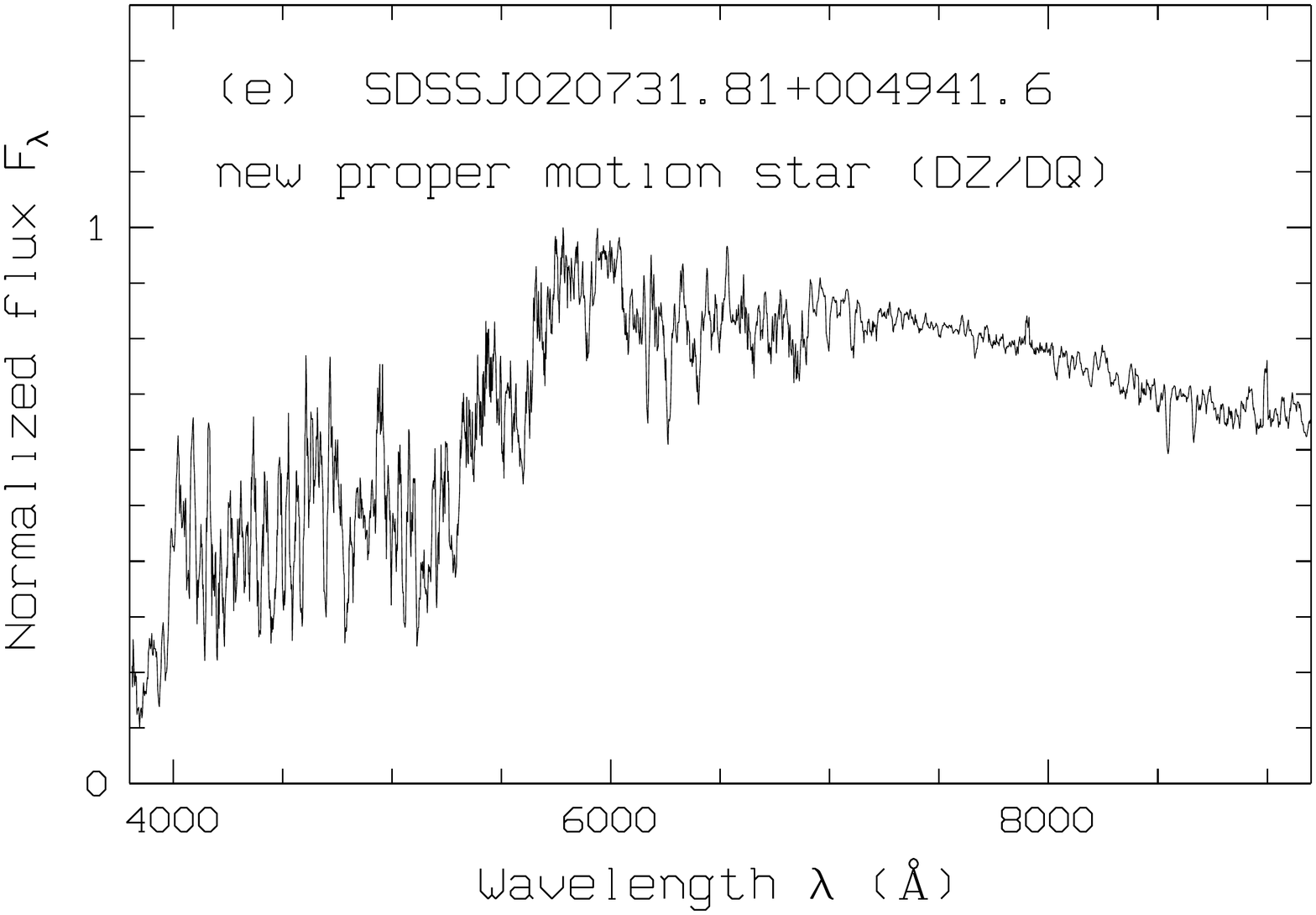} \=
\includegraphics[bb=45 80 580 780,scale=0.178,angle=270,clip]{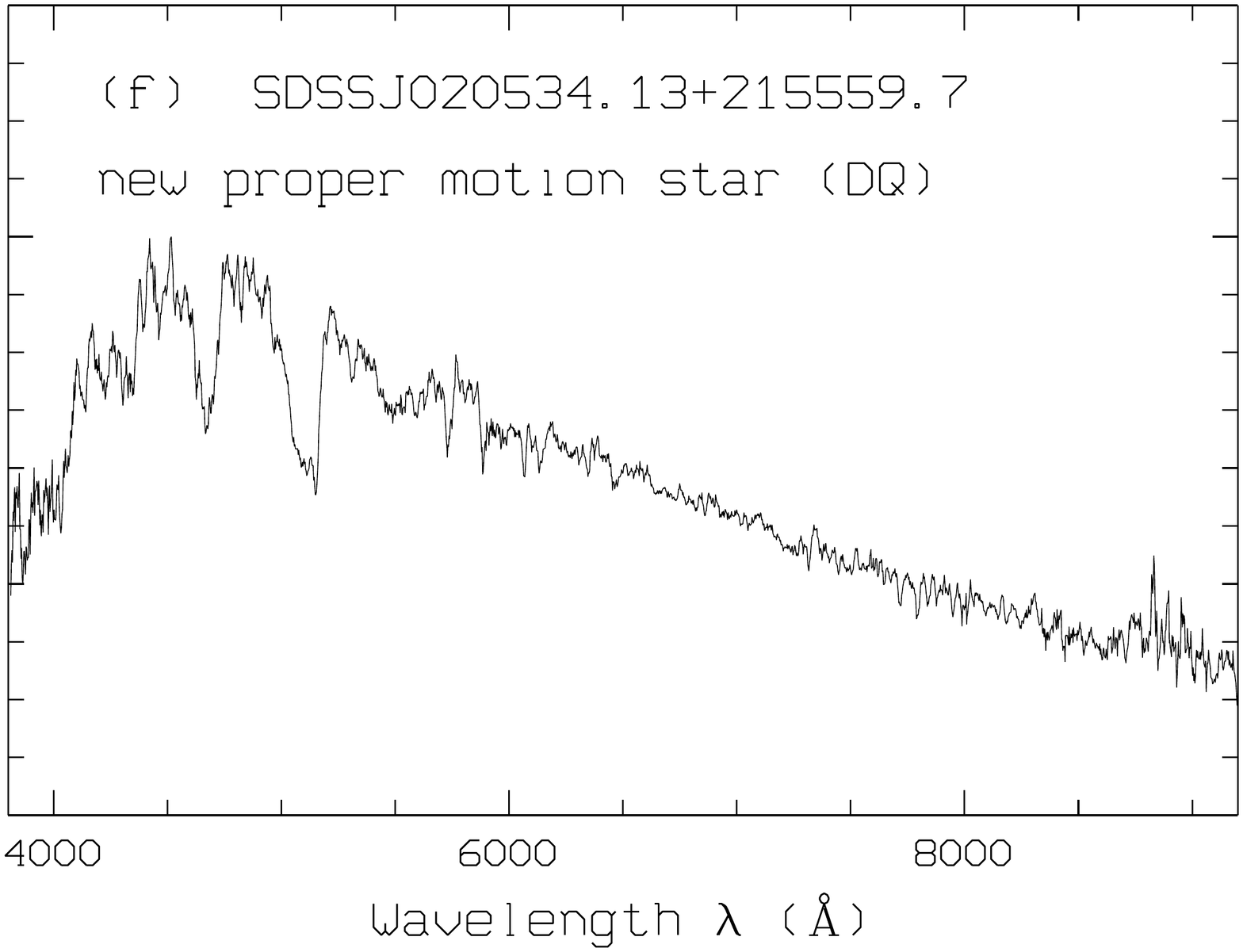} \=
\includegraphics[bb=45 80 580 780,scale=0.178,angle=270,clip]{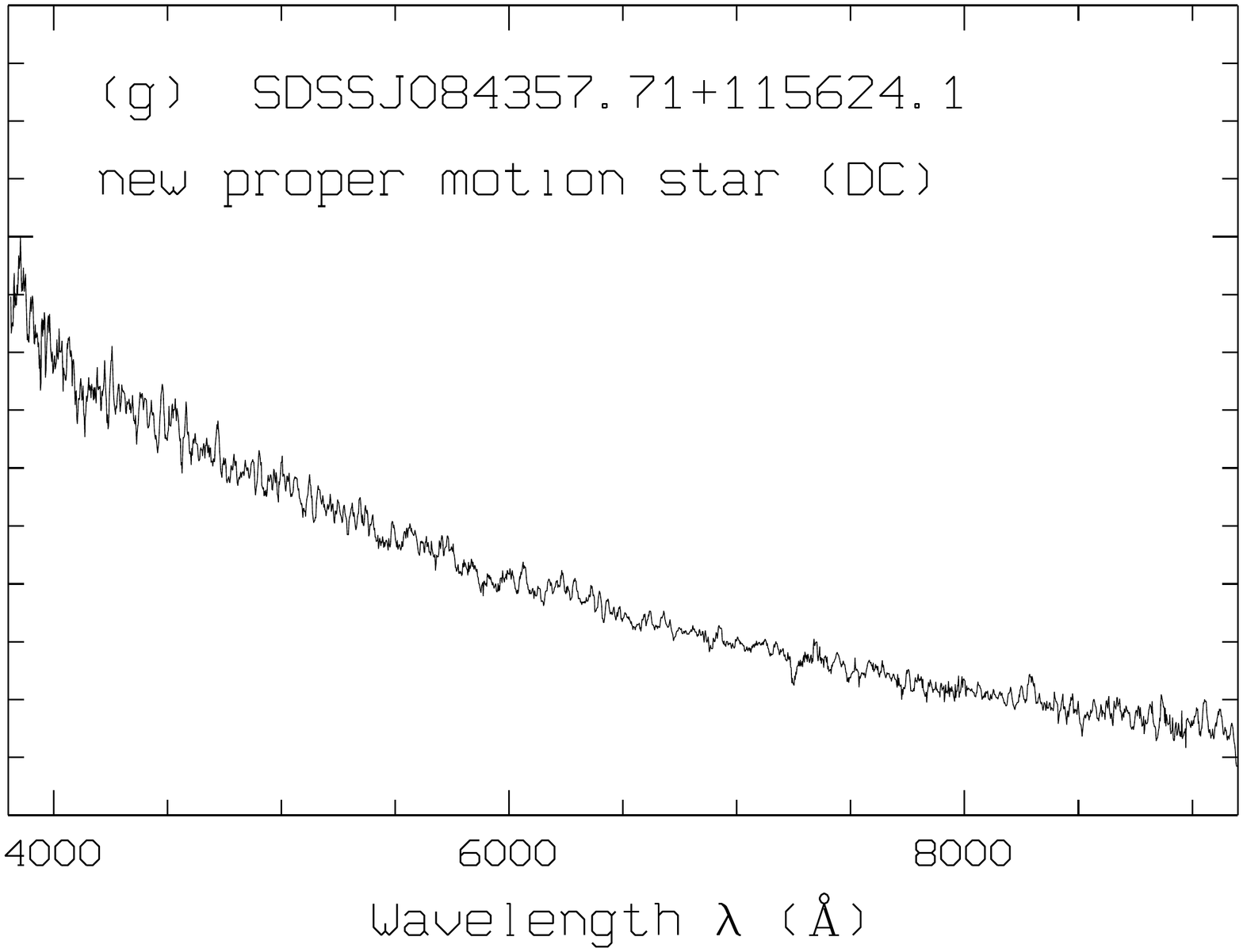} \=
\includegraphics[bb=45 80 580 780,scale=0.178,angle=270,clip]{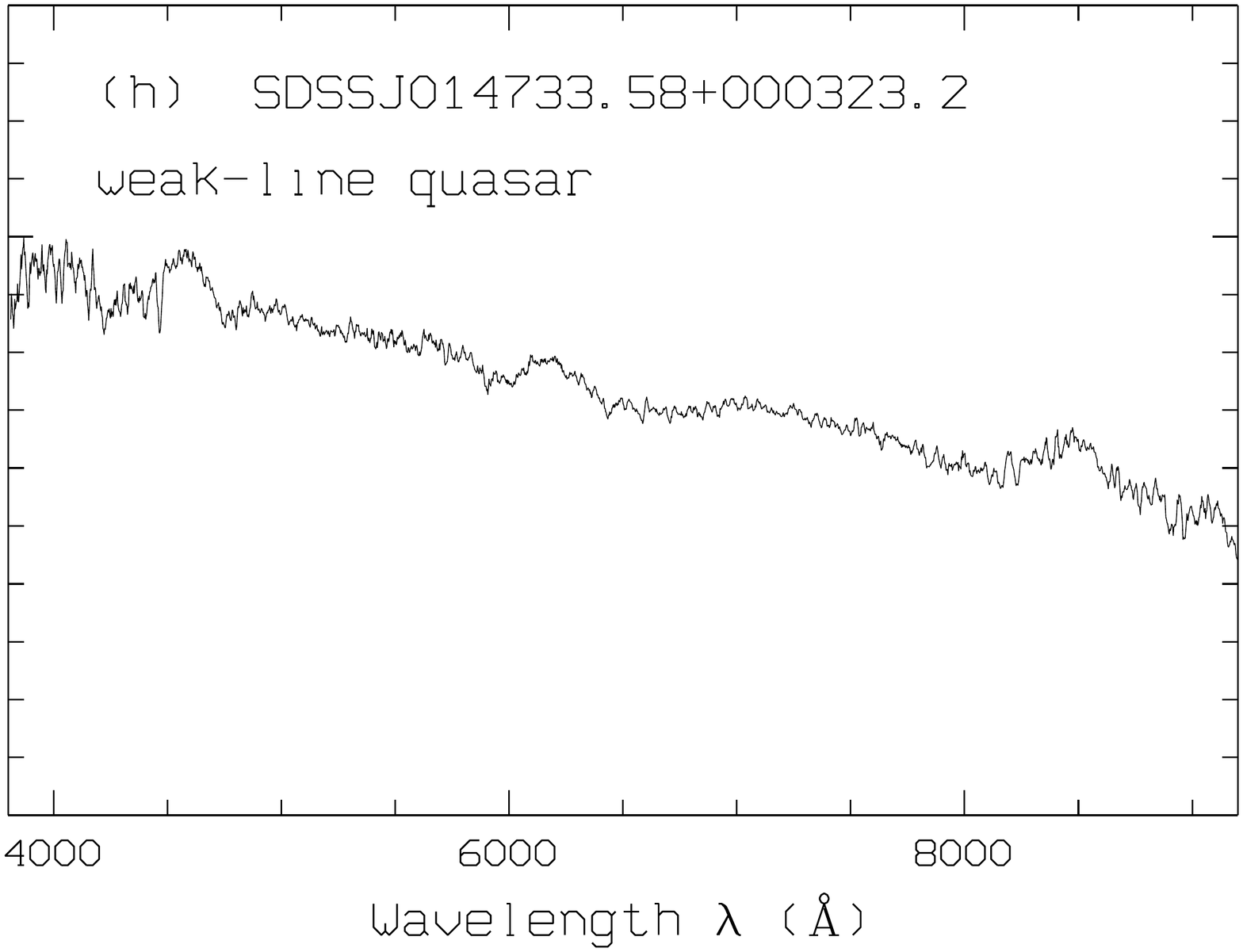} \\
\end{tabbing}
\caption{Examples for unusual spectra which were confirmed (panels 
{\bf a} to {\bf g}) or not confirmed (panel {\bf h}), respectively, as 
rare types of stellar contaminants in the initially 
selected quasar sample.}
\label{fig:stellar-cont}
\end{figure*}

%
\section{Selection and classification of unusual spectra}\label{sec:selection}
%

\subsection{Selection of spectra from the SOMs}\label{subsec:selection}

Originally, our search for unusual spectra was focused on unusual 
FeLoBAL quasars. The longest-wavelength lines of
the strongest UV iron multiplet, UV\,1, ($\lambda \le 2600$\AA) is
shifted into the spectral window of the SDSS spectra for $z\ga 0.6$.
This defines the lower limit for our quasar selection. The upper limit
is set at $z\la 4.3$ by the demands that {\it (a)} the \ion{C}{iv} line 
should be well within the spectral window and {\it (b)} the number of 
quasars per $z$ bin must not be too small. Therewith, the quasar sample 
is restricted to the
range  $z = 0.6 \ldots 4.3$. Altogether, 99\,151 spectra in this 
$z$ range were analysed with the Kohonen method.

For each of the 37 resulting SOMs, the corresponding grid of spectra was 
produced in the form of an ``icon map'', where the spectral resolution was
reduced by an order of magnitude. Outlier spectra are clustered at the 
edges and corners of the SOMs (Sect.\,\ref{sec:kohonen}, Figs.\,\ref{fig:z-map} 
and \ref{fig:u-map}) and are thus easy to select by means of the
visual inspection of the icon maps. The selected quasar sample
is expected to contain various spectral types.
We note, however, that the selection procedure 
{\it (a)} is  certainly not free of a subjective bias,
{\it (b)} does not use a quantitative criterion, and
{\it (c)} is thus incomplete.
The latter is, of course, a direct consequence of the term
"unusual" not being defined {\it a priori}.
 
We selected 1575 spectra of 1530 objects, 41 objects from this sample 
have more than one spectrum in DR7. The selected spectra were first 
corrected for Galactic foreground extinction and then individually checked
to estimate $z$ and object type. The Milky Way extinction curve from Pei
(\cite{Pei92}) was used for the extinction correction with $E(B-V)$
obtained from the ``Galactic Dust Extinction
Service''\footnote{http://irsa.ipac.caltech.edu/applications/DUST/} 
of the NASA/IPAC Infrared Science Archive.

While obtaining the redshift from a typical quasar spectrum is mostly
straightforward, the situation is different for highly unusual spectra. 
If prominent emission lines are present and 
clearly identified, the redshift is estimated as usual by comparing 
the measured line centres with a catalogue of their rest-frame wavelengths.
The spectral lines were selected manually to account for such effects as
noise, broad absorption lines, or artefacts from bright night-sky lines.
If the spectrum is not dominated by BAL troughs and emission lines are only 
rudimentarily present, or not at all, we tried to estimate the
redshift by fitting the continuum to that of the SDSS quasar composite
spectrum (VandenBerk et al. \cite{VandenBerk01}).
Redshift estimation was approached with 
particuliar caution
for the BAL quasars without clearly identified emission lines but
with many narrow absorption troughs or overlapping troughs. Here, we basically 
follow the approach outlined by Hall et al. (\cite{Hall02}). In several cases,
redshifts could be estimated only roughly. Dubious as well as uncertain redshifts
were flagged: $f_{z} = 1$ for likely but not certain, 2 for
uncertain due to the lack of redshift indicators, and 3 for uncertain due to low
S/N, respectively. For example, $z$ could not be estimated
for three objects with featureless blue spectra, which were flagged therefore 
$f_{z} = 2$.

\subsection{Contamination}\label{subsec:contaminants}

The procedure of individually investigating each spectrum
was repeated in three consecutive, independent runs. At the end, 
1014 of the selected 1530 objects (66\%) were included in the (preliminary) 
catalogue of unusual quasars,
8\% turned out to be stars, another 8\% were classified as galaxies, 
and 5\% were rejected because of too low S/N and/or errors.
About 13\% were rejected as (nearly) normal quasars. Some of these quasars 
were positioned on the SOM in the area of unusual spectra because the
assigned redshift from the SDSS pipeline was wrong. Others were simply located in 
transition regions between unusual and usual spectra but turned out to be, more
or less, usual when individually inspected.
Objects displaying strong stellar continua and typical low-$z$ absorption and/or
narrow emission lines without broad components 
were simply classified as galaxies and subsequently rejected. 
Among the remaining 1014 catalogue entries, there are 18 objects for which
$z$ is very uncertain or not determined at all ($f_{z} > 1$).

Most of the rejected stars were of late spectral type, a few others are
early-type stars and a small fraction were identified with exotic types.
While normal stars were easily recognised based on their typical spectral features
and rejected from the database, there remains a risk that some (a few) rather 
unusual stellar spectra were not correctly identified, e.g., rare types of
white dwarfs (WD; Schmidt et al. \cite{Schmidt99},\cite{Schmidt07}; 
Szkody et al. \cite{Szkody04}; Carollo et al. \cite{Carollo06}).
Fig.\,\ref{fig:stellar-cont}\,{\bf a}-{\bf g} show peculiar spectra that were
classified as quasars by SDSS but rejected later on in the present study
either during the inspection of the spectra or as a result of proper motion 
determinations (see below):
{\bf (a)} the superimposition of the spectra of a normal quasar and a late-type star 
(\object{SDSS J014349.15+002128.3}), 
{\bf (b)} the WD-M star binary \object{SDSS J100658.64+121133.9} 
(Heller et al. \cite{Heller09}),
{\bf (c)} a member of the new class of magnetic WD binaries 
with extremely low mass transfer rates 
(\object{SDSS J103100.55+202832.1}; Schmidt et al. \cite{Schmidt07}),
{\bf (d)} a rare magnetic cataclysmic variable with extreme
cyclotron features (\object{SDSS J132411.57+032050.4}; 
Szkody et al. \cite{Szkody04}),
{\bf (e)} the newly discovered proper motion star 
\object{SDSS J020731.81+004941.6}\footnote{According to
the SDSS explorer, a quasar with $z=1.1165$ in DR7 and $z=2.3273$ in DR8, 
respectively.} (Table\,\ref{tab:pm}). In a first attempt, we had
classified SDSS J020731.81+004941.6 as a low-redshift FeLoBAL quasar
(similar to
\object{SDSS J112526.12+002901.3} and \object{SDSS J112828.31+011337.9} from 
Hall et al. \cite{Hall02}). Its significant proper motion suggests
that \object{SDSS J020731.81+004941.6} is a possible extreme DZ or carbon 
star,
{\bf (f)} the newly discovered proper motion DQ WD 
\object{SDSS J020534.13+215559.7}, and
{\bf (g)} the featureless blue spectrum of a newly discovered 
proper motion star (\object{SDSS J084357.71+115624.1}).

Finally, \object{SDSS J014733.58+000323.2} (Fig.\,\ref{fig:stellar-cont}\,{\bf h})
is an example of a spectrum that was erroneously rejected in our first analysis
attempt. The object was classified as a stellar mass black-hole candidate
by Chisholm et al. (\cite{Chisholm03}). Though there is a blue continuum with 
apparent broad emission-line components, these lines do not fit the positions
of the typical broad lines in quasar spectra. Our referee, Dr. Patrick B. Hall, 
pointed out that there is \ion{C}{iv}$\lambda\lambda$ 1548.20,1550.77 absorption
at z=1.8815. This high redshift clearly proves that SDSS J014733.58+000323.2
is a quasar. The closer inspection of the spectrum leads to
its classification as a slightly reddened ($E(B-V) = 0.07$ for SMC-like dust) 
weak-line quasar at $z = 1.98$ without \ion{C}{iii}] emission and with a 
shift between the \ion{C}{iv} and \ion{Mg}{ii} emission lines.

Stellar contamination remains a serious problem even after the individual
inspection of all spectra. Various studies have used 
astrometric information for the quasar selection 
(e.g., Sandage \& Luyten \cite{Sandage67}; Kron \& Chiu \cite{Kron81}; 
Meusinger et al. \cite{Meusinger02},\cite{Meusinger03}; 
Kaczmarczik et al. \cite{Kaczmarczik09}; Lang et al. \cite{Lang09}). 
Quasars are sufficiently distant to able to neglect there having any
measurable absolute proper motion (pm) in presently available position data.
Therefore, absolute pm are considered helpful data to discriminate quasars from 
(nearby) stars. We cross-correlate our quasar sample
with the PPMXL catalogue (R\"oser et al. \cite{Roeser10}). This catalogue 
is the largest collection (of nearly one billion objects) of pm 
in the International Celestial Reference System (ICRS), which is primarily 
realized by the {\it Hipparcos} catalogue. The PPMXL aims to be complete down
to $V \sim 20$, while our quasar sample extends to fainter magnitudes 
(mean g magnitude $\langle g \rangle = 20.02\pm 1.41$). 
With a search radius of 5$\arcsec$, we identified 918 objects (91\%) 
from our sample in the PPMXL (mean position difference
$0\farcs3$). We express the probability of significant non-zero pm
in terms of a simple pm index, which is defined as the total pm $\mu$ in units
of the pm error $\epsilon$:
\begin{equation}
I_{\rm  pm} = \sqrt{(\mu_\alpha \cos \delta)^2+\mu_\delta^2}\ \Big/
                    \sqrt{\epsilon_{\mu_\alpha \cos \delta}^2+\epsilon_{\mu_\delta}^2}.
\end{equation} 
For the 918 identified objects, we found a mean value of
$\langle I_{\rm pm} \rangle = 2.46$ and a median at 0.98; 
90\% of the objects have  $I_{\rm  pm} < 2.5$. 
We restricted the further analysis to the 95 objects with 
$I_{\rm  pm} > 2.5$. One third of this subsample have large pm
$\mu > 150$\ arcsec\ yr$^{-1}$, which are likely due to these objects beeing
fakes (R\"oser et al. \cite{Roeser10}). We re-investigated the spectra of
all 95 pm candidates and searched for possible entries in the 
SIMBAD\footnote{http://simbad.u-strasbg.fr/simbad/} and 
NED\footnote{http://ned.ipac.caltech.edu} database. For only one object, 
\object{SDSS J020534.13+215559.7}, did the spectrum turn out to be
characteristic of a DQ WD (Fig.\,\ref{fig:stellar-cont}).
The star was not found in any stellar database. For 81 objects, the 
spectra could be assigned to quasars without doubt, another 9 objects 
are likely quasars. Another four selected pm candidates have 
relatively featureless spectra, one of them, 
\object{SDSS J153939.1+274438}, is a known radio source with
dominant blazar characteristics (Massaro et al. \cite{Massaro09}).
The large AGN fraction among the pm-selected
objects obviously means that the accuracy of the available pm data
is insufficient to efficiently select stellar contaminants in the
magnitude range of our sample. The trend of increasing 
$I_{\rm  pm}$ with fainter magnitudes 
(e.g., $\langle I_{\rm  pm} \rangle = 1.12$ for $g<20$ compared 
to $\langle I_{\rm  pm} \rangle = 4.10$ for $g>20$) and vice versa
(e.g., $\langle g \rangle = 19.7$ for $I_{\rm pm} \le 2.5$ compared to
$\langle g \rangle = 21.0$ for $I_{\rm pm} > 2.5$) indicates that the 
pm errors are systematically underestimated at faint magnitudes.

\begin{figure}[htbp]
\vspace{6cm}
\rotatebox{270}{\includegraphics[bb=592 0 10 790,scale=0.3,clip]{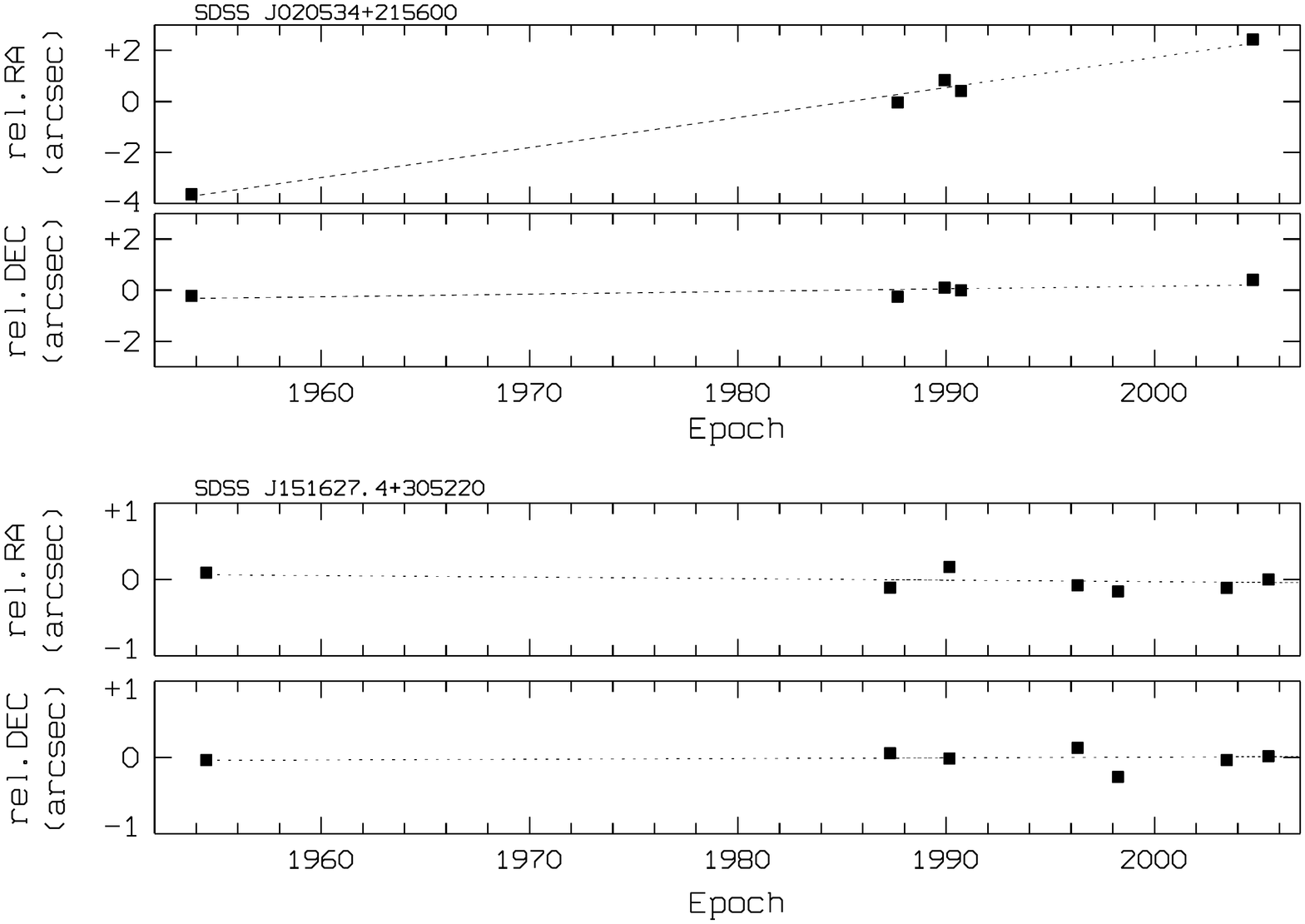}}
\caption{Two examples for estimating proper motions of uncertain quasar
candidates: the  DQ WD \object{SDSS J020534.13+215559.7} (top) and the quasar 
\object{SDSS J151627.40+305219.7} (bottom).
} 
\label{fig:pm}
\end{figure}

In a second step, we therefore used all available multi-epoch positions
from different sky surveys to determine an improved pm.
The longest time baseline (with epochs between 1950 and 2002)
was provided by the SuperCOSMOS Sky Surveys 
(SSS; Hambly et al.~\cite{hambly01}), including the measurements of 
overlapping Schmidt plates. In the case of problems with the SSS 
measurements on the first epoch Palomar Observatory Sky Survey
(POSS1) plates, we used the APM measurements of these POSS1 plates
(McMahon et al. \cite{mcmahon00}). New data (for epochs between 
1998 and 2009) came from two data releases of the SDSS
(DR7, Abazajian et al.~\cite{Abazajian09};
DR8, Aihara et al.~\cite{aihara11}; with 30-70 epochs in 
the equatorial stripe), the eighth data release of
UKIDSS\footnote{The UKIDSS project is defined in 
Lawrence et al.~(\cite{lawrence07})}, and for the brightest objects, from
the Carlsberg Meridian Catalog (\cite{cmc06}),
and the Two Micron All Sky Survey
(2MASS; Skrutskie et al.~\cite{skrutskie06}).
Finally, we included the epoch 2010 positions of the objects (if measured)
in the WISE (Wide-field Infrared Survey Explorer; 
Wright et al. \cite{Wright10}) preliminary data release.
The pm obtained from simple linear fitting of all positions 
over time were not only judged in terms of their formal errors
(typically smaller than the PPMXL errors, if available), but 
also by the influence of the less-accurate Schmidt plate data.
In particular, we did not trust any formally significant pm
if the fit was dominated by only one POSS1 position and could not be 
confirmed after excluding this plate from the fit.
Examples of the proper motion fits are shown in Fig.\,\ref{fig:pm}.

\begin{figure*}[htbp]
\begin{tabbing}
\includegraphics[bb=45 00 495 780,scale=0.178,angle=270,clip]{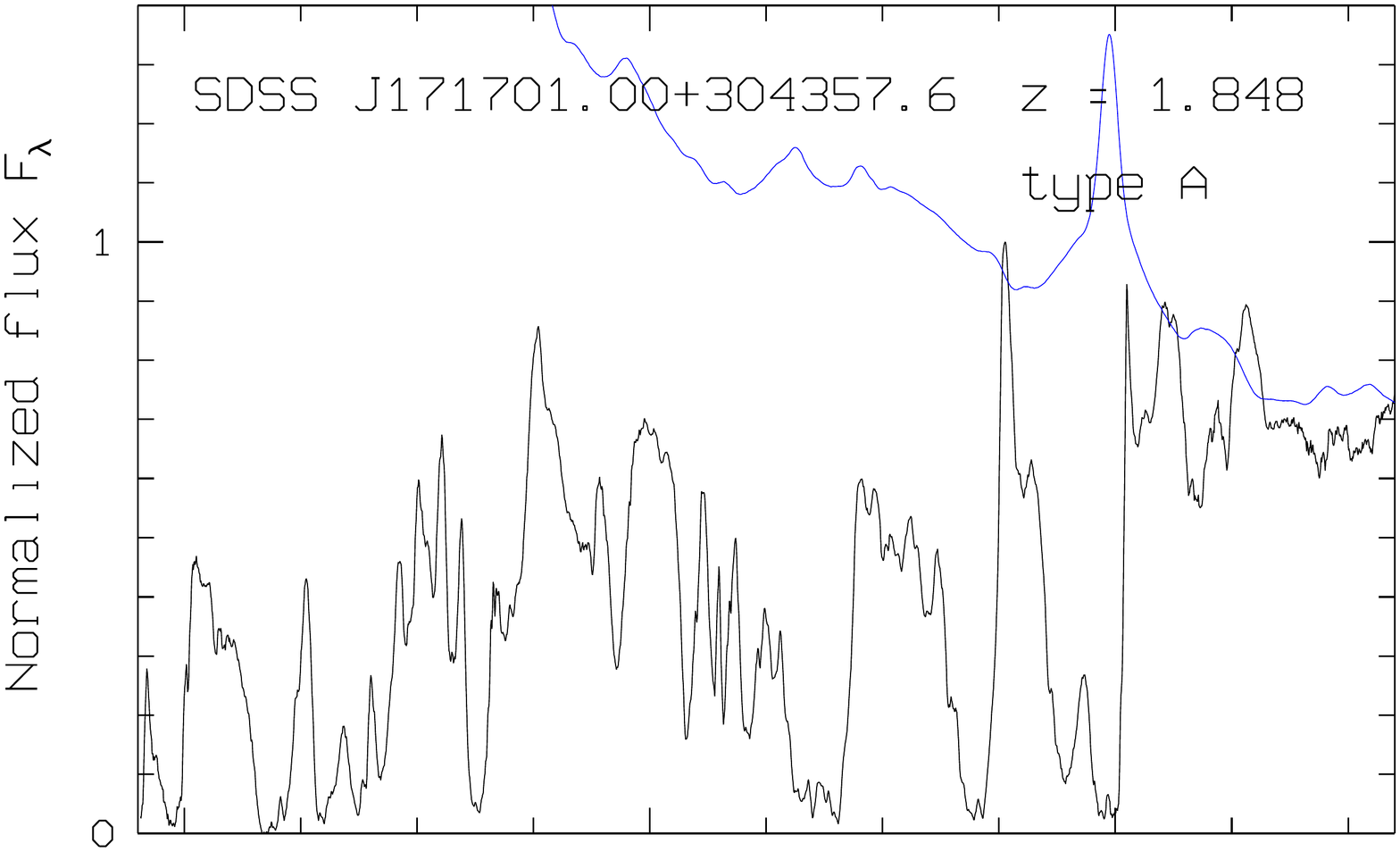} \=
\includegraphics[bb=45 80 495 780,scale=0.178,angle=270,clip]{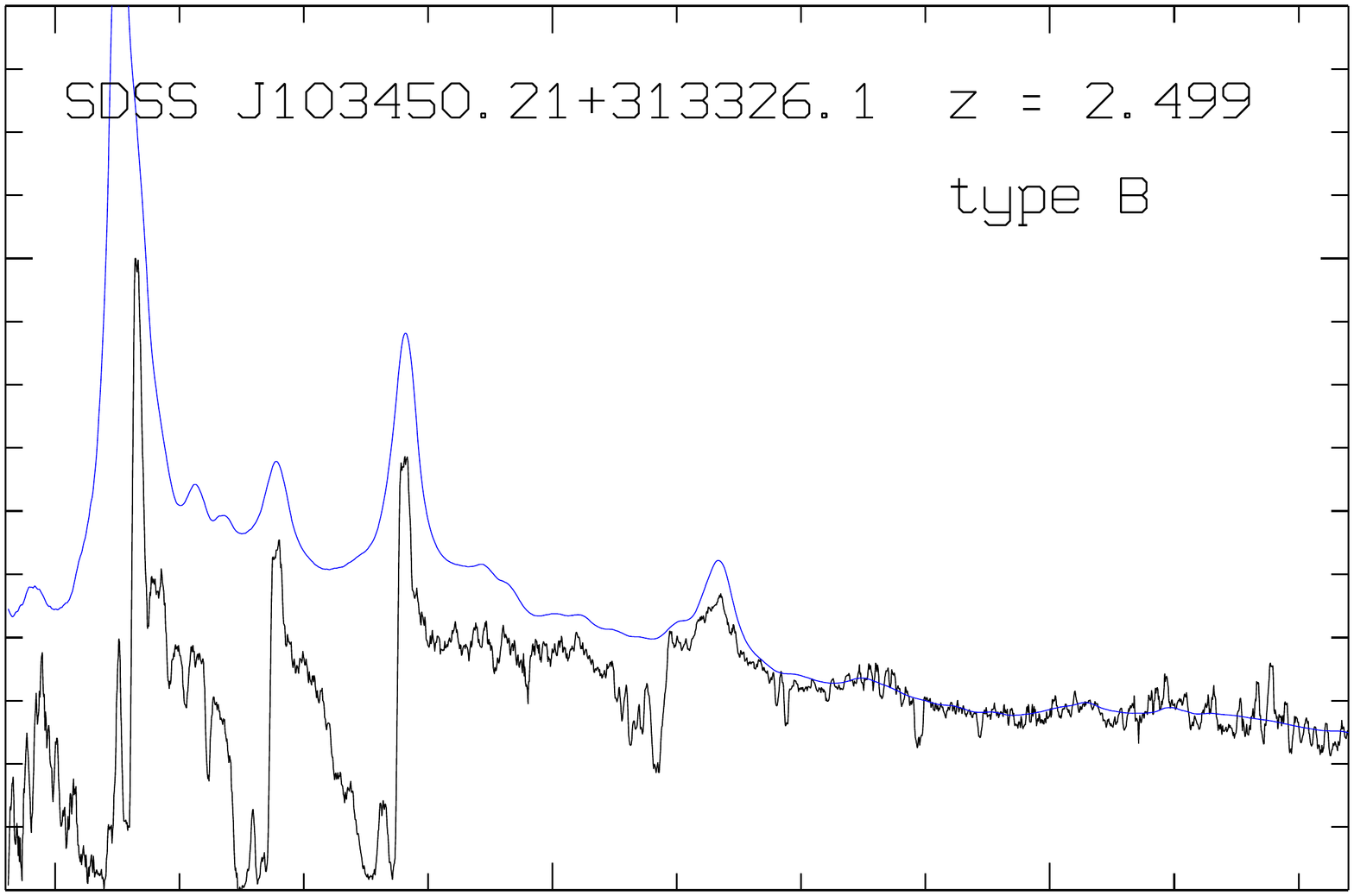} \=
\includegraphics[bb=45 80 495 780,scale=0.178,angle=270,clip]{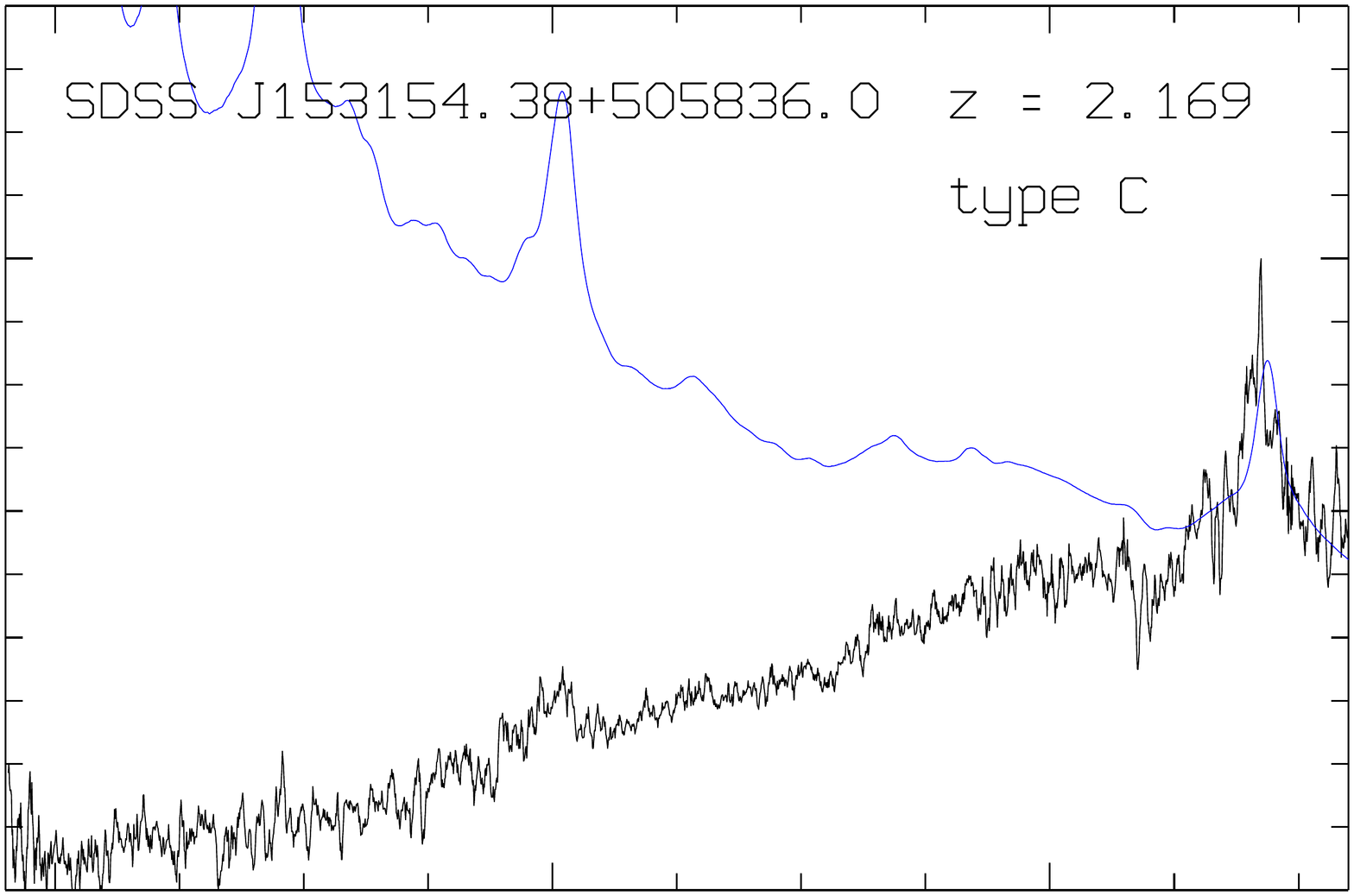} \=
\includegraphics[bb=45 80 495 780,scale=0.178,angle=270,clip]{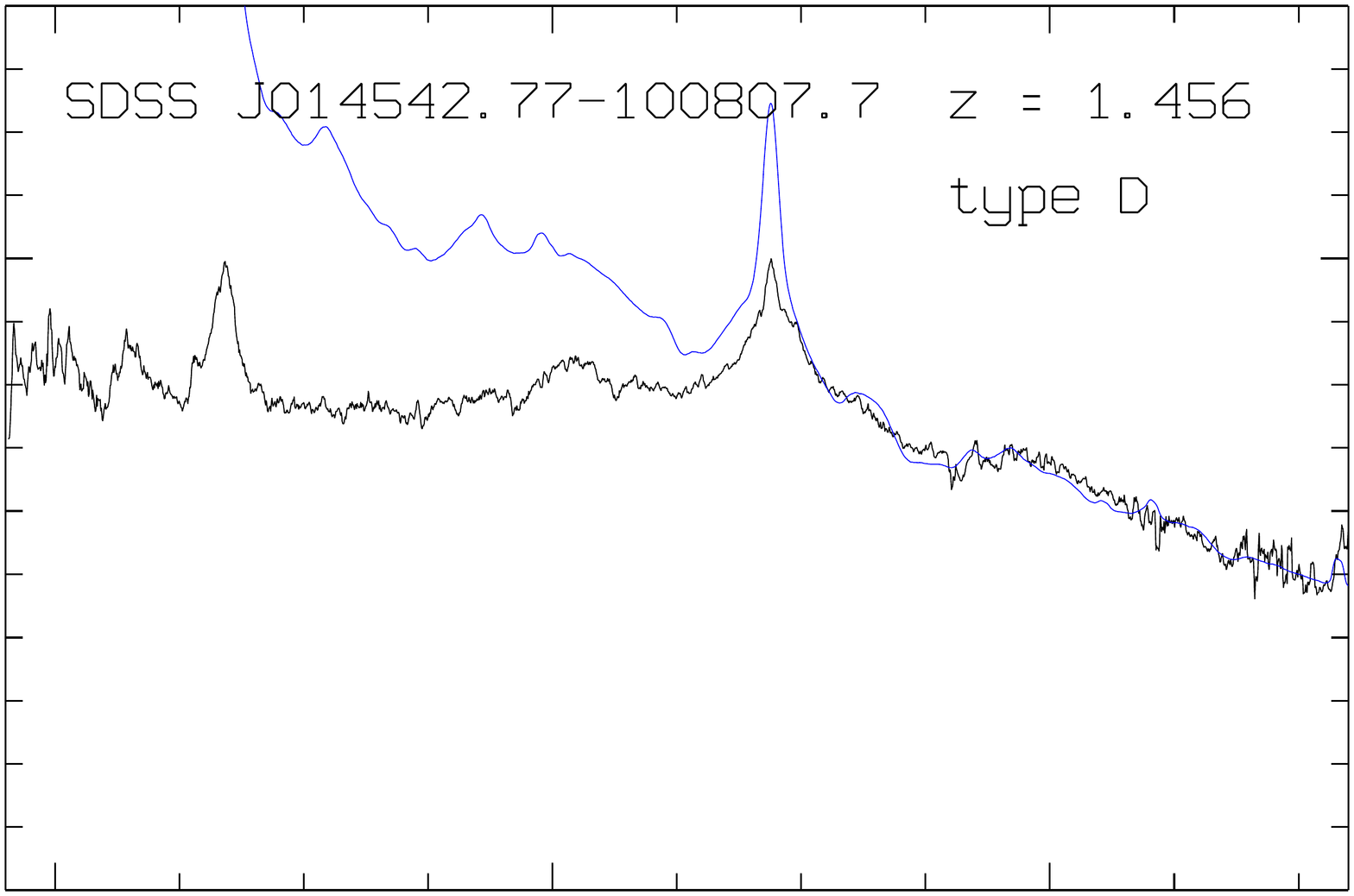} \\
\includegraphics[bb=45 00 580 780,scale=0.178,angle=270,clip]{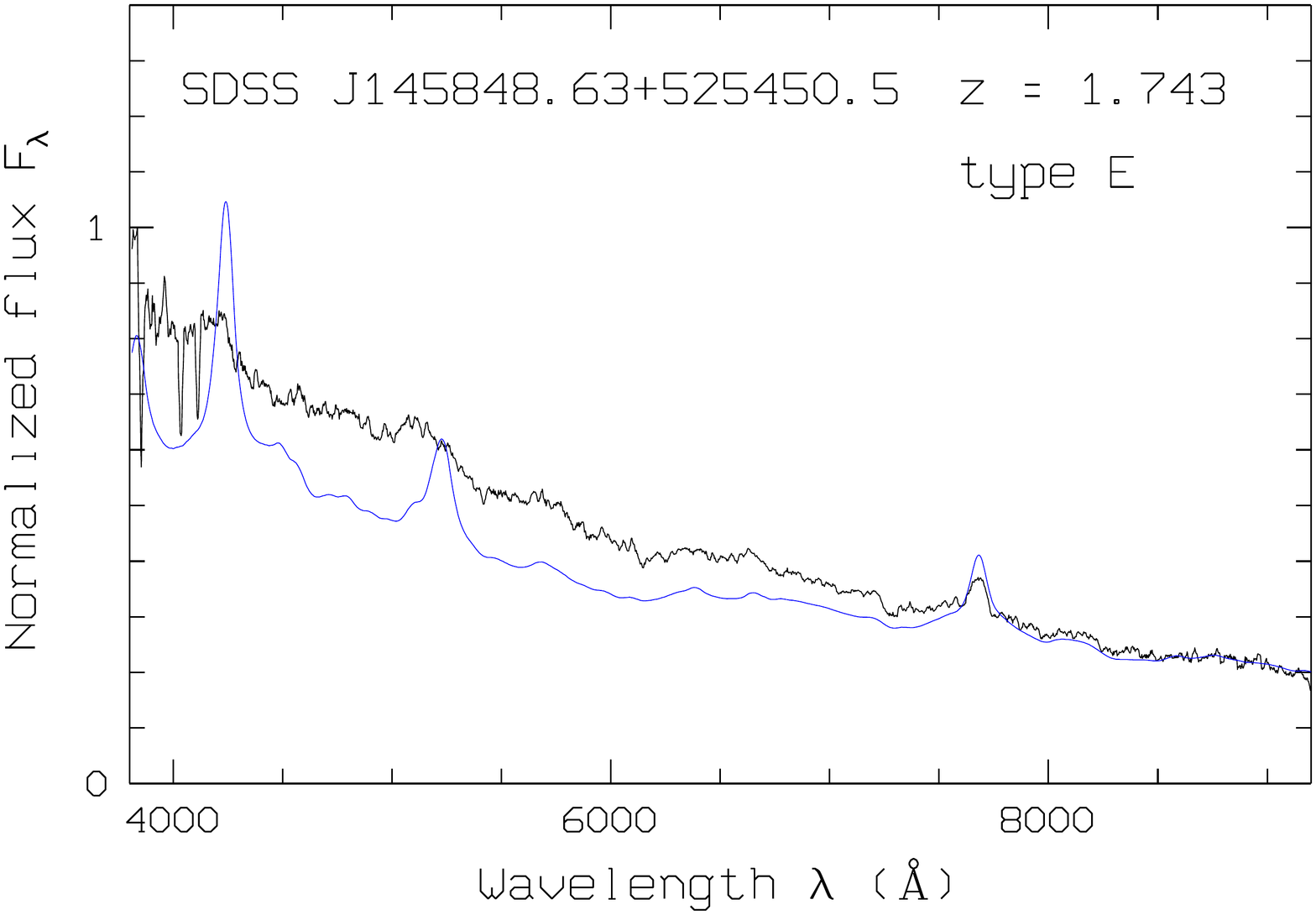} \=
\includegraphics[bb=45 80 580 780,scale=0.178,angle=270,clip]{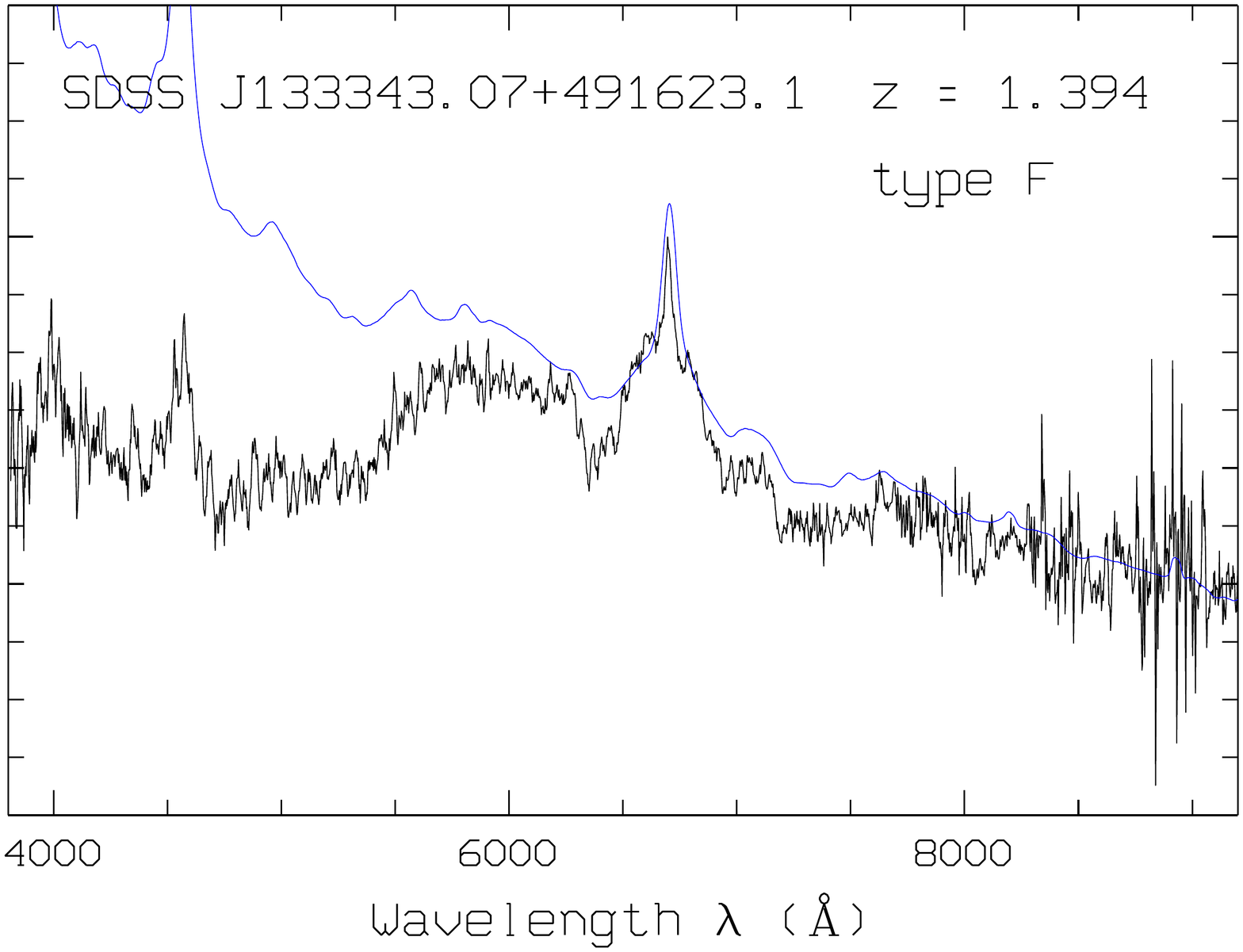} \=
\includegraphics[bb=45 80 580 780,scale=0.178,angle=270,clip]{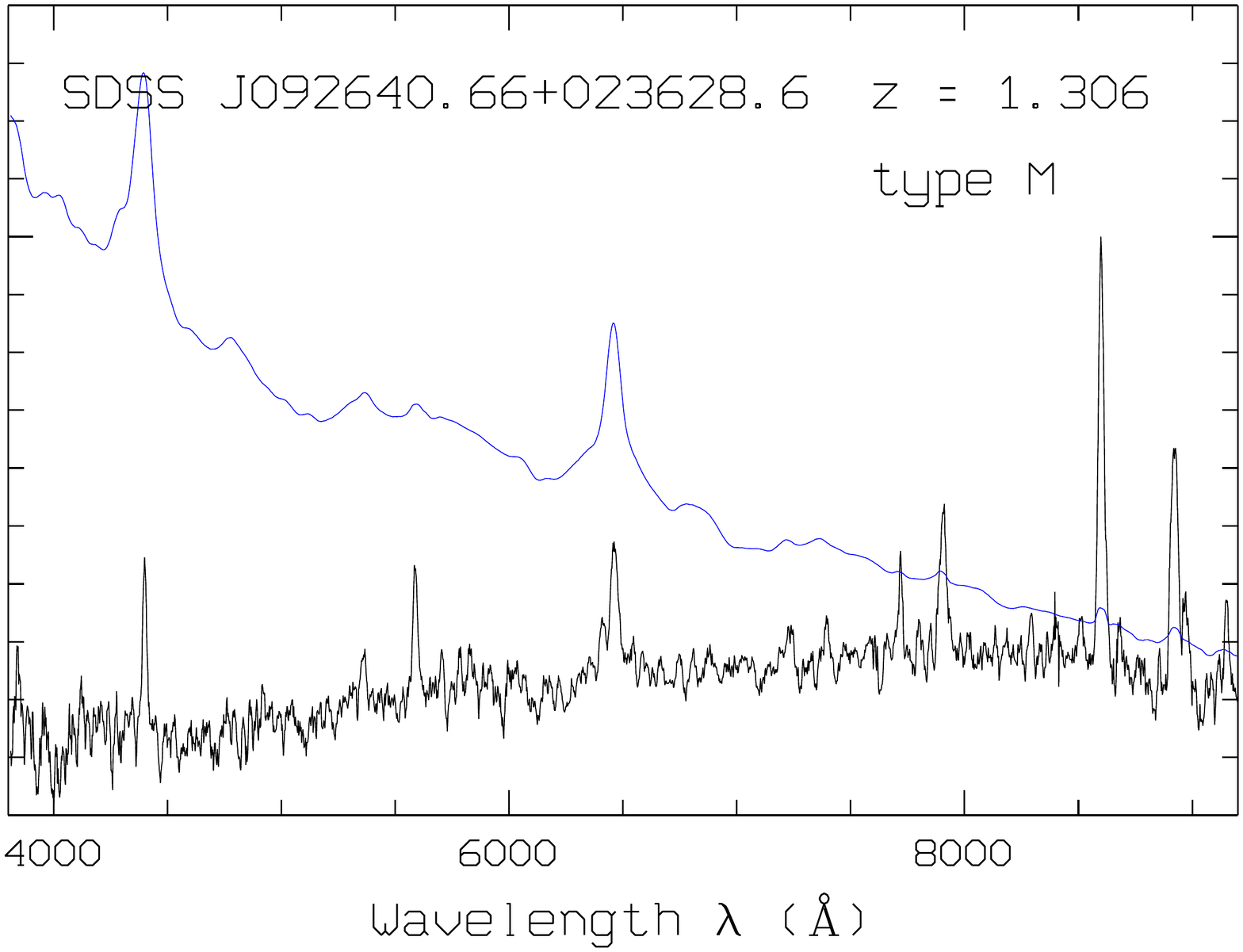} \=
\includegraphics[bb=45 80 580 780,scale=0.178,angle=270,clip]{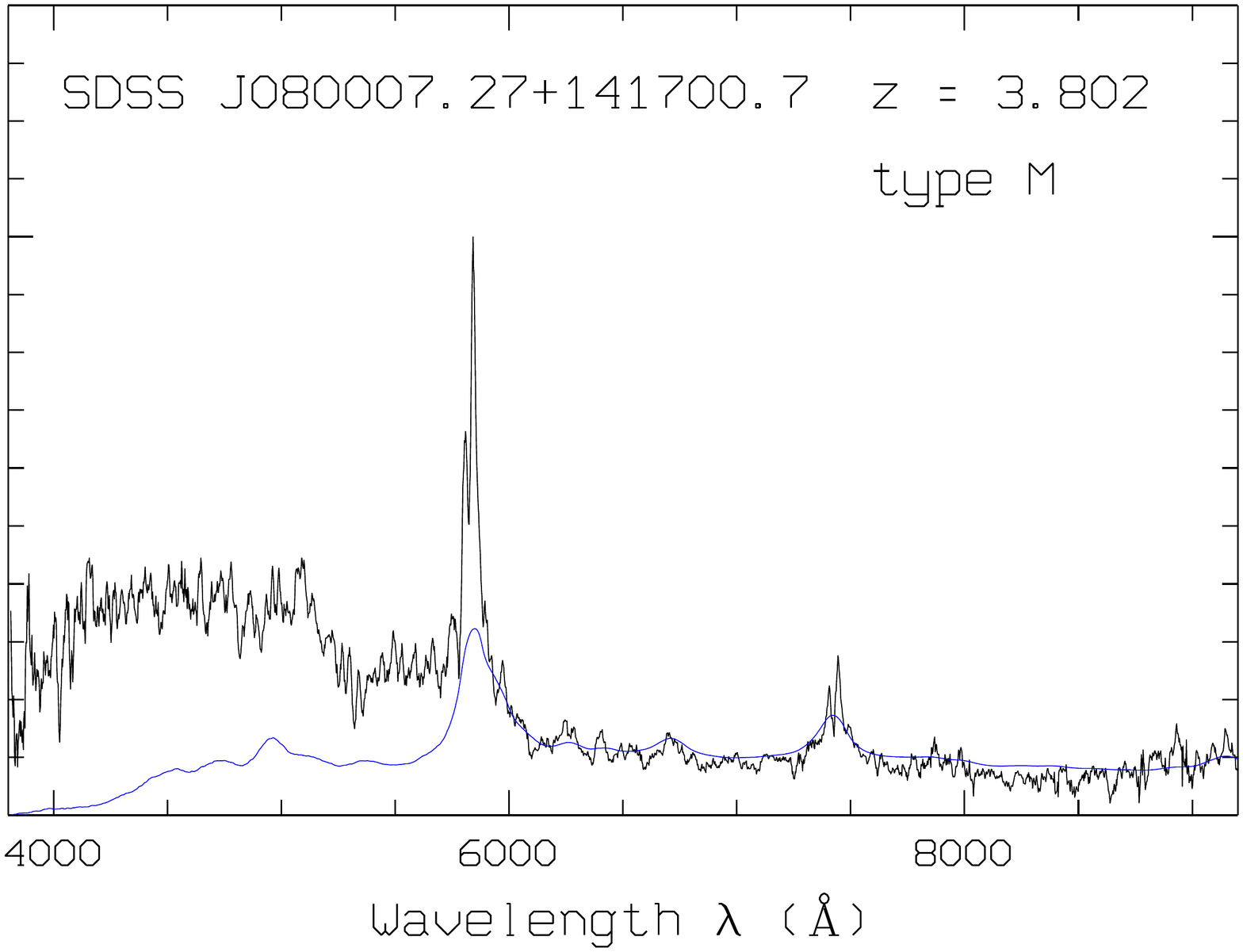} \\
\end{tabbing}
\caption{Example spectra for the unusual quasar types A to M. 
For comparison, the SDSS quasar composite spectrum (VandenBerk \cite{VandenBerk01})
is shown, arbitrarily normalized to each object spectrum at the red end.}
\label{fig:examples_groups}
\end{figure*}

The procedure was applied to objects from three subsamples: 
{\it (a)} the 95 pm candidates from the PPMXL, 
{\it (b)} the two ``mysterious'' objects from Hall et al. (\cite{Hall02}) 
and 15 possibly related objects (Sect.\,\ref{subsect:myst}),
and {\it (c)} an additional 9 objects that had
either featureless spectra or spectral features we were unable
to interpret. Among these 121 objects are in particular all those
with $f_{z} = 2$. As expected, most quasar candidates have zero
proper motions that we were able to measure very accurately if many epochs
were available. However, for seven candidates we found significant
proper motions (Table\,\ref{tab:pm}).
With the exception of \object{SDSS J144354.73+242906.6}, the spectra
of all these objects are blue and suggest that their classification as WDs
with the spectral types given in the last column of Tab.\,\ref{tab:pm}.
The spectra of \object{SDSS J134913.51+205646.9} and 
\object{SDSS J144354.73+242906.6} are noisy; the former 
is probably a DC and the latter may be a late-type star, but the S/N
is too low for a reliable type estimation to be possible.
Another two unusual quasar candidates, \object{SDSS J032907.24+002438.4}
and \object{SDSS J033716.08+000041.7}, turned out to be galaxies. These
9 entries were removed so that our catalogue finally contains 1005 entries,
among them 14 objects
with $f_{z} \ge 2$ that were excluded from the
statistical investigations described in Sect.\,\ref{sec:groups}.

\begin{table}[h]
\caption{Rejected quasar candidates with significant proper motions
from the present study ($N_{\rm e}$: number of epochs, $T_{\rm WD}$: WD spectral type).}
\centering
\begin{tabular}{lrrrl}                
\hline\hline                                   
\ \ \ \ \ \ \ \ SDSS J   & $\mu_\alpha\cos\delta$  \ \  & $\mu_\delta$ \ \ \ \ \
& $N_{\rm e}$ & $T_{\rm WD}$\\ 
                       &  (mas/yr)               \ \  & (mas/yr)  \   & \\ 
\hline
$020534.13+215559.7$ & $ 109.0\pm7.2$ & $ +10.3\pm5.6 $ & 5 & Q\\ 
$020731.81+004941.6$ & $  29.2\pm1.0$ & $  -6.2\pm1.5 $ &39 & Z\\ 
$024058.80-003934.5$ & $ -10.4\pm0.9$ & $  -7.5\pm1.2 $ &80 & C\\ 
$084357.71+115624.1$ & $  -9.4\pm8.0$ & $ +45.8\pm6.0 $ & 9 & C\\ 
$100149.22+144123.8$ & $-345.5\pm1.5$ & $  -2.3\pm3.5 $ & 7 & C\\ 
$134913.51+205646.9$ & $ -68.4\pm8.7$ & $ -17.6\pm3.8 $ & 5 & ?\\ 
$144354.73+242906.6$ & $-107.7\pm6.2$ & $-112.8\pm7.1 $ & 6 & ?\\ 
\hline       
\end{tabular}
\label{tab:pm}
\end{table}

As emphasised above, our method of selecting unusual quasars is not
aimed at a complete sample. Nevertheless, to get an idea of the
completeness we considered the sample of the unusual BAL quasars from 
Hall et al. (\cite{Hall02}). Among these 23 objects, four were classified 
by the SDSS DR7 as {\sc unknown (spec\_cln=0)}, and for another three the
spectra were unavailable for download. Hence, 16 quasars from this
sample are expected to be selected by our approach.
Only one Hall quasar is missed,
\object{SDSS J$121441.42-000137.8$}, a BAL quasar with possible 
relatively strong \ion{Fe}{iii} absorption.

\subsection{Categories of unusual quasar spectra}\label{subsec:classification}

According to the different areas populated by unusual quasar spectra 
in the Kohonen maps, a classification into essentially four categories
was immediately suggested. The corresponding dominant spectral features are 
{\it (a)} strong BALs, 
{\it (b)} red continua, 
{\it (c)} weak or absent emission lines, and 
{\it (d)} strong optical or UV iron emission. 
As a result of evaluating all selected 
spectra, we decided to subdivide the first category into two types: unusual BAL 
structures (mostly LoBALs) and more or less normal BAL quasars (mostly strong HiBALs).  
Moreover, we found that the category of red quasars contains a substantial
fraction of spectra where the continuum is significantly red or reddened at UV
wavelengths (e.g., $\lambda \la 3000$\AA), but seem to reasonably fit the SDSS quasar 
composite spectrum at longer wavelengths  
(see also Hall et al. \cite{Hall02}, their section 6.3.1). 
Therefore, we split this category 
into the two types of (pure) red continua and UV-red continua, though it was
unclear {\it ab initio} whether they represent distinct types. Finally, a
relatively small number of objects could not be adequately sorted into one of these six
types. These quasars form a small and quite inhomogeneous group, including
those that have spectra with
exceptionally blue continuum, unusual line profiles, unusual continuum shape
(perhaps caused by an error in the flux calibration), or quasars with narrow
emission lines but lacking substantial broad emission-line components (type 2 quasar candidates).

Therewith, this paper deals with seven categories or types of unusual quasar spectra
(Fig.\,\ref{fig:examples_groups}):
\vspace{-0.2cm}
\begin{itemize}
\item[A] - Unusual BAL quasars (mostly LoBAL quasars);
\item[B] - Quasars with strong but more or less usual BAL structures (mostly HiBALs);
\item[C] - Quasars where the whole continuum is significantly redder than the SDSS composite;
\item[D] - As type C, but strong reddening is obvious only in the UV;
\item[E] - Quasars with weak or absent emission lines;
\item[F] - Quasars with strong iron emission;
\item[M] - Miscellaneous.
\end{itemize}
Since some of the selected spectra appear to be very complex, we decided to build
the classification on the subjective impression of the overall picture rather 
than on quantitative criteria.

The transitions between the various categories are smooth. Many spectra show a mixing of 
characteristic features of two or three types (see e.g., Fig.\,\ref{fig:examples_groups}).
Therefore we allocated to each quasar up to three out of the seven possible types 
$T_{\rm i}, \ i = 1\ldots3$ ($T_1 \ne T_2 \ne T_3$) ), where the 
significance decreased with $i$.
Forty-one percent of the quasars have more than one classification and 8\% have more than two.
In particular, there is a strong coupling between unusual BALs, red 
continua, and strong iron emission. Table\,\ref{tab:catalogue} lists all
three types. In the rest of the paper, we consider the type of highest
significance ($T = T_1$) only.

\subsection{The catalogue}\label{subsec:catalogue}

\begin{table*}[htbp]
\caption{Classification and properties of unusual SDSS quasars. 
(Only the first five rows are shown here; 
the full table is available electronically only.) 
}
\begin{flushleft}
\begin{tabular}{rcccccccrrrccrl}
\hline\hline
Nr           &
SDSS J       &
$f_{\rm c}$  &
$z$          &
$f_z$  &
$i_0$        &
$M_{\rm i}$  &
$f_{\rm r}$  &
$F_{1.4}$    &
$R_{\rm i}$  &
$\chi^2$     &
$T_1$        &
$T_2$        &
$T_3$        &
Comment\\
\hline
1 &$000009.38+135618.4$& 1 & 2.233 & 1 & 18.17 &$-27.45$& 0  &... &... &  2 & 5 &  6 &... & \ion{Fe}{iii}\ em?\\
2 &$000326.67+001157.1$& 1 & 1.845 & 0 & 18.95 &$-26.23$& 0  &... &... & 18 & 4 &... &... & \\
3 &$000625.25+154625.8$& 1 & 1.527 & 1 & 19.58 &$-25.16$& 0  &... &... & 77 & 3 &... &... & ao(\ion{Mg}{ii});strong\,[\ion{Ne}{v}];noisy\\
4 &$000728.45-042345.5$& 1 & 1.198 & 0 & 18.12 &$-26.04$& 0  &... &... & 15 & 6 &  1 &... & \ion{Fe}{iii}\ em; see 2215(H02)\\
5 &$000920.01+005618.4$& 1 & 2.387 & 0 & 20.20 &$-25.56$& 0  &... &... & 10 & 4 &  5 &... & \\
\hline
\end{tabular}\\
\end{flushleft}
\label{tab:catalogue}
\end{table*}

The results of our visual inspection of the spectra of the 1005 unusual SDSS quasars
are given in Table\,\ref{tab:catalogue}. The catalogue was matched
with both the QCDR7 (Schneider et al. \cite{Schneider10}) and the 08Jul16 version
of the catalogue from the FIRST Survey (Becker et al. \cite{Becker95}). 
The vast majority (98.3\%) of the quasars have entries in the QCDR7
(quasar catalogue flag $f_{\rm c} > 0$). The agreement between the redshifts 
from the present study and those from the QCDR7 is generally 
good.\footnote{For the quasars with certain redshifts ($f_{z} = 0$),
we found six with deviations $>10 $\%.
These objects are indicated by $f_{\rm q} = 2$. The deviations were larger than
50\% for \object{SDSS J093437.53+262232.6} and \object{SDSS J131524.00+041734.4}.} 
As expected, the discrepancies are larger when the redshifts 
from the SDSS pipeline are considered: 98 (41) quasars with deviations 
$>10 (50)$\%. The 18 objects
missed in the QCDR7 have normal absolute magnitudes and redshifts
($\overline{M}_{\rm i} = -25.5,\ \bar{z} =$ 1.84) but their redshifts are mostly uncertain
(6/7/4/0 objects have  $f_{z}$ = 0/1/2/3). For these objects, the entries in cols.
6 and 8 of the catalogue were taken from the SDSS DR7 Explorer.

The full catalogue is available only in electronic form. Table\,\ref{tab:catalogue} 
lists the first five entries for guidance regarding its content and form:\\
Column (1): The running catalogue number.\\
Column (2): The SDSS J2000 equatorial coordinates for the quasar (taken from the
  spectrum fits header).\\
Column (3): The QCDR7 catalogue flag ($f_{\rm c}$ = 0: not in QCDR7, 1: in QCDR7 with 
  $z$ deviation $\le 10$\%, 2: in QCDR7 with $z$ deviation $> 10$\%).\\
Columns (4) and (5): The redshift $z$ and the redshift flag $f_{z}$
  from the present study ($f_{z} = 0$: certain, 1: likely but not certain, 2:
  lack of certainly identifiable redshift indicators,  3: low signal-to-noise).\\
Column (6): The SDSS apparent i band magnitude corrected for Galactic foreground extinction.\\
Column (7): The absolute i band magnitude, computed as described
  by Kennefick \& Bursick (\cite{Kennefick08}) for a spectral index $\alpha_\nu =-0.5$.\\
Columns (8): The FIRST radio detection flag from the QCDR7 (0 for non-detection).\\ 
Column (9): The FIRST 1.4 GHz peak flux $F_{1.4}$ (mJy).\\
Column (10): The radio loudness parameter $R_{\rm i}$ computed from $F_{1.4}$ and the 
  extinction-corrected SDSS i band magnitude following Ivezi\'c et al. (\cite{Ivezic02}).\\ 
Column (11): The peculiarity index $\chi^2$ of the spectrum (Sect.\,\ref{subsec:peculiarity}).\\
Columns (12) to (14): The type classifications from the present study 
  (Sect.\,\ref{subsec:classification}) in numerical format (1=A, 2=B, ...) for 
  computational reasons. The priority decreases from $T_1$ to $T_3$.\\
Column (15): Comments on spectral peculiarities. Abbreviations:  
\vspace{-0.2cm}
\begin{itemize}  
 \item abs: absorption
 \item ao: associated absorption    
 \item bel: blueshifted emission line
 \item bq: blue quasar  
 \item em: emission
 \item dlas: damped Ly\,$\alpha$ absorption system  
 \item fg: foreground    
 \item lst: longwards-of-systemic trough
 \item mnt: many narrow troughs  
 \item myst?: related or possibly related to the two mysterious objects 
       from Hall et al. (\cite{Hall02})  
 \item nlq: narrow-line quasar  
 \item nt: narrow troughs  
 \item ot: overlapping troughs  
 \item sdvdpc: spatially distinct velocity-dependent partial covering
       of the continuum source
 \item see 0810(H02): see object SDSS J081024.75+480615.5   
  \hspace{2.1cm} from Hall et al. (\cite{Hall02})
 \item uc: unusual continuum
 \item ulp: unusual line profile
\end{itemize}


%
\section{Mean properties of the various unusual quasar types}\label{sec:groups}
%

The number of quasars in the seven types  are listed in
Table\,\ref{tab:groups} along with some mean properties discussed below 
in this section.

\begin{table}[hhh]
\caption{
Mean properties of the unusual quasar types. 
}
\begin{flushleft}
\begin{tabular}{lrrcclll}
\hline\hline
   $T$         
   & $N$  \
   & $\chi^2$  
   & $z$ 
   & $M_{\rm i}$
   & $E_{B-V}^{\rm (intr)}$
   & \ \ $M_{\rm i}^{\rm (cor)}$
   & $f_{\rm RL}$ \\
\hline
A  & 215 &  78.8& 2.02 & $-26.3$   & 0.18 & $-27.6$& 0.26 \\
B  & 209 &  12.7& 2.92 & $-27.0$   & 0.09 & $-27.9$& 0.11 \\
C  & 116 &  61.3& 1.96 & $-25.9$   & 0.38 & $-28.5$& 0.31 \\
D  & 150 & 115.6& 1.47 & $-25.9$   & 0.20 & $-27.0$& 0.30 \\
E  & 185 &   9.3& 2.14 & $-27.0$   & 0.00 & $-27.0$& 0.26 \\
F  & 112 &  28.5& 1.40 & $-26.2$   & 0.06 & $-26.5$& 0.16 \\
M  &  18 &  13.0& 1.53 & $-25.7$   &\ \ \ ... &\ \ \ \ ...& 0.33 \\
&&&&&&&\\
all& 1005&  48.8& 2.06 & $-26.4$   & $0.15^a$ & $-27.4^a$  & 0.23 \\
\hline
\end{tabular}
\end{flushleft}
\tiny{
{\bf Notes.} $^{(a)}$ Types A to F only.
}
\label{tab:groups}
\end{table}

\subsection{Peculiarity index}\label{subsec:peculiarity}
 
As a  quantitative measure of the deviation of a given spectrum from the SDSS quasar
composite spectrum (VandenBerk \cite{VandenBerk01}), we computed
\begin{equation}\label{eq:chi_2}
\chi^2 = \frac{1}{N \sigma^2} \sum_{\rm i = 1}^{N}
\big[F_{\rm n}(\lambda_{\rm i})-F_{\rm comp, n}(\lambda_{\rm i})\big]^2,
\end{equation}
where $F_{\rm n}$ and $F_{\rm comp,n}$ are the normalised spectra of the quasar and the
composite, respectively, and $\sigma$ is the average noise of  $F_{\rm n}$. 
The wavelengths $\lambda_{\rm i}$ refer to the observer frame 
and the index $i$ indicates the pixel number. 
The quasar spectrum is normalised to the integrated flux
with an integration interval from 4000\AA\ to 9000\AA\ for $z<2.29$ and 
from $1216\cdot(1+z)$\AA\ to 9000\AA\ for $z\ge 2.29$, respectively.  
The SDSS quasar composite spectrum was shifted to the redshift $z$
and normalised to match the red end of $F_{\rm n}(\lambda)$ at $\lambda$ = 8600-9100\AA.
The noise $\sigma$ was derived from the difference spectrum
$D_{\rm n}(\lambda) = F_{\rm n}(\lambda) - \tilde{F}_{\rm n, s10}(\lambda)$, 
where $\tilde{F}_{\rm n, s10}(\lambda)$ is the smoothed version of $F_{\rm n}(\lambda)$ 
using a 20 pixel boxcar filter.
To reduce the effect of residuals from narrow spectral features, five rest-frame wavelength 
intervals of $\sim50$\AA\ width were selected where the contribution from
emission lines is small (pseudo-continuum windows; see Sect.\,\ref{subsec:composites}).
The average standard devation of $D_{\rm n}(\lambda)$ in these parts of the spectrum 
measured in the observer frame is taken as a proxy for $\sigma$. Care has been taken to
exclude the regions around the strongest telluric emission lines at $\lambda 5577$\AA\ and
$\lambda 6300$\AA. 

At least for the statistical comparison of subsamples, $\chi^2$ can be taken as a
useful integral measure of the ``peculiarity'' of the spectrum. 
The individual values cover the broad range from $\sim 1$ to 5\,000 with a mean value 
of $\sim 50$. 
For comparison, the 15 unusual Hall quasars (Hall et al.
\cite{Hall02}) cover the range from 35 to 1089 with a mean value of 261. When we restrict
the selection to $\chi^2>35$, we have 296 quasars corresponding to an unusual
quasar fraction of $f_{\rm uq} = 296/99151 = 0.0029$ in perfect agreement with 
$23/8000=0.0029$ for the selection by Hall et al.. However, if we choose the 
threshold so that our selected sample has the same mean value of $\chi^2$ as the
Hall sample, our selection is much less efficient with $f_{\rm uq} = 0.0012$.

The peculiarity index $\chi^2$ is particularly sensitive to deviations from the 
composite spectrum over wide wavelength intervals, i. e., owing to intrinsic
reddening or an intrinsically red continuum. It is thus unsurprising that the
largest values are measured for the red/UV-red quasars of the types C and D, but 
also type A. (However, type C spectra frequently suffer from strong
noise, i.e., large values for $\sigma$). For weak-line quasars, on the other hand,
the deviations are restricted mainly to the small wavelength intervals of the
lines and thus $\chi^2$ is generally small. The unusual BAL quasars of type A
have significantly larger values of $\chi^2$ than their more normal relatives
of type B, in agreement with their stronger apparent intrinsic reddening
(see below).

\begin{figure*}[htbp]
\begin{tabbing}
\includegraphics[bb=23 220 548 800,scale=0.52,clip]{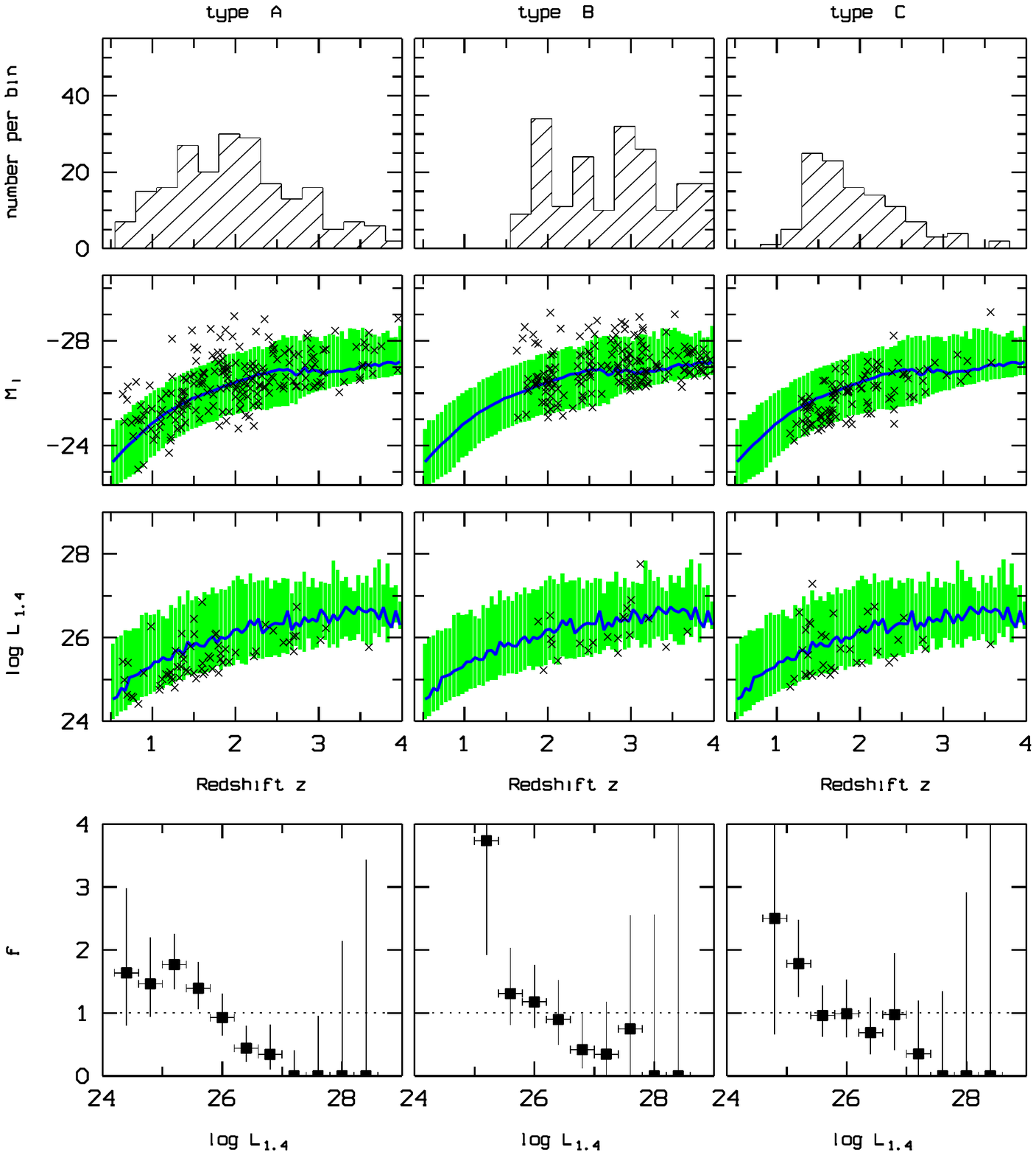}\hfill \=
\includegraphics[bb=84 220 548 800,scale=0.52,clip]{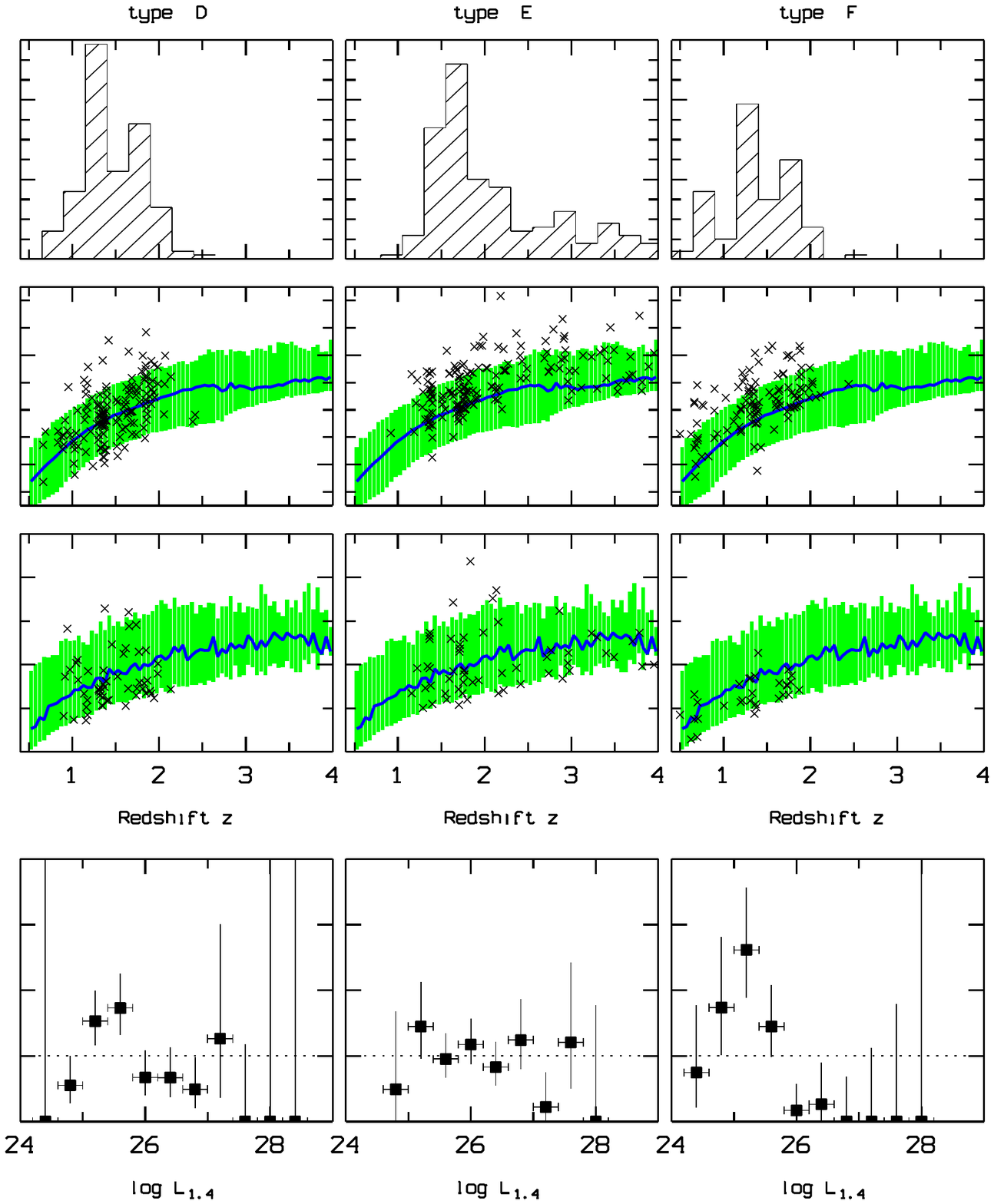}\\
\end{tabbing}
\caption{{\it Top}: Redshift distribution for types A to F.
{\it Second and third row:} absolute i magnitude and
1.4 GHz luminosity, respectively, as function of redshift
compared with QCDR7.
{\it Bottom}: Relative distribution of the 1.4 GHz luminosity. 
}
\label{fig:luminosity-z}
\end{figure*}

\begin{figure*}[bhtp]   
\begin{tabbing}
\includegraphics[bb=10 30 430 190,scale=0.68,angle=270,clip]{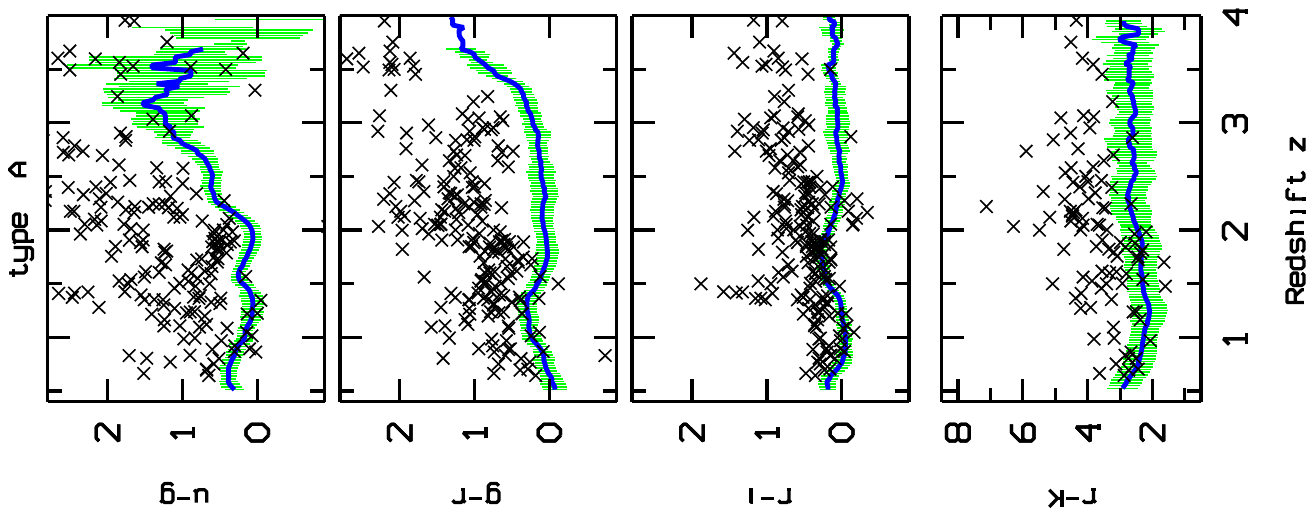}\hfill \=
\includegraphics[bb=10 70 430 190,scale=0.68,angle=270,clip]{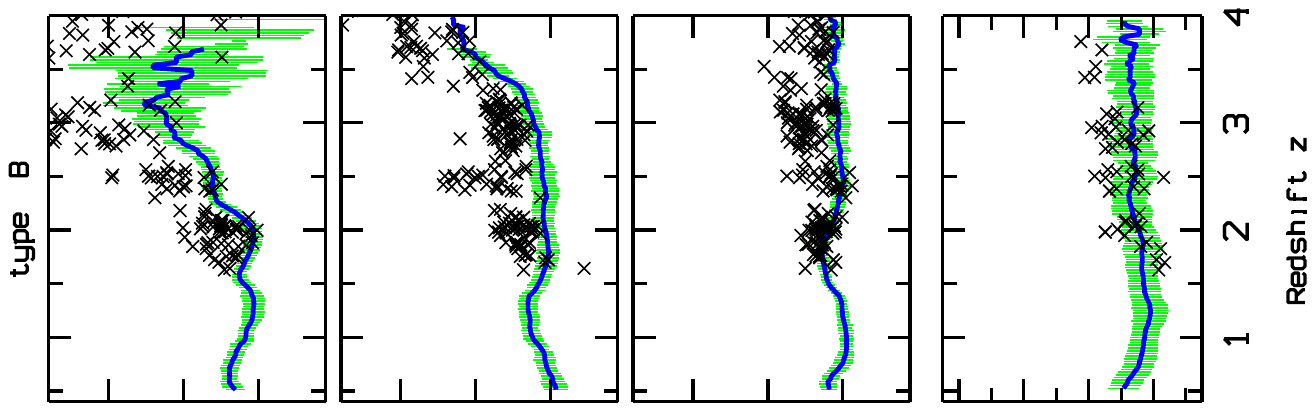}\hfill \=
\includegraphics[bb=10 70 430 190,scale=0.68,angle=270,clip]{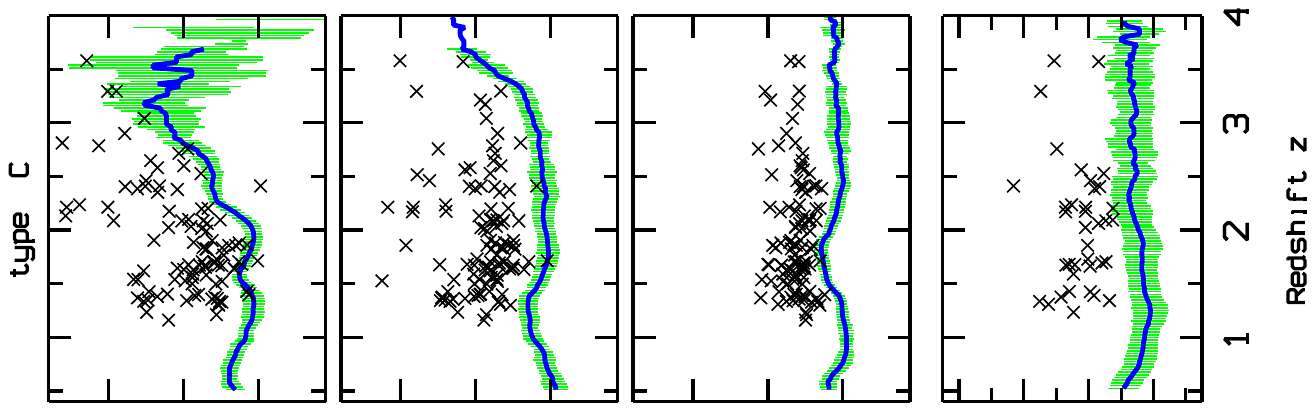}\hfill \=
\includegraphics[bb=10 70 430 190,scale=0.68,angle=270,clip]{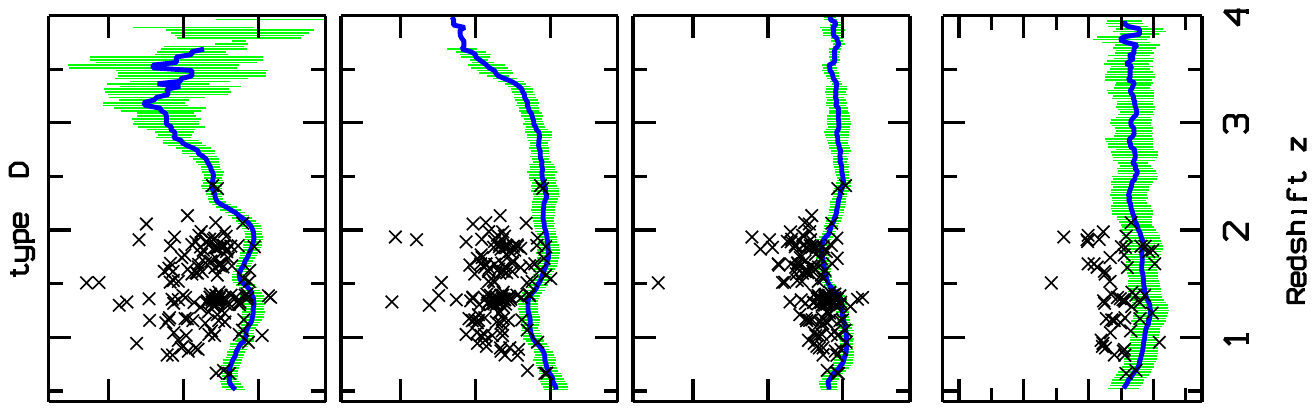}\hfill \=
\includegraphics[bb=10 70 430 190,scale=0.68,angle=270,clip]{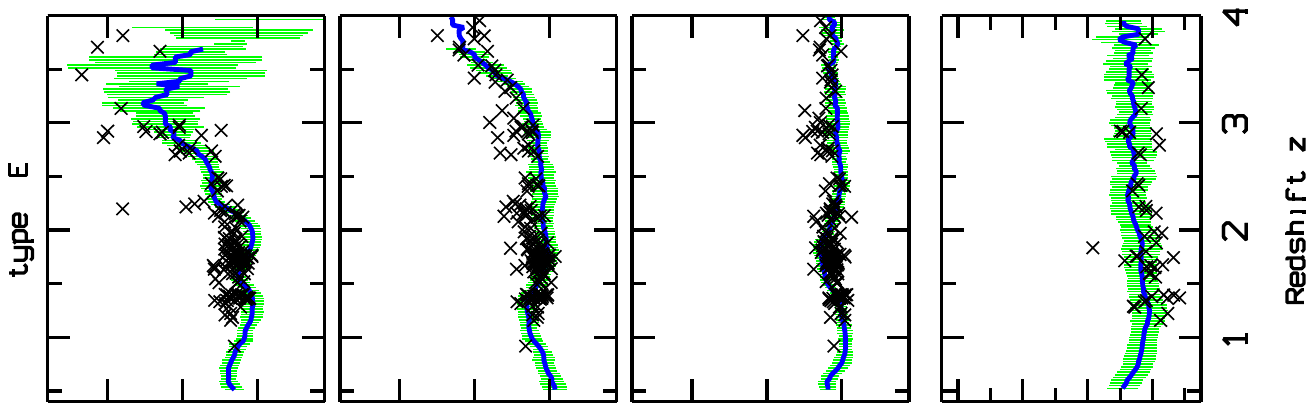}\hfill \=
\includegraphics[bb=10 70 430 190,scale=0.68,angle=270,clip]{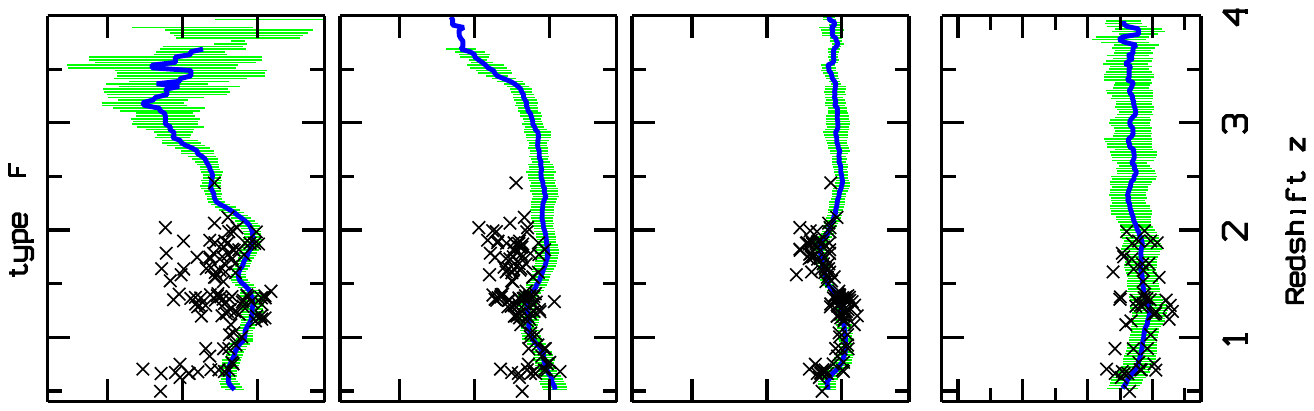}\hfill \\
\end{tabbing}
\caption{Colour indices (extinction corrected)
as function of redshift for types A to F (top to bottom) in comparison 
with the SDSS DR7 quasar catalogue (single values and median relations).}
\label{fig:colour-indices}
\end{figure*}

\subsection{Redshifts and absolute magnitudes $M_{\rm i}$}\label{subsec:redshifts}

The first row of Fig.\,\ref{fig:luminosity-z} shows the histograms of the $z$ 
distributions. As can be seen, the different types cover different redshift
intervals. This is simply due to the limited spectral window of the observations in
combination with the characteristic spectral features being tied to
special wavelength intervals. For example, the change in the continuum slope
for the type D quasars frequently appears around rest-frame wavelengths between 2000\AA\ 
to 3000\AA\ and is thus best observed for $z\sim 1.5$. At higher $z$, 
the wavelength of the turnover is shifted towards and beyond the red edge of the
spectral window and the spectrum appears to be red (see also Fig.\,\ref{fig:examples_groups}).
Strong Fe emission in the UV is restricted to the same wavelength interval.
Consequently, quasars of types D and F cover similar $z$ intervals.

In the second row of Fig.\,\ref{fig:luminosity-z}, the absolute magnitudes
$M_{\rm i}$ are plotted as a function of $z$ (crosses), along with the median relation and
$1\sigma$ deviations for the QCDR7 quasars. Owing to the strong inherent
redshift bias, the $M_{\rm i}$ distribution of any type cannot
be compared with that from the whole QCDR7. For the same reason, the 
$M_{\rm i}$ distributions of the various quasar types cannot be compared with
each other. Therefore, we constructed for each type corresponding comparison samples from
the QCDR7 that have identical $z$ distributions. For every quasar (redshift $z$,
flag $f_{z} <2$), we randomly selected one quasar from the
QCDR7 with a redshift in the interval $z-0.05 \ldots z+0.05$. This procedure was
performed 100 times to create 100 different comparison samples. 

To test whether our unusual quasars and their comparison samples 
represent the same quasar population with respect to $M_{\rm i}$,
we applied the two-tailed two-sample Kolmogorov-Smirnov (KS) test
(e.g., Siegel \& Castellan \cite{Siegel88}).
The null hypothesis $H0$, that the members of a given type of unusual quasars
have the same $M_{\rm  i}$ distribution as the quasars in a comparison 
sample from QCDR7, is tested against the alternative 
$H1$, that the $M_{\rm i}$ distributions of the two samples are different.
The KS test uses the maximum difference 
$D = \max_i |S_i - S_{i,\,{\rm comp}}|$ between the cumulative
distributions $S_{\rm i}$ from the two samples. For the two-tailed test,
$H0$ has to be rejected at a chosen level $\alpha$ if $D$ is so large
that the probability $p$ of its occurrence is $p < \alpha$.
Since we have 100 comparison samples, we performed 100 tests per type. 
Adopting $\alpha = 0.05$, $H0$ had to be rejected for all comparison
samples of types E and F. For the other types the null hypothesis has to be rejected for 
only 45\% of the tests for type C and for $\le 10$\% for the types A, B,
and D.

According to Fig.\,\ref{fig:luminosity-z},
the quasars of types E and F tend to be more luminous than normal. 
We applied a one-tailed KS test with the alternative hypothesis
$H1$ that unusual quasars are more luminous than usual quasars.
Again, $H0 (\alpha=0.05)$ had to be rejected in favour of
$H1$ for the types E and F.  We conclude that both the weak-line quasars and the
strong iron emitters are more luminous in the optical than typical quasars 
from QCDR7. Just et al. (\cite{Just07}) created a sample of 32 of the most 
luminous quasars from the SDSS DR3 quasar catalogue adopting a minimum
luminosity at $M_{\rm i} = -29.28$. Our whole sample includes three quasars
brighter than this threshold; all three are of type E, among them the 
ultra-luminous quasar \object{SDSS J152156.5+520238} with $M_{\rm i} = -30.17$.

It is conceivable that our selection process for spectra with 
weak emission lines or with strong iron emission might have
preferentially selected quasars with higher spectral S/N, which will be the more luminous
ones at each redshift. We estimated S/N for the unusual quasar spectra as well
as for the spectra of $\sim 10^4$ comparison quasars. The noise was measured
in defined areas of the spectrum as described in Sect.\,\ref{subsec:peculiarity}, 
the signal was given by the mean flux in the same areas. We indeed found a higher
average S/N for the unusual quasars namely 11.3 and 15.6 for types E and F, respectively,
compared to 8.1 and 9.2 for the corresponding comparison samples. This is, however,
not completely unexpected because of the trend of S/N with 
apparent magnitude. In the next step, we created new comparison 
samples where for each unusual quasar one comparison quasar was identified 
with similar $z$ (as before) {\it and} with similar S/N. For either type E and F,
we found that the absolute magnitudes $M_{\rm i}$ of the unusual quasars of
types E and F are on average about 0.4 mag brighter than the comparison
samples of normal quasars. The KS test confirms that the differences are significant.

Among all six types, the red quasars of type C have the lowest
optical luminosities, as expected.
We performed the one-tailed KS test to check whether 
the red quasars are significantly underluminous and found that this is unlikely
to be the case.

\subsection{Radio luminosity and loudness}\label{subsec:radio_luminosity}

The radio detection fraction, $f_{\rm RD}$, of our whole sample is 0.29, compared to 0.08 for
the quasars from the QCDR7 in the same redshift range ($0.5 \le z \le 4.4$).
A similar difference is seen for the fraction $f_{\rm RL}$ of radio-loud
quasars\footnote{Using the criterion $R_{\rm i} > 1$  as definition for radio loudness; see 
Ivezi\'c et al. (\cite{Ivezic02}).}, 
which is 0.22 in our sample, compared to 0.07 for the QCDR7 quasars. 
This could imply that independently of colour selection, unusual quasars are
detectable over a much longer time span at the FIRST level than normal quasars. 
Alternatively, these significant differences may be caused by the
SDSS having also targeted FIRST sources for spectroscopy when their optical
colours did not meet the quasar selection criteria. The FIRST target
flag is indeed set for 21\% of the quasars in our sample, compared to 5\% in the
whole quasar catalogue. 

To test whether the excess radio-detected unusual quasars were only discovered 
through FIRST targeting, we considered the numbers $N_{\rm F}$ of quasars with the 
FIRST target flag set and $N_{\rm C}$ of quasars with their colour target flag set.
We found $N_{\rm F}/N_{\rm C} = 0.33$ for the unusual quasars compared to
0.07 for the QCDR7. A similar result was found when we compared the number $N_{\rm Fs}$
of quasars selected solely by the FIRST selection (but not the colour selection)
to the number $N_{\rm Cs}$ of quasars without FIRST counterparts but 
selected solely based on their colours, namely $N_{\rm Fs}/N_{\rm Cs} = 0.22$ for
the unusual quasars and 0.01 for the QCDR7.
Moreover, when we binned the unusual quasars into intervals of $\chi^2$, 
it could be clearly seen that the fraction of quasars with
their FIRST target flag set increases with the mean peculiarity index 
(Tab.\,\ref{tab:chi2}). Both $N_{\rm F}/N_{\rm C}$ and $N_{\rm Fs}/N_{\rm Cs}$ 
rise strongly with  $\chi^2$ and reach values of $\sim 1$ for the
objects showing the strongest mean deviation from the SDSS quasar composite 
spectrum. We could identify no physical reason for such a trend and concluded therefore
that the increase in the fraction $f_{\rm RD}$ of quasars with FIRST
detections towards the more peculiar spectra is more likely a selection bias: 
a large fraction of the unusual quasars were not selected by the colour selection
criterion but were targeted by SDSS just because they had been detected
as FIRST radio sources. 

A similar trend, but weaker, was also observed
for the fraction $f_{\rm RL}$ of the radio-loud quasars. On the other hand, 
no indication of this trend is seen for the ratio $f_{\rm RL}/f_{\rm RD}$.
The fraction of radio-loud among the radio-detected quasars for our 
whole sample is 0.76, compared to 0.90 for the QCDR7 sample.  This means
that our selection criterion of peculiar spectra did not induce a
substantial bias towards quasars that are more active at radio 
frequencies than normal.

\begin{table}[hhh]
\caption{
Radio detection fraction for different spectral peculiarities. 
}
\begin{flushleft}
\begin{tabular}{lrccccc}
\hline\hline
 $\chi^2$  
 & $N$  
 & $N_{\rm F}/N_{\rm C}$ 
 & $N_{\rm Fs}/N_{\rm Cs}$
 & $f_{\rm RD}$
 & $f_{\rm RL}$ 
 & $f_{\rm RL}/f_{\rm RD}$ \\
\hline
$   0\ldots 5$   & 248  & 0.17  & 0.09 & 0.19 &  0.18 &  0.94 \\
$   5\ldots 20$  & 333  & 0.20  & 0.15 & 0.23 &  0.16 &  0.72 \\
$  20\ldots 50$  & 184  & 0.34  & 0.22 & 0.28 &  0.23 &  0.82 \\
$  50\ldots 100$ & 123  & 0.69  & 0.41 & 0.48 &  0.38 &  0.80 \\
$ 100\ldots 200$ &  61  & 0.73  & 0.62 & 0.48 &  0.39 &  0.83 \\
$> 200        $  &  42  & 1.16  & 0.91 & 0.55 &  0.31 &  0.57 \\
\hline
\end{tabular}
\end{flushleft}
\label{tab:chi2}
\end{table}

On the other hand, it is interesting to study whether our lower radio loudness fraction
is significant and how it depends on the type properties. For this purpose,
we converted the FIRST radio flux densities $F_{\rm 1.4}$ to the specific 
luminosities $L_{\rm 1.4}$ emitted at 1.4 GHz (restframe) via
\begin{equation}
\log L_{\rm 1.4} = 23.08 + 2\,\log\,D_{\rm L} + \log\,F_{\rm 1.4}-(1+\alpha)\,\log\,(1+z),
\end{equation}
where $L_{\rm 1.4}$ and $F_{\rm 1.4}$ are given in W\,Hz$^{-1}$ and mJy,
respectively, 
$D_{\rm L}$ is the luminosity distance in Gpc, and $\alpha$ is the spectral
index ($F_\nu \propto \nu^{\,\alpha}$), where we assumed that $\alpha = -0.5$.
The results are shown in Fig.\,\ref{fig:luminosity-z} for the quasars of each
type (crosses). The (normal) quasars from the QCDR7 are represented by the median
relation and the area populated by 80\% of the quasars closest to the median
on either side.

As discussed above for the absolute magnitudes, comparison samples
with the same $z$ distribution are needed to check whether the radio luminosities
in our sample differ from those of normal QCDR7 quasars.
Such comparison samples were constructed in the same way as for $M_{\rm i}$, with
the only exception that now only the subsamples of the radio-detected quasars
($F_{1.4} \ge 1$\,mJy) were considered. We again first performed the two-tailed
KS test with the null hypothesis $H0$, that there is no difference between the radio
luminosities of the unusual quasars of a given type and the comparison sample
from the QCDR7, and the alternative $H1$, that both groups have different
radio properties.
We found that $H0$ has to be rejected for all comparison samples of type A
and for 80\% of the comparison samples of type F. For the other types,
the fraction of rejections is 20\% or less.    
Since the fraction of radio-loud quasars tends to be smaller for the unusual
quasars, we also applied one-tailed tests with $H1$: the radio luminosities of
the unusual quasars are lower. We found that $H0$ has to be rejected in favour of
$H1$ for the unusual BAL quasars of type A ($p = 0.01$)
and the strong Fe-emitting quasars ($p = 0.04$) of type F. No such
firm conclusion can be drawn for the other types.   

The bottom row of Fig.\,\ref{fig:luminosity-z} shows the distribution of the
ratio $f = N(L_{1.4})/N_{\rm QC}(L_{1.4})$ where $N(L_{1.4})$ is the differential
radio luminosity distribution for the corresponding type with equidistant 
$\log L_{1.4}$ intervals and  $N_{\rm QC}(L_{1.4})$
is the distribution averaged over 10 comparison samples from QCDR7,
where both distributions are normalised.
The vertical bars indicate the propagation of the counting errors of $N(L_{1.4})$ and 
$N_{\rm QC}(L_{1.4})$, which were identified with the Poisson limits 
at a confidence level of 0.84\footnote{Corresponding to Gaussian statistics 1$\sigma$}.
To compute the upper and lower Poisson levels, the approximate equations given 
by Gehrels (\cite{Gehrels86}) were applied.
It can be clearly seen that strong BALs are not among the most radio-loud quasars.
The fraction of BAL quasars drops by a factor
of at least three between $\log\,L_{1.4} = 25$ and 27. This decline in the fraction
of BAL quasars with increasing radio power confirms previous results from Becker et al.
(\cite{Becker00}) for a smaller sample of BAL quasars and Shankar et al.
(\cite{Shankar08}) for a larger sample based on the catalogue of BAL quasars
from Trump et al. (\cite{Trump06}). Shankar et al. (\cite{Shankar08}) argue that
such a trend fits well within a simple geometric model. 
Together with the results in Tab.\,\ref{tab:chi2}, the increase in $f$ towards
lower $L_{1.4}$ for type A indicates that the real fraction of unusual 
BAL quasars might be considerably larger than in present samples, where these
quasars have instead been targeted mainly as FIRST radio sources.

Finally, we note that there is a similar trend of $f (\log\,L_{1.4})$ for types B and C.
This trend seems to also exist for type F but is not statistically significant.
 
\begin{figure*}[htbp]
\begin{tabbing}
\includegraphics[bb=30 00 520 780,scale=0.24,angle=270,clip]{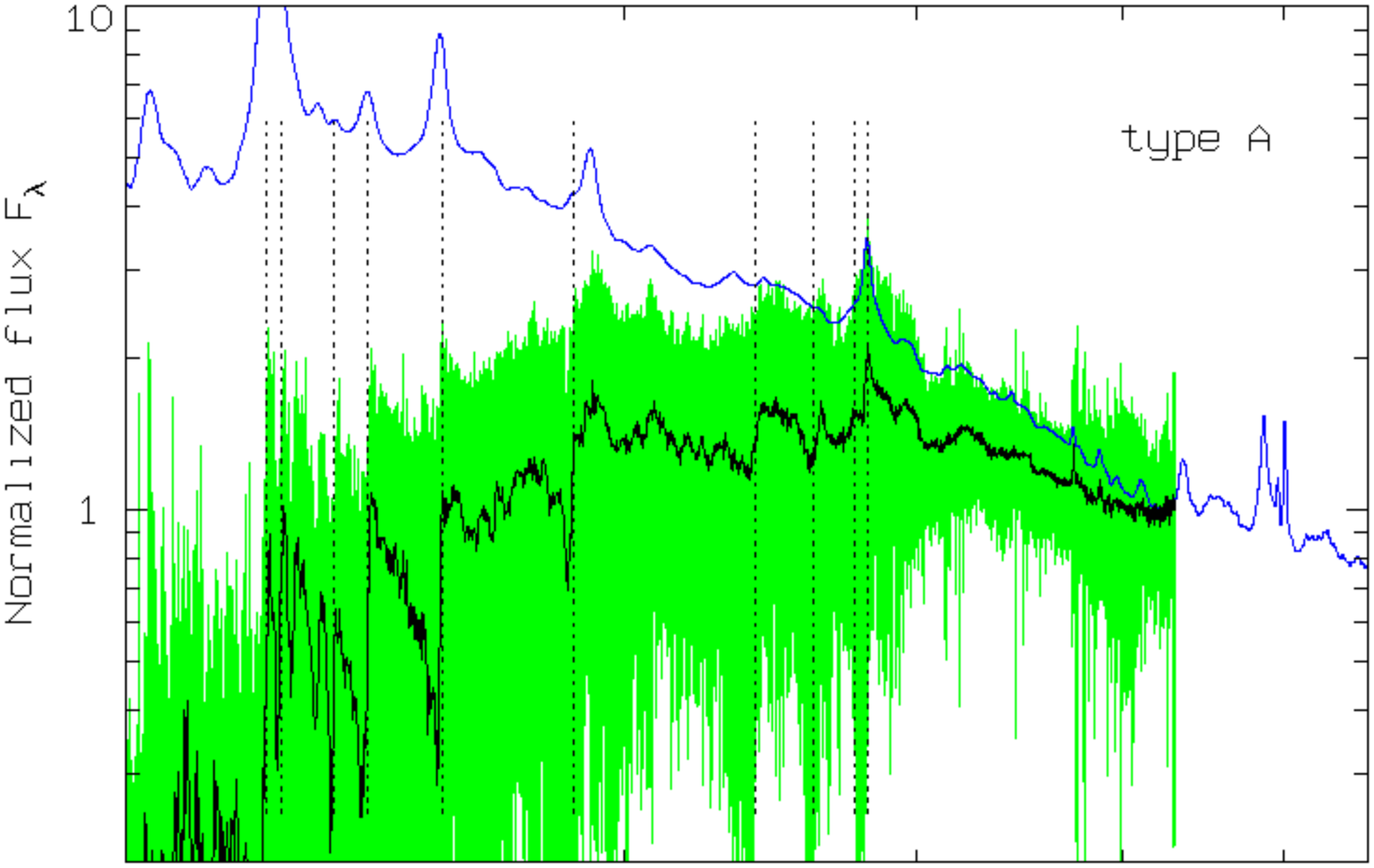}\hfill \=
\includegraphics[bb=30 80 520 780,scale=0.24,angle=270,clip]{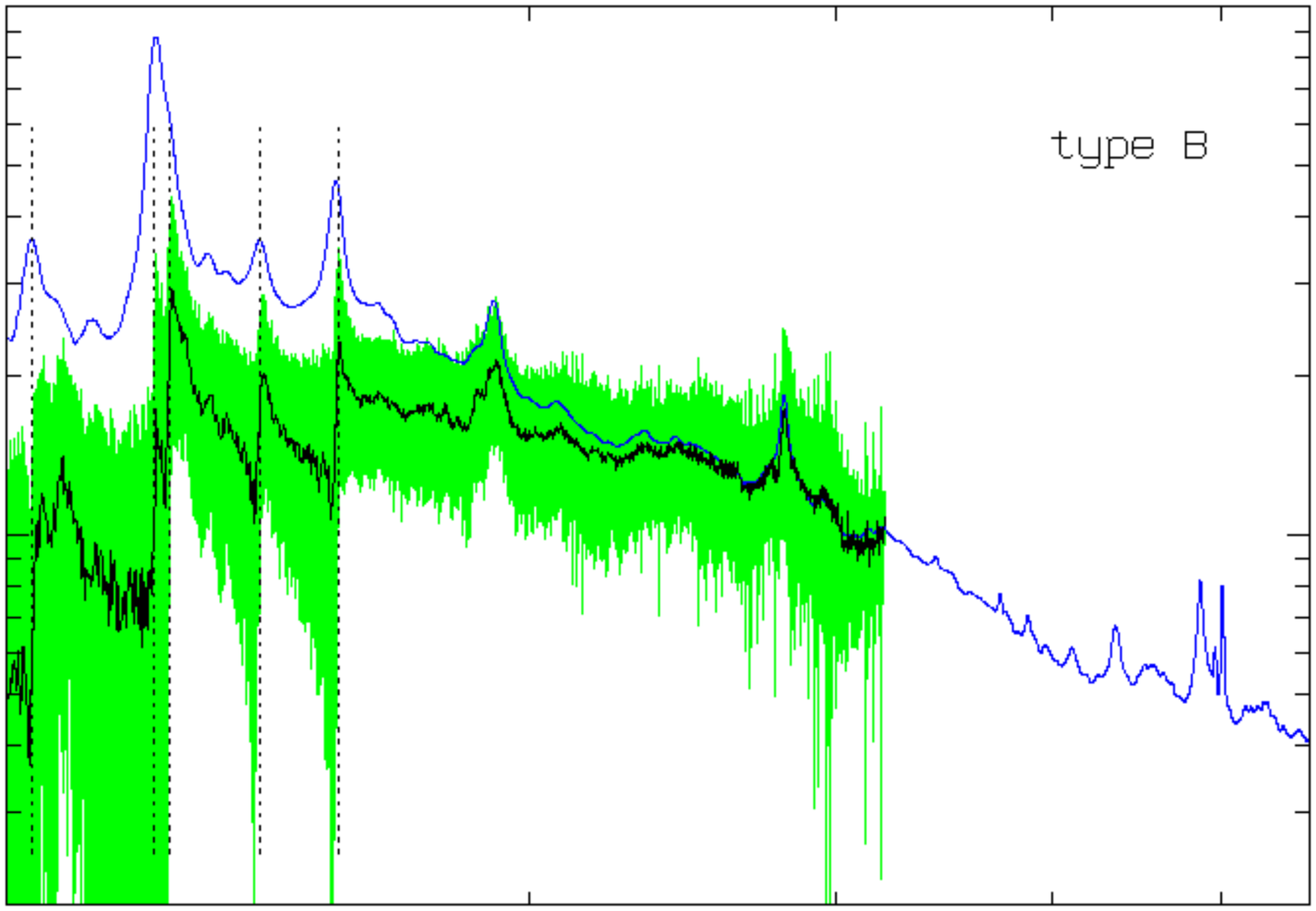}\hfill \=
\includegraphics[bb=30 80 520 780,scale=0.24,angle=270,clip]{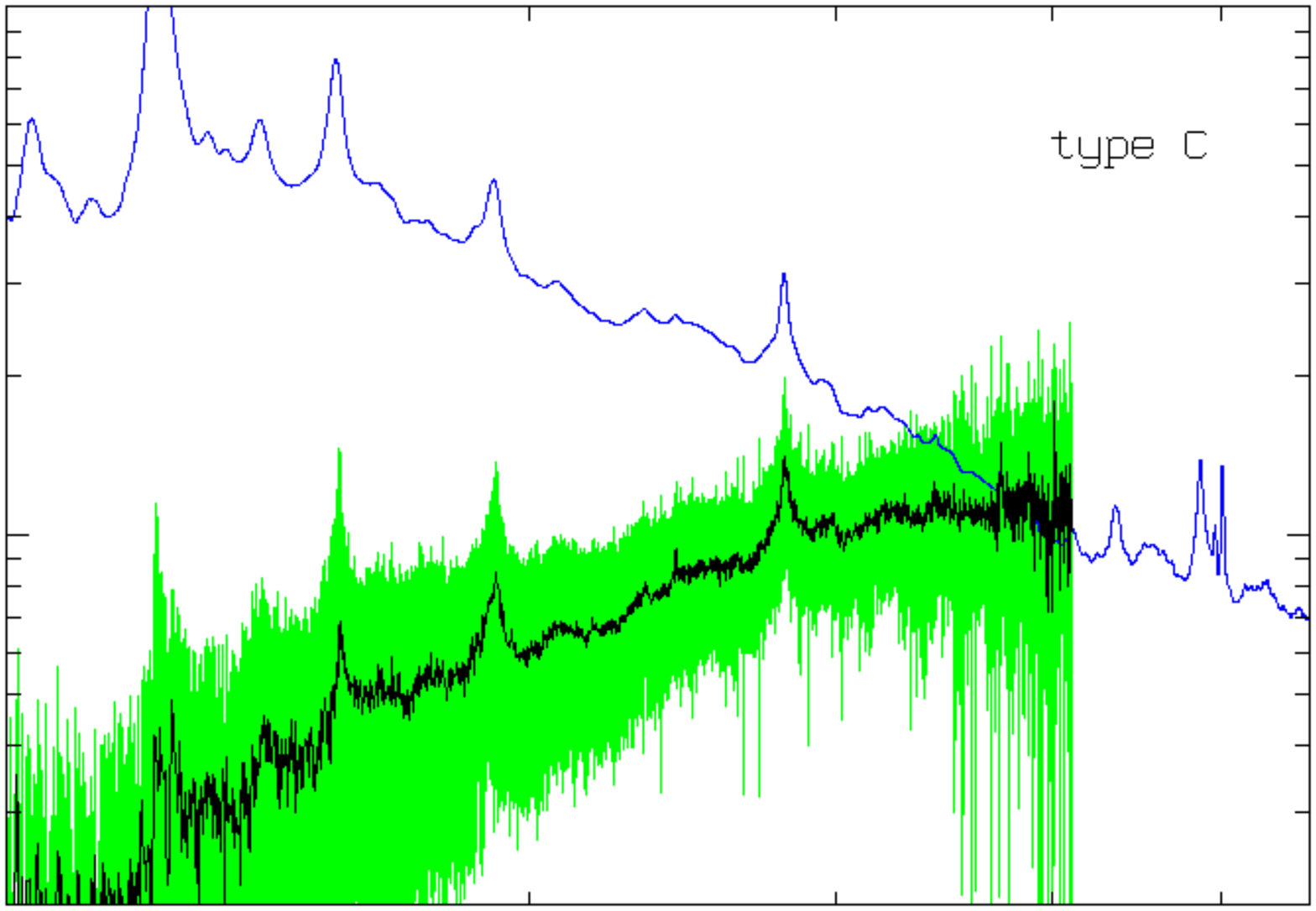}\hfill \\
\includegraphics[bb=50 00 580 780,scale=0.24,angle=270,clip]{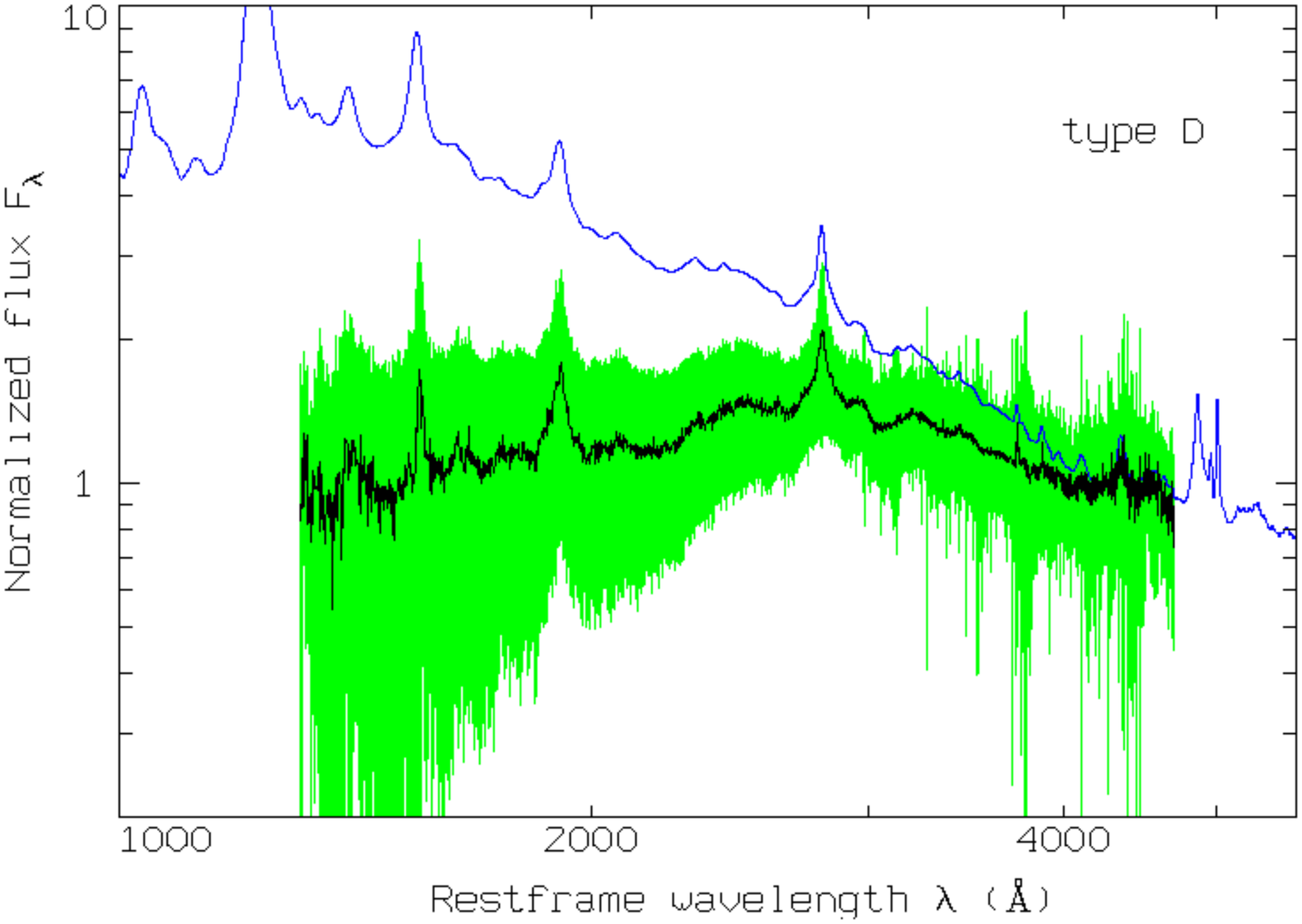}\hfill \=
\includegraphics[bb=50 80 580 780,scale=0.24,angle=270,clip]{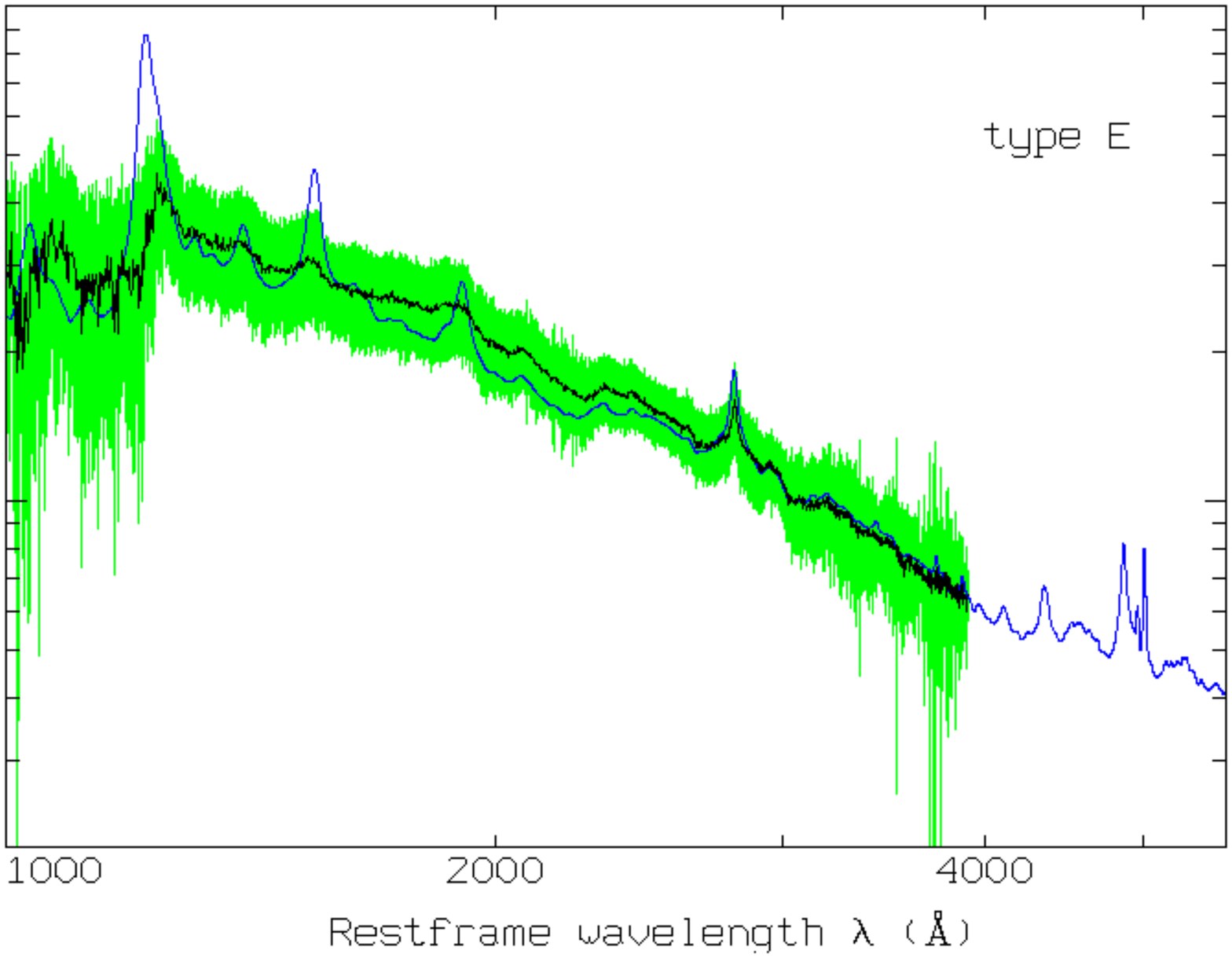}\hfill \=
\includegraphics[bb=50 80 580 780,scale=0.24,angle=270,clip]{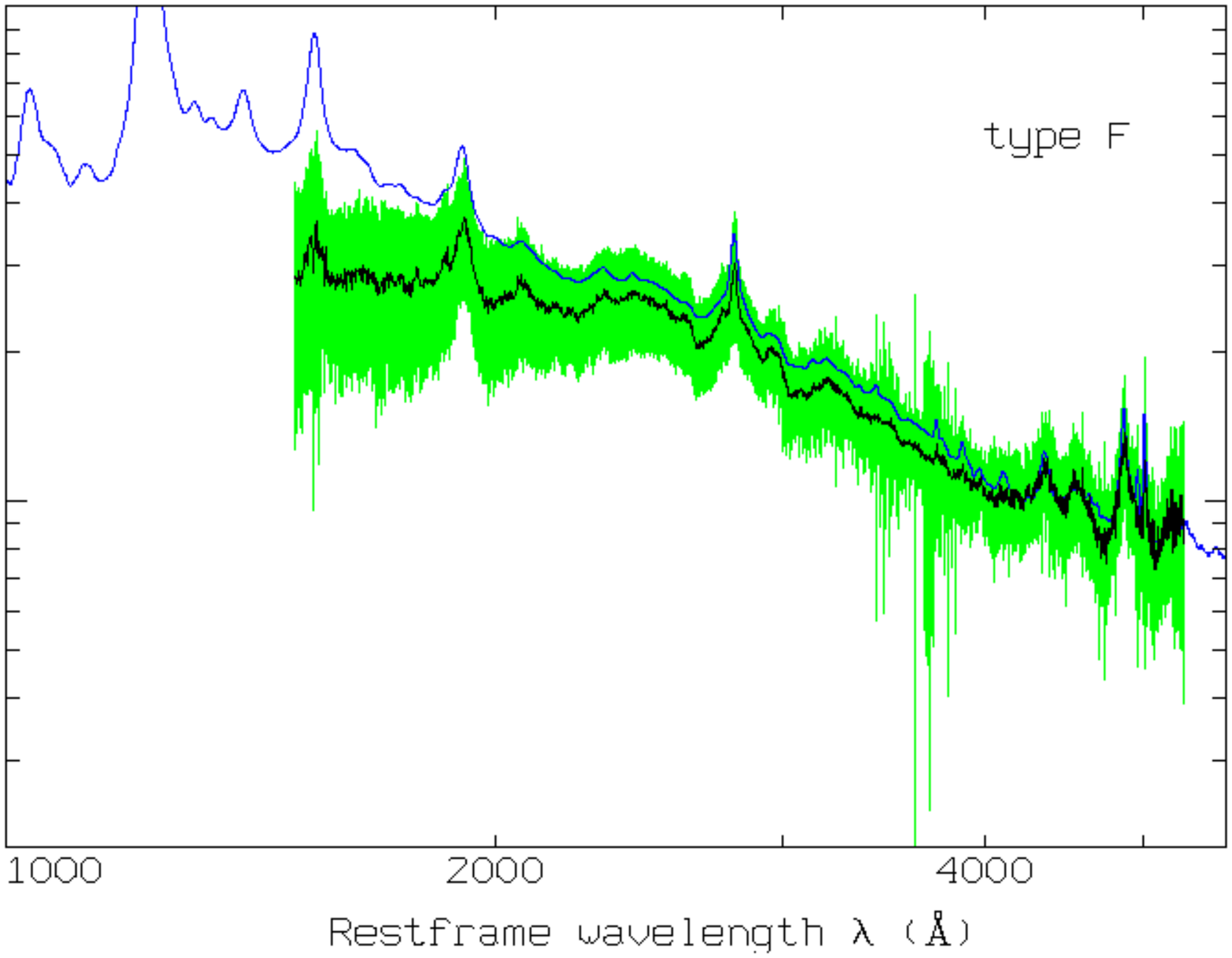}\hfill \\
\end{tabbing}
\caption{Arithmetic median composite spectra for types A to F.
For comparison the  SDSS quasar composite spectrum from 
VandenBerk et al. (\cite{VandenBerk01}) is shown,
arbitrarily normalized at the red end. The dashed vertical lines indicate the
strongest absorption lines for types A and B.}
\label{fig:median-composite}
\end{figure*}

\subsection{Optical colours}\label{subsec:colours}

Systematic trends of the SED can be illustrated
by average spectra but also by the distribution of the colour indices. 
We first discuss the colours. 

Since the colour indices of quasars vary with $z$, it is useful to study 
colour-redshift diagrams rather than colour-colour diagrams. Such colour-$z$ 
diagrams are displayed in Fig.\,\ref{fig:colour-indices} for $u-g, g-r$, $r-i$,
and $r-K$, where $u,g,r,i$ are the SDSS magnitudes from the QCDR7 corrected for Galactic
foreground extinction and $K$ is the K band magnitude from the 2MASS survey
(Skrutskie et al. \cite{skrutskie06}). About 60\% of our quasars have 2MASS measurements.
We do not see significant differences between the subsamples with
or without 2MASS counterparts. As in Fig.\,\ref{fig:luminosity-z}, the 
quasars from the QCDR7 are plotted for comparison (median and 1$\sigma$
deviation).

In general, BAL quasars are known to be significantly redder than those without
BALs, and LoBAL quasars are even redder 
(Sprayberry \& Foltz \cite{Sprayberry92}; 
Reichard et al. \cite{Reichard03};
Gibson et al. \cite{Gibson09}).
It can be clearly seen  in Fig.\,\ref{fig:colour-indices} that types A to D tend to have
redder colours than the whole SDSS quasar population with strongest deviations
for type A. The colours of type C and D quasars are by definition redder than normal. 
For type E, the colour-redshift diagrams resemble those of the whole quasar population.
The diagrams for type F quasars indicate a moderate reddening of $u-g$ and $g-r$, while
$r-i$ and $r-K$ are approximately the same as for normal quasars.  

As shown in Sect.\,\ref{subsec:redshifts}, types A to D have
the same absolute magnitudes as normal quasars. However, if their red colours are 
due to intrinsic reddening by dust and/or gas, the intrinsic luminosities of these 
quasars appear to be higher than normal. In Sect.\,\ref{subsec:reddening}, we
estimate the intrinsic absorption and perform the KS test for the 
correspondingly corrected $M_{\rm i}$.

\subsection{Composite spectra}\label{subsec:composites}

Composite spectra are particularly useful for determining the average spectral 
properties of quasar samples  (Francis et al. \cite{Francis91}; VandenBerk et al.
\cite{VandenBerk01}; Richards et al. \cite{Richards03}). The general 
procedure of constructing quasar composites is the following:
after the redshifts are determined, the individual
(foreground) extinction-corrected spectra are rebinned to a common wavelength
scale in their restframe, and the quasars are sorted by redshift.
Then, generating composites requires essentially two steps: 
{\it (a)} normalisation of the spectra and
{\it (b)} combining the normalised spectra. Both steps can be performed in
different ways. 

\begin{figure*}[htbp]
\begin{tabbing}
\includegraphics[bb=30 00 520 780,scale=0.24,angle=270,clip]{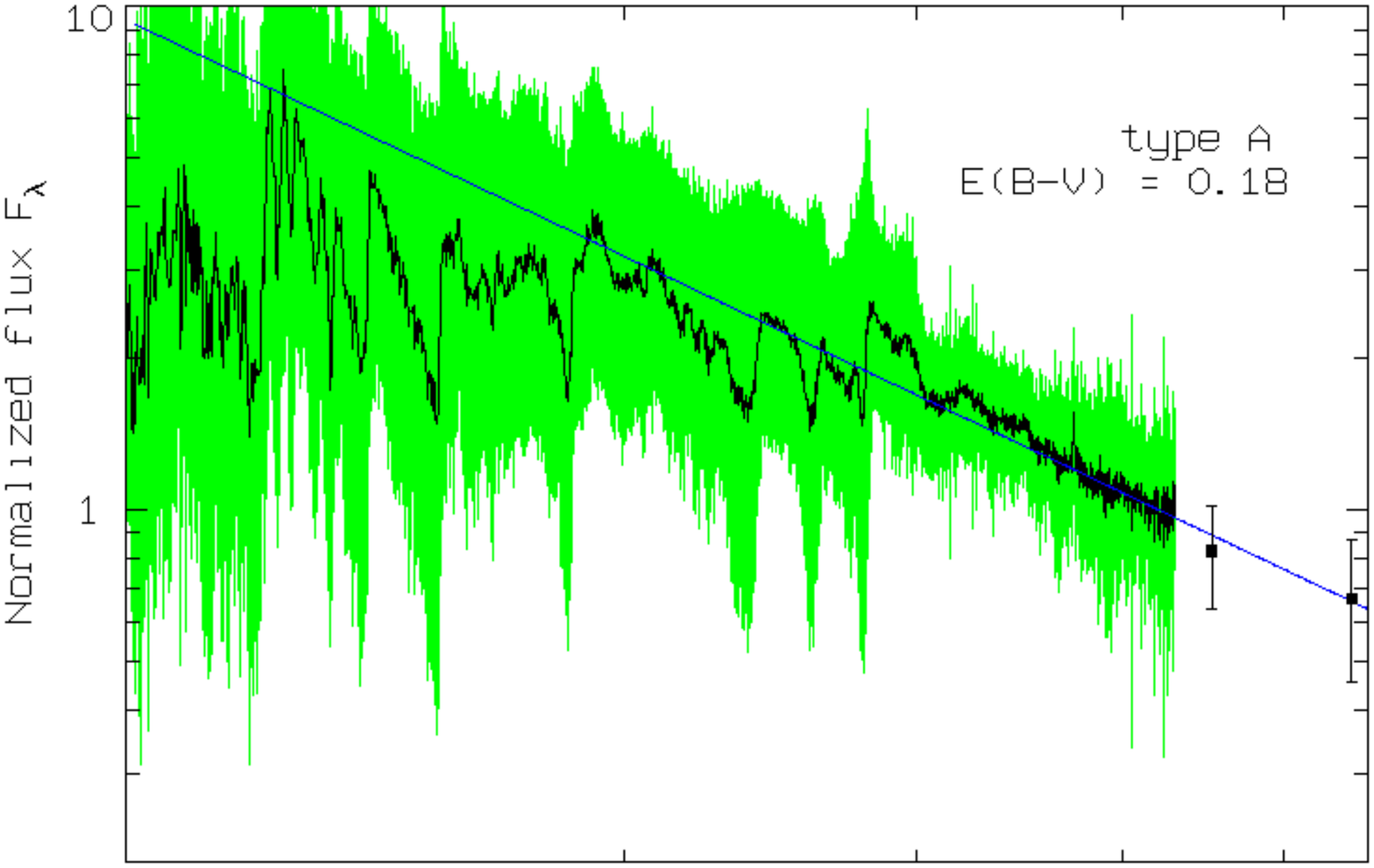}\hfill \=
\includegraphics[bb=30 80 520 780,scale=0.24,angle=270,clip]{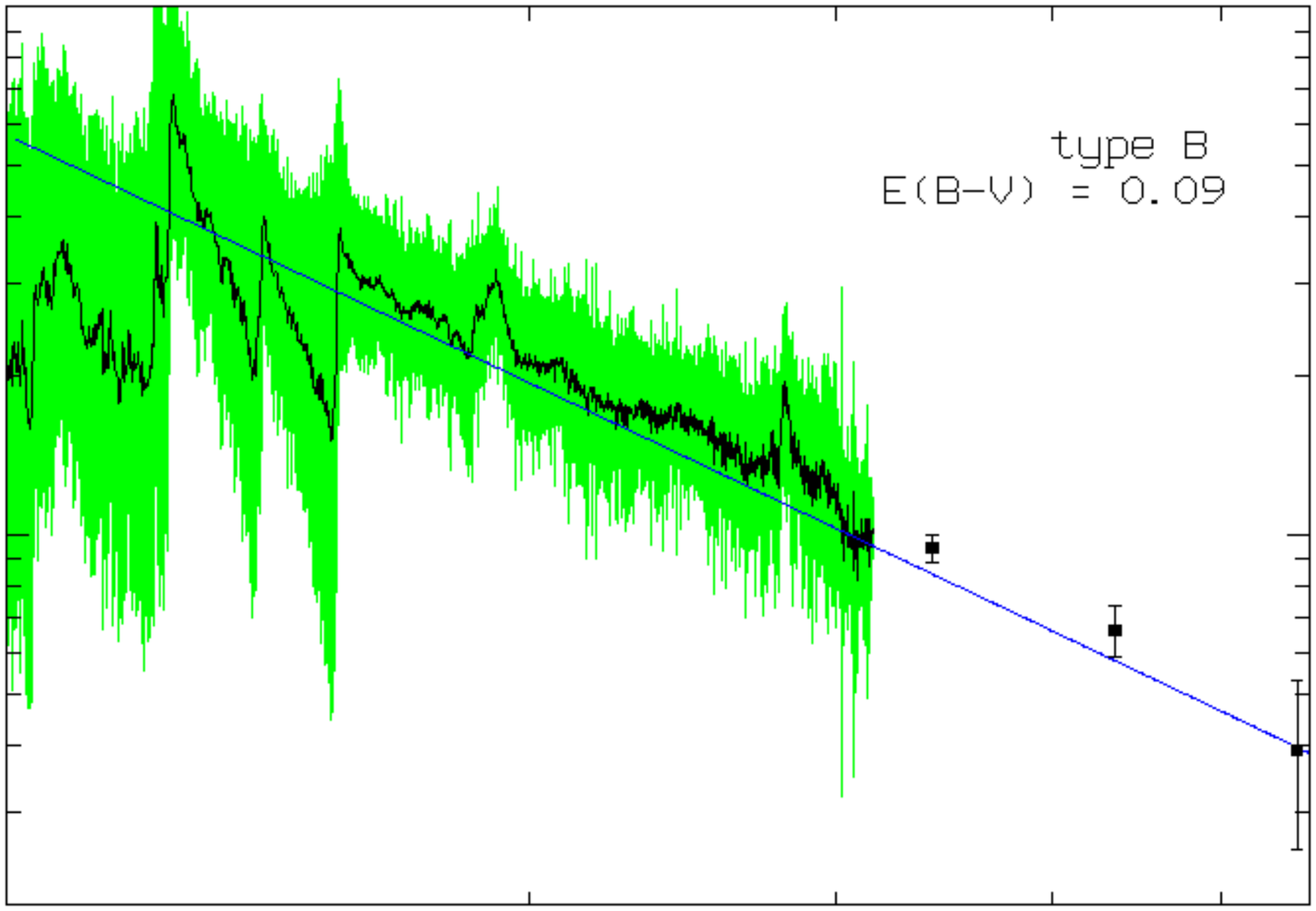}\hfill \=
\includegraphics[bb=30 80 520 780,scale=0.24,angle=270,clip]{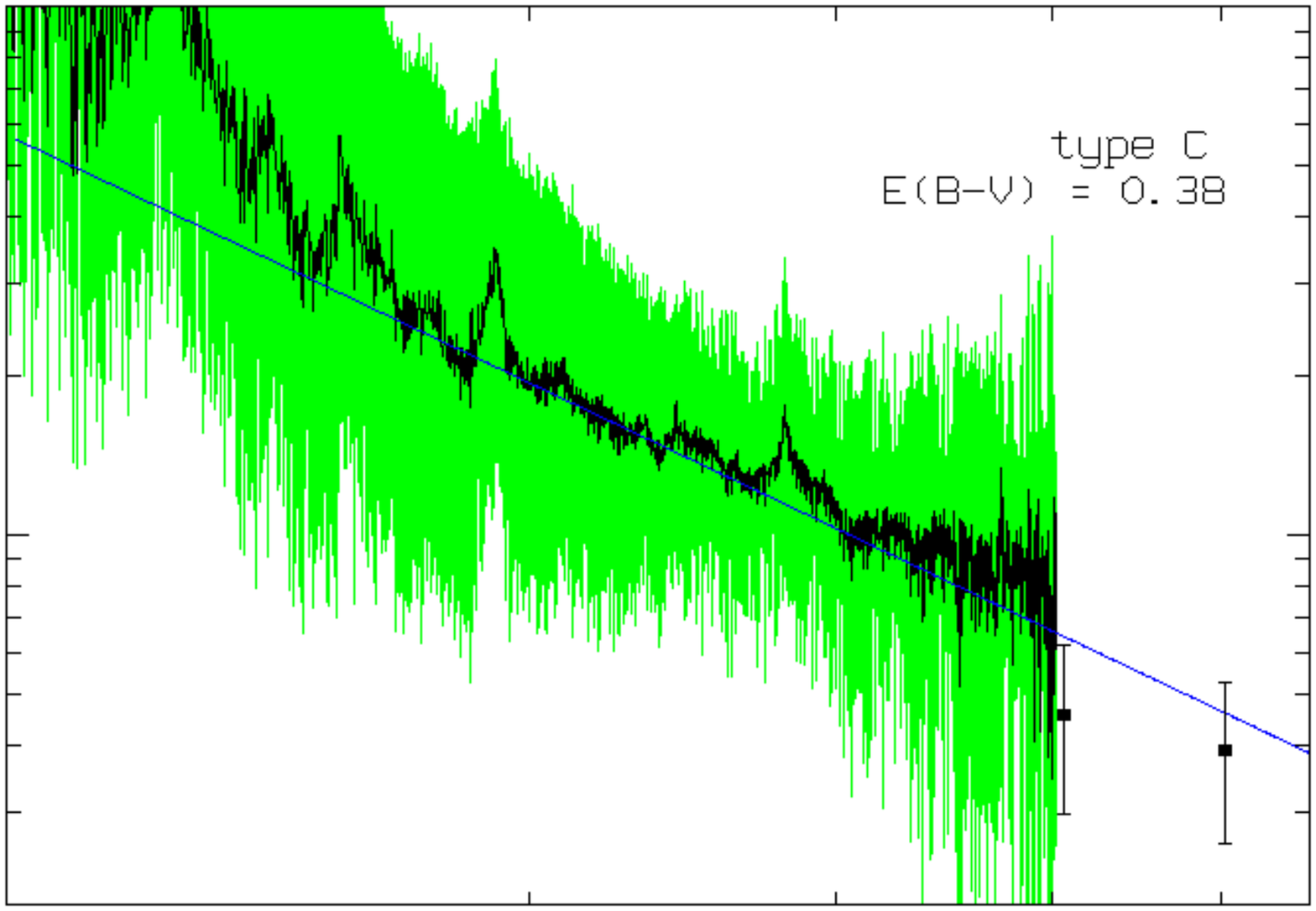}\hfill \\
\includegraphics[bb=50 00 580 780,scale=0.24,angle=270,clip]{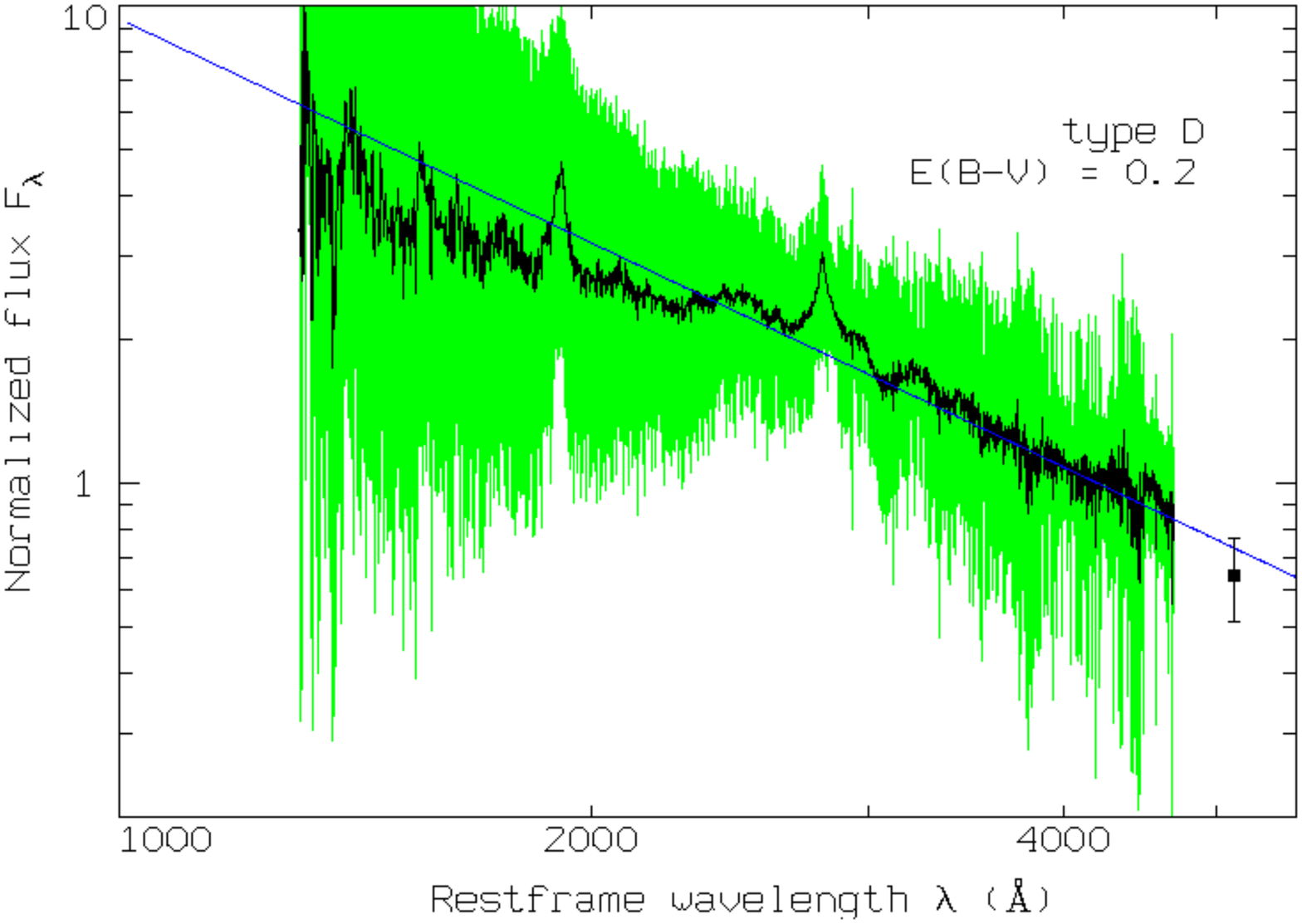}\hfill \=
\includegraphics[bb=50 80 580 780,scale=0.24,angle=270,clip]{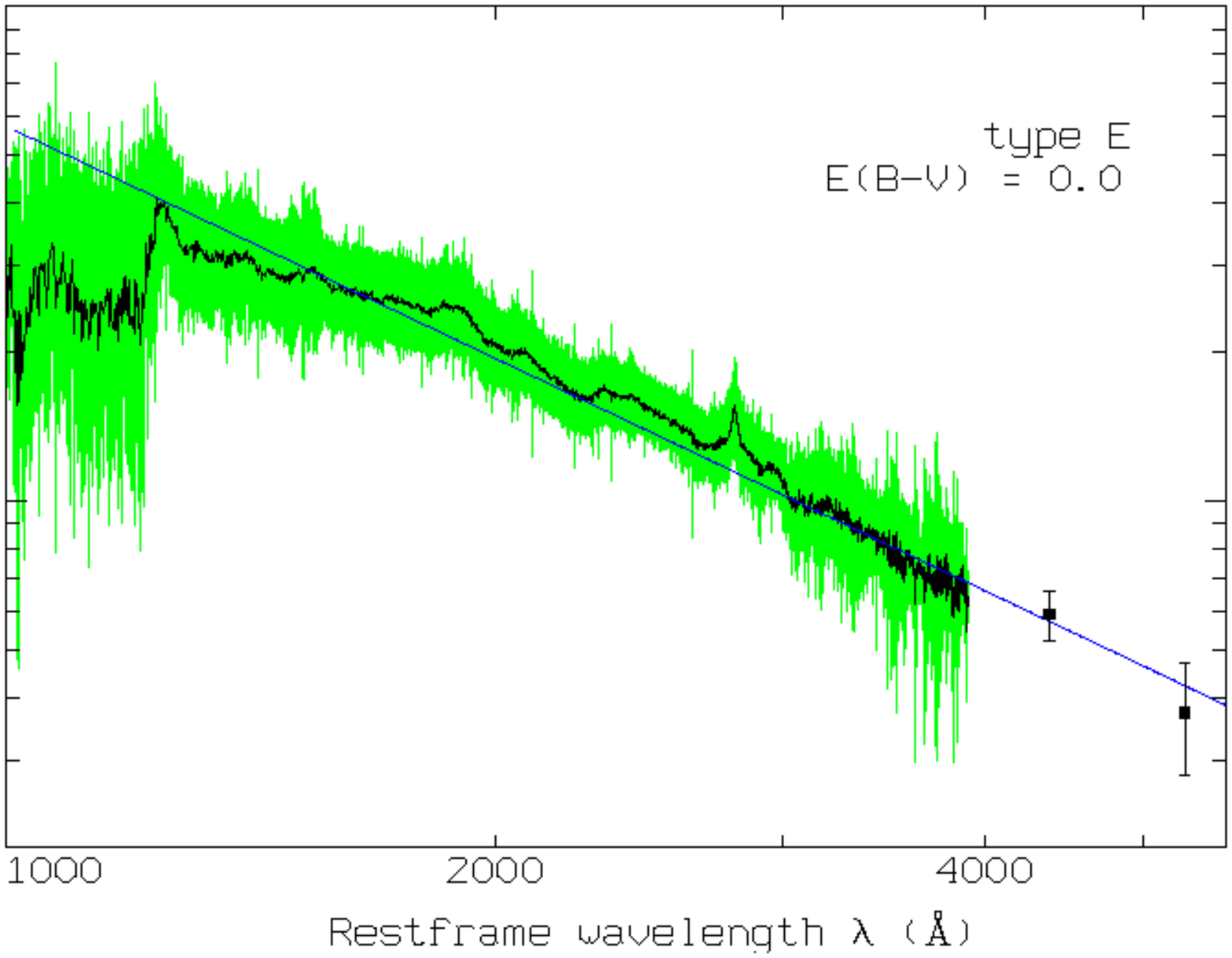}\hfill \=
\includegraphics[bb=50 80 580 780,scale=0.24,angle=270,clip]{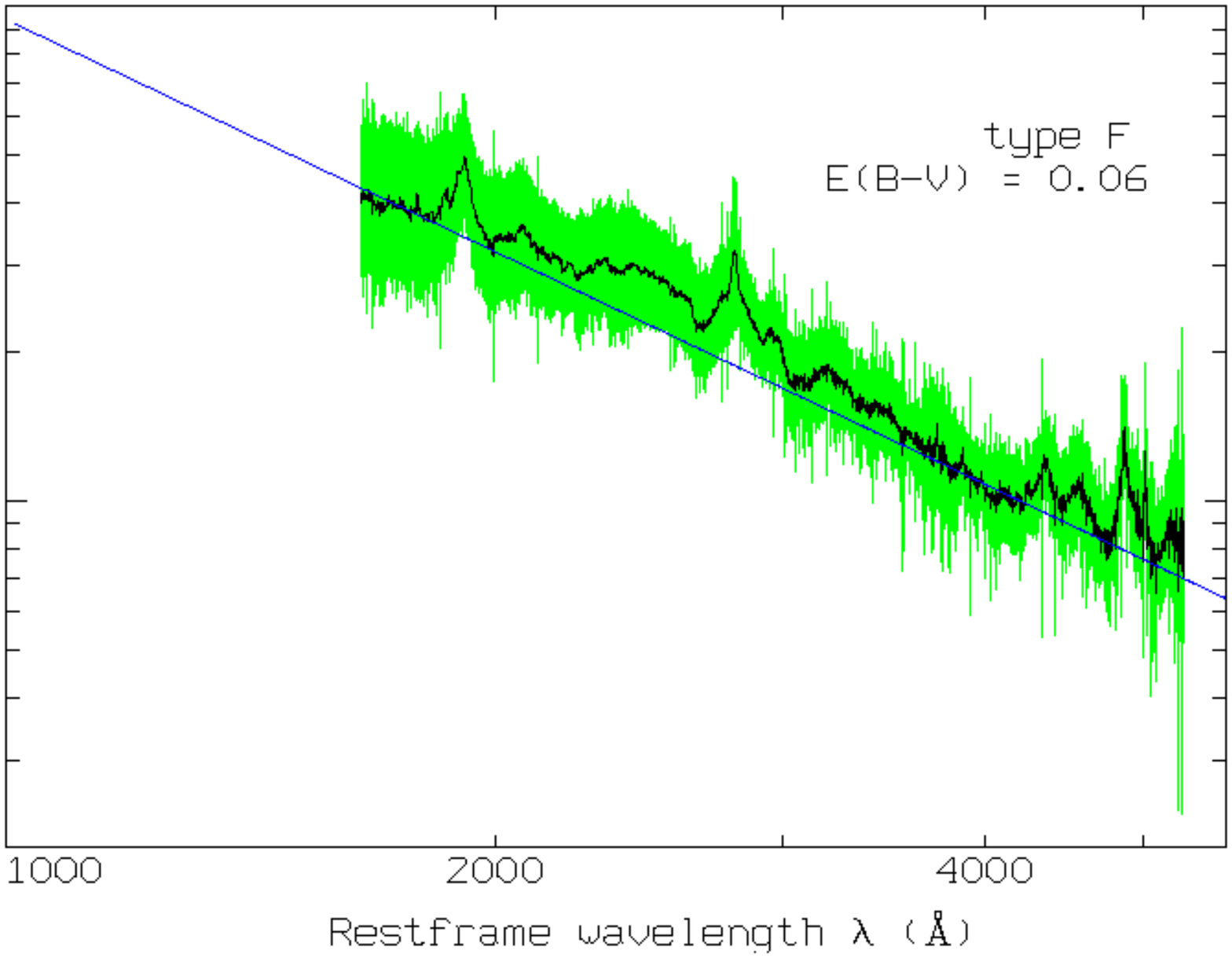}\hfill \\
\end{tabbing}
\caption{Geometric mean composite spectra corrected for intrinsic reddening. The SDSS spectra were
combined with JHK fluxes form 2MASS (squares with error bars). For comparison, the power-law 
continuum fit to the geometric mean SDSS composite from VandenBerk et al. (\cite{VandenBerk01}) is shown.}
\label{fig:geometric-mean-composite}
\end{figure*}

The normalisation is not entirely trivial because the spectra cover different
rest-frame wavelength intervals. To create global composite spectra of normal SDSS
quasars, Vanden Berk et al. (\cite{VandenBerk01}) started with the arbitrarily scaled
spectrum of the lowest redshift quasar (with measured [\ion{O}{iii}]\,$\lambda$5007
line emission), and scaled each subsequent spectrum by the overlap of the preceding
average spectrum (i.e., by its mean flux density in the wavelength interval of the overlap).
The same procedure was applied in our previous study (Meusinger et al. \cite{Meusinger11}).
Here, however, we followed a slightly different approach. The main reason is that,
for our present sample, the flux density of the overlapping interval can be strongly
modified by the spectral peculiarities for which we search. Hence, we decided to restrict
the normalisation to a few, relatively narrow continuum windows. Though there is
no ideal continuum window where a suppression by BALs
and/or the contribution from emission lines can be completely ignored, usable
pseudo-windows exist however at 3540-3600\AA, 3030-3090\AA\ 
(Tsuzuki et al. \cite{Tsuzuki06}), and 2000-2300\AA\
(Sameshima et al. \cite{Sameshima11}). We started by scaling the lowest-redshift spectra.
If spectra existed with $z<1.5$, they were arbitrarily normalised to an average flux density
$F_\lambda = 1$ over the wavelength interval $\lambda =$ 3540-3600\AA. The thus normalised
spectra were combined to a first composite $\vec{C}_1$. Next, the spectra in the redshift
range $z=$1.5-1.94 were scaled to the flux density of $\vec{C}_1$ at $\lambda =$
3030-3090\AA\ and the normalised spectra were co-added together with the normalized spectra
from the previous step to create a second composite $\vec{C}_2$. In the same way, the continuum
window at $\lambda =$ 2200-2230\AA\ was used to extend the composite $\vec{C}_3$ to
$z=$3. To include quasars at higher redshift, we introduced another
pseudo-continuum window at $\lambda =$ 1650-1700\AA\ where the spectra with $z>3$ were 
scaled to the flux density of $\vec{C}_3$.

Combining the normalised spectra to a composite requires choosing between mean
or median and arithmetic or geometric averaging. Following Vanden Berk et al.
(\cite{VandenBerk01}), we used combining techniques to create {\it (a)} the arithmetic median,
which preserves the relative fluxes along the spectra, and {\it (b)} the geometric mean spectrum,
which preserves the global shape of a power-law continuum. The arithmetic median spectrum was
computed in the same way as in our previous study (Meusinger et al. \cite{Meusinger11}):
All normalised (restframe) spectra $\vec{S}_i$ of a given type were inserted into a
2D image, one spectrum per row with 1 \AA\ binning on the horizontal axis. The arithmetic
median was then computed by averaging over each column using the procedure average/image
from the ESO-MIDAS package\footnote{http://www.eso.org/sci/software/esomidas} with the
median average option. A similar approach to the previous section
was used for the geometric mean, where however the 2D image was build from the
$\log\,\vec{S}_{i}$ instead of $\vec{S}_{i}$
to compute the mean value $\langle \log \vec{S} \rangle_{i}$.  
The geometric mean spectrum $\vec{S}_{\rm gm}$ and the geometric standard deviation
$\sigma_{\rm g}$ are given by
\begin{equation}\label{eq:geom-mean}
\vec{S}_{\rm gm} = \Bigg[\prod_{i=1}^{N} \vec{S}_{i}\Bigg]^{1/N}
                 = 10^{\, \langle \log \vec{S}_{i} \rangle}, \quad
\sigma_{\rm g}   = 10^{\, \sqrt{\langle [\log (\vec{S}_{i}/\vec{S}_{\rm gm})]^2\rangle}}.
\end{equation}
The resulting spectra are shown on a log-log scale in 
Fig.\,\ref{fig:median-composite} for the arithmetic median and in 
Fig.\,\ref{fig:geometric-mean-composite} for the geometric mean (extinction-corrected, 
see below), respectively, except for type M where a composite makes no 
sense as this subsample is too small and very heterogeneous.
The spectra of different types have different normalisations owing to their
different wavelength coverages. The $1\sigma$ variation is shown
by the shaded area. The variation in the flux density across the spectrum is
expected to reflect the spectrum-to-spectrum differences caused by differences in the
continuum shapes and in the properties of the emission and/or absorption lines
(see Vanden Berk et al. \cite{VandenBerk01}).

The average rest-frame spectra differ significantly from type to type, as well as
from the Vanden Berk composite of ``normal'' SDSS quasars. 
The defining spectral features are clearly indicated in the median spectra.
The median spectrum of the unusual BAL quasars of type A is remarkably
dominated by the typical absorption troughs of FeLoBAL quasars (see Hall et al.
\cite{Hall02}; their table 1), of which the strongest are indicated by the dashed
vertical lines:
Ly\,$\alpha$, 
\ion{N}{v} $\lambda$1334,
\ion{C}{ii} $\lambda$1334,
\ion{Si}{iv} $\lambda$1398, 
\ion{C}{iv} $\lambda$1550, 
\ion{Al}{iii} $\lambda$1860, 
\ion{Fe}{ii} $\lambda$2400, 
\ion{Fe}{ii} $\lambda$2600, 
\ion{Fe}{ii} $\lambda$2750, and
\ion{Mg}{ii} $\lambda$2800\footnote{Most of these lines are actually blends of two
or several lines. However, in many cases throughout this paper it is adequate
to treat these blends as single lines.}. 
On the other hand, the spectrum-to-spectrum variations for type A are huge. 
This reflects the particularly high degree of diversity among these spectra.
As expected, the dominating spectral features in the median spectrum for type B 
are the absorption troughs from the high ionisation lines
\ion{O}{iv} $\lambda$1033, 
Ly\,$\alpha$, 
\ion{N}{iv} $\lambda$1240,
\ion{Si}{iv} $\lambda$1398, and 
\ion{C}{iv} $\lambda$1550.
There are no stringent indications for BALs in the median spectra of the other 
types, though these features are present in the spectra of some members of these types
(Sect.\,\ref{subsec:classification}).

\subsection{Intrinsic reddening and corrected absolute magnitudes}\label{subsec:reddening}

The general trend of BAL quasars to show stronger UV reddening than non-BAL quasars and of 
LoBAL quasars to be even redder (see Sect.\,\ref{subsec:colours}) is clearly 
indicated by the type composites in Fig.\,\ref{fig:median-composite}. 
The type F composite indicates significant reddening shortwards of
$\ion{Mg}{ii}$. For type E, we found that quasars with $z \la 2$ tend to have steeper continua
than normal quasars (e.g., Fig.\,\ref{fig:examples_groups}), but that the spectra are
slightly shallower at higher $z$. This yields an average spectrum for type E with a
continuum that reasonably fits the SDSS composite of normal quasars.
The quasars of the types C and D are by definition redder than normal quasars.

We assumed that the red continua are due to reddening by dust related to the
quasar and/or the host galaxy and estimated the mean intrinsic reddening 
$\langle E^{\rm (int)}_{\rm B-V} \rangle$ of each type 
from its geometric mean spectrum 
\begin{equation}\label{int-ext}
\log \vec{S}_{\rm gm,\, 0} = \log \vec{S}_{\rm gm} 
+ 0.4\,\langle E^{\rm (int)}_{B-V} \rangle [R_{V} + Q_{\lambda}],
\end{equation}
where $\vec{S}_{\rm gm}$ is the observed geometric mean from
Eq.\,(\ref{eq:geom-mean}) and  $\vec{S}_{\rm gm,\, 0}$ is the 
emitted geometric mean spectrum before dust extinction.  
As has been shown in several previous studies 
(e.g., Reichard et al. \cite{Reichard03}), SMC dust provides 
acceptable fits for the intrinsic reddening in quasars.
Hence, we adopted the SMC extinction curve 
with $R_{V}=2.93$ and $Q(\lambda) = E_{\lambda-V}/E_{B-V}$ from Pei (\cite{Pei92}) .
We assumed that $\vec{S}_{\rm gm,\, 0}$ can be identified with the geometric
mean spectrum of usual quasars. Vanden Berk et al. (\cite{VandenBerk01}) found that a
single power-law adequately fits the continuum between Ly\,$\alpha$ and H$\beta$ with
$\alpha_\nu = -0.44$.
We used $\langle E^{\rm (int)}_{B-V}\rangle$ in Eq.\,(\ref{int-ext}) as a free
parameter to fit the continuum of the corrected spectrum of each type to the 
power-law with $\alpha_\nu = -0.44$ and identified the best-fit value with the mean
reddening parameter of this type.

This approach is hampered by the UV ($\lambda \la 3000$\AA)
continua being strongly affected by the unusual absorption and/or
emission features, perhaps with the exception of type E. 
On the other side, the composite spectra do not cover a wide wavelength 
interval longwards of 3000\AA, except for type F.
To stretch the useful wavelength interval for the continuum fit,
we combined the SDSS spectra with the fluxes derived from JHK magnitudes given
in the 2MASS catalogue (Skrutzki et al. \cite{skrutskie06}). The percentage of
quasars with 2MASS magnitudes varies between 45\% for type B and 70\% for
type A with a mean value of 58\%. For each object identified in 2MASS, the JHK
fluxes were normalised with the same normalisation factors used for the
corresponding spectrum. The fluxes derived from the 2MASS magnitudes generally 
agree with the extrapolation of the composite spectra towards longer 
wavelengths. 

The extinction-corrected geometric mean spectra are shown in 
Fig.\,\ref{fig:geometric-mean-composite}. The mean reddening parameters are
given in Tab.\,\ref{tab:groups}. The values are larger than those found
in previous studies 
(e.g., $E_{B-V}$ = 0.08 (0.02) for LoBALs (HiBALs) according to
Reichard et al. \cite{Reichard03}),  
which is most likely due to differences between the quasar samples.
The contributions from host galaxies in the optical are 
obviously suppressed by the strong fluxes from the bright quasars. 

As can be seen in Fig.\,\ref{fig:geometric-mean-composite}, 
the rest-frame optical colours of the unusual quasar types are 
essentially the same as for the SDSS quasar composite (for a similar 
result see , e.g., Zhang et al. \cite{Zhang10}), while they tend to be redder in
the UV. In particular, the UV flux of the unusual BAL quasars of 
type A is substantially lower than the normal BAL quasars
of type B. This is likely caused by quasars for which the UV flux
is strongly suppressed by BAL troughs and/or overlapping troughs (Sect.\,\ref{so}). 
Owing to these features, the fit to the type A composite is particularly uncertain.
Types B and F are well-fitted by the SDSS continuum.
No reasonable fit can be achieved for type D, where the slope of
the dereddened continuum becomes shallower shortwards of \ion{Mg}{ii}.
Since a substantial fraction of red quasars also show BALs 
(Sect.\,\ref{subsec:classification}), the combined effect of broad-line
absorption may have some influence on the average continuum, but there are
no strong indications of BALs in the type D composite spectrum. An extinction
curve steeper than the SMC curve would be another explanation.
Finally, the type E geometric mean spectrum fits
the SDSS composite at longer wavelengths without dereddening but 
flattens approximately below the \ion{C}{iv} line. This may be an indication 
for differences between the weak-line quasars of high and low redshift. 

In the next step, we used the mean reddening parameters from Tab.\,\ref{tab:groups} 
to estimate the absolute magnitudes statistically corrected for intrinsic dust 
extinction adopting the approximation
\begin{equation}\label{corr-int-ext}
\langle M_{\rm i}^{\rm (cor)}\rangle 
\approx \langle M_{\rm i}\rangle - 
\langle E_{B-V}^{\rm (int)}\rangle\cdot[R_{V}-\langle Q(\lambda_{\rm i, eff}^{\rm (rf)})\rangle],
\end{equation} 
where the angular brackets symbolise the ensemble average and
$\lambda_{\rm i, eff}^{\rm (rf)} = \lambda_{\rm i, eff}/(1+z)$ is the 
rest-frame wavelength corresponding to the effective wavelength of the i band with
$\lambda_{\rm i, eff} = 7461$\AA\ (Schneider et al. \cite{Schneider10}).
We again adopted the SMC extinction curve from Pei (\cite{Pei92}). As can be seen
from the results listed in Tab.\,\ref{tab:groups},
this correction is significant. The one-sided KS (Sect.\,\ref{subsec:redshifts}) 
applied to the corrected $M_{\rm i}^{\rm (cor)}$ inferred that all six quasar types A to F
are on average more luminous then the comparison samples of normal quasars. 
The finding of the weaker \ion{C}{iv} emission of BAL quasars compared to that of
non-BAL quasars (Gibson et al. \cite{Gibson09}) thus appears as a manifestation of the
Baldwin-effect (Baldwin \cite{Baldwin77}; Osmer et al. \cite{Osmer94}) relation between 
the \ion{C}{iv} emission line strength and the UV luminosity.

\subsection{Intervening absorbers}\label{subsec:foreground}

Finally, we checked by eye the spectra of all quasars for narrow
absorption lines distinct from the quasar lines and at lower $z$ than the 
quasar. One of the easiest species to detect is the \ion{Mg}{ii} 
$\lambda\lambda$2796,2804 doublet, which is frequently in combination with   
\ion{Fe}{ii} $\lambda\lambda$2587,2600,
\ion{Fe}{ii} $\lambda\lambda$2374,2382,
\ion{Fe}{ii} $\lambda$2344, and
\ion{Mg}{i} $\lambda$2853.  
While the origin of the absorbing gas remains unclear, a direct association with
galaxies is suggested by several studies (e.g., Bahcall \& Spitzer
\cite{Bahcall69};
Steidel et al. \cite{Steidel97};
Bowen \& Chelouche \cite{Bowen11}; 
Kacprzak et al. \cite{Kacprzak11}). 
Dust embedded in foreground absorber systems along the line of sight is expected to
absorb and scatter the UV photons from the quasar and redden the quasar spectrum.  
Detection of the quasar reddening related to absorber systems has been reported
e.g. by York et al. (\cite{York06}) and M\'enard et al. (\cite{Menard08}) for
\ion{Mg}{ii}, by Wild et al. (\cite{Wild06}) for \ion{Ca}{ii}, and by
Vladilo et al. (\cite{Vladilo08}) for damped Ly$\alpha$ absorber systems. 
While the extinction curves of the absorbers are generally considered to be 
consistent with that of the SMC, there are probably also cases of Milky Way-like 
dust extinction that can strongly affect the quasar spectrum in a narrower
wavelength interval owing to the broad extinction bump around 2175\AA\
(Jiang et al. \cite{Jiang11}).

\begin{table}[hhh]
\caption{
Unusual quasars with structures close to the sightline
on the SDSS images.
}
\begin{flushleft}
\begin{tabular}{ccclrc}
\hline\hline
   SDSS J            & $z$& $T$ & Object$^{\, a}$     & AL$^{\, b}$& Ref.\\
\hline
$024230.65-000029.7$ & 2.486& B & NGC\,1068      & no   & \\
$075518.31+245432.3$ & 1.335& C & gal            & no   & \\
$081959.80+535624.2$ & 2.235& C & gal            & yes  & 1\\
$090334.94+502819.3$ & 3.579& C & gal group      & yes  & 2\\
$103410.26+344200.7$ & 2.503& B & ?              & yes  & \\
$103443.66+411120.4$ & 2.064& F & gal            & no   & \\
$104513.86+171952.2$ & 3.679& E & gal            & ?    & \\
$112851.83+062315.3$ & 1.503& A & gal            & ?    & \\
$135048.67+433406.2$ & 1.765& E & gal            & no   & \\
$143445.95+422516.7$ & 1.407& F & gal            & no   & \\
$144002.24+371058.5$ & 1.409& A & gal            & no   & \\
$151150.23+140506.4$ & 2.156& E & gal group      & no   & \\
$164847.85+374006.3$ & 1.328& C & gal            & no   & \\
\hline
\end{tabular}
\end{flushleft}
\tiny{
{\bf Notes.}
$^{(a)}$ Foreground object: gal=galaxy, ?=unclear;
$^{(b)}$ Absorption line system in SDSS spectrum

{\bf References.}
(1) Inada et al. (\cite{Inada10});
(2) Johnston et al. (\cite{Johnston03})
}
\label{tab:fg_sdss}
\end{table}

\begin{figure}[hbtp]
\includegraphics[bb=45 0 570 770,scale=0.335,angle=270,clip]{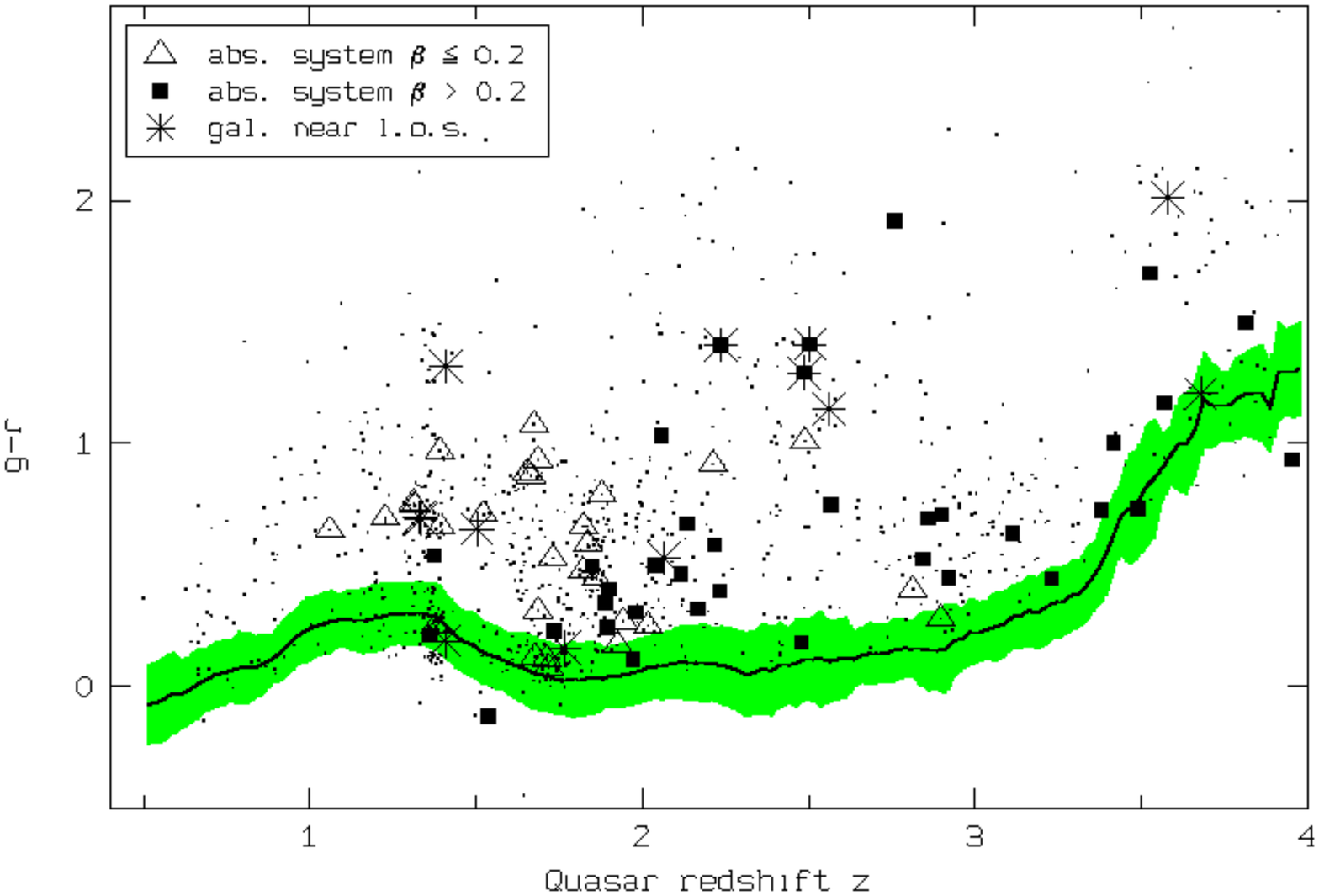}
\caption{
Colour indices $g-r$ (foreground extinction-corrected) as function of redshift
for all types (dots).
Other symbols: quasars with indications for foreground systems.
}
\label{fig:gr_z}
\end{figure}

We identified 76 absorber systems in 66 quasar spectra with a mean absorber
redshift $\langle z_{\rm abs} \rangle = 1.44\pm0.45$. Among them are 8 quasars with two and
one quasar (\object{SDSS J082747.14+425241.1}) with three different absorber systems.
The redshift differences between the absorber and the quasar, expressed by
the velocity parameter $\beta = v/c = (R^2-1)/(R^2+1)$, where 
$R = (1+z_{\rm q})/(1+z_{\rm abs})$, covers 
the range $\beta = 0.02$ to 0.72 (mean value 0.28). 
The fraction of quasars with registered absorption systems strongly depends on 
the spectral type. It is lowest for type A (3\%) and highest for type E 
(13\%). This is almost certainly a selection bias related to the intrinsic
complexity of the type A spectra. 
For any given spectral type, on the other hand, quasars with or without 
these narrow-line absorber systems have on average the same peculiarity
parameters $\chi^2$.

Figure\,\ref{fig:gr_z} shows the colour index $g-r$ (corrected for Galactic
extinction) versus $z$ for all unusual quasars (as in Fig.\,\ref{fig:colour-indices}),
where those showing indications of foreground galaxies
are marked by symbols. Since absorbers with small $\beta$ may be associated
with the quasar, we split the sample into two groups of
29 quasars with $\beta_{\rm max} \le 0.2$ and 37 with $\beta_{\rm max} > 0.2$. 
As outflows from quasars usually have $\beta < 0.2$ (Foltz et al. \cite{Foltz83}),
the latter group (with $\langle \beta \rangle = 0.41\pm0.14$) is expected 
to represent a clean sample of quasars with spectroscopically identified
foreground galaxies. The majority of our quasars with foreground galaxies are redder
than the normal quasars from the QCDR7. Hence, we cannot exclude that a fraction of
the red quasars (types 3 and 4) were selected as unusual owing to properties 
related to the foreground rather than the quasar itself (as is indeed the case for the
two gravitationally lensed quasars \object{SDSS J081959.80+535624.2} and
\object{SDSS J090334.94+502819.3}). However, we did not find any significant
differences between the reddening properties of these quasars in our
sample showing evidence of foreground galaxies close to the line of sight
and those which do not. 

In the Appendix A, we present the results of a simple search for
extended structures close to the line of sight on direct images.

%
\section{Very peculiar quasar spectra}
%

We did not identify any fundamentally new type of unusual quasar spectra.
However, our selection substantially enlarges the sample of known unusual quasars and 
thus enables statistical studies of the various peculiarity types and searches for
relations between them. Representative examples of our 7 types were shown in
Fig.\,\ref{fig:examples_groups}. The detailed investigation of individual spectra 
is beyond the scope of this paper. However, it is worth commenting on a 
few very peculiar and interesting spectra, including another three mysterious objects 
and another dozen possibly related quasars.

\subsection{Unusual BALs and strong Fe emission}\label{so}

\begin{figure*}[hbtp]	
\begin{tabbing}
\includegraphics[bb=40 35 510 785,scale=0.23,angle=270,clip]{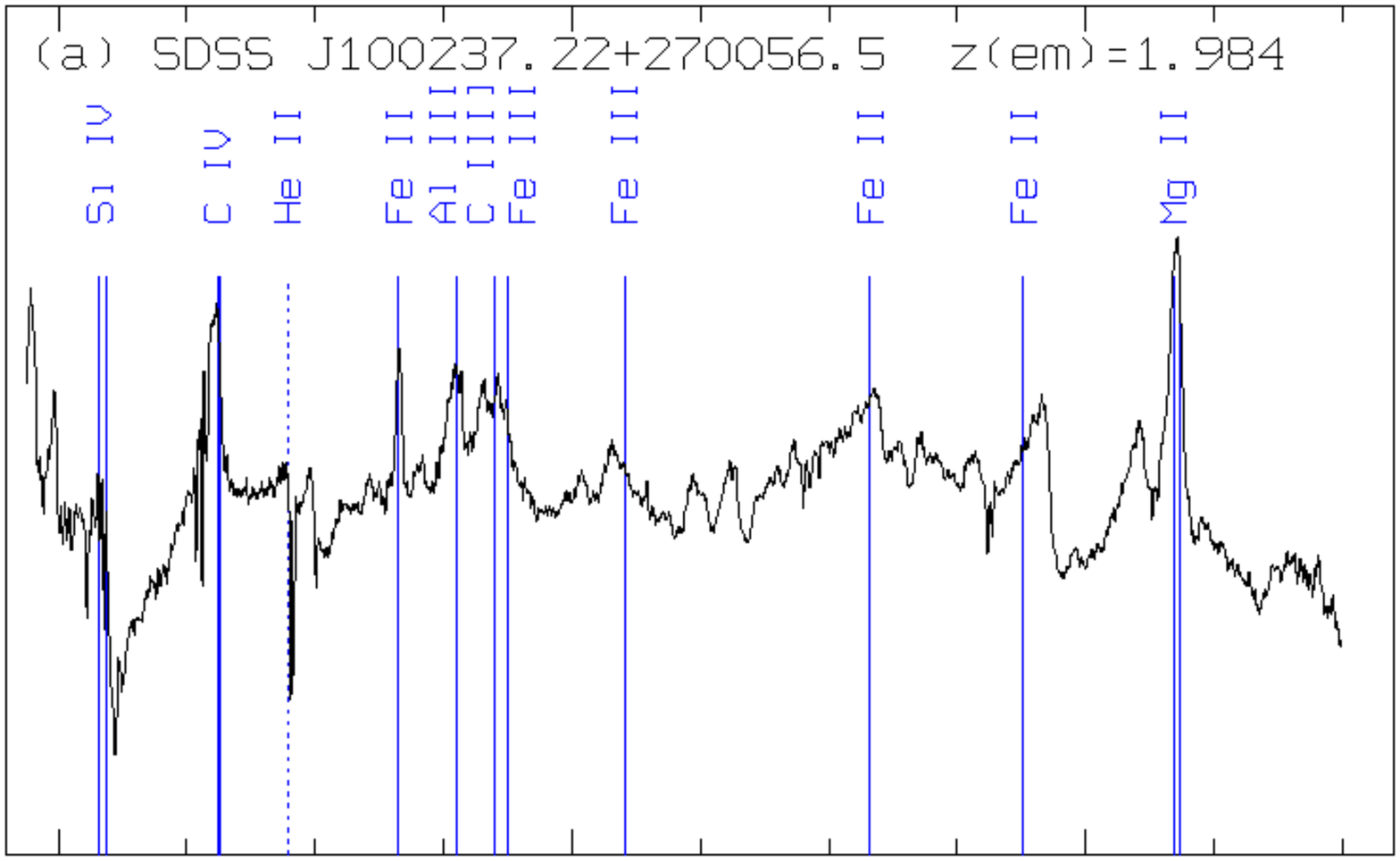}\hfill \=
\includegraphics[bb=40 35 510 785,scale=0.23,angle=270,clip]{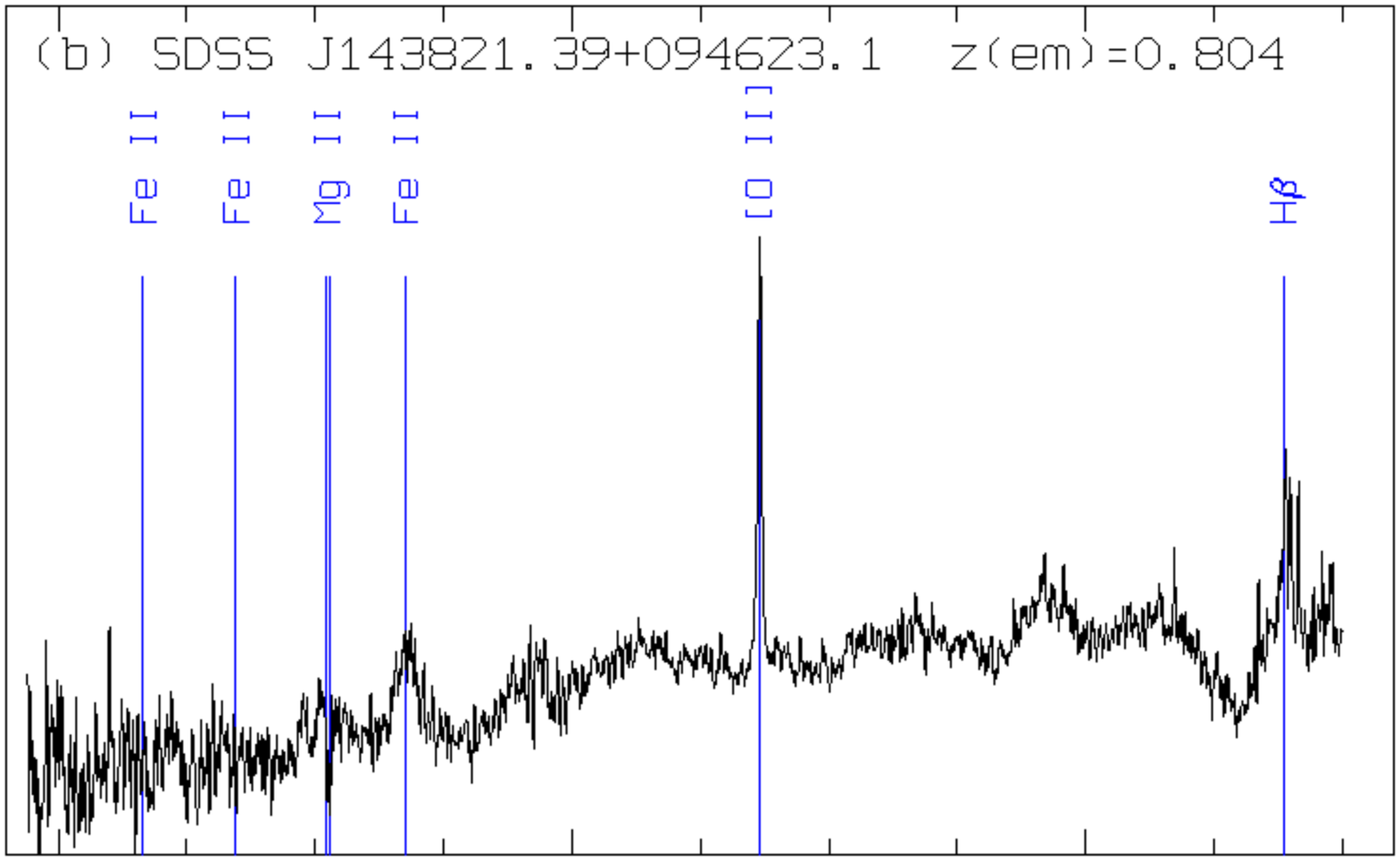}\hfill \=
\includegraphics[bb=40 35 510 785,scale=0.23,angle=270,clip]{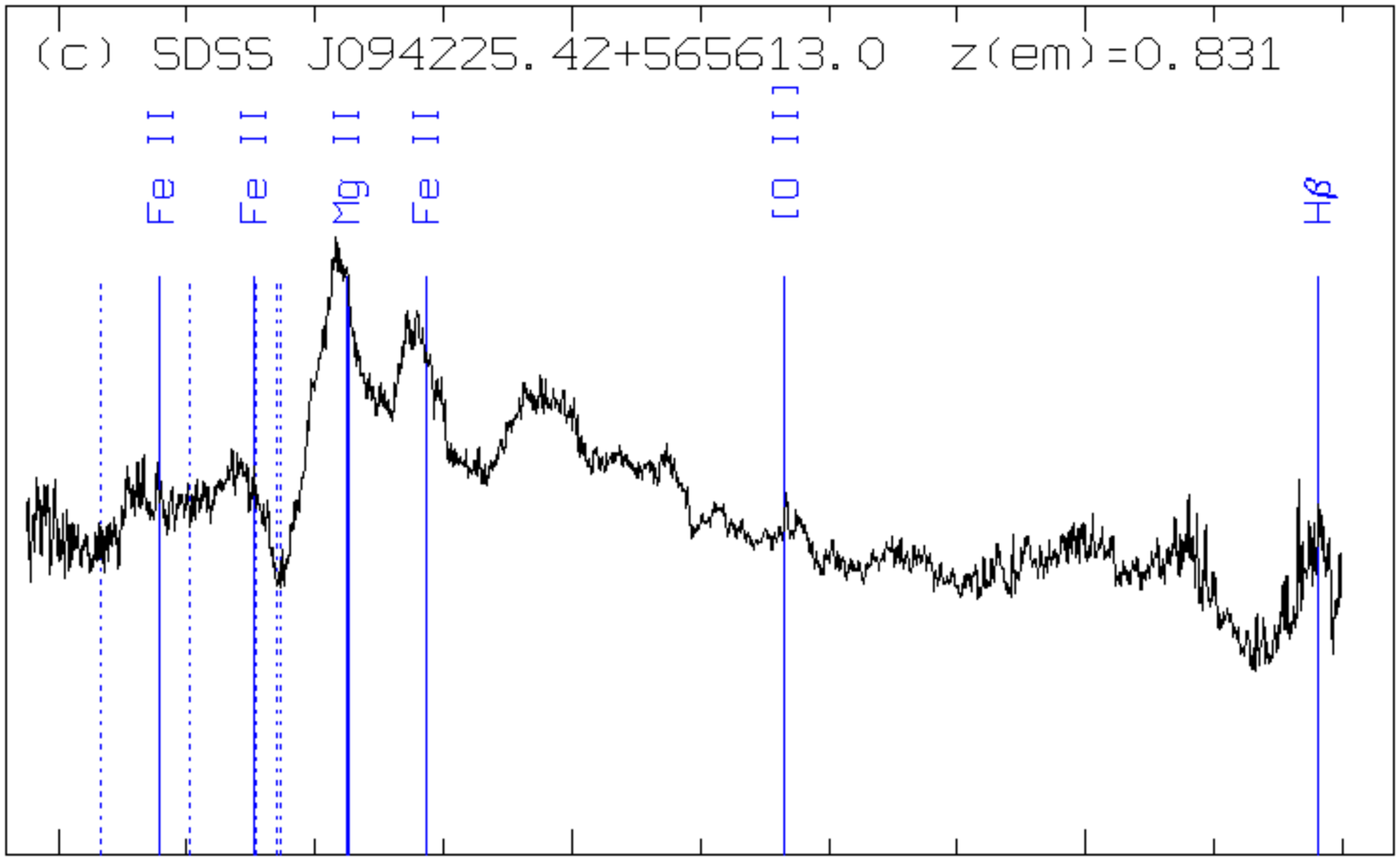}\hfill \\
\includegraphics[bb=40 35 510 785,scale=0.23,angle=270,clip]{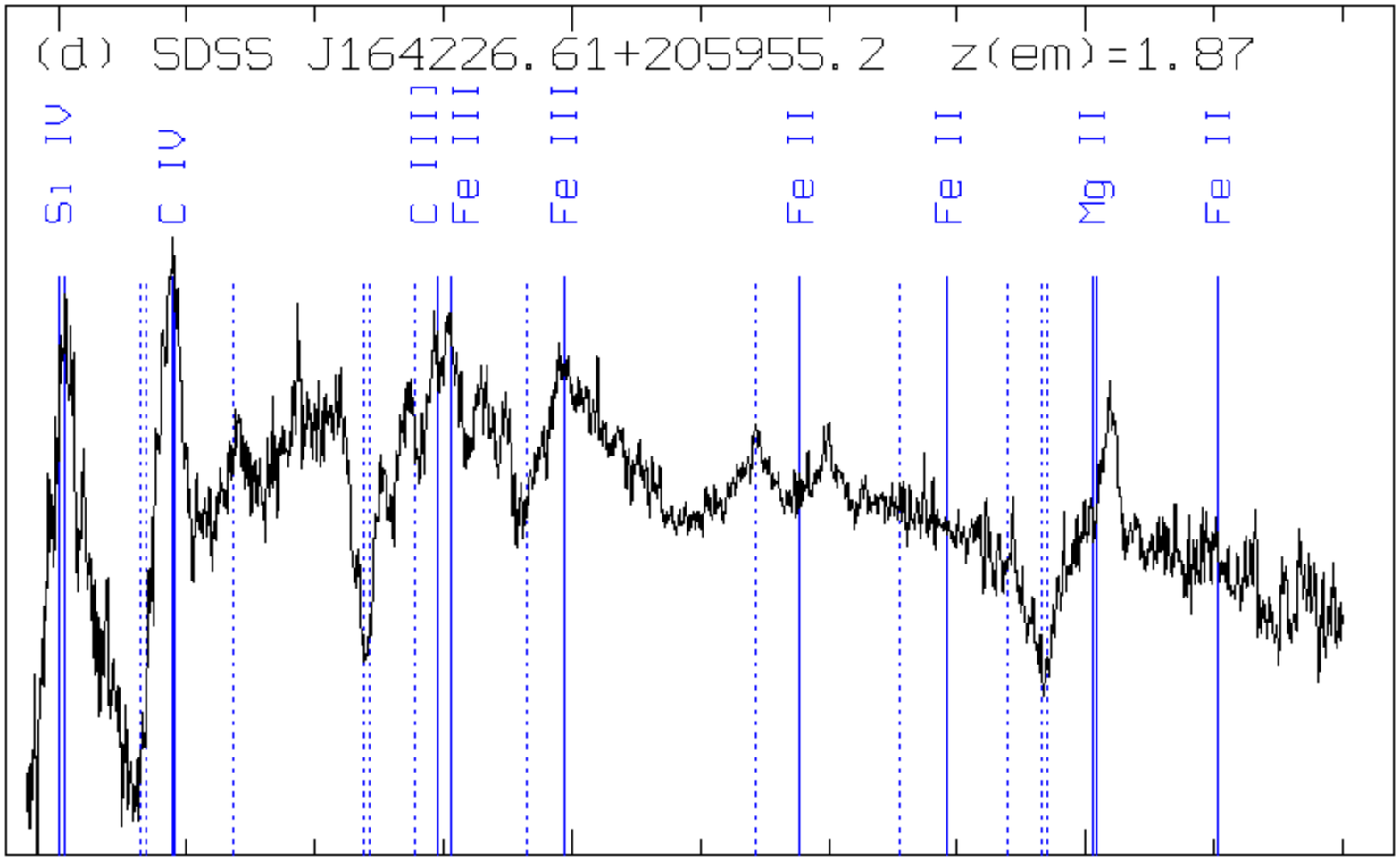}\hfill \=
\includegraphics[bb=40 35 510 785,scale=0.23,angle=270,clip]{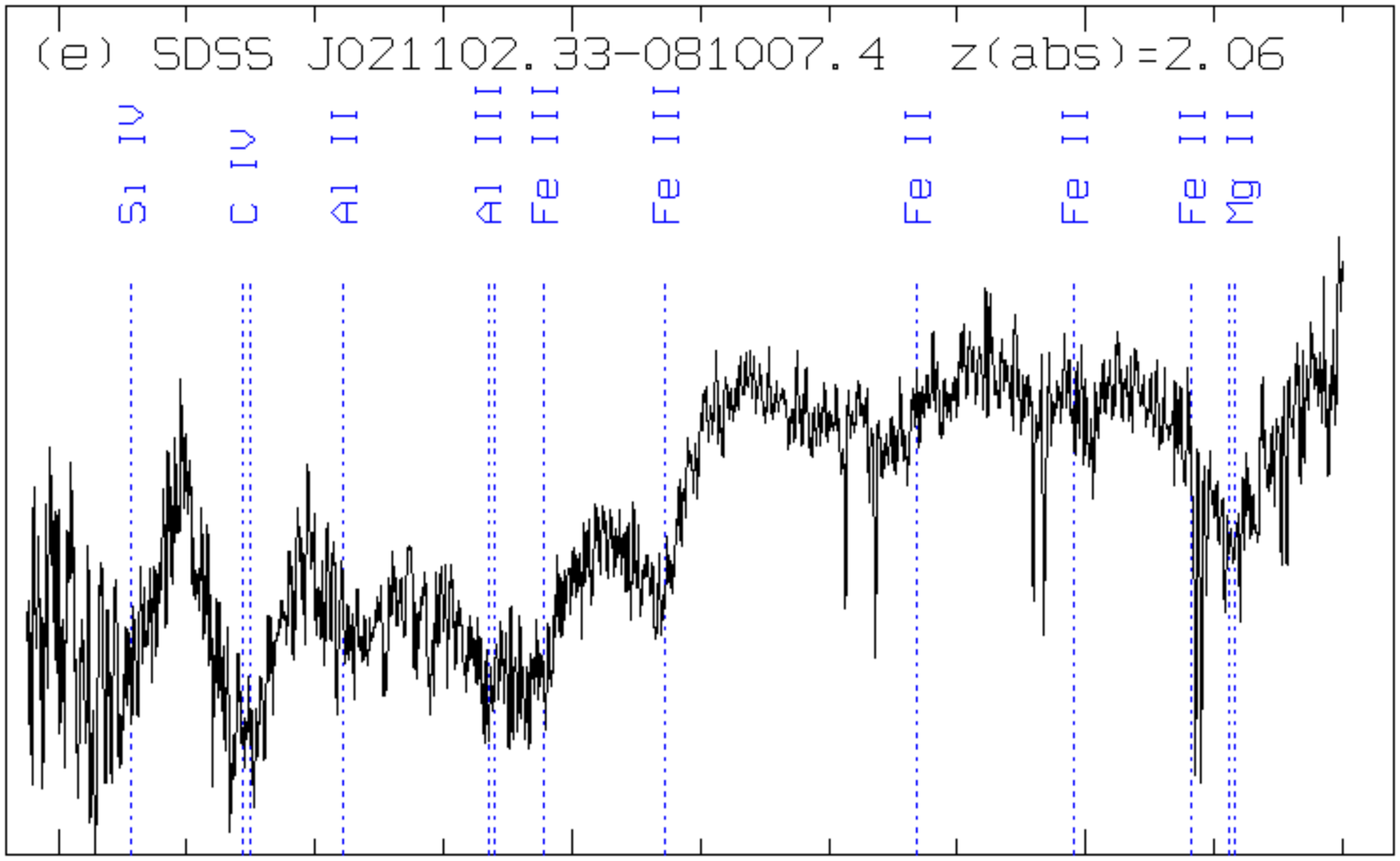}\hfill \=
\includegraphics[bb=40 35 510 785,scale=0.23,angle=270,clip]{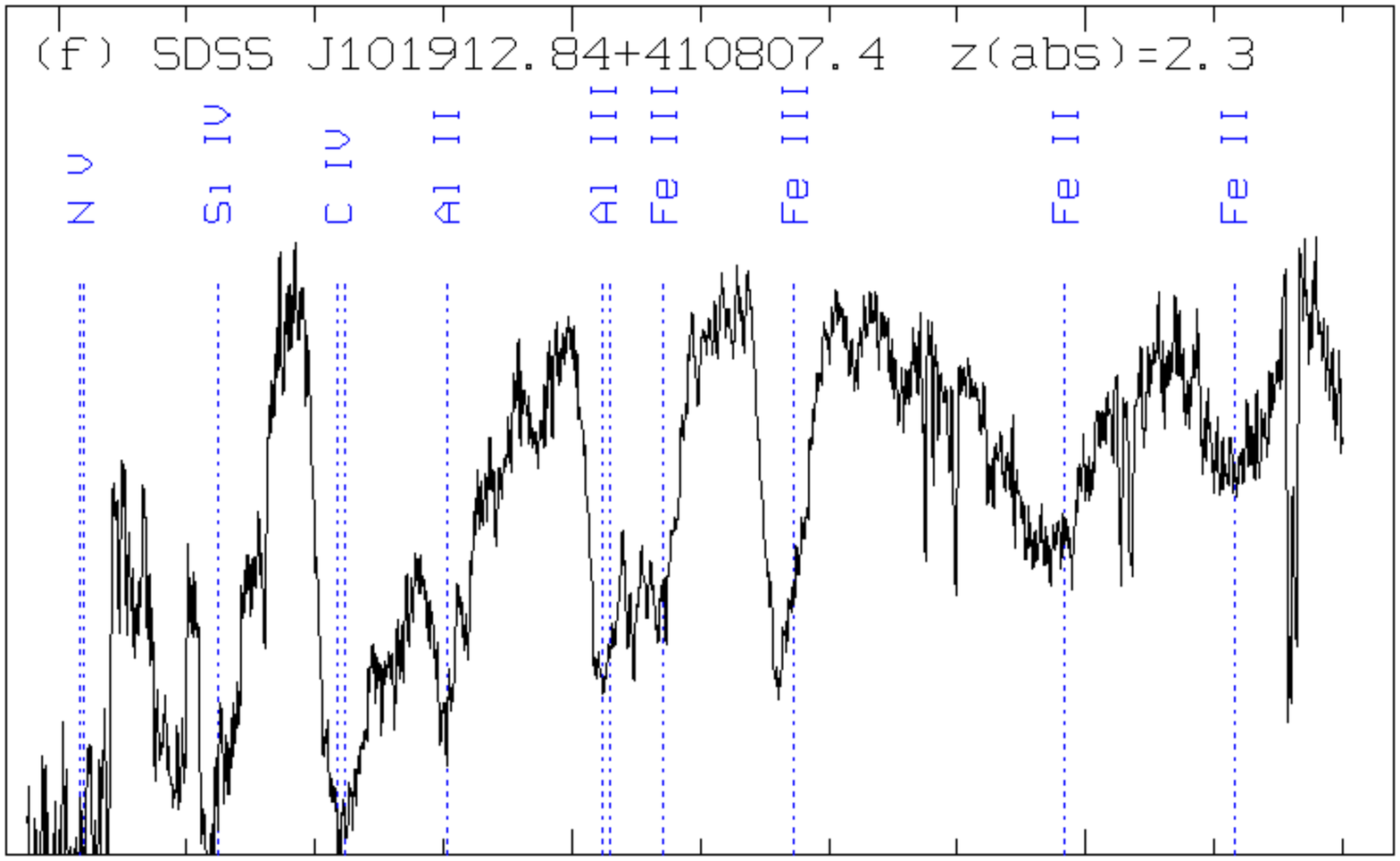}\hfill \\
\includegraphics[bb=40 35 510 785,scale=0.23,angle=270,clip]{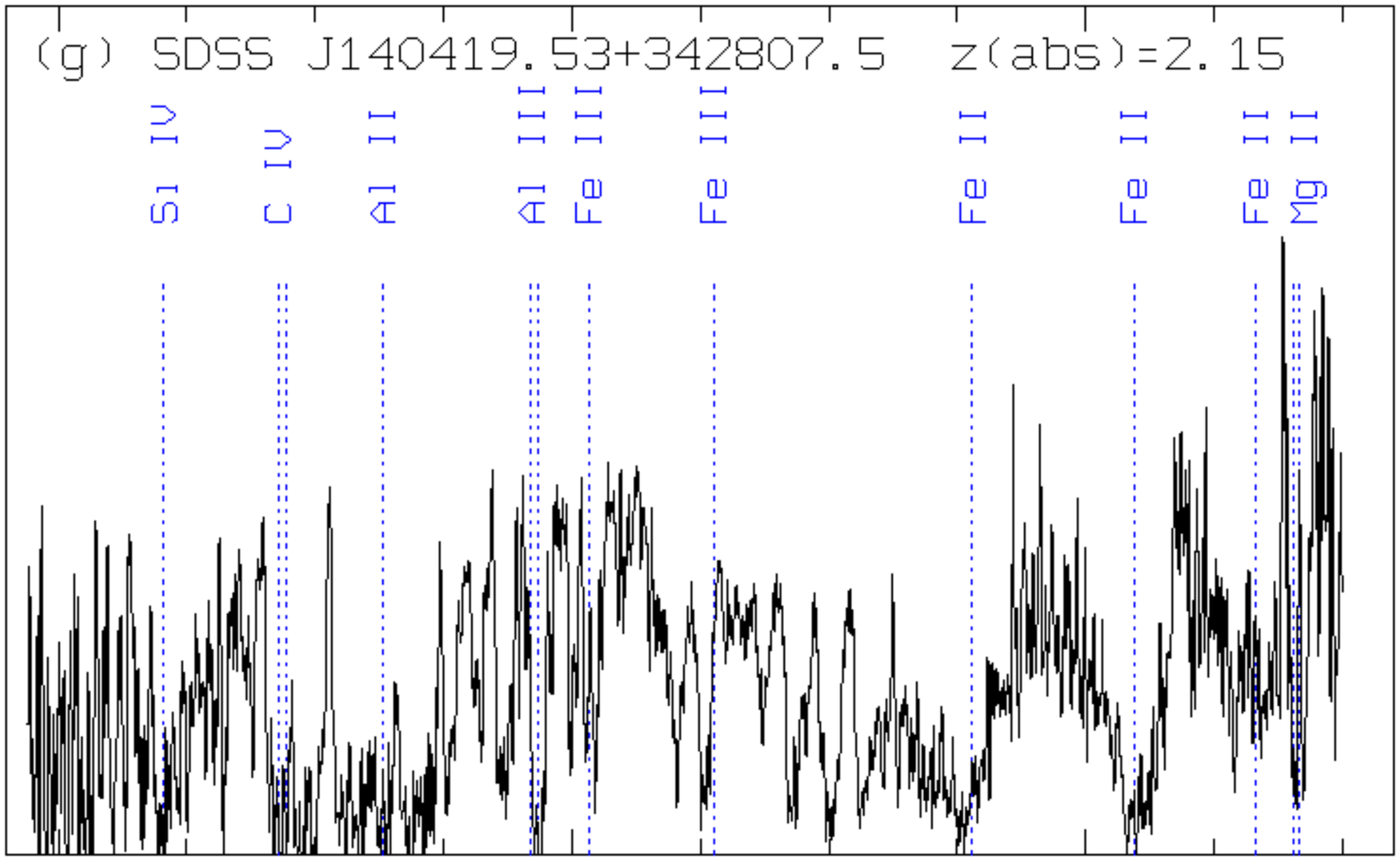}\hfill \=
\includegraphics[bb=40 35 510 785,scale=0.23,angle=270,clip]{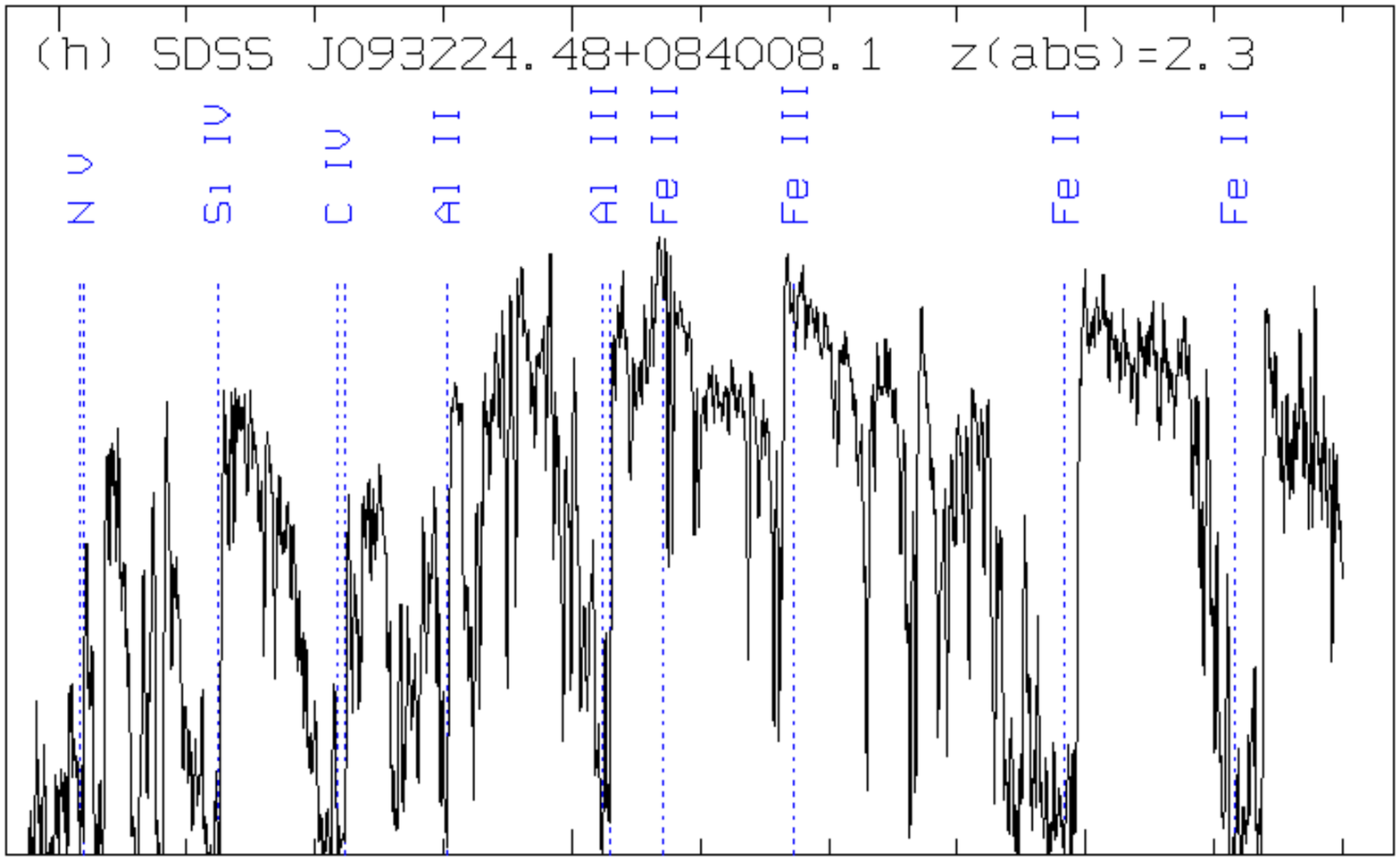}\hfill \=
\includegraphics[bb=40 35 510 785,scale=0.23,angle=270,clip]{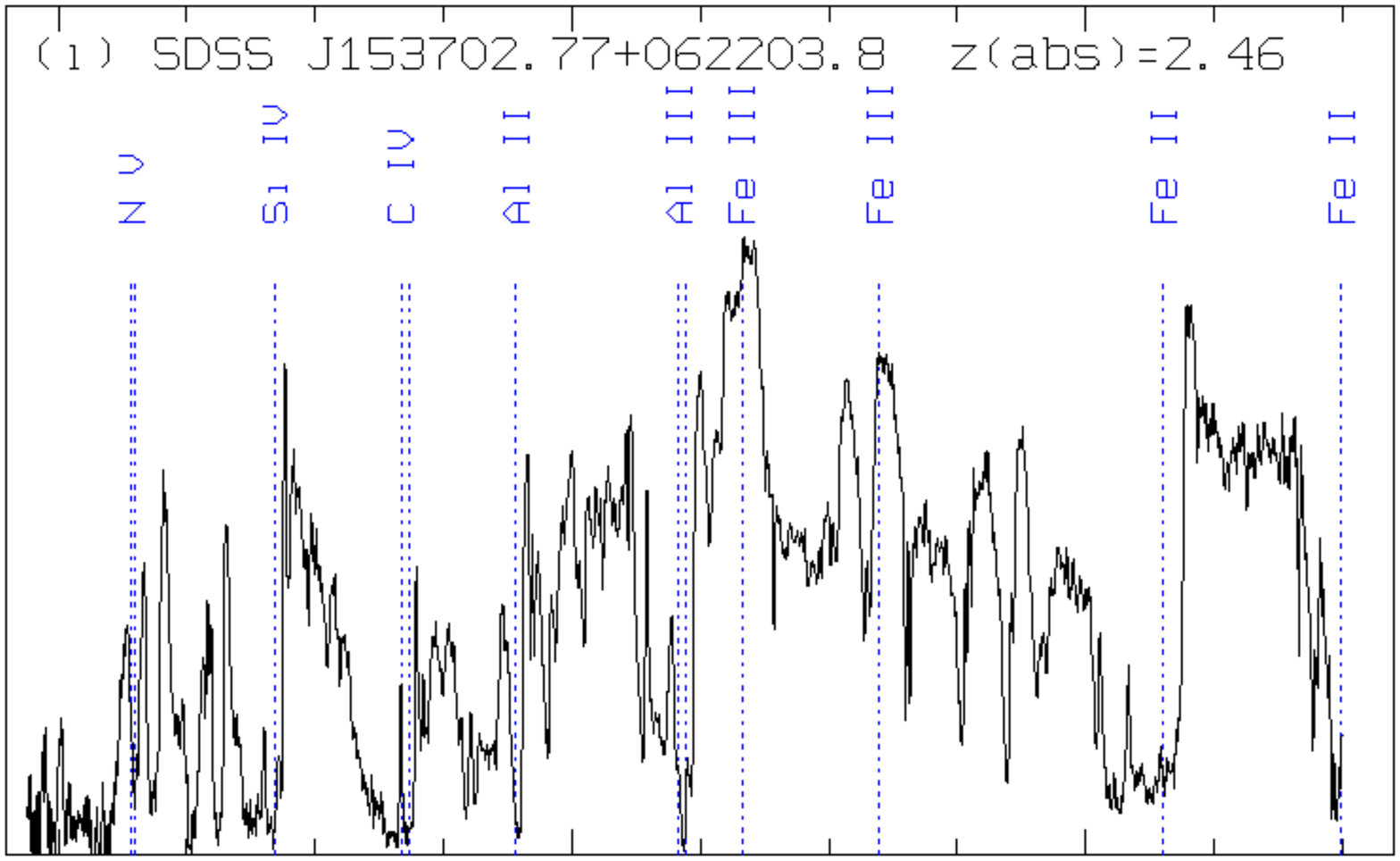}\hfill \\
\includegraphics[bb=40 35 590 785,scale=0.23,angle=270,clip]{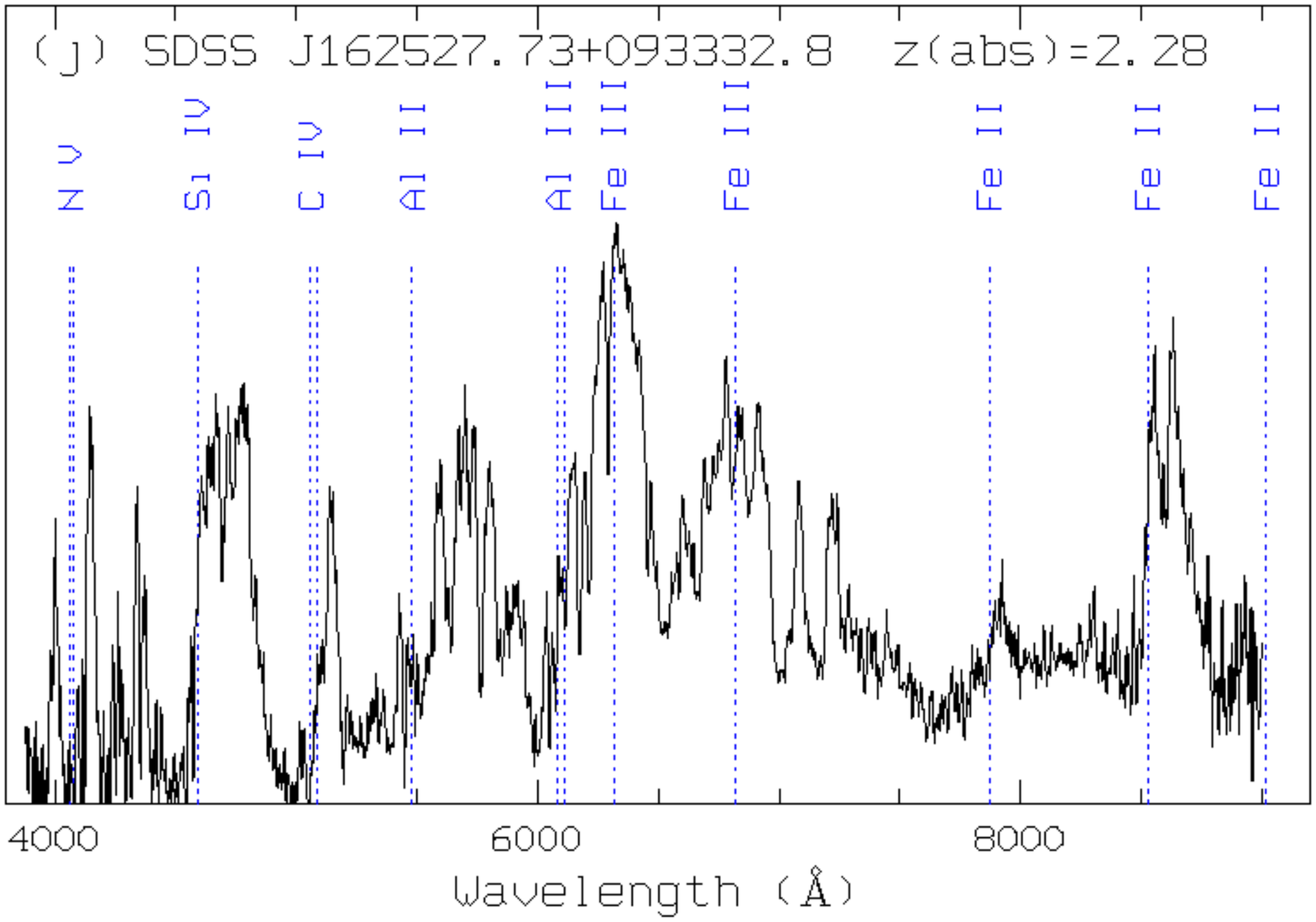}\hfill \=
\includegraphics[bb=40 35 590 785,scale=0.23,angle=270,clip]{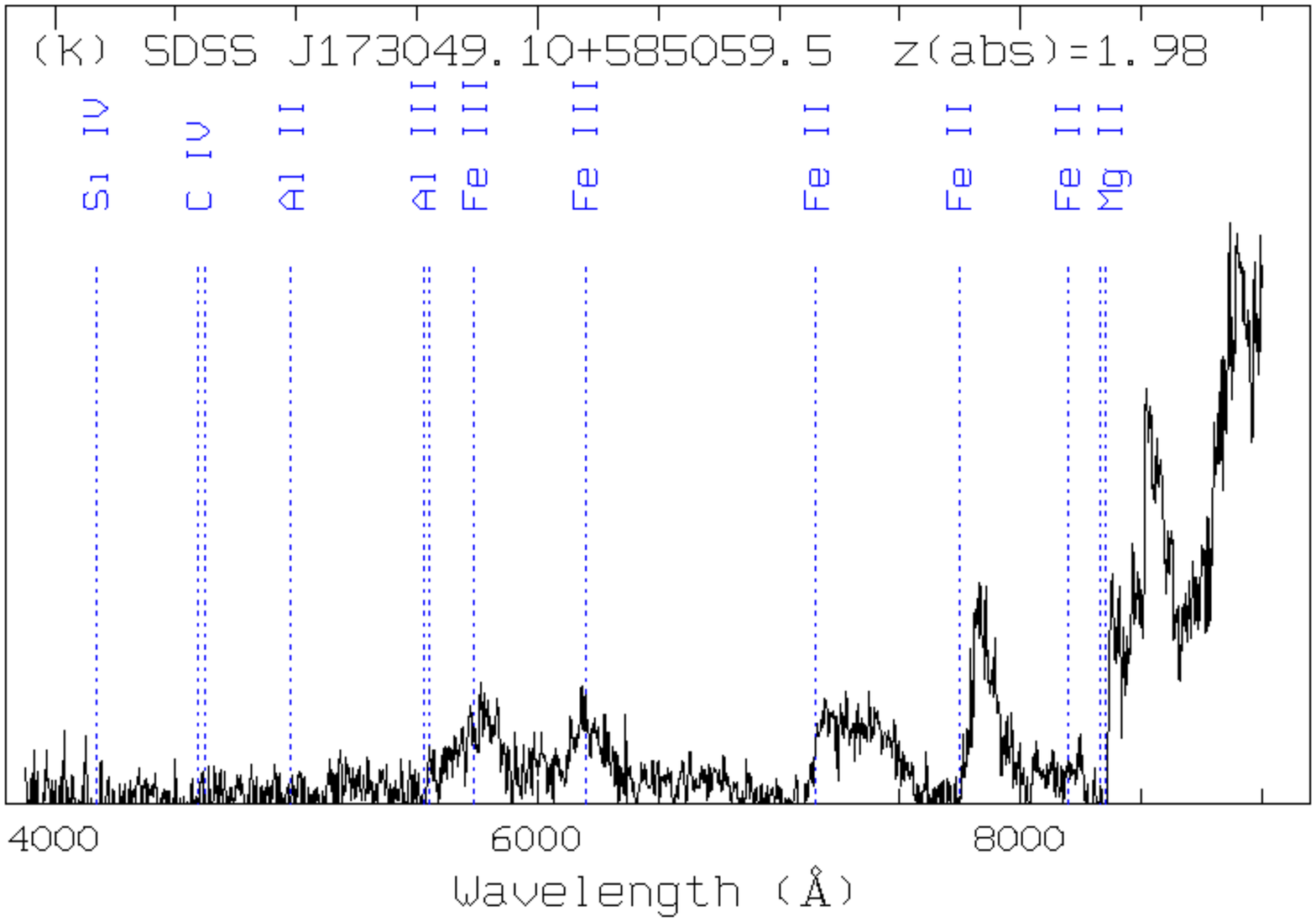}\hfill \=
\includegraphics[bb=40 35 590 785,scale=0.23,angle=270,clip]{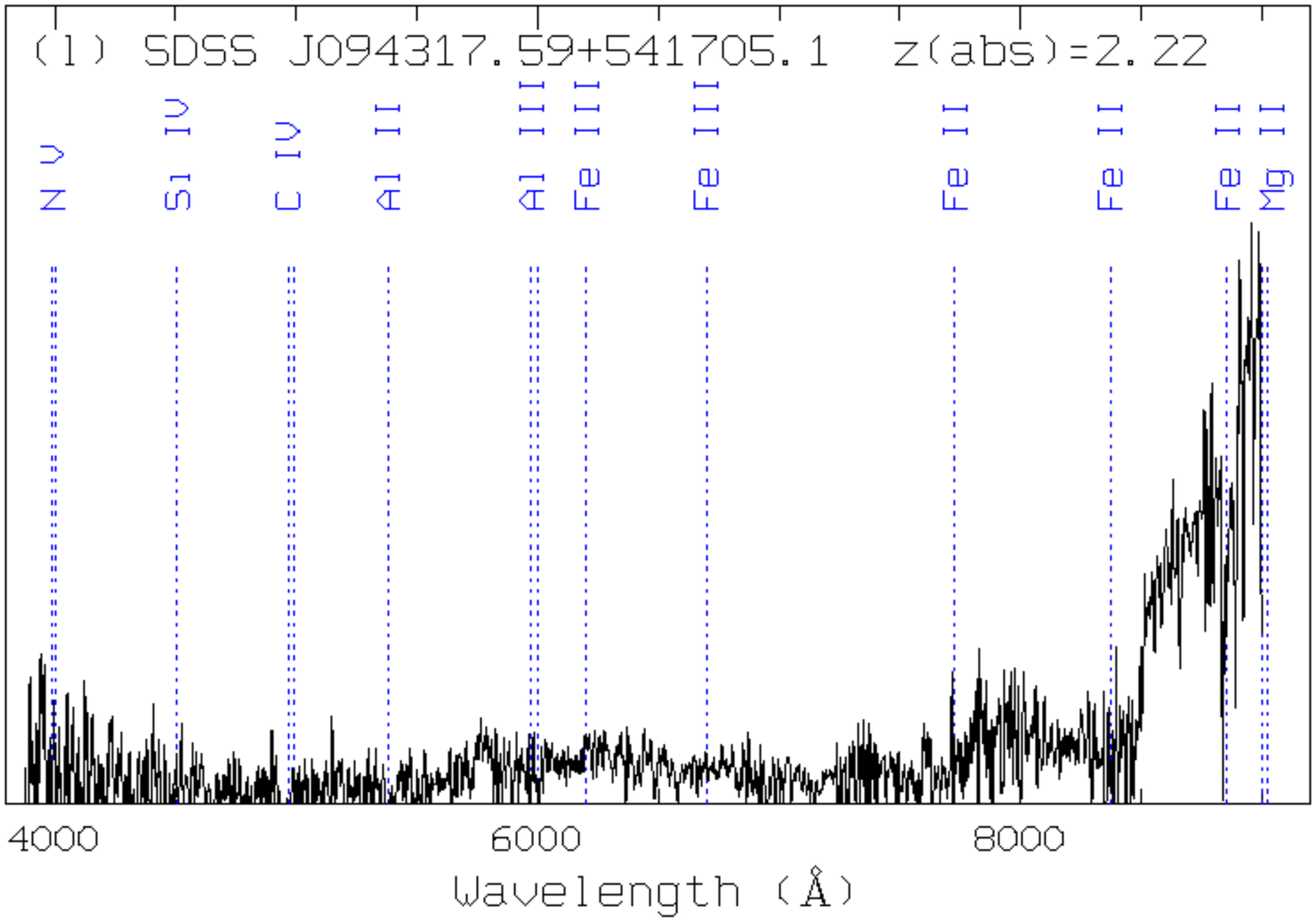}\hfill \\
\end{tabbing}
\caption{Examples of particularly unusual SDSS quasars: strong UV iron emission ({\it
top}), \ion{Fe}{iii} absorption ({\it second row}), BAL quasars with many troughs
({\it third row}), and with overlapping troughs ({\it bottom}).}
\label{fig:so}
\end{figure*}

A selection of SDSS quasars with particularly unusual spectra caused by 
extreme BAL features or/and strong iron emission is presented in 
Fig.\,\ref{fig:so}. Each panel shows the normalised flux $F_\lambda(\lambda)$
over the observed wavelength interval 3900\AA\ to 9000\AA,
the object name is given at the top of the panel.
Here we briefly discuss these 12 spectra from top to 
bottom and left to right.  

The first three panels show quasars with strong iron emission. 
In the first panel, we present
\object{SDSS J100237.22+270056.5}\footnote{A similar object 
(not shown here) is \object{SDSS J124244.36+624659.1}.}
which exhibits emission from the multiplets
\ion{Fe}{ii}  $\lambda$1785 (UV67,UV191),
\ion{Fe}{iii} $\lambda$1926 (UV34),
\ion{Fe}{iii} $\lambda$2070 (UV48), 				
\ion{Fe}{ii}  $\lambda$2400 (UV2), 	      
\ion{Fe}{ii}  $\lambda$2600 (UV1), 
and
\ion{Fe}{ii}  $\lambda$2750 (UV62,UV63).
The positions of these lines, as well as the positions of 
the typical quasar emission lines 
(\ion{Si}{iv}  $\lambda$1400,    
\ion{C}{iv}   $\lambda\lambda$1542.2,1550.8, 
\ion{Al}{iii} $\lambda\lambda$1854.7,1862.8,
\ion{C}{iii}] $\lambda$ 1908.7,
\ion{Mg}{ii}  $\lambda\lambda$2796.3,2803.5),
are marked by the vertical solid lines.  
There is a high-velocity outflow at $z_{\rm abs} = 1.724$ seen 
in \ion{C}{iv}, \ion{Al}{ii}, \ion{Al}{iii}, and \ion{Mg}{ii}.
The absorption line redward of \ion{C}{iv}, close to \ion{He}{ii} $\lambda$1640,
is probably a narrow intervening system.
\object{SDSS J100237.2+270057} is one of the most luminous quasars in
our sample with $M_{\rm i} = -28.94$.

The spectrum in panel {\bf (b)} shows a quasar (\object{SDSS J092211.56+365120.2})
where the emission from \ion{Fe}{ii} UV1 and UV2 and \ion{Fe}{iii} UV48 
is stronger than the (weak) broad emission lines \ion{C}{iv} and \ion{Mg}{ii}.  
Strong iron emission is known to be correlated with the occurrence of BAL 
troughs (Boroson \& Meyers \cite{Boroson92b}; Zhang et al. \cite{Zhang10}), 
as can be seen in many of our spectra of types A or F 
(Sect.\,\ref{subsec:classification}). A nice example is  \object{SDSS J094225.42+565613.0}
in panel {\bf (c)}. The positions of absorption troughs at $z_{\rm abs} = 0.735$ are
marked by the dashed lines. If we adopt the systemic redshift of \object{SDSS J094225.42+565613.0}
from the (weak) [\ion{O}{ii}] $\lambda$3728 emission line, the \ion{Mg}{ii} emission 
line is either blueshifted or absorbed at the red side. 
In Tab.\,\ref{tab:catalogue}, the remark ``lst?'' is made, indicating possible 
but not certain longwards-of-systemic absorption (see Hall et al. \cite{Hall02}).
The \ion{Mg}{ii} emission of \object{SDSS J094225.42+565613.0} may instead be
affected in this case by broad \ion{Fe}{ii} absorption on its red side.

As a consequence of the overlapping \ion{Fe}{ii} troughs, the continuum of 
\object{SDSS J094225.42+565613.0} is depressed shortwards of \ion{Mg}{ii}
down to the lowest observed wavelengths. Interestingly, the \ion{Fe}{ii} UV2 
trough appears to be deeper than UV1. As outlined by Hall et al. (\cite{Hall02}), the only
way of explaining an increase in the absorption strength when troughs overlap is
spatially distinct velocity-dependent partial covering (sdvdpc) of the continuum
source, provided that the troughs are saturated (as is usually the case even
when the absorption is not black). Hall et al. emphasise that these objects
``may not be all that rare, just difficult to recognise'', though the only quasar
known to them to definitely exhibit sdvdpc was FBQS 1408+3054. 
From the inspection of our sample, we found at least another eight
candidates probably showing that effect (Table \ref{tab:sdvdpc}). 

\begin{table}[htbp]
\caption{
Candidates for spatially distinct velocity-dependent partial covering (sdvdpc) 
of the continuum source by \ion{Fe}{ii} UV troughs.
}
\begin{flushleft}
\begin{tabular}{cccl}
\hline\hline
   SDSS J       & $T$  & $z$   & Remark  \\
\hline
$015151.58-093215.3$ & A  & 1.413 & ot(sdvdpc)                \\
$082350.53+244653.1$ & C  & 1.766 & ot(sdvdpc)? 	      \\
$094225.42+565613.0$ & A  & 0.831 & ot(sdvdpc?,lst?); sFe(opt) \\
$101912.84+410807.4$ & A  & 2.450 & nt(sdvdpc); \ion{Fe}{iii} BAL?     \\
$135246.37+423923.5$ & A  & 2.000 & ot(sdvdpc)  	      \\
$142010.28+604722.3$ & A  & 1.349 & ot(sdvdpc)  	      \\
$150848.80+605551.9$ & A  & 1.525 & ot(sdvdpc?) 	      \\
$162527.73+093332.8$ & A  & 1.413 & ot(sdvdpc)  	      \\
\hline
\end{tabular}
\end{flushleft}
{\tiny
{\bf Notes.} Remarks see Table 3.\\
}
\label{tab:sdvdpc}
\end{table}

In the second row of Fig.\,\ref{fig:so}, we give three  
examples of LoBAL quasars with possible strong \ion{Fe}{iii} absorption.
\object{SDSS J$021102.33-081007.4$} resembles the two possible 
\ion{Fe}{iii} BAL quasars \object{SDSS J$014905.28-011404.9$} and 
\object{SDSS J$081024.75+480615.4$} from Hall et al. (\cite{Hall02}).
It appears to have \ion{Fe}{iii} UV34, UV48 troughs (which probably overlap),
in addition to troughs from \ion{Si}{iv}, \ion{C}{iv}, \ion{Mg}{ii}, 
and weak \ion{Fe}{ii} UV1, UV2 absorption. Even more pronounced are the iron 
absorption troughs (plus troughs from other elements) 
seen in \object{SDSS J$101912.84+410807.4$}. For both spectra,
and also for all following quasars shown in Fig.\,\ref{fig:so}, 
it seems impossible to clearly identify any emission line, thus only
absorption redshifts can be given. The BAL quasars can show, of course, two or 
more absorption systems at different $z_{\rm abs}$. The values of 
$z_{\rm abs}$ given in Fig.\,\ref{fig:so} usually refer to the 
centres of the strongest system.

The third row shows another three examples for FeLoBAL 
quasars with many narrow troughs, similar to the three high-$z$ quasars 
of this category discussed by Hall et al. (\cite{Hall02}). Although the spectra
appear to be very complex, a large number of absorption features can be identified.
Only a selection of the strongest troughs is labelled here. We again did not
try to estimate the emission redshift since emission lines can be mimicked 
by overlapping absorption troughs that make detailed line identification
quite difficult. As for the previously known objects of this type, 
no absorption trough reaches zero flux, indicating that there is a
partial coverage of the continuum source by the absorbers and/or a contribution
from scattered light (see Hall et al. \cite{Hall02}).

The SDSS has discovered several FeLoBAL quasars with sharply declining flux near
\ion{Mg}{ii}, which is interpreted as being caused by overlapping iron troughs 
(Hall et al. \cite{Hall02}). The bottom row of Fig.\,\ref{fig:so} exhibits 
three examples where the absorption strength appears to increase from left to right. 
Here, the redshifts refer to the red edges of the absorption troughs.
Only the longest wavelength lines are marked for the Fe multiplets.
\object{SDSS J162527.73+093332.8} displays both many narrow troughs at
$\lambda \la 7200$\AA\ (observed) and overlapping \ion{Fe}{ii} UV1 and UV2 
at longer wavelengths (\ion{Fe}{ii} UV2 is deeper than UV1, see above).  

\object{SDSS J173049.10+585059.5} was already described  
by Hall et al. (\cite{Hall02}). With the exception of two narrow wavelength  
intervals around \ion{C}{iii}] $\lambda$1908.7, there is almost no flux 
shortwards of the onset of \ion{Fe}{ii} UV2 at 7000\AA\ (observed) down
to the shortest wavelengths. A similar spectrum, but even more extreme, is that of
\object{SDSS J094317.59+541705.1} at the right-hand side. The flux drops 
rapidly at $\sim 9000$\AA\ (observed) towards shorter wavelengths and is 
below the detection threshold in the SDSS g and u bands, i.e. shortwards of the
\ion{C}{iv} line. The object is detected in 2MASS with $K = 14.25$. The 2MASS colours
correspond to a slowly decreasing flux with increasing wavelengths 
from 1.2 to 2.2 \,$\mu$m. At first glance, the SDSS spectrum suggests a high
redshift quasar at $z > 6$. However, a higher-quality Keck spectrum presented by
Urrutia et al. (\cite{Urrutia09}) clearly indicates the \ion{Mg}{ii}
emission line, along with the typical absorption features of a FeLoBAL 
at $z=2.224$.

Given that \object{SDSS J162527.73+093332.8} is more luminous than the 
other two overlapping trough quasars 
\object{SDSS J173049.10+585059.5} and \object{SDSS J094317.59+541705.1}
($M_{\rm i} = -26.27$ compared  to $-24.65$ and $-24.75$),
all three spectra are most likely very similar. The main 
difference is a slightly higher redshift of \object{SDSS J162527.73+093332.8}, 
hence the 3000\AA\ jump is shifted out of the spectral window. The very
red optical-to-near infrared colours of \object{SDSS J162527.73+093332.8}
(which is one of the reddest SDSS quasars; $r-K = 7.17$) indicate that its 
flux must strongly increase longwardss of the red edge of the SDSS spectrum.
Lower-luminosity versions of these quasars at $z \ga 2.3$ are
strongly biased against being included in optical magnitude-limited samples.

\subsection{Mysterious and possibly related objects}\label{subsect:myst}

\begin{table*}[htbp]
\caption{
Mysterious objects and possibly related quasars.
}
\begin{flushleft}
\begin{tabular}{lcllccccrrrl}
\hline\hline
Name &
Panel in &
\ \ \ $z$   & 
Ref.  &
\ $T$  & 
$M_{\rm i}$ &
$R_{\rm i}$ &
$E_{B-V}^{\rm (intr)}$ &
$\mu_\alpha\cos\delta$ &
$\mu_\delta$ \ \ \ \ &
$N_{\rm e} $ &
\ pm \\
&Fig.12&&&&(mag)&&(mag)&(mas/yr)&(mas/yr)&&\\
\hline
\multicolumn{5}{l}{(a) Mysterious objects:}  \\
SDSS J$010540.75-003313.9^{~a}$ &(c)& 1.179$^{~b}$&  1 &...&$-26.49$& 1.19 &0.18&$  8.1\pm2.1$&$-4.1\pm2.3$&40&pmx?$^{~d}$\\
SDSS J$085502.20+280219.6$      &(f)& 1.511	  &    &  D&$-25.55$& ...  &0.15&$ -8.1\pm1.0$&$ 1.5\pm2.0$&13&pmx?$^{~d}$\\
FBQS J$105528.80+312411.3$      &(a)& 0.497	  &  1 &  F&$-25.13$& 1.16 &0.45&$ -3.1\pm2.7$&$-2.2\pm3.5$& 7&no pm \\
SDSS J$130941.35+112540.1^{~a}$ &(h)& 1.362$^{~b}$&  3 &...&$-26.62$& 0.72 &0.20&$ -2.2\pm5.2$&$-1.2\pm2.6$&10&no pm \\
VPMS J$134246.24+284027.5^{~a}$ &(i)& 1.300	  &  4 &...&$-25.36$& 0.89 &0.12&$  3.7\pm5.8$&$-0.5\pm5.9$&22&no pm \\
SDSS J$145045.56+461504.2$      &(g)& 1.894	  &  3 &  D&$-26.48$& 1.17 &0.30&$  3.5\pm5.7$&$ 1.8\pm1.8$& 6&no pm \\
SDSS J$160827.08+075811.5$      &(e)& 1.182$^{~b}$&    &  D&$-27.57$& 1.18 &0.18&$  1.5\pm1.2$&$ 1.6\pm1.9$&11&no pm \\
SDSS J$161836.09+153313.5^{~a}$ &(d)& 1.358$^{~b}$&    &...&$-26.02$& 1.22 &0.10&$  1.4\pm2.1$&$-9.0\pm2.1$& 8&pmy?$^{~d}$\\
SDSS J$220445.27+003141.8$      &(b)& 1.353	  &  1 &  D&$-27.81$& 0.49 &0.22&$ -1.6\pm1.3$&$-1.5\pm1.6$&23&no pm \\
\hline
\multicolumn{5}{l}{(b) Possibly related objects:}\\
SDSS J$073816.91+314437.0^{~a}$ &(q)& 2.01	  &  2 &...&$-26.80$& 0.80 &0.13&$ -0.1\pm1.7$&$-0.6\pm3.4$& 11&no pm \\
SDSS J$075437.85+422115.3$ 	&(s)& 1.964	  &    &  C&$-26.17$& 1.29 &0.27&$ 31.4\pm6.2$&$27.3\pm8.0$& 12&pm?$^{~e}$\\
SDSS J$091613.59+292106.2$ 	&(l)& 1.143$^{~b}$&    &  D&$-25.64$& ...  &0.18&$ -0.4\pm1.1$&$-3.4\pm1.4$& 13&no pm \\
SDSS J$091940.97+064459.9$ 	&(r)& 1.351$^{~b}$&    &  A&$-26.02$& 2.05 &0.27&$     ...   $&$    ...   $&...& ...  \\
SDSS J$101723.04+230322.1$ 	&(o)& 1.794	  &    &  A&$-27.18$& 1.30 &0.20&$  4.6\pm2.6$&$ 0.8\pm6.8$&  7&no pm \\
SDSS J$110511.15+530806.5$ 	&(t)& 1.936	  &    &  D&$-26.63$& ...  &0.50&$  4.8\pm2.6$&$-1.7\pm2.4$&  6&no pm \\
SDSS J$120337.91+153006.6$      &(n)& 1.238$^{~b}$&    &  A&$-26.62$& 0.55 &0.12&$     ...   $&$    ...   $&...& ...  \\
SDSS J$134951.93+382334.1$      &(j)& 1.094$^{~b}$&    &  A&$-25.20$& 1.12 &0.30&$  2.4\pm3.6$&$-9.3\pm5.2$&  7&no pm \\
SDSS J$151627.40+305219.7$ 	&(p)& 1.846	  &    &  D&$-27.79$& 0.43 &0.10&$ -1.8\pm1.5$&$ 0.5\pm2.0$& 11&no pm \\
SDSS J$152438.79+415543.0$      &(m)& 1.227$^{~b}$&    &  A&$-26.00$& 1.01 &0.22&$ -2.7\pm6.4$&$-1.5\pm6.8$& 14&no pm \\
SDSS J$215950.30+124718.4$ 	&(k)& 1.516	  & & D&$-26.39$&...$^{~c}$&0.50&$-11.5\pm2.3$&$ 2.0\pm6.0$&  9&pmx?$^{~d}$\\
\hline
\end{tabular}
\end{flushleft}
\tiny{
{\bf Notes.} $^{(a)}$ not in Tab.\,\ref{tab:catalogue};
\ $^{(b)}$ redshift based on [\ion{O}{ii}];
\ $^{(c)}$ not in FIRST survey area;
$^{(d)}$ ``pmx?'' and ``pmy?'' indicate small ($\sim$10 mas/yr) proper motion components
$\mu_{\alpha}\cos{\delta}$ or $\mu_{\delta}$ respectively. These results are formally significant 
but doubtful, as caused by probably erroneous POSS1 positions. 
$^{(e)}$  ``pm?'' indicates formally significant proper motions in both components that are
likely caused by a faint object close to the line of sight.\\
{\bf References.} 
(1) Hall et al. (\cite{Hall02});
(2) Hall et al. (\cite{Hall04});
(3) Plotkin et al. (\cite{Plotkin08});
(4) Meusinger et al. (\cite{Meusinger05})
}
\label{tab:mysts}
\end{table*}

\begin{figure*}[hbtp]   
\begin{tabbing}
\includegraphics[bb=53 00 500 770,scale=0.20,angle=270,clip]{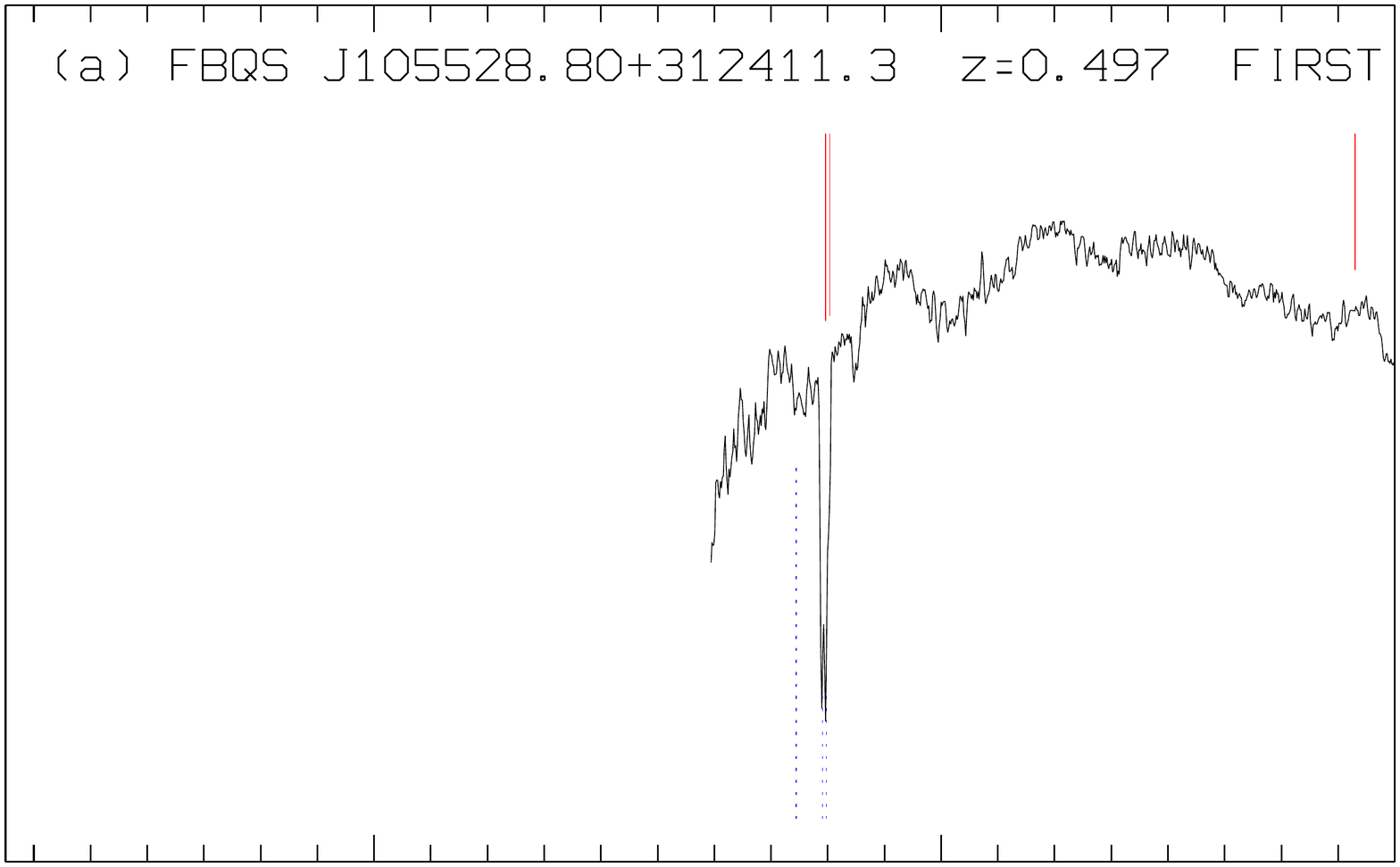}\hfill \=
\includegraphics[bb=53 20 500 770,scale=0.20,angle=270,clip]{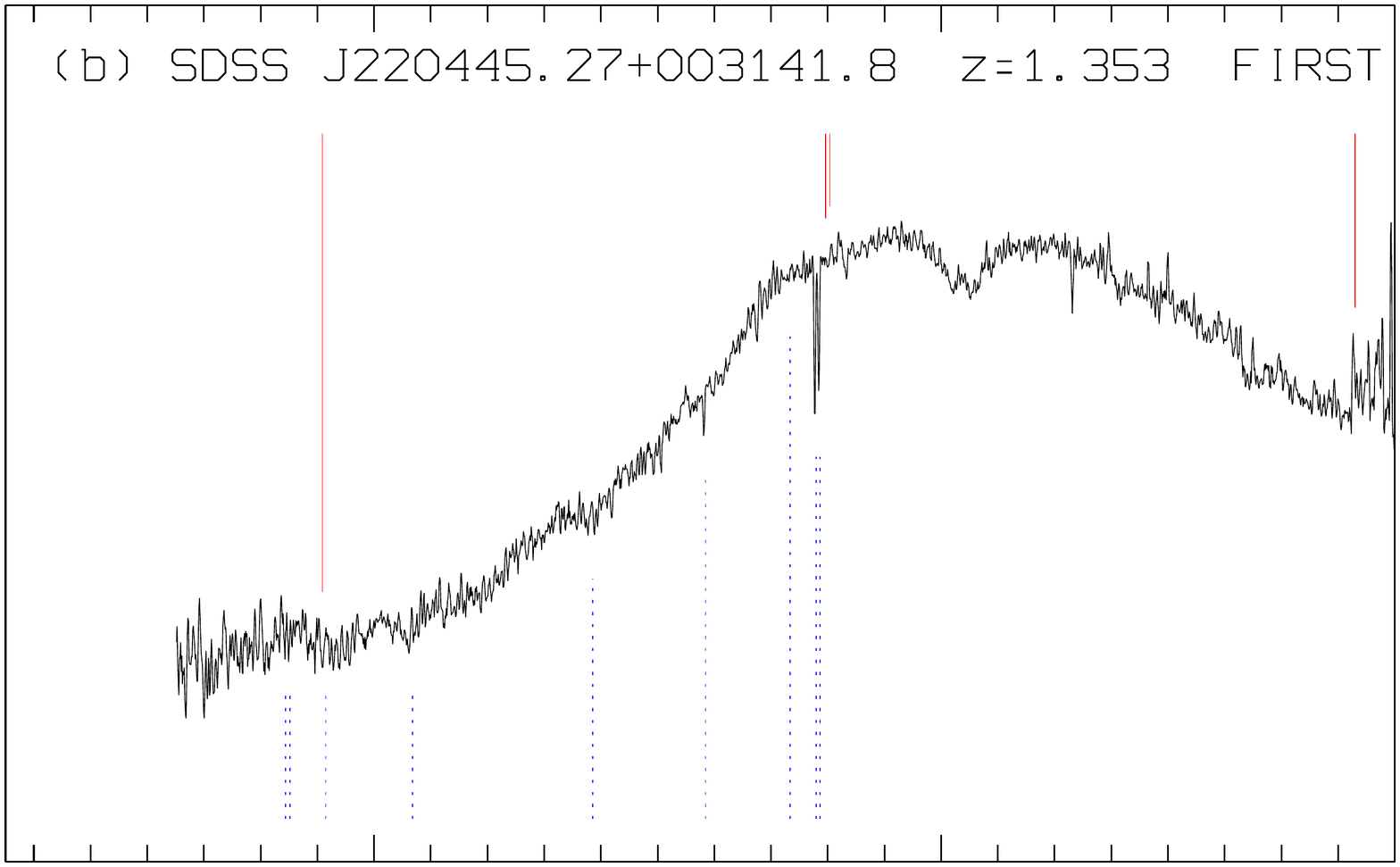}\hfill \=
\includegraphics[bb=53 20 500 770,scale=0.20,angle=270,clip]{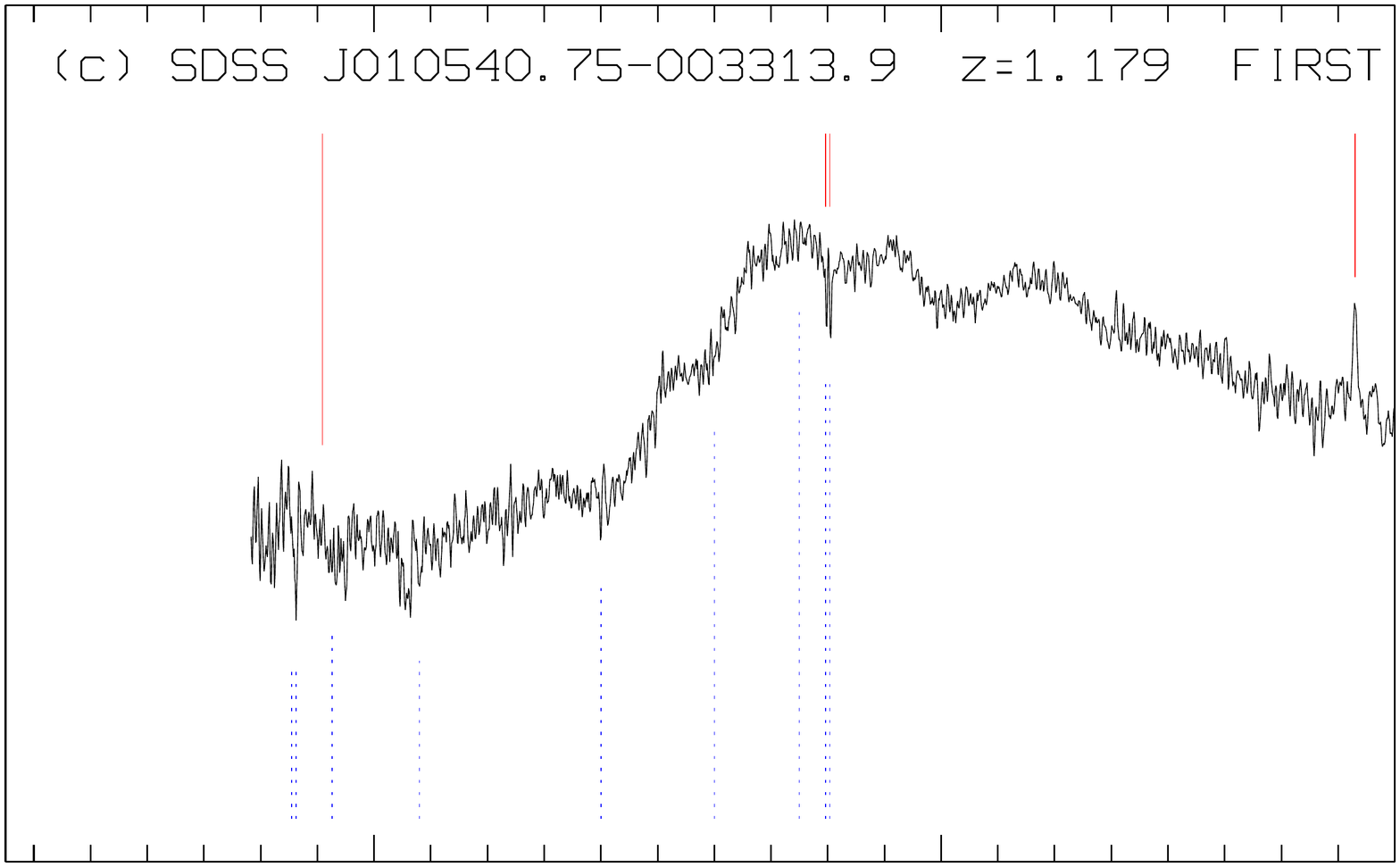}\hfill \\
\includegraphics[bb=53 00 500 770,scale=0.20,angle=270,clip]{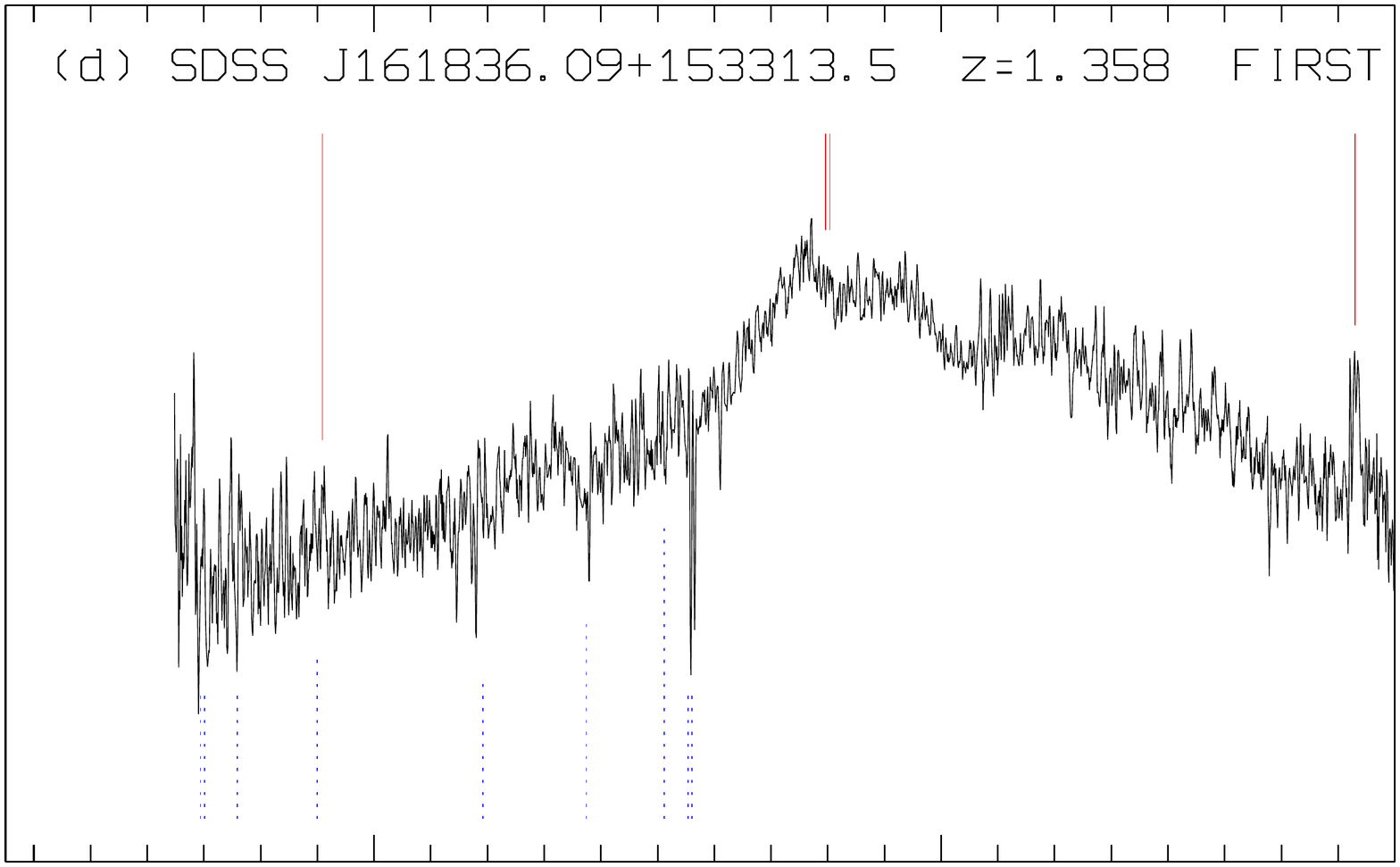}\hfill \=
\includegraphics[bb=53 20 500 770,scale=0.20,angle=270,clip]{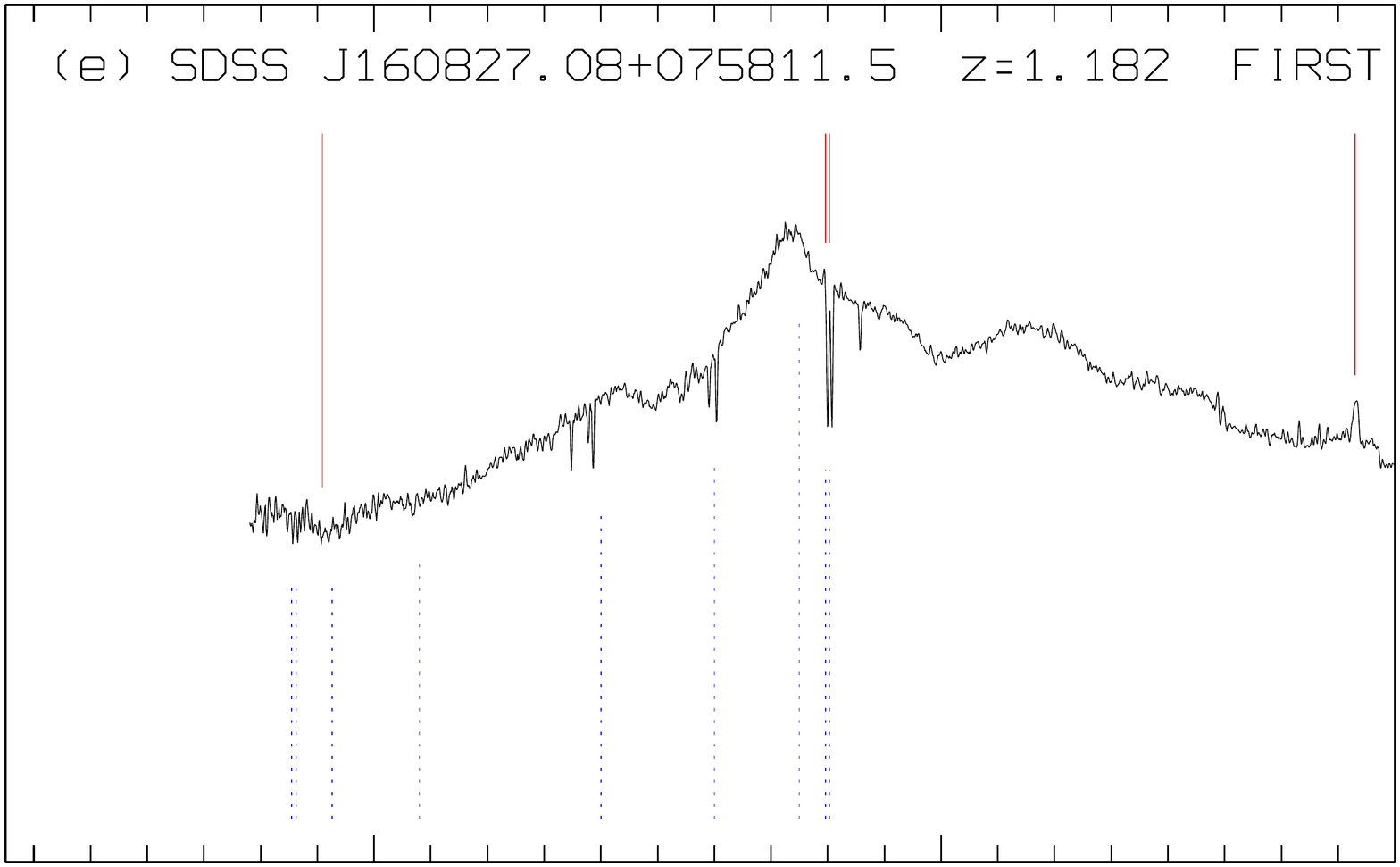}\hfill \=
\includegraphics[bb=53 20 500 770,scale=0.20,angle=270,clip]{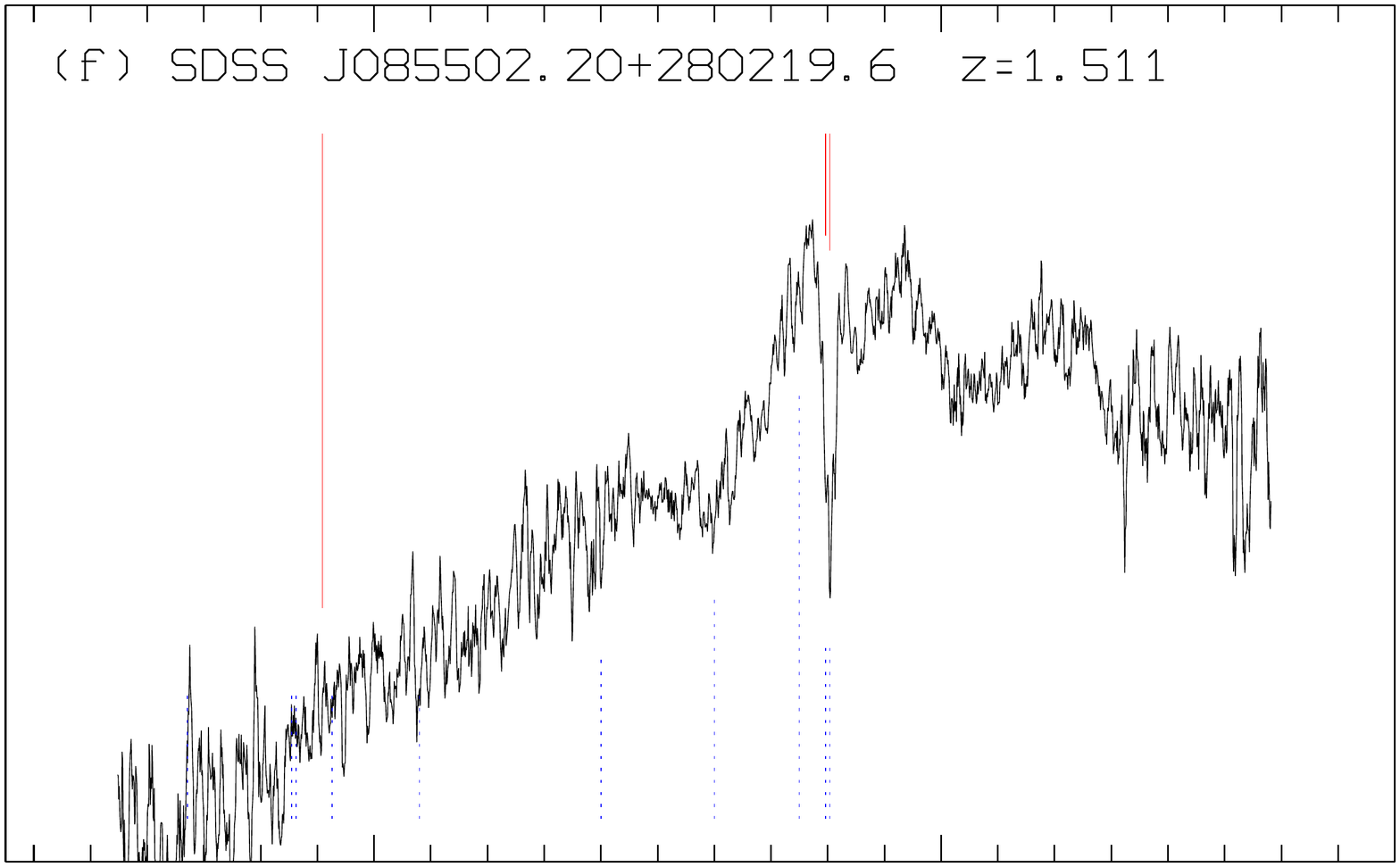}\hfill \\
\includegraphics[bb=53 00 500 770,scale=0.20,angle=270,clip]{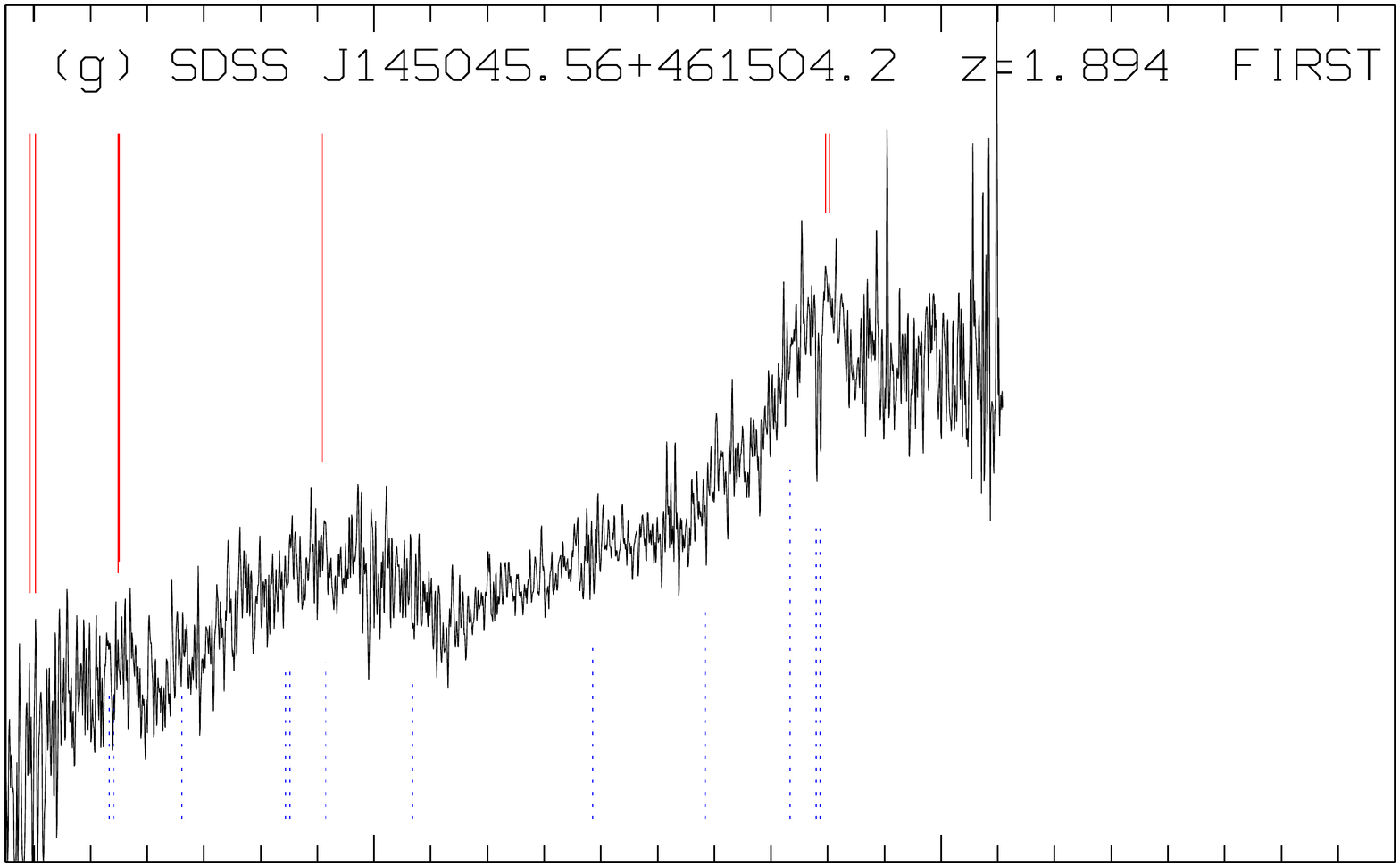}\hfill \=
\includegraphics[bb=53 20 500 770,scale=0.20,angle=270,clip]{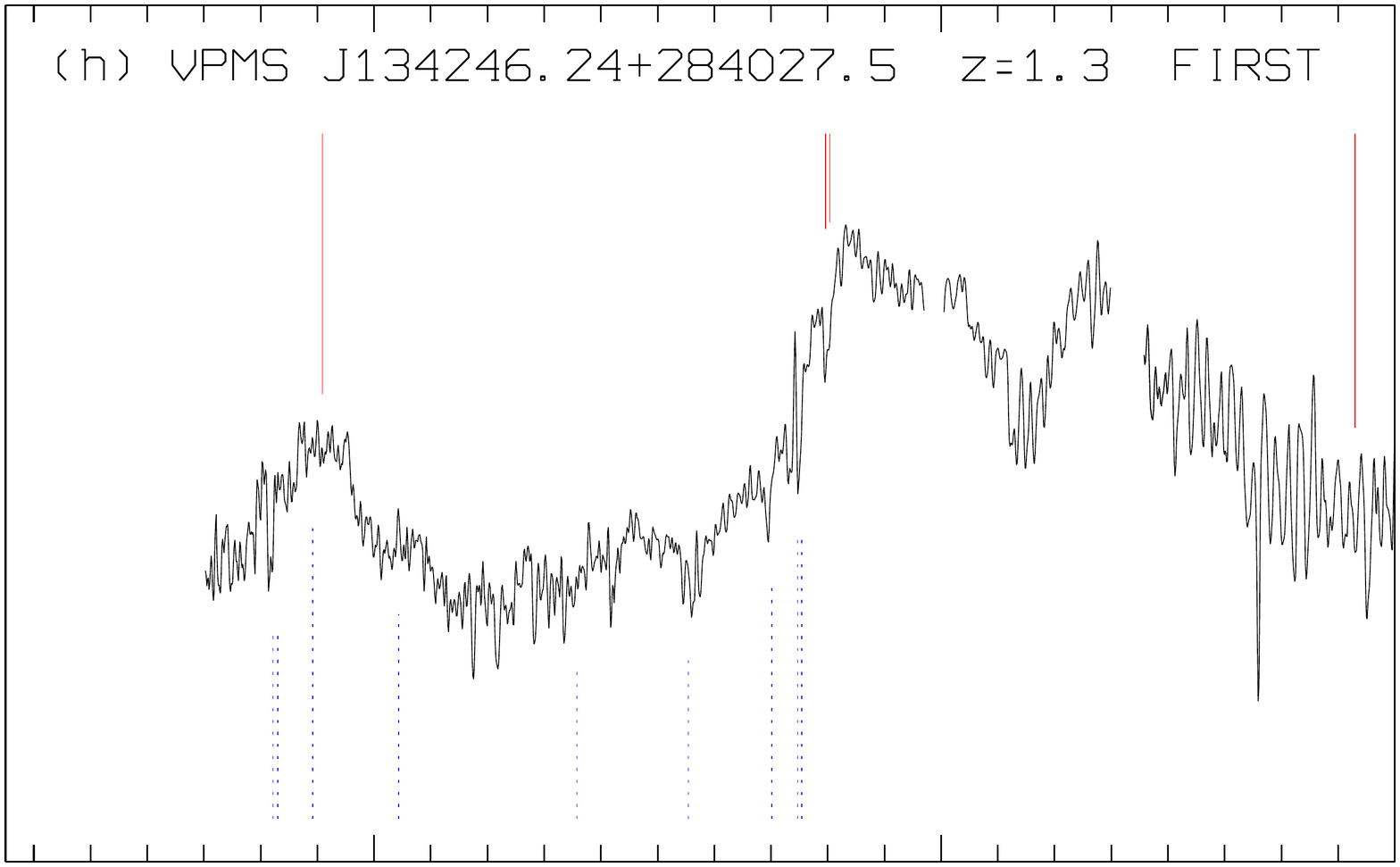}\hfill \=
\includegraphics[bb=53 20 500 770,scale=0.20,angle=270,clip]{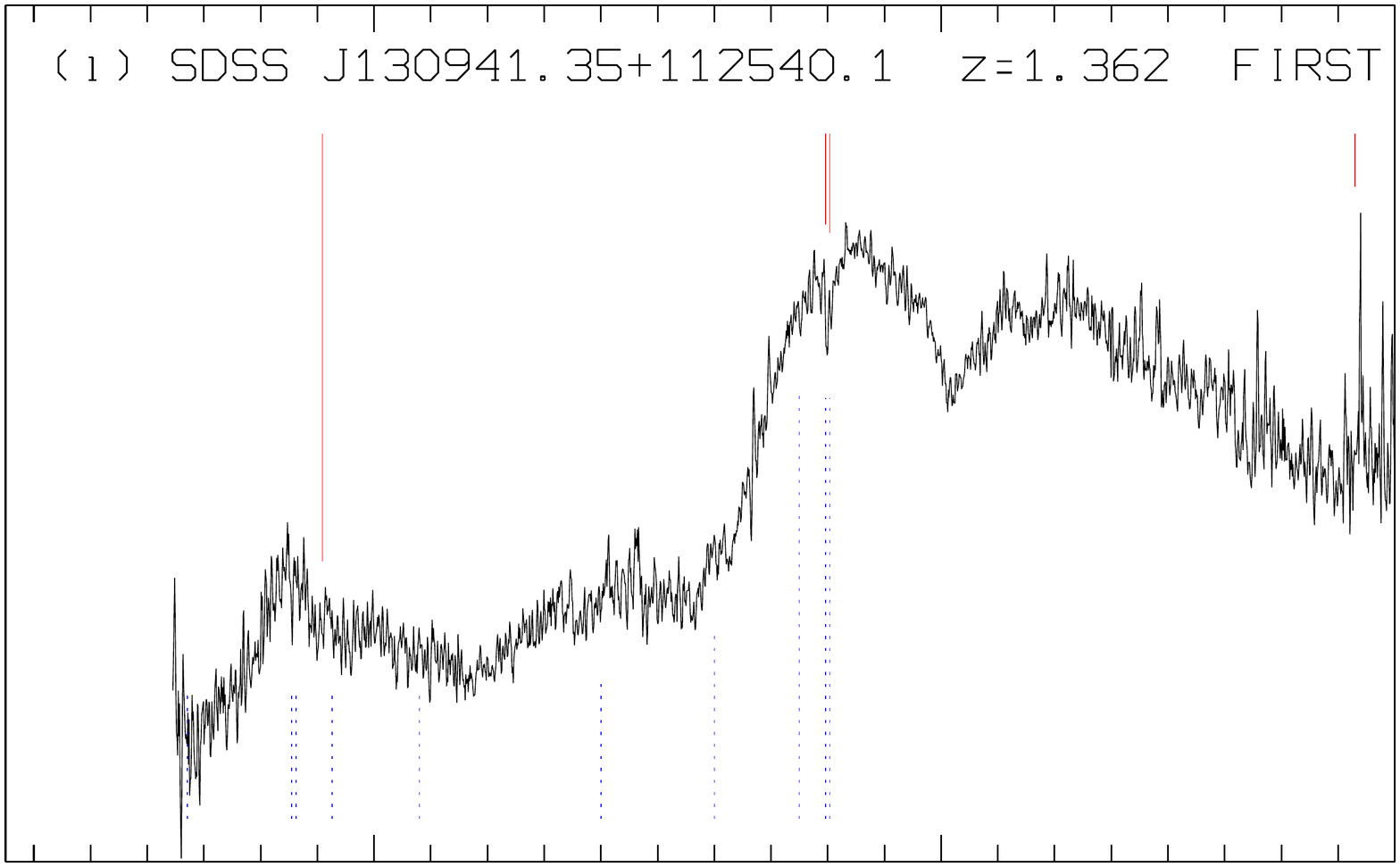}\hfill \\
\includegraphics[bb=53 00 500 770,scale=0.20,angle=270,clip]{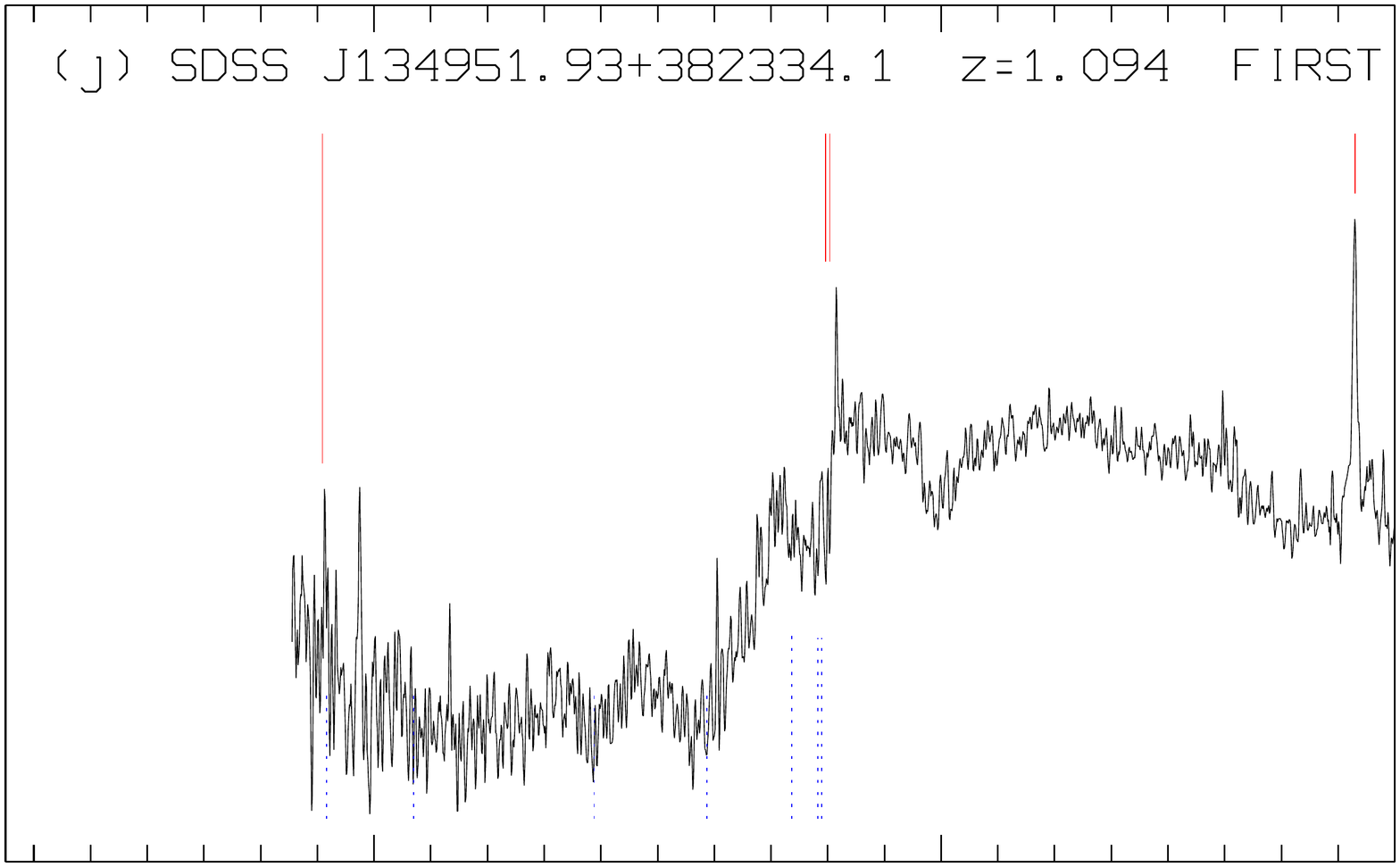}\hfill \=
\includegraphics[bb=53 20 500 770,scale=0.20,angle=270,clip]{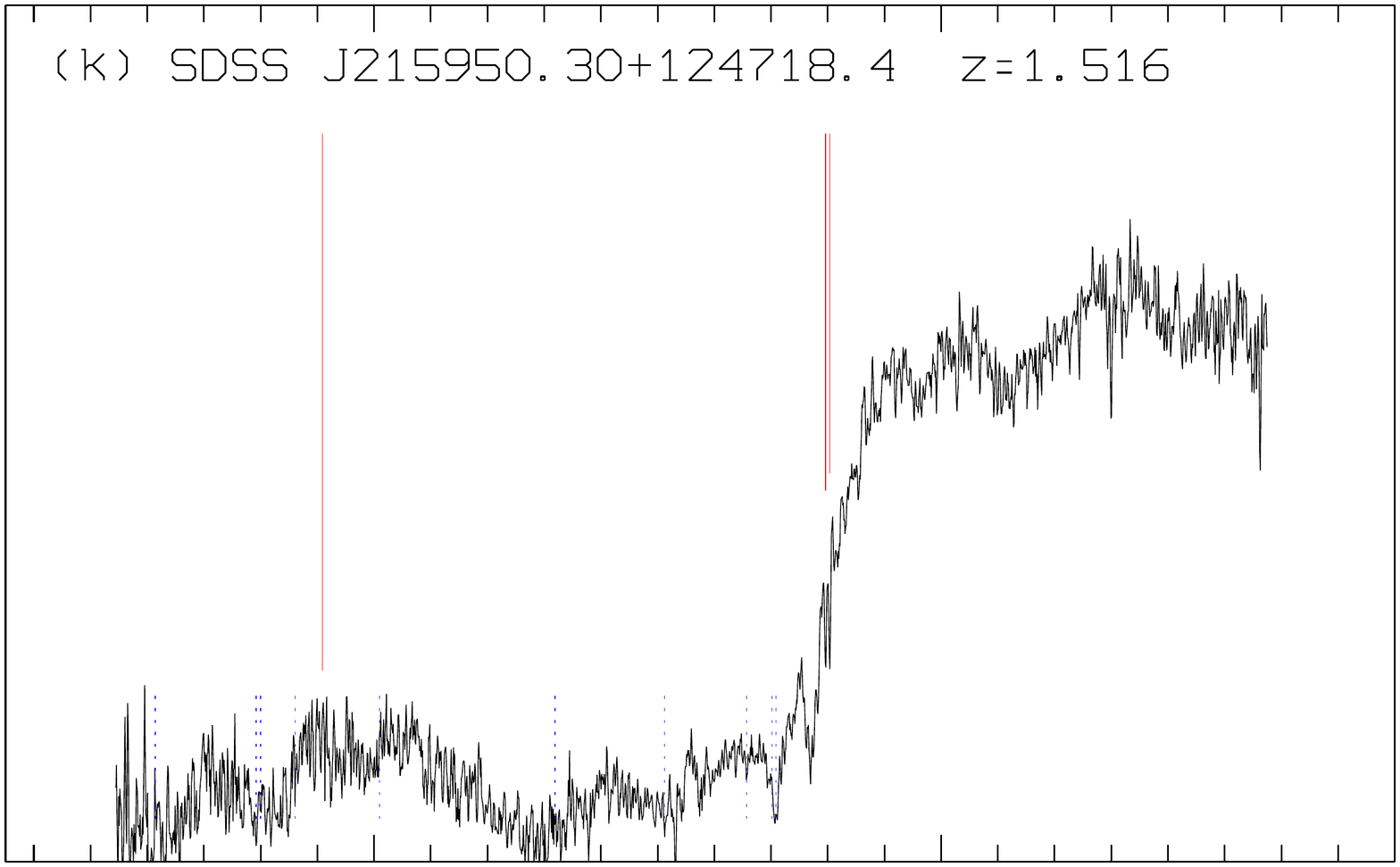}\hfill \=
\includegraphics[bb=53 20 500 770,scale=0.20,angle=270,clip]{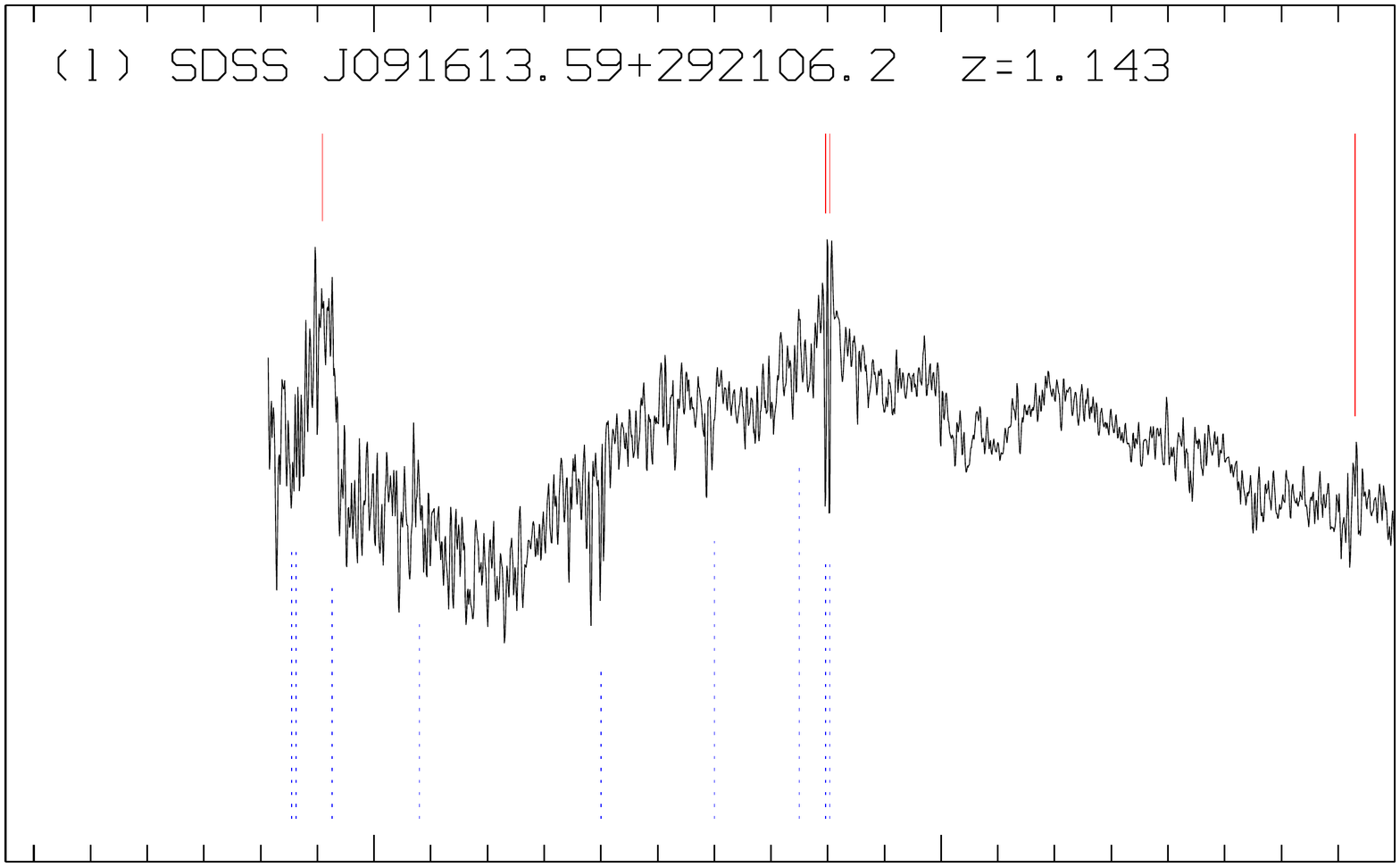}\hfill \\
\includegraphics[bb=53 00 500 770,scale=0.20,angle=270,clip]{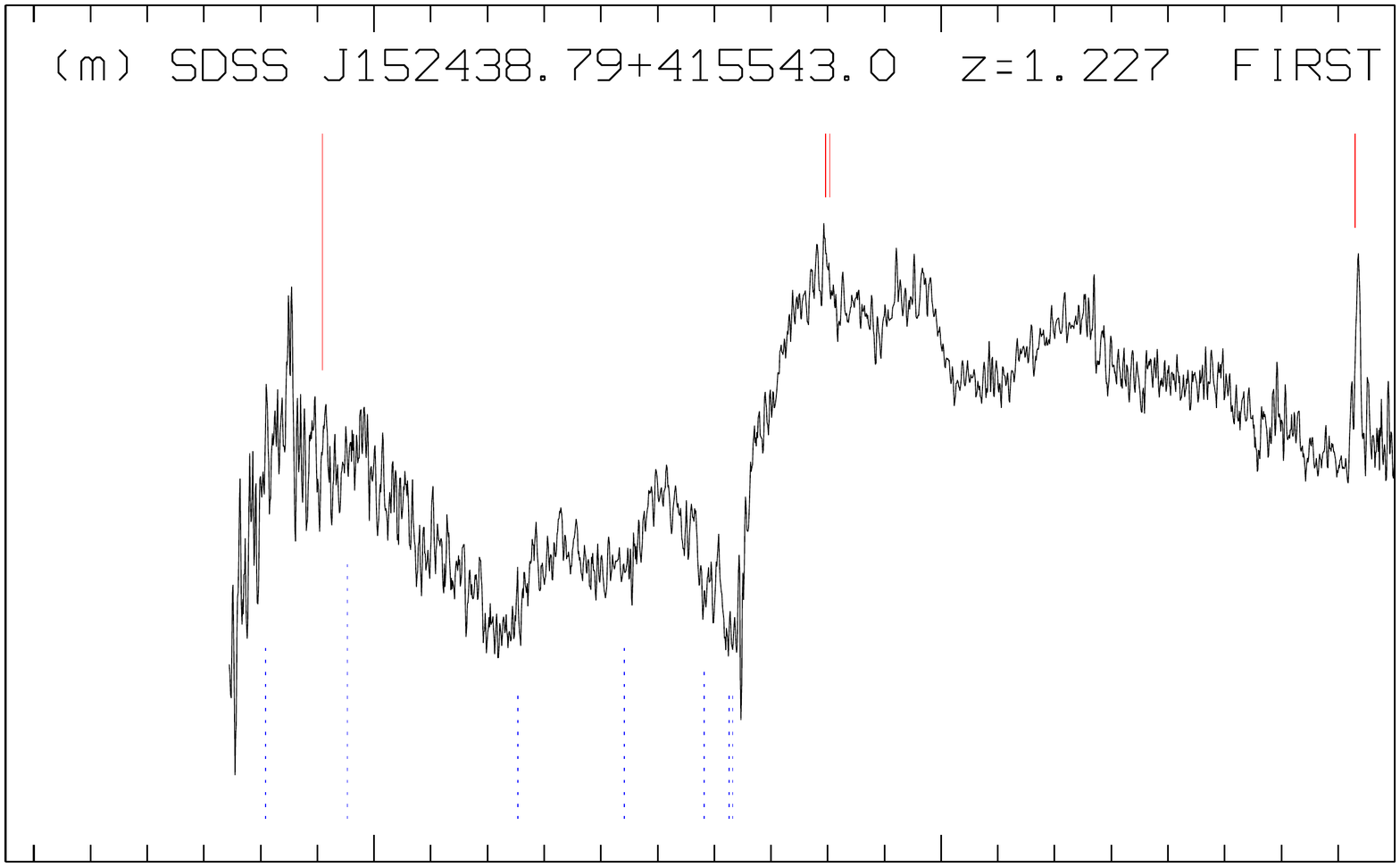}\hfill \=
\includegraphics[bb=53 20 500 770,scale=0.20,angle=270,clip]{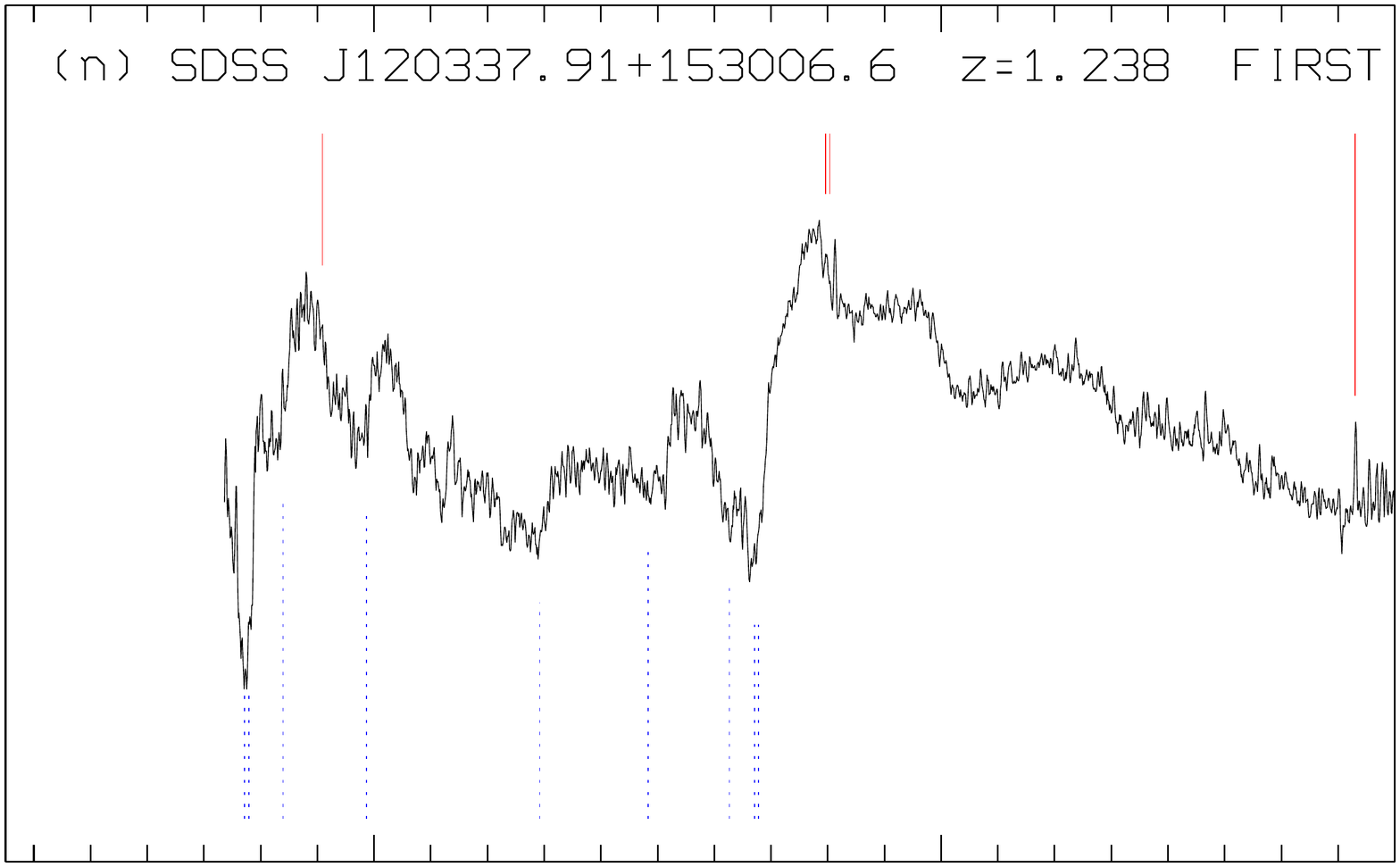}\hfill \=
\includegraphics[bb=53 20 500 770,scale=0.20,angle=270,clip]{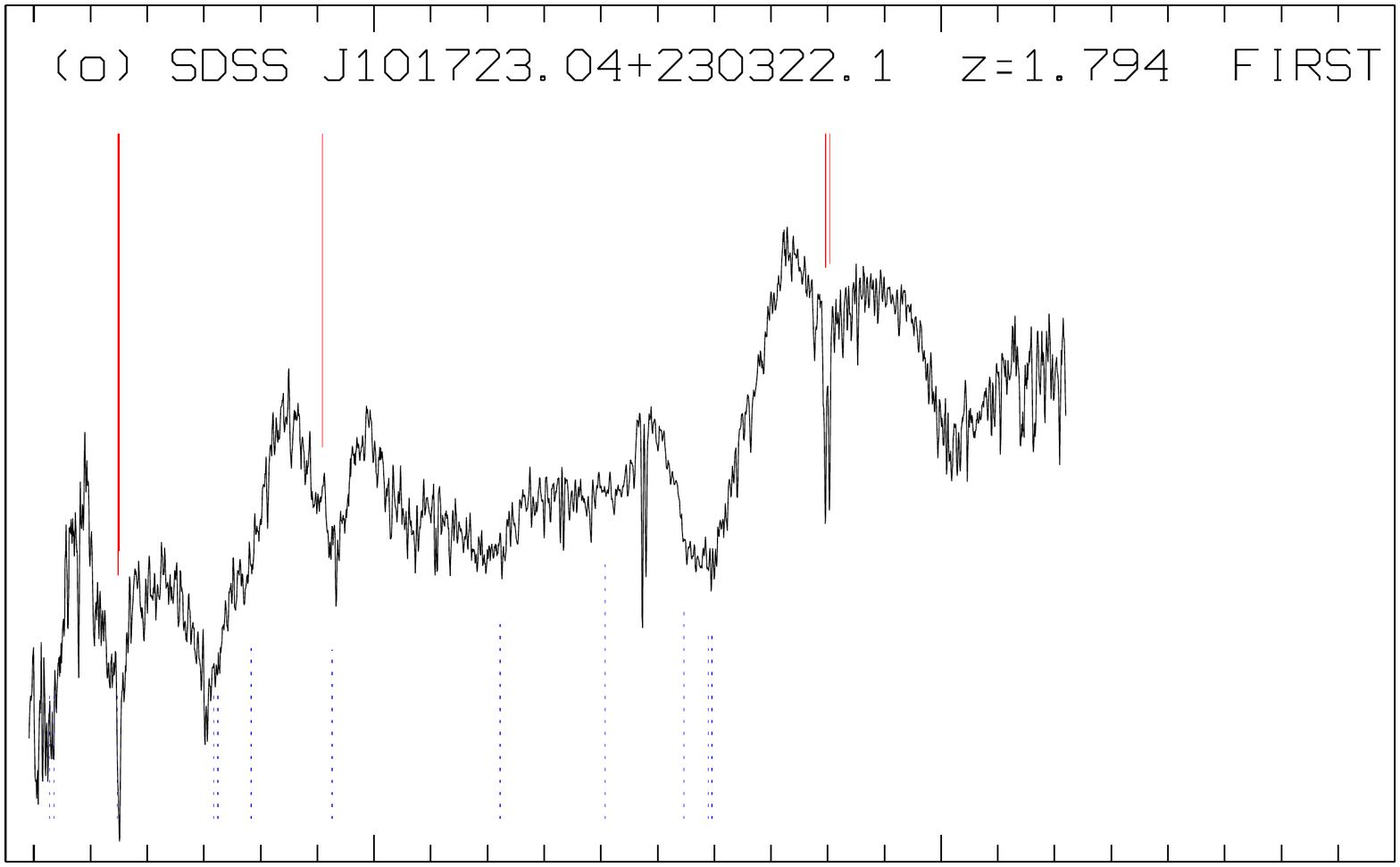}\hfill \\
\includegraphics[bb=53 00 500 770,scale=0.20,angle=270,clip]{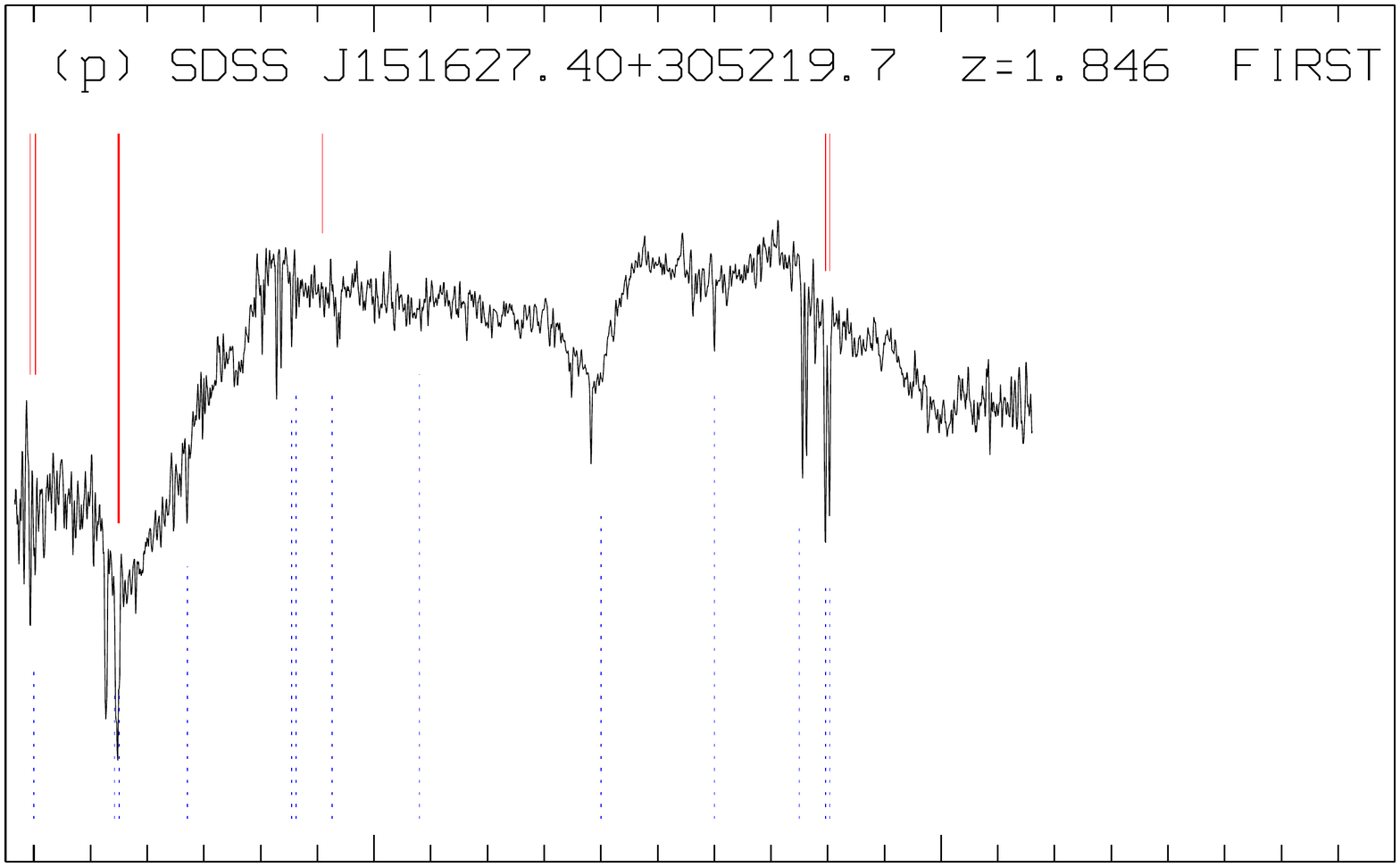}\hfill \=
\includegraphics[bb=53 20 500 770,scale=0.20,angle=270,clip]{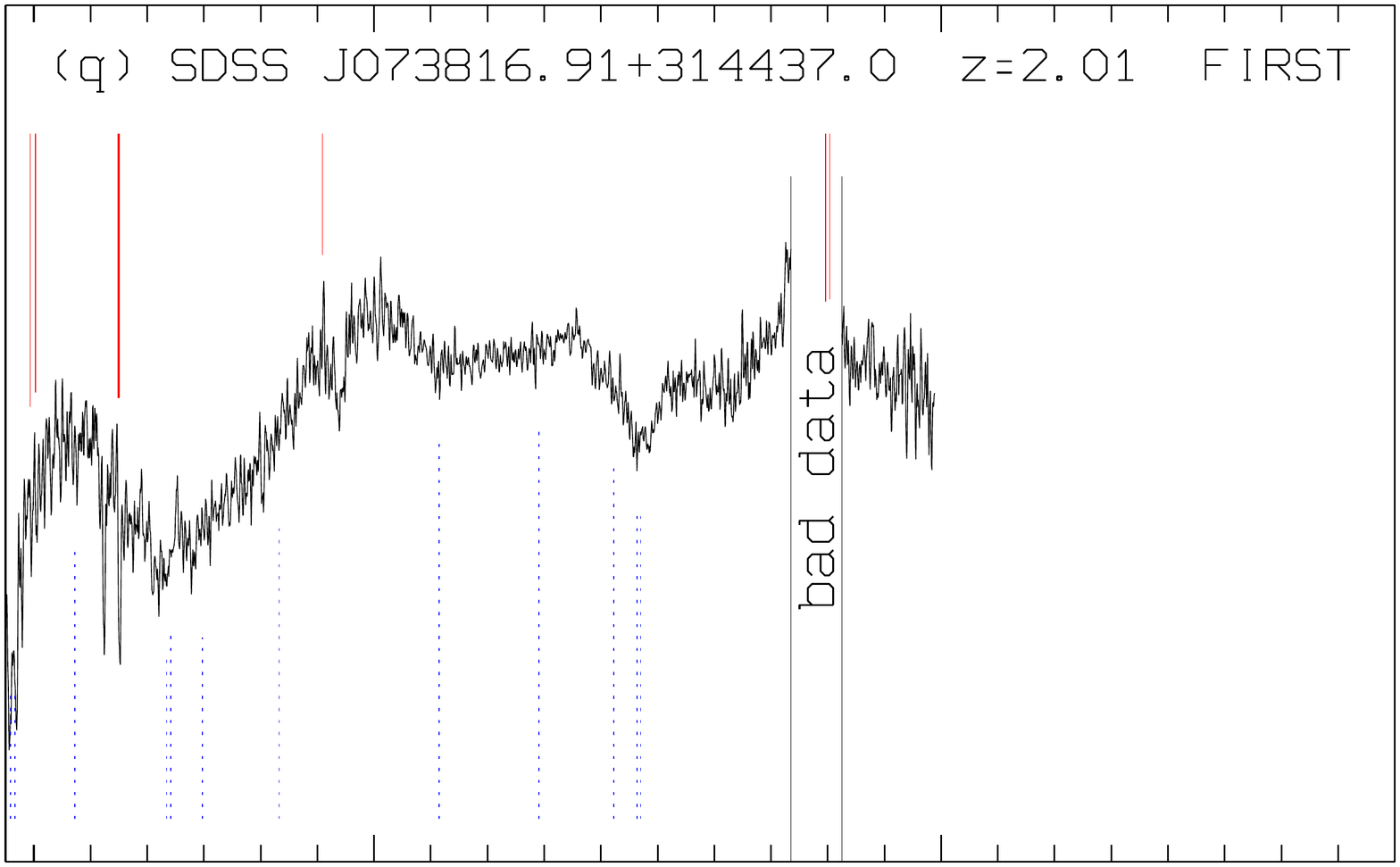}\hfill \=
\includegraphics[bb=53 20 500 770,scale=0.20,angle=270,clip]{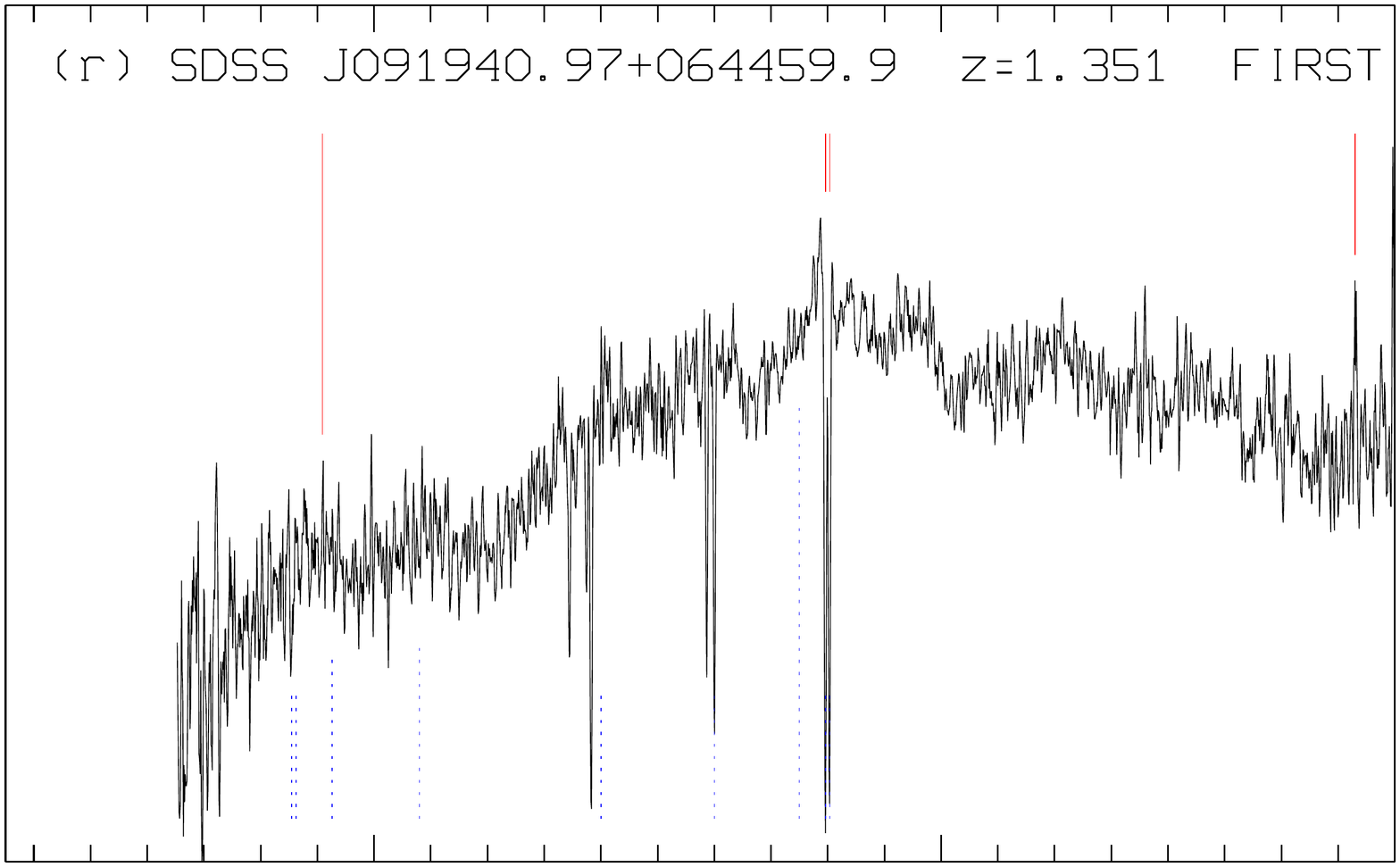}\hfill \\
\includegraphics[bb=53 00 570 770,scale=0.20,angle=270,clip]{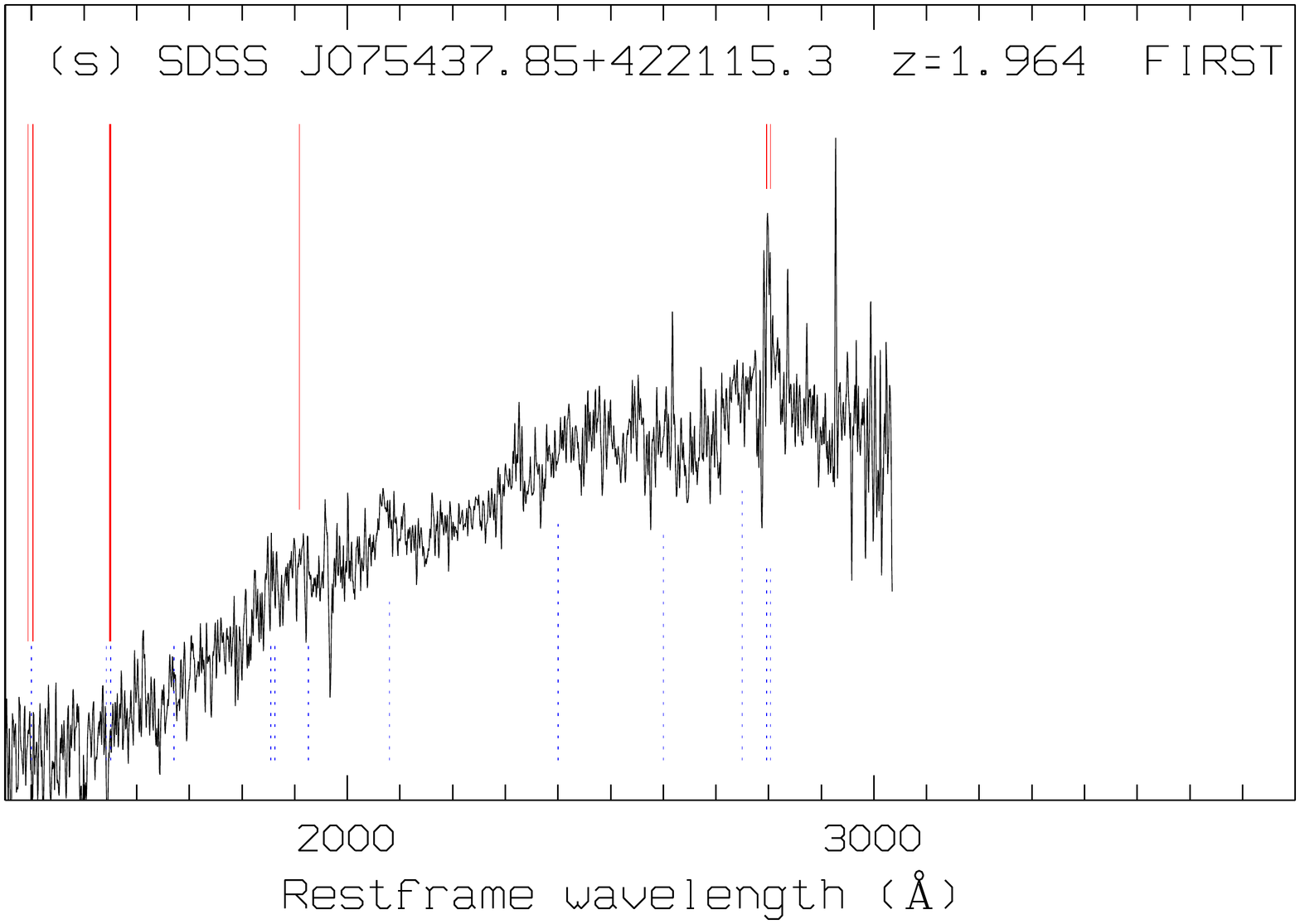}\hfill \=
\includegraphics[bb=53 20 570 770,scale=0.20,angle=270,clip]{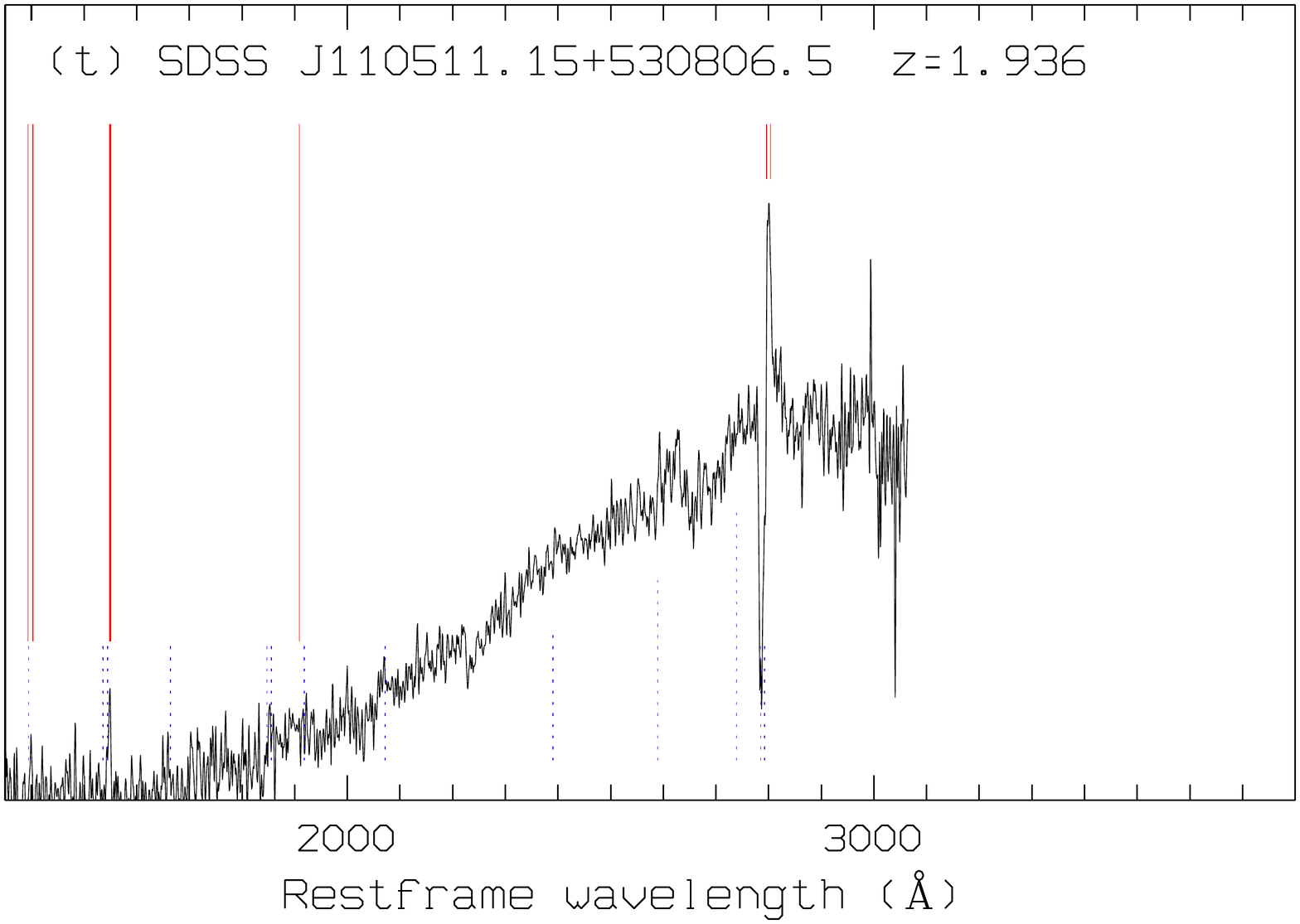}\hfill \=
\includegraphics[bb=53 20 570 770,scale=0.20,angle=270,clip]{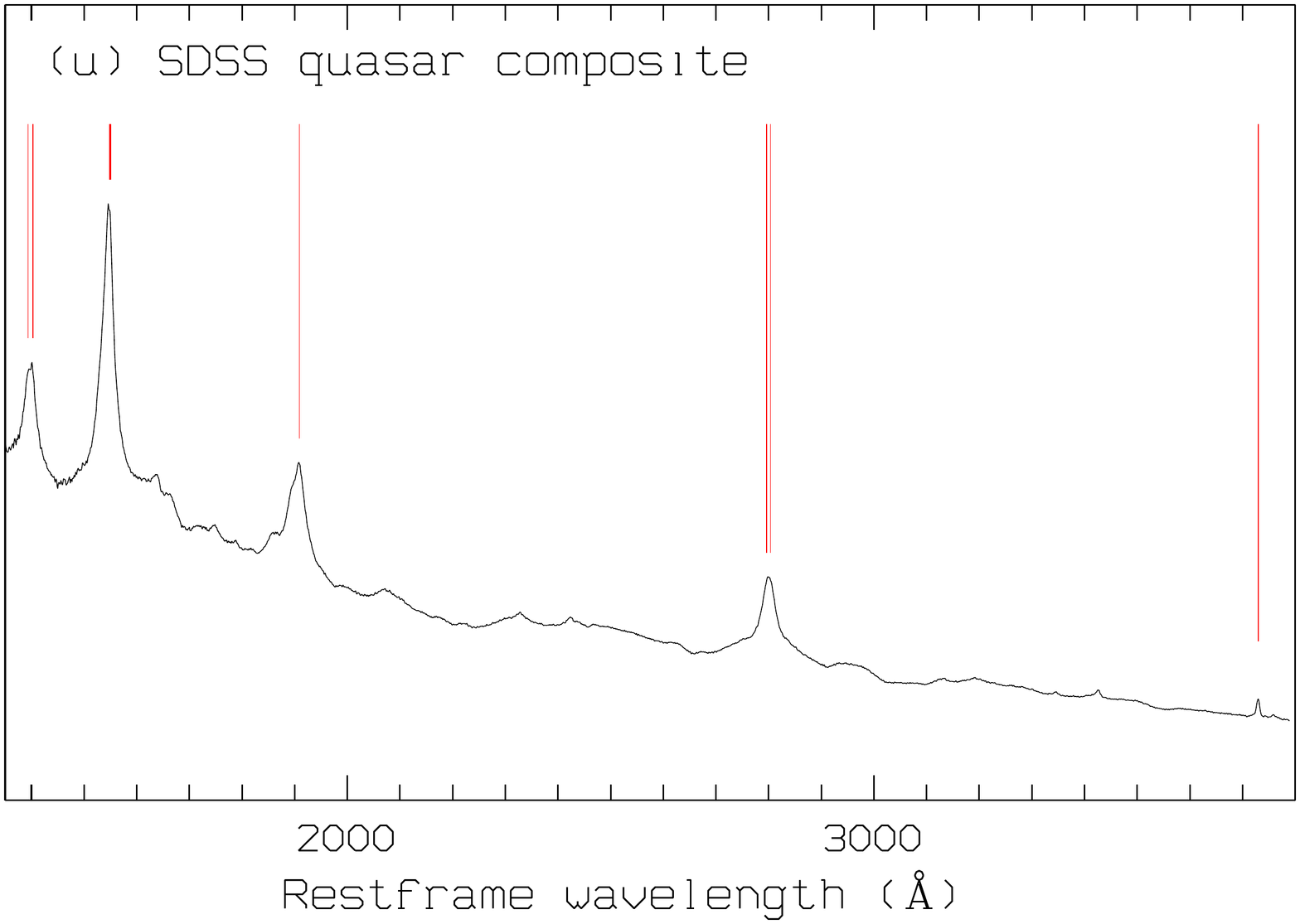}\hfill \\
\end{tabbing}
\caption{Partial spectra of 9 mysterious objects and 11 possibly related quasars.
Radio sources detected in FIRST are labeled. Vertical bars: positions 
of typical strong quasar emission (solid) and absorption (dashed) lines, 
respectively.}
\label{fig:mysts}
\end{figure*}

Hall et al. (\cite{Hall02}) discussed at length two objects discovered by the
SDSS with spectra classified by these authors as ``mysterious'':
\object{SDSS J$010540.75-003313.9$} and \object{SDSS J220445.27+003141.8}
(Fig.\,\ref{fig:mysts}{\bf b},{\bf c}). The most prominent spectral features are:
{\it (a)} a lack of substantial typical quasar emission lines
(except broad \ion{Fe}{ii} emission and [\ion{O}{ii}] $\lambda$3730 in
the case of \object{SDSS J$010540.75-003313.9$});
{\it (b)} a blue continuum longwards of $\sim 3200$\AA;
{\it (c)} a dip at $\sim 3000$\AA;
{\it (d)} a continuum drop-off shortwards of
\ion{Mg}{ii} $\lambda$2800 which appears too steep to be 
caused by dust reddening; and
{\it (e)} no obvious BAL troughs. 
Both objects show associated \ion{Mg}{ii} absorption lines, 
are unresolved FIRST radio sources, and are much more luminous than any 
galaxy. Hall et al. discussed various viable explanations for these spectra
and concluded that none of them are particularly satisfactory.
These strange objects may be very unusual BAL quasars with partial covering 
of different regions of the continuum source as a function of velocity, 
probably in combination with moderate reddening.

Regardless of the physical reason for these mysterious spectra, they almost
certainly represent quasar types that are extremely rare in presently
available samples. Hall et al. (\cite{Hall02}) were aware of only two objects 
with probably similar spectra, the low-$z$  quasars \object{FBQS 1503+2330} 
($z=0.40$) and \object{FBQS 1055+3124} ($z=0.49$) from the FIRST Bright 
Quasar Survey (White et al. \cite{White00}).
In a search for BL\,Lac objects, Plotkin et al. (\cite{Plotkin08})
revealed another two FIRST sources with SDSS spectra similar to the two 
Hall et al. objects
\object{SDSS J130941.35+112540.1} and \object{SDSS J145045.56+461504.2}.
Neither of these two new objects ultimately survived as BL\,Lac
candidates during the manual inspection by these authors.
The first similar object that was not targeted as a radio
source was \object{VPMS J134246.24+284027.5}, which was discovered during
a variability and (zero-) proper
motion selection of quasar candidates (Meusinger et al. \cite{Meusinger05}). 
All seven of these objects have counterparts in the FIRST images with radio fluxes
on the mJy level and moderate or small radio-loudness parameters.
 
Browsing through our sample (Tab.\,\ref{tab:catalogue}) yields another three 
comparable SDSS objects and about one dozen possibly related objects with spectral
properties that are partly similar to those of the mysterious objects.
In addition, we checked a subsample of the \textsc{\texttt{unknowns}} 
(\textsc{\texttt{spec\_cln}}=0) in the DR7. Only $\sim 10^3$ spectra 
of relatively high S/N were clustered in one single SOM.
The resulting icon map was used solely for selecting spectra that resemble
those of the mysterious objects mentioned above. Two additional objects of
this type were found:
the mysterious object \object{SDSS J010540.75$-$003313.9} from Hall et al.
(\cite{Hall02}) and \object{SDSS J073816.91+314437.0}, which was discussed
by Hall et al. (\cite{Hall04}).\footnote{Another highly peculiar
object from this SOM is \object{SDSS J164941.87+401455.9}. This spectrum has
a similar type to \object{SDSS J033810.85+005617.6} from Hall et al. (\cite{Hall02}).}

\begin{figure*}[hbtp]   
\begin{tabbing}
\includegraphics[bb=53 00 500 770,scale=0.20,angle=270,clip]{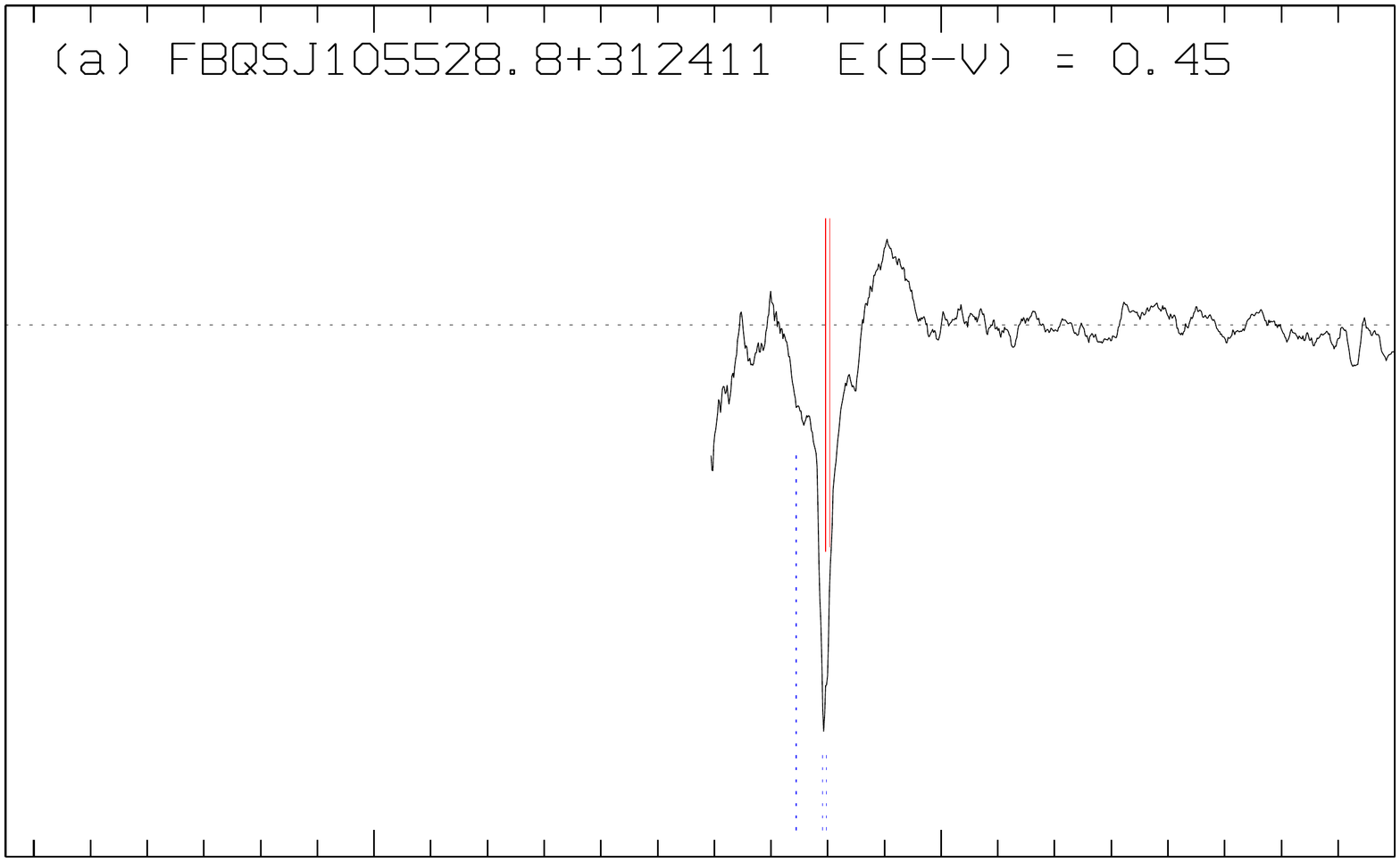}\hfill \=
\includegraphics[bb=53 20 500 770,scale=0.20,angle=270,clip]{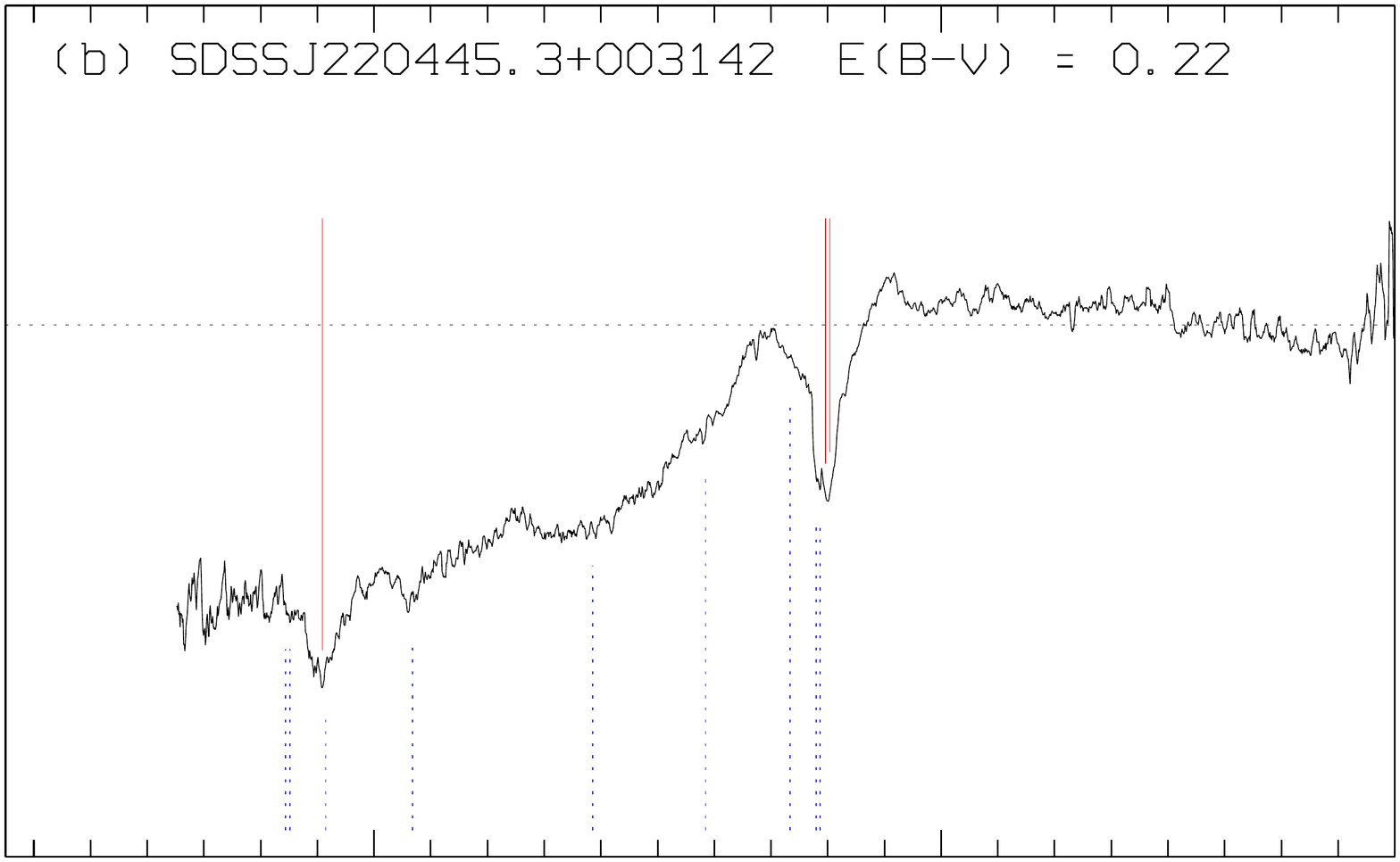}\hfill \=
\includegraphics[bb=53 20 500 770,scale=0.20,angle=270,clip]{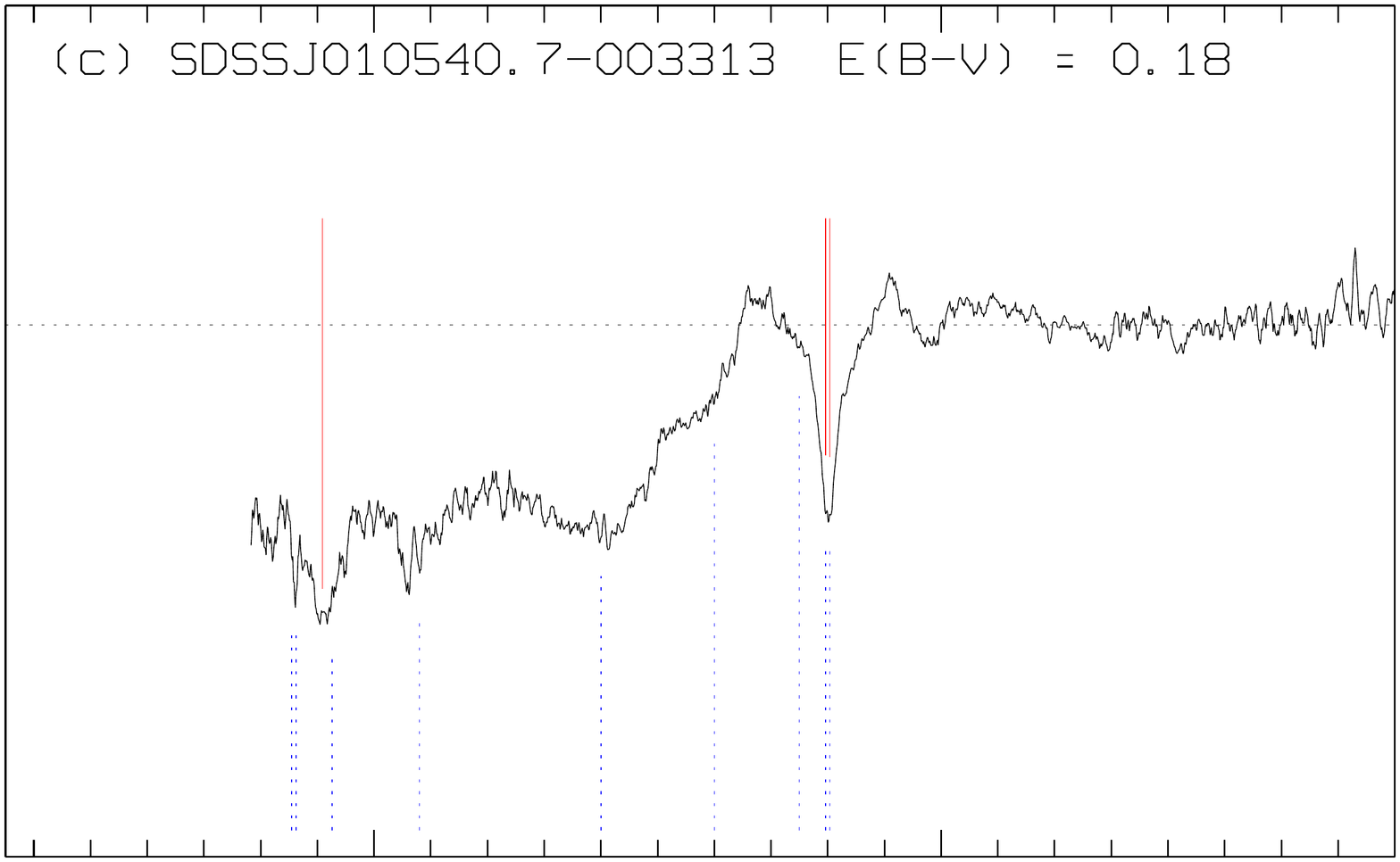}\hfill \\
\includegraphics[bb=53 00 500 770,scale=0.20,angle=270,clip]{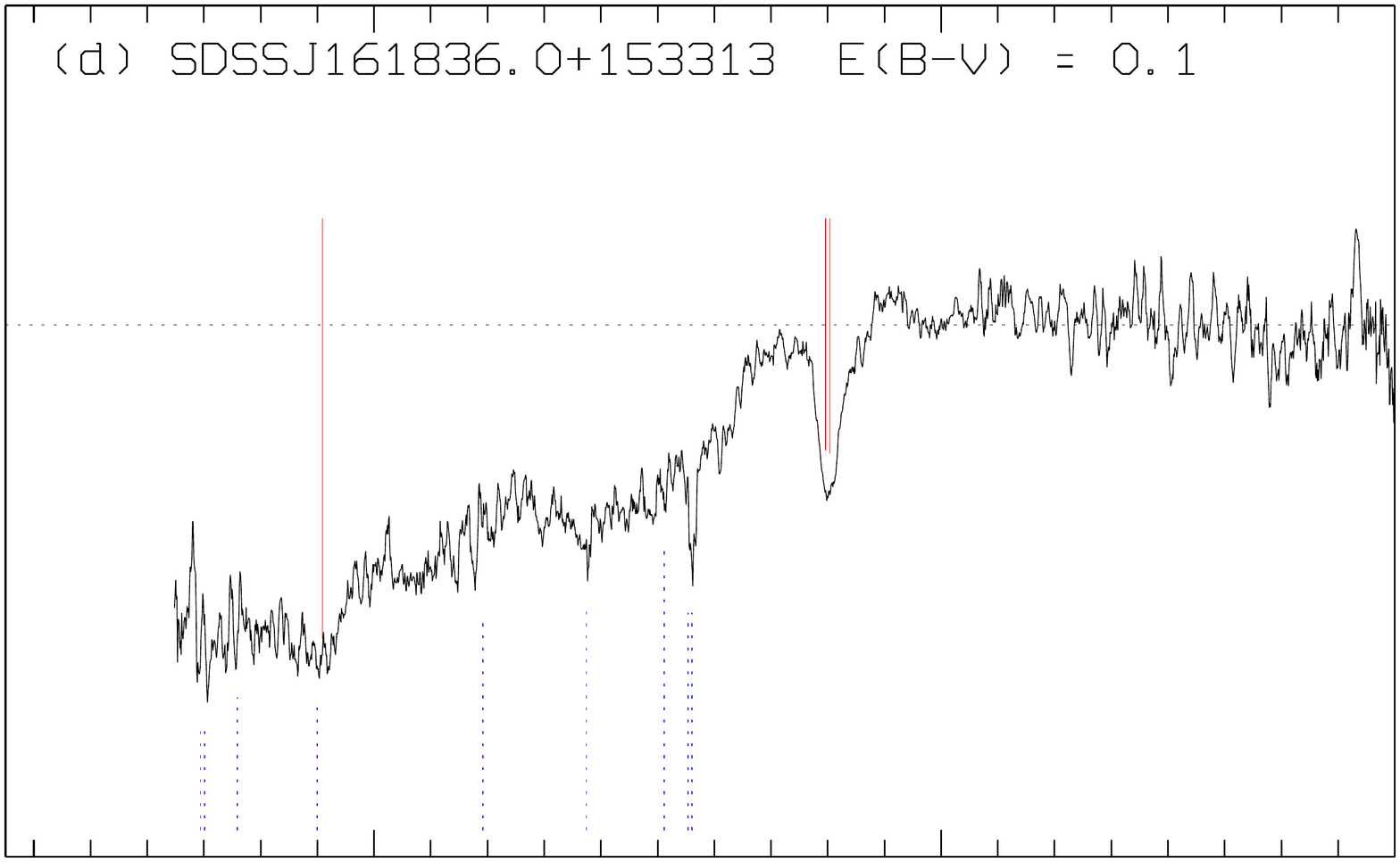}\hfill \=
\includegraphics[bb=53 20 500 770,scale=0.20,angle=270,clip]{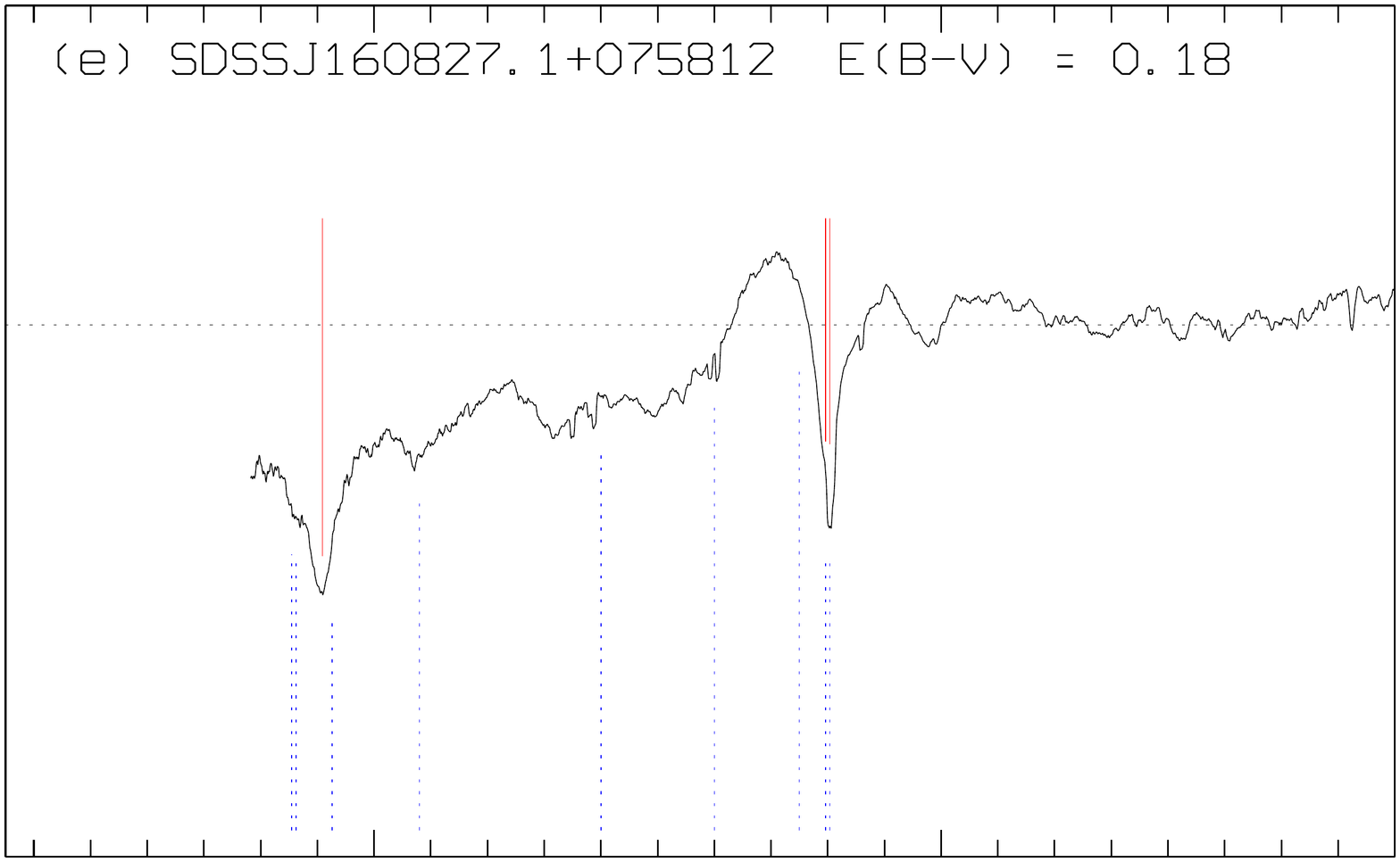}\hfill \=
\includegraphics[bb=53 20 500 770,scale=0.20,angle=270,clip]{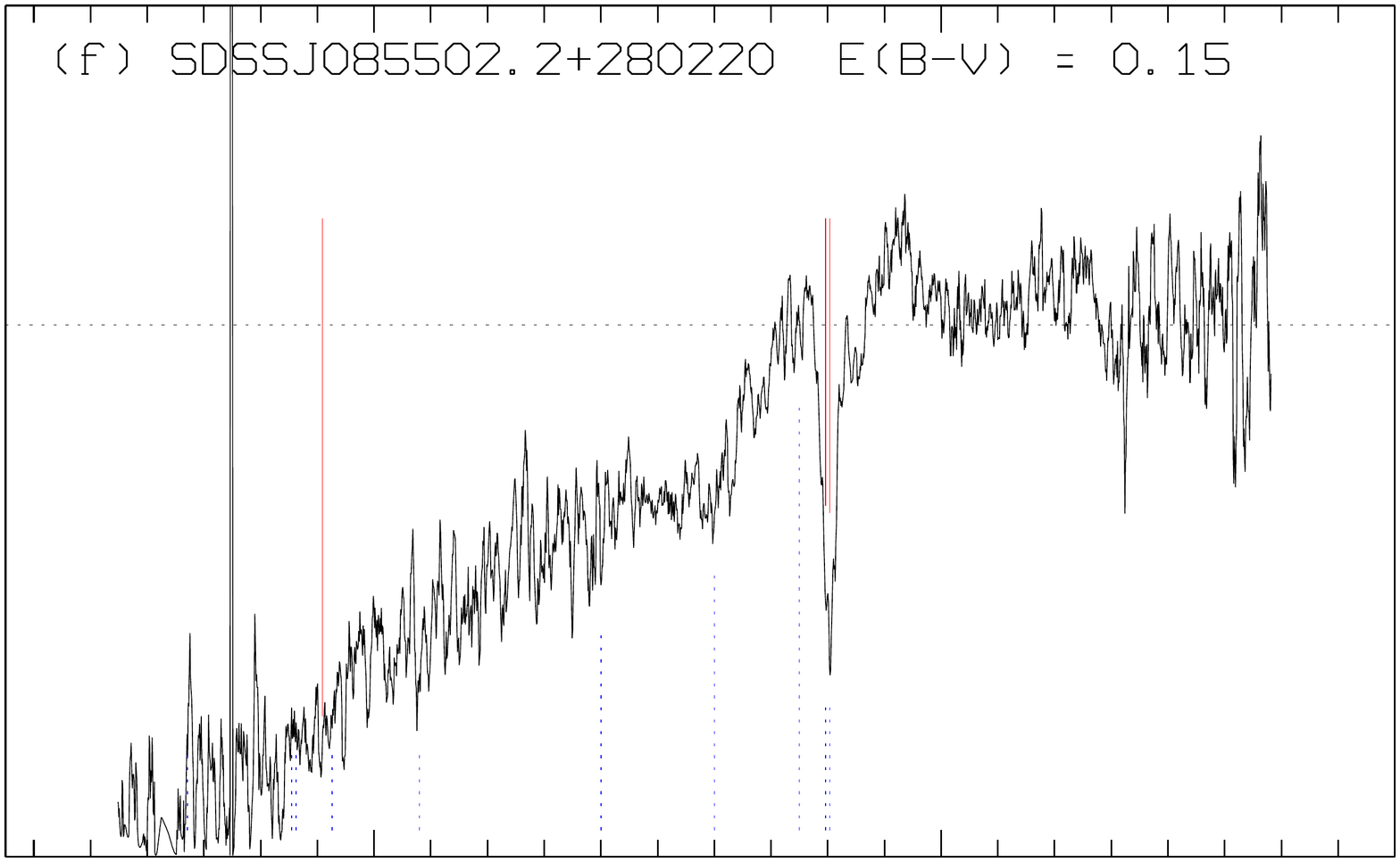}\hfill \\
\includegraphics[bb=53 00 500 770,scale=0.20,angle=270,clip]{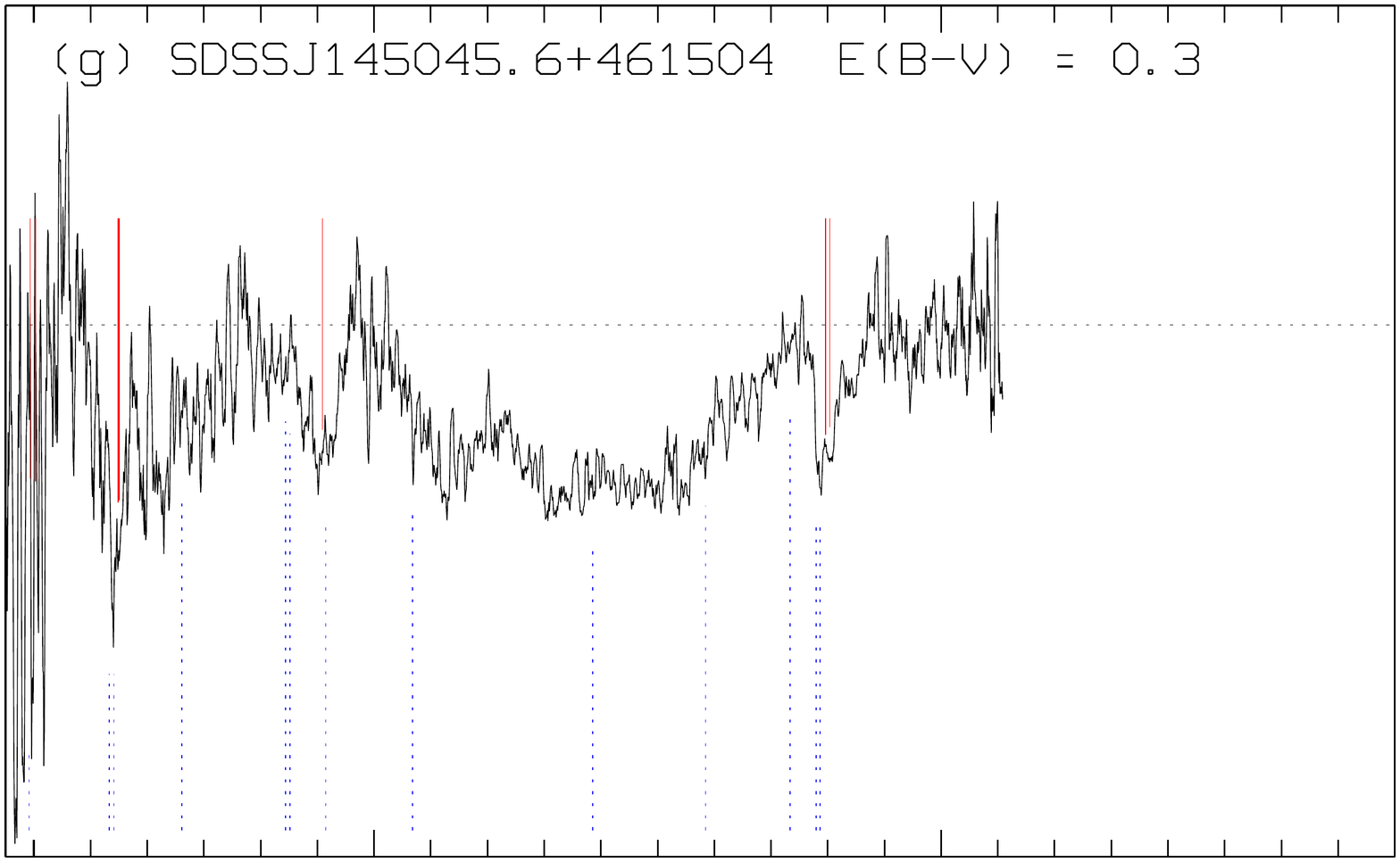}\hfill \=
\includegraphics[bb=53 20 500 770,scale=0.20,angle=270,clip]{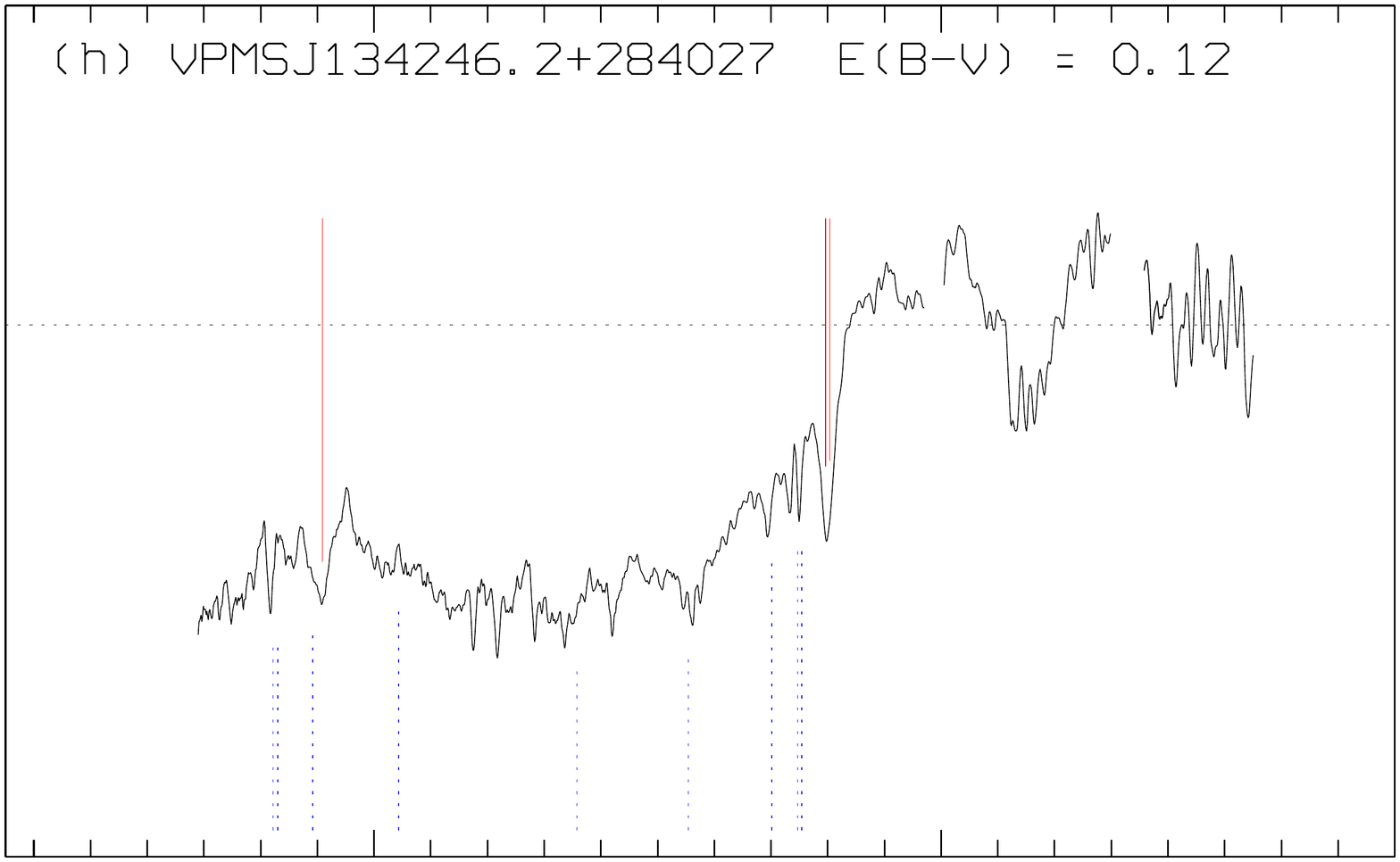}\hfill \=
\includegraphics[bb=53 20 500 770,scale=0.20,angle=270,clip]{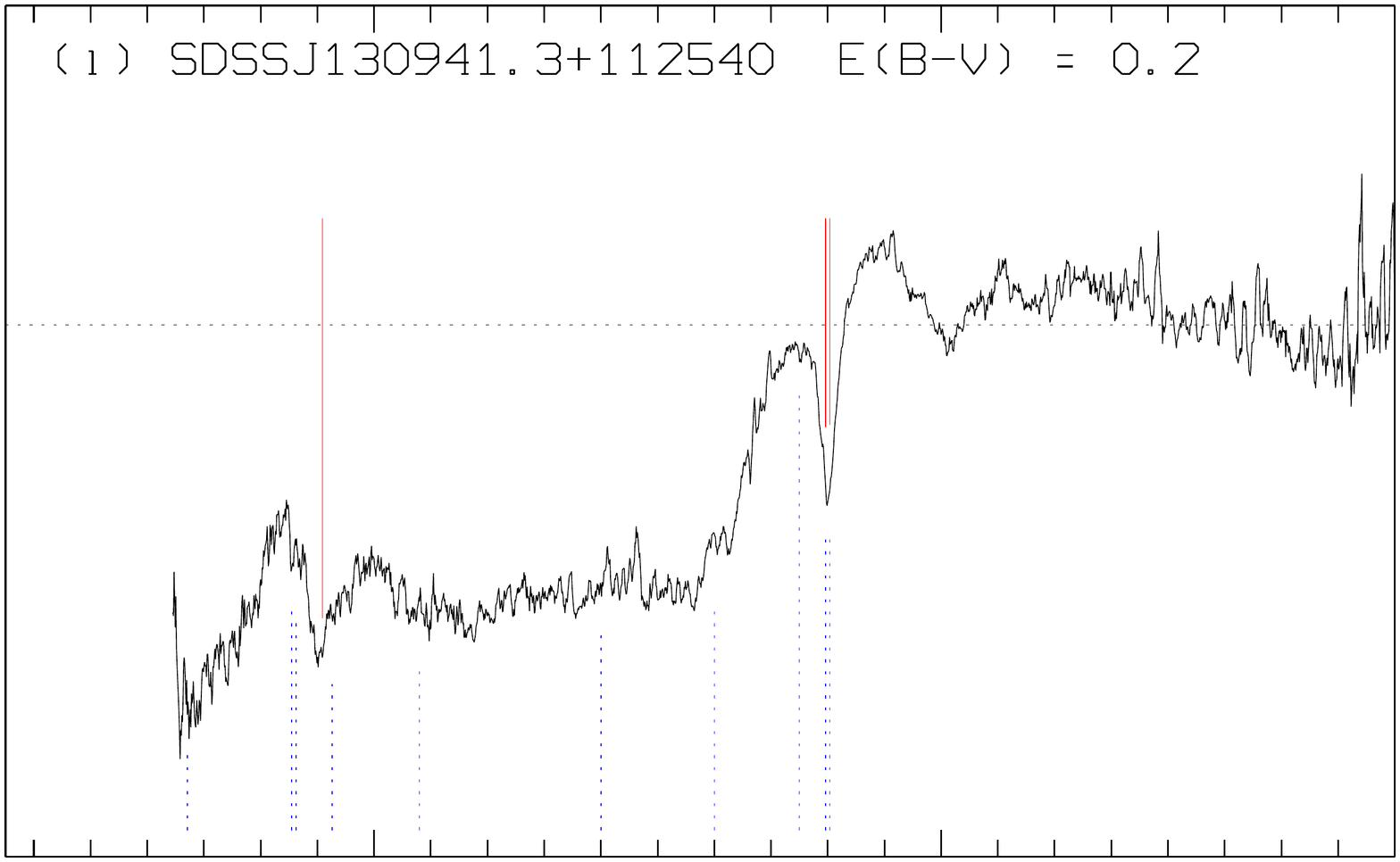}\hfill \\
\includegraphics[bb=53 00 500 770,scale=0.20,angle=270,clip]{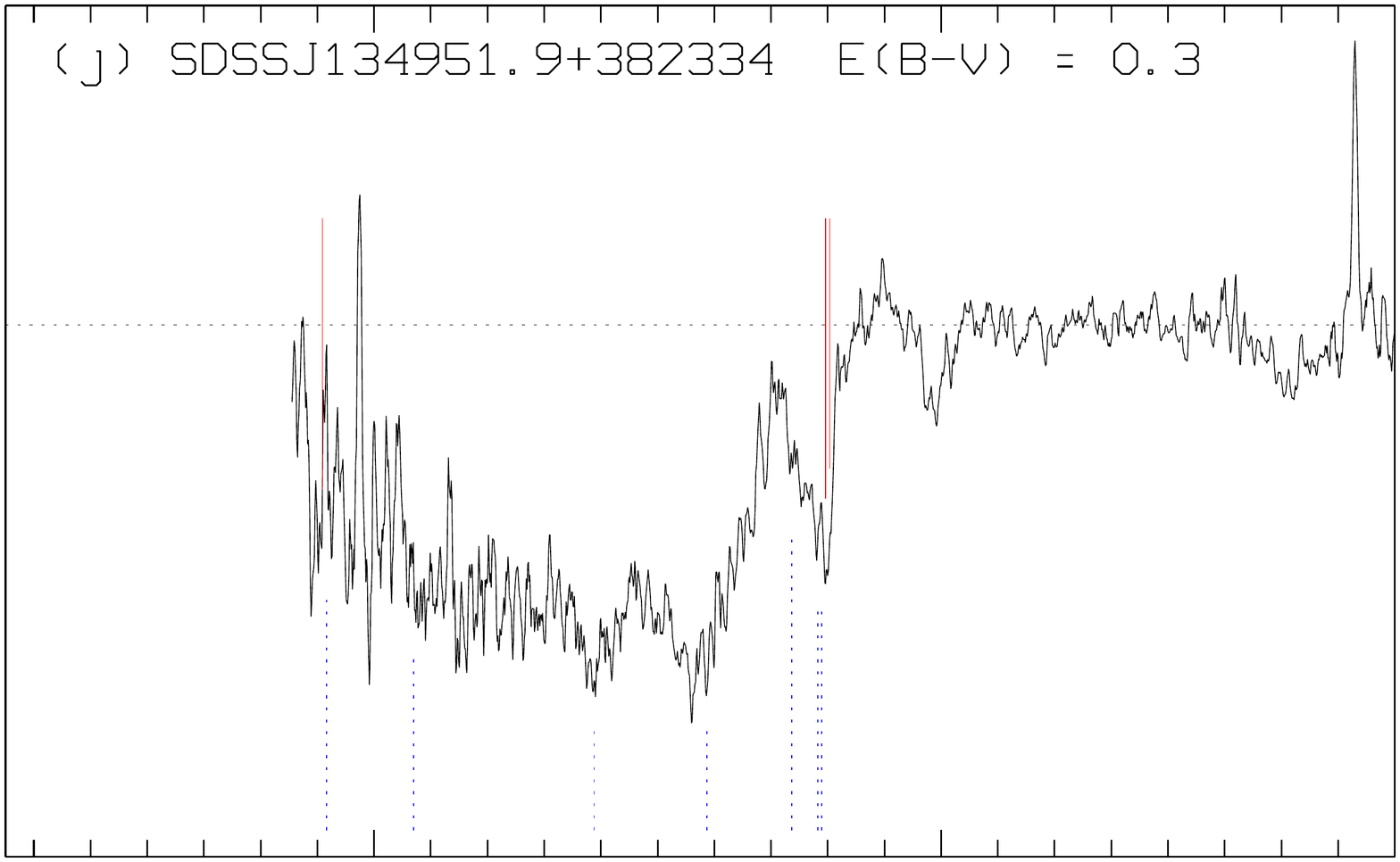}\hfill \=
\includegraphics[bb=53 20 500 770,scale=0.20,angle=270,clip]{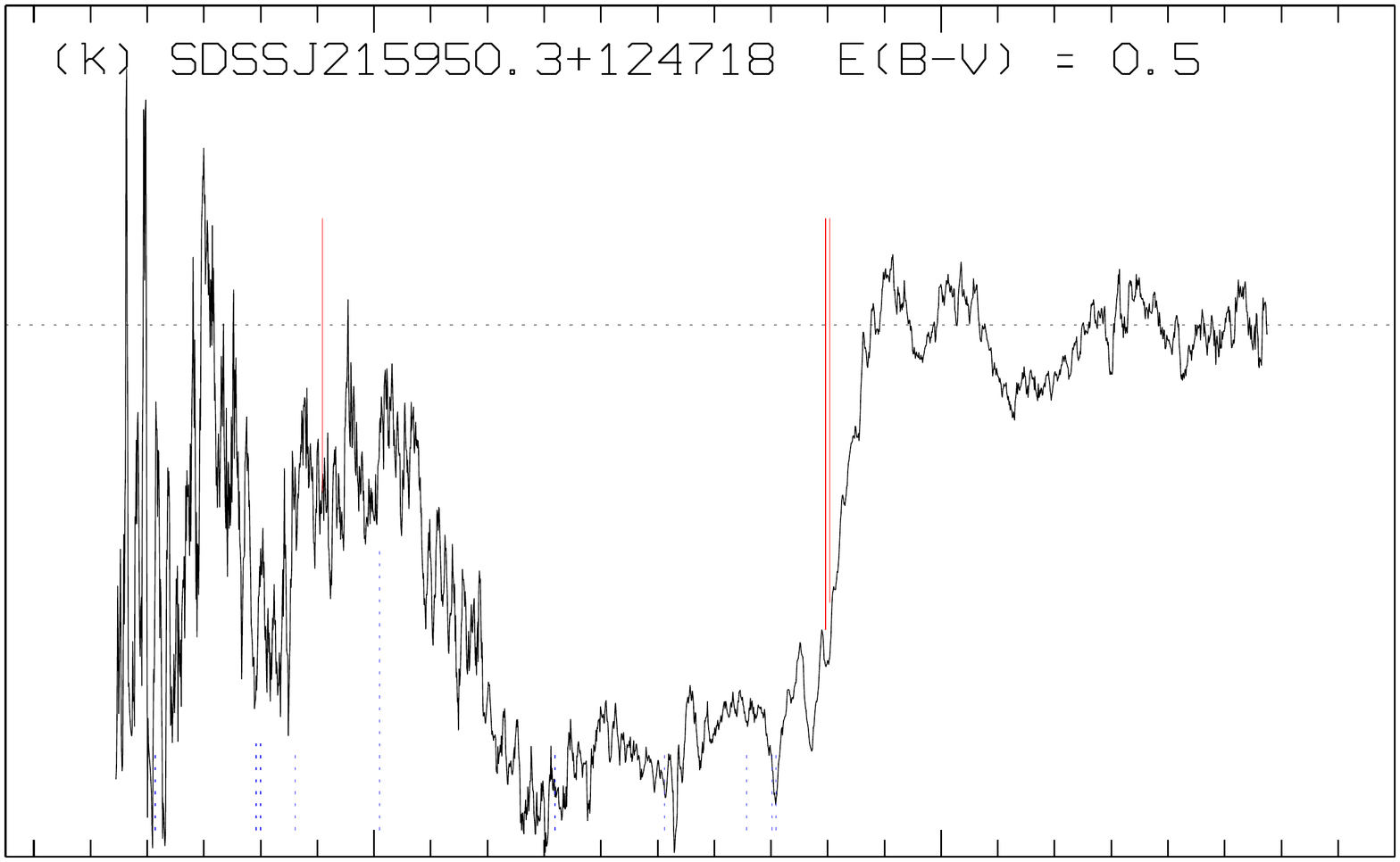}\hfill \=
\includegraphics[bb=53 20 500 770,scale=0.20,angle=270,clip]{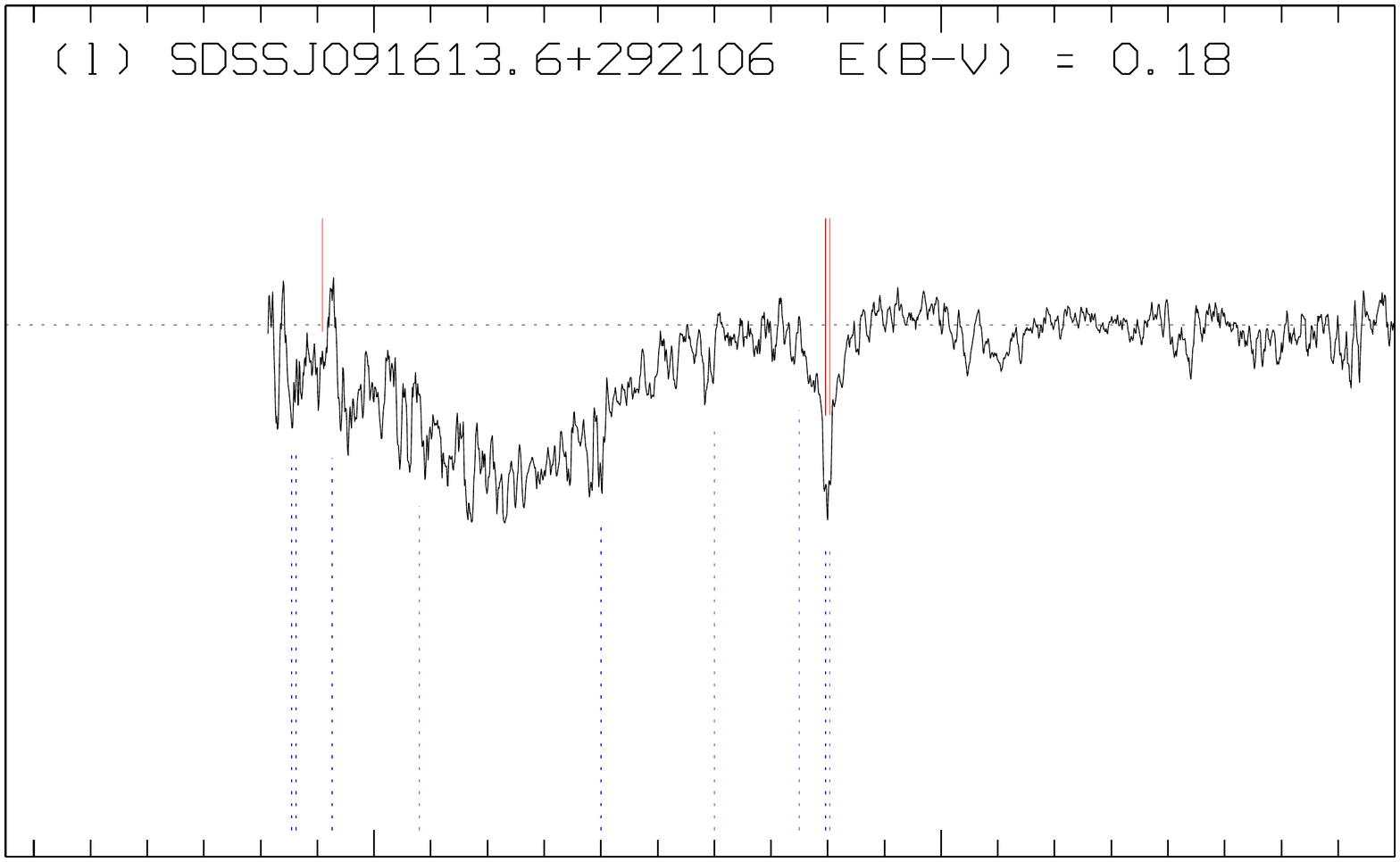}\hfill \\
\includegraphics[bb=53 00 500 770,scale=0.20,angle=270,clip]{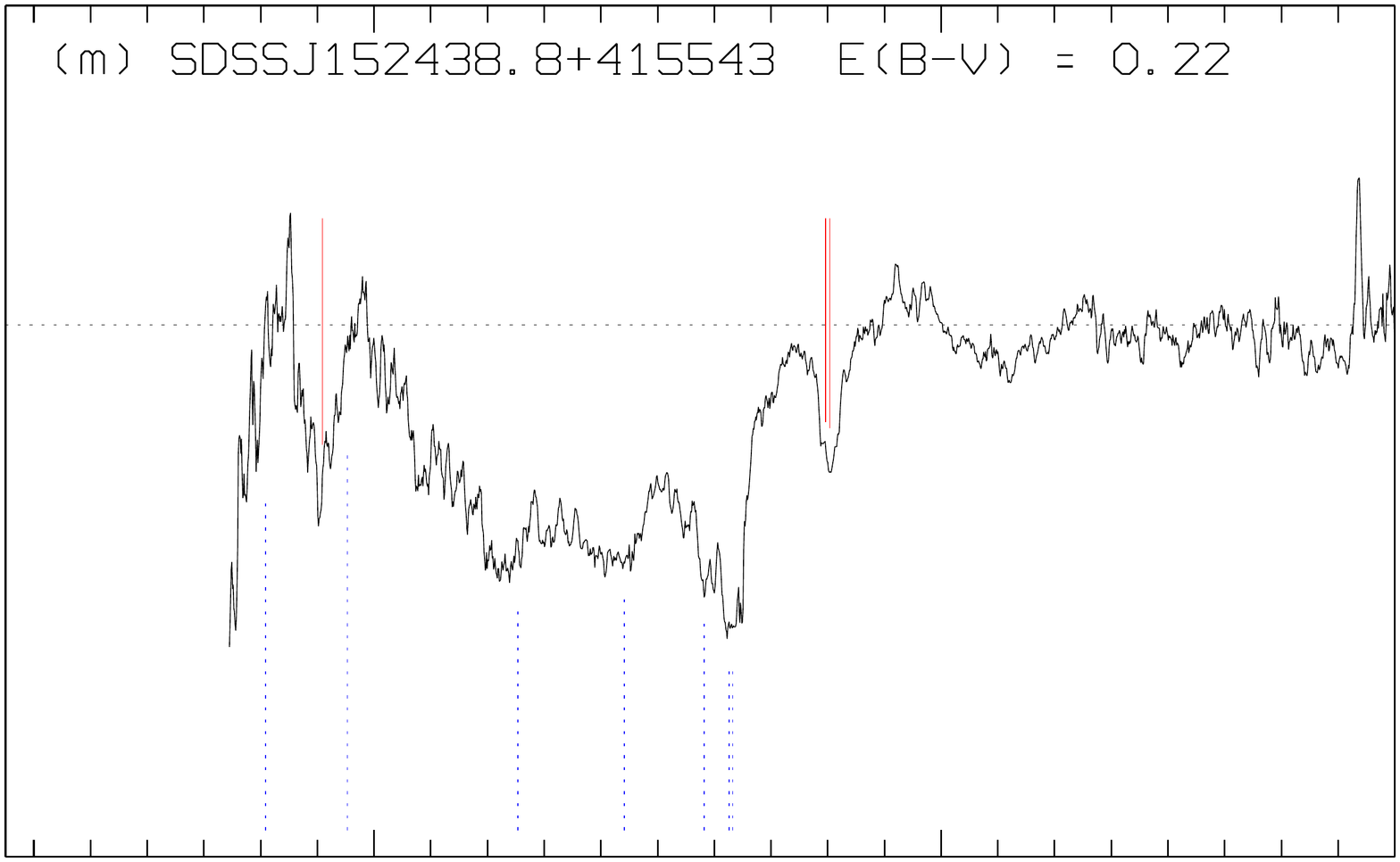}\hfill \=
\includegraphics[bb=53 20 500 770,scale=0.20,angle=270,clip]{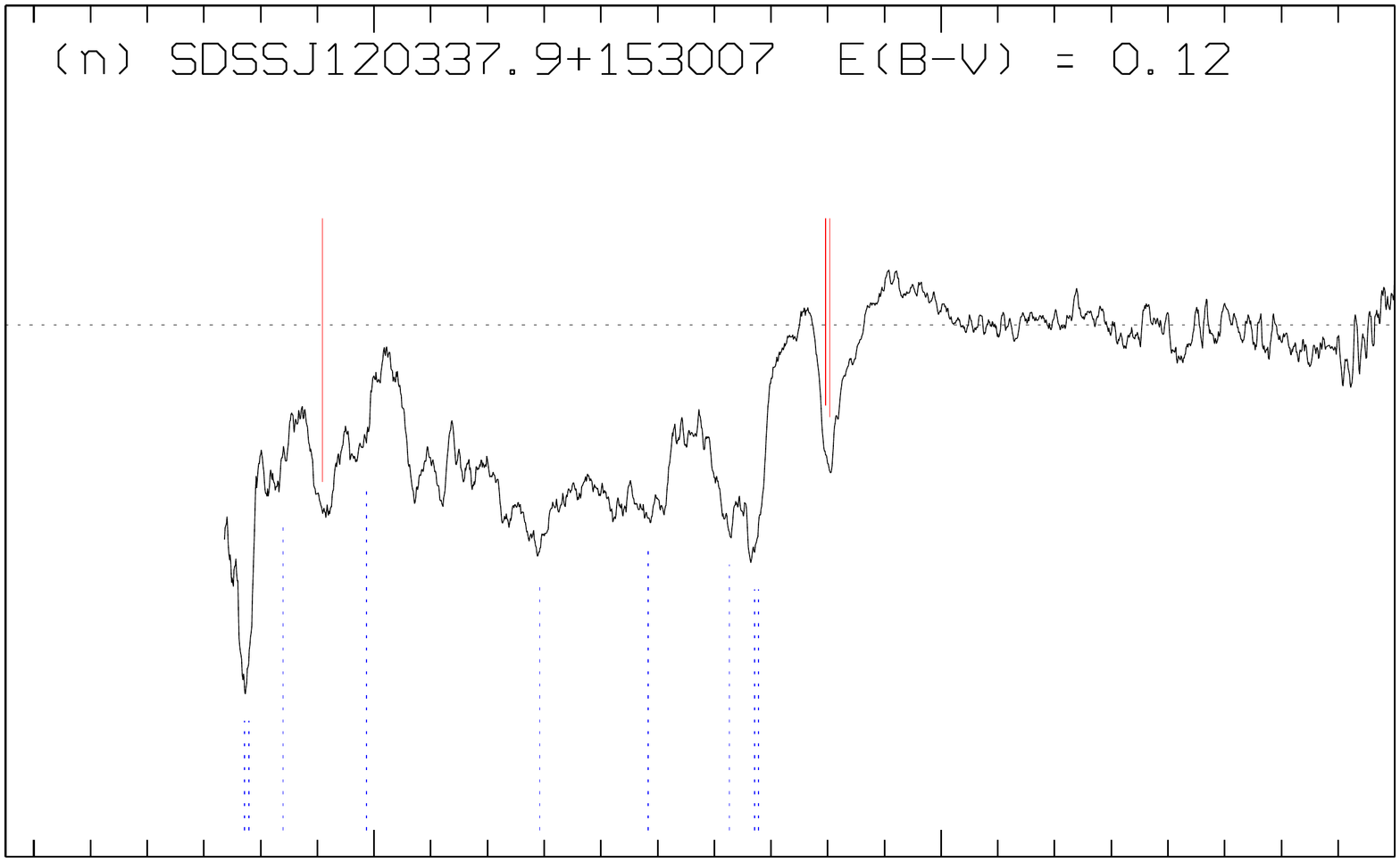}\hfill \=
\includegraphics[bb=53 20 500 770,scale=0.20,angle=270,clip]{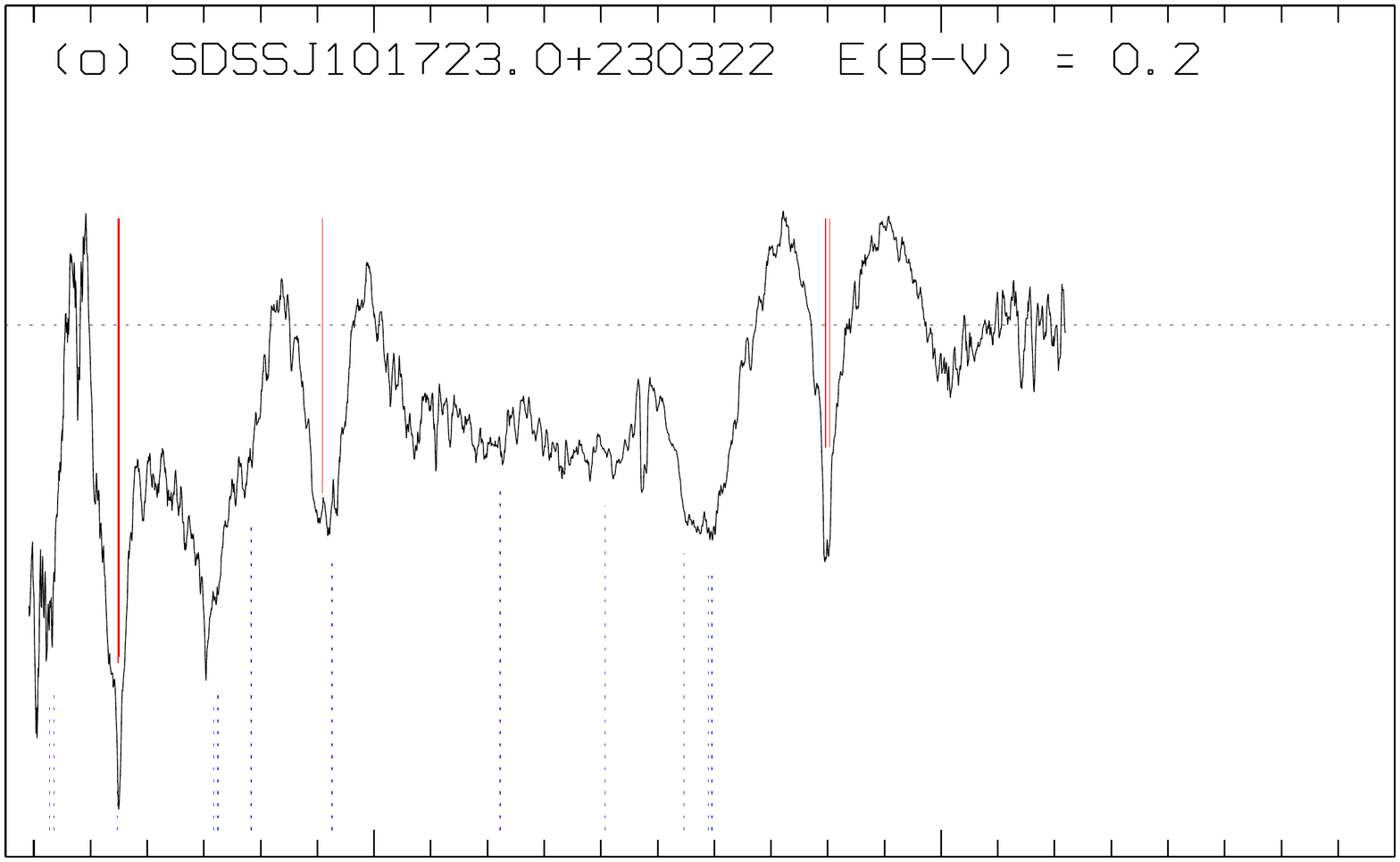}\hfill \\
\includegraphics[bb=53 00 500 770,scale=0.20,angle=270,clip]{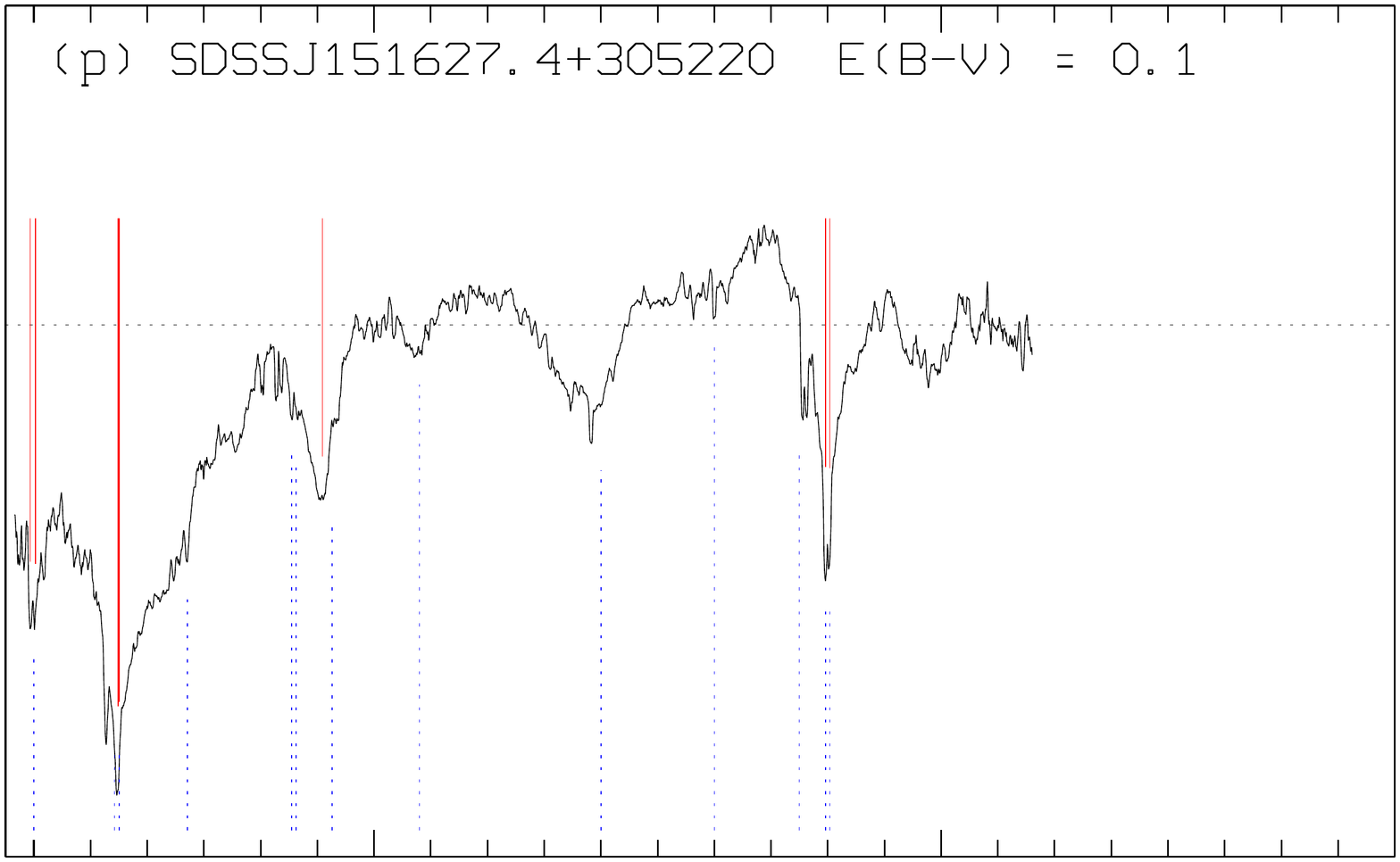}\hfill \=
\includegraphics[bb=53 20 500 770,scale=0.20,angle=270,clip]{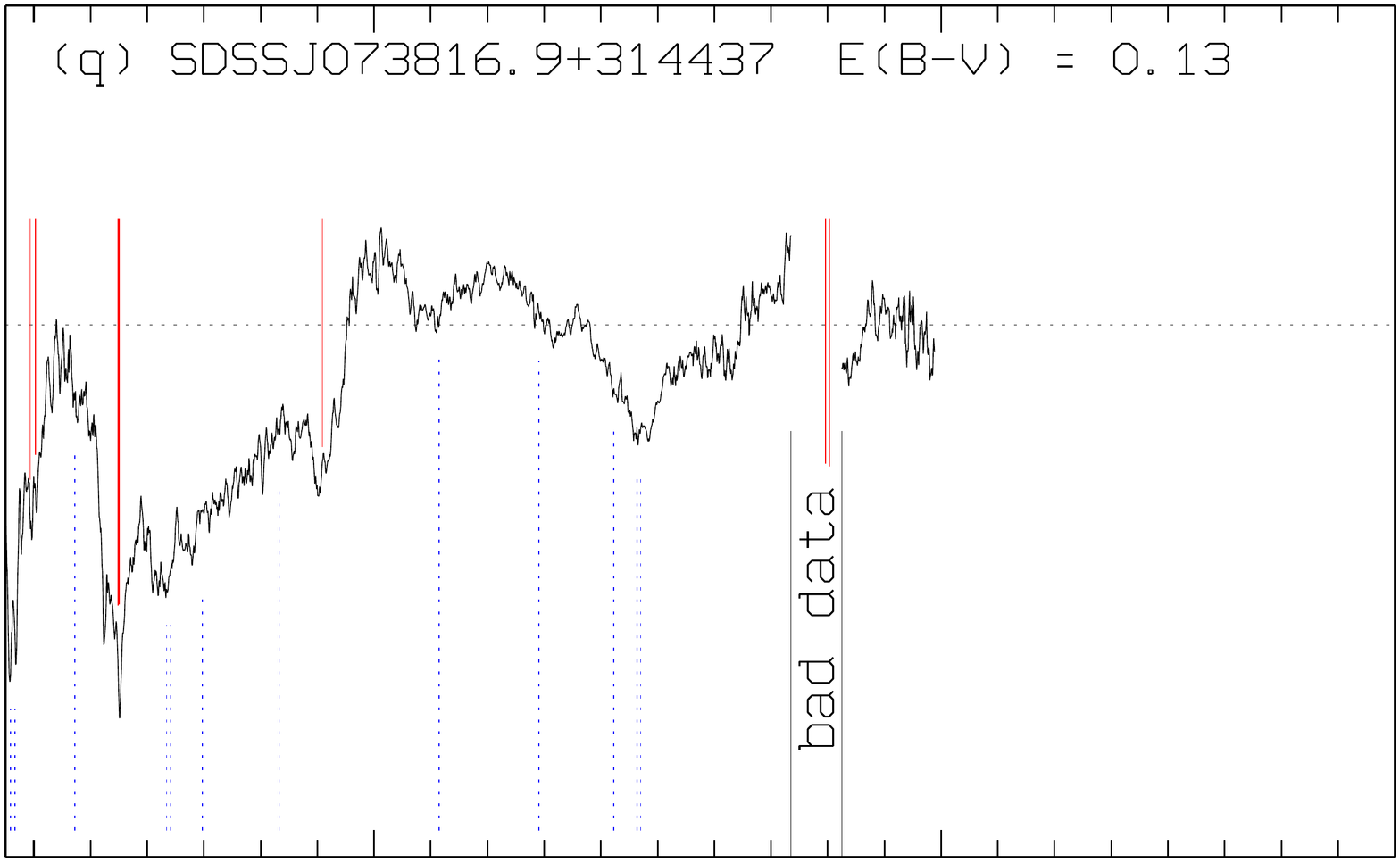}\hfill \=
\includegraphics[bb=53 20 500 770,scale=0.20,angle=270,clip]{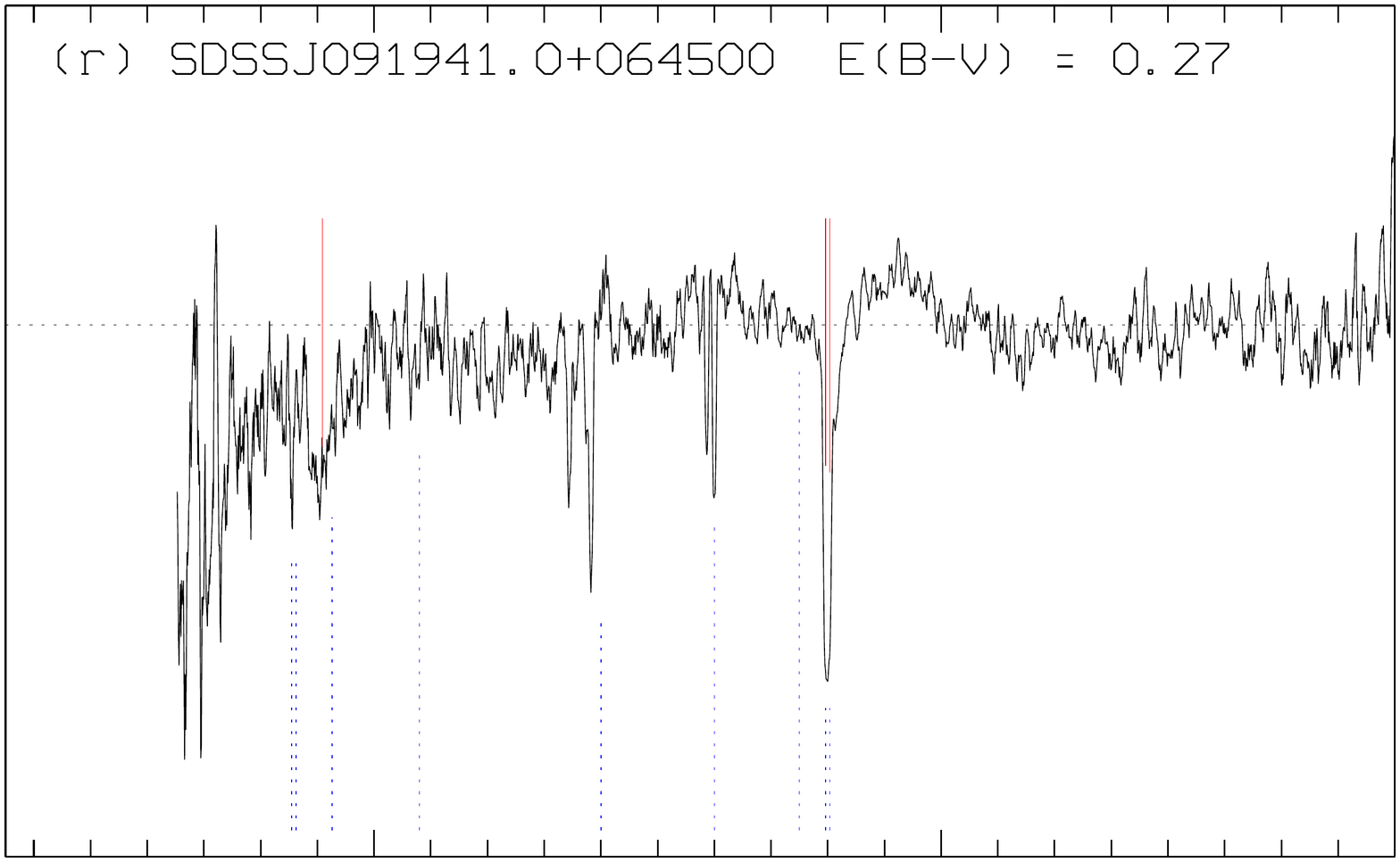}\hfill \\
\includegraphics[bb=53 00 570 770,scale=0.20,angle=270,clip]{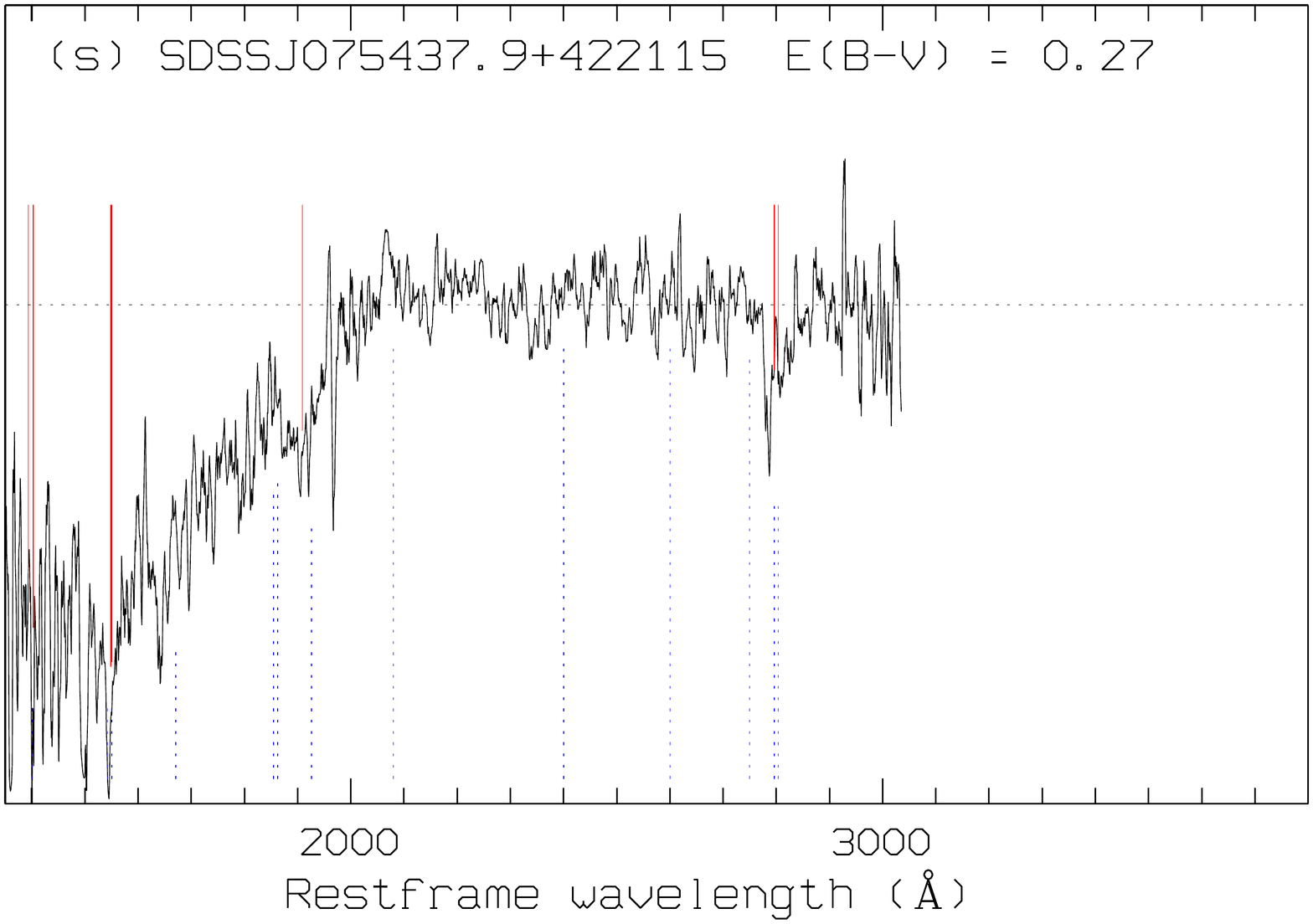}\hfill \=
\includegraphics[bb=53 20 570 770,scale=0.20,angle=270,clip]{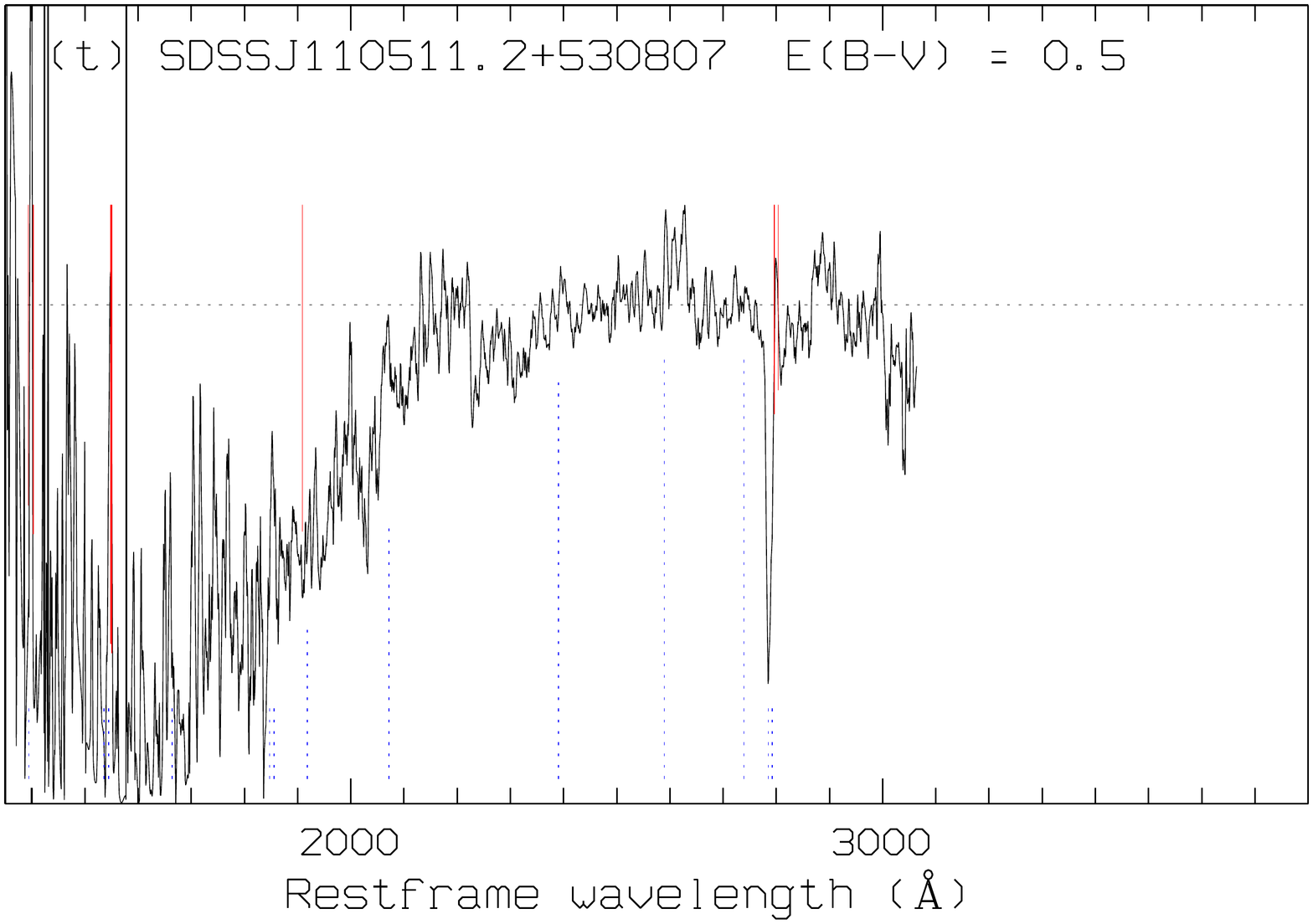}\hfill \=
\includegraphics[bb=53 20 570 770,scale=0.20,angle=270,clip]{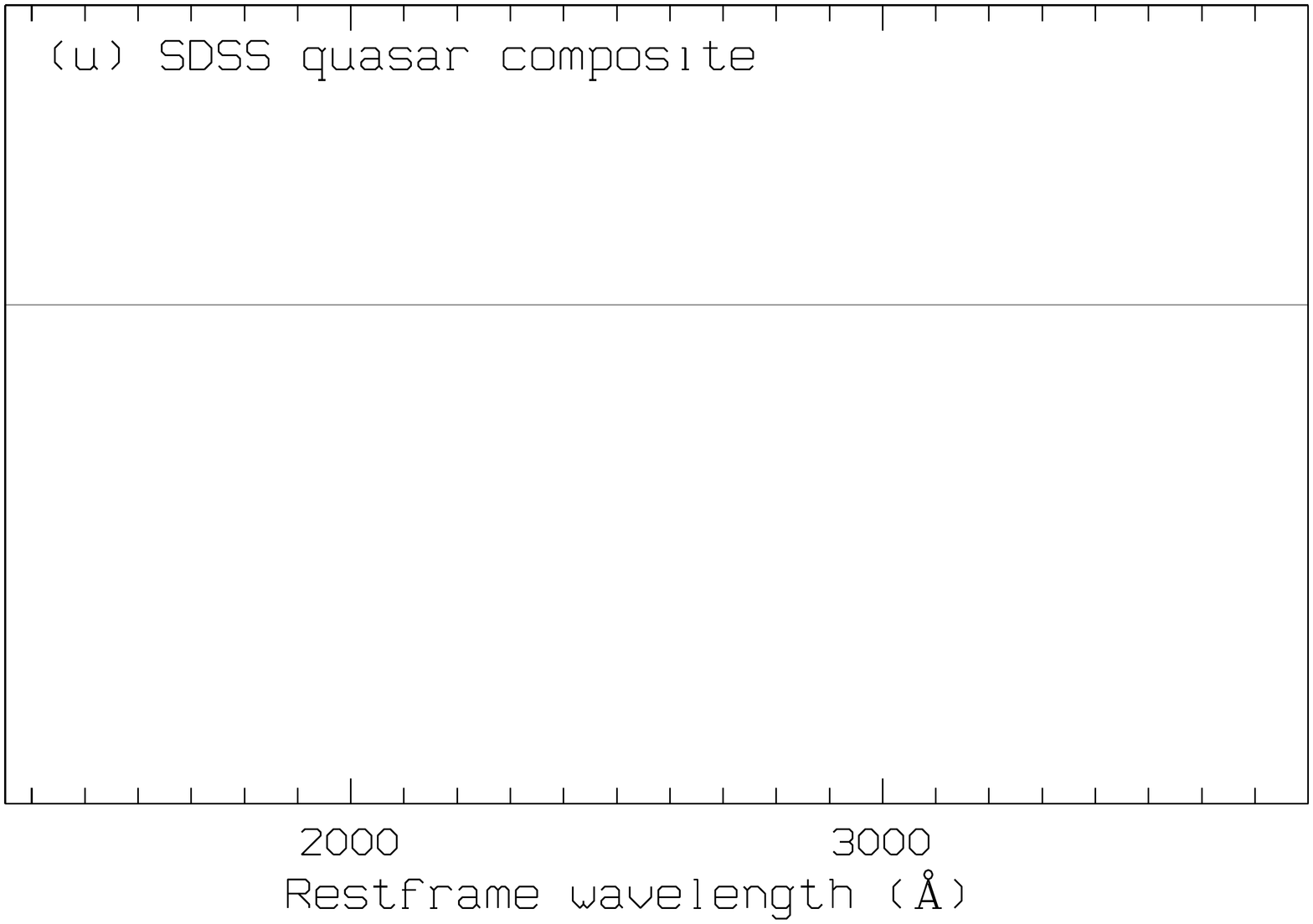}\hfill \\
\end{tabbing}
\caption{As Fig.\,\ref{fig:mysts} but for the ratio of the dereddened spectra to the
SDSS quasar composite.}
\label{fig:ratio_mysts}
\end{figure*}

\begin{figure*}[hbtp]   
\begin{tabbing}
\includegraphics[bb=53 00 500 770,scale=0.20,angle=270,clip]{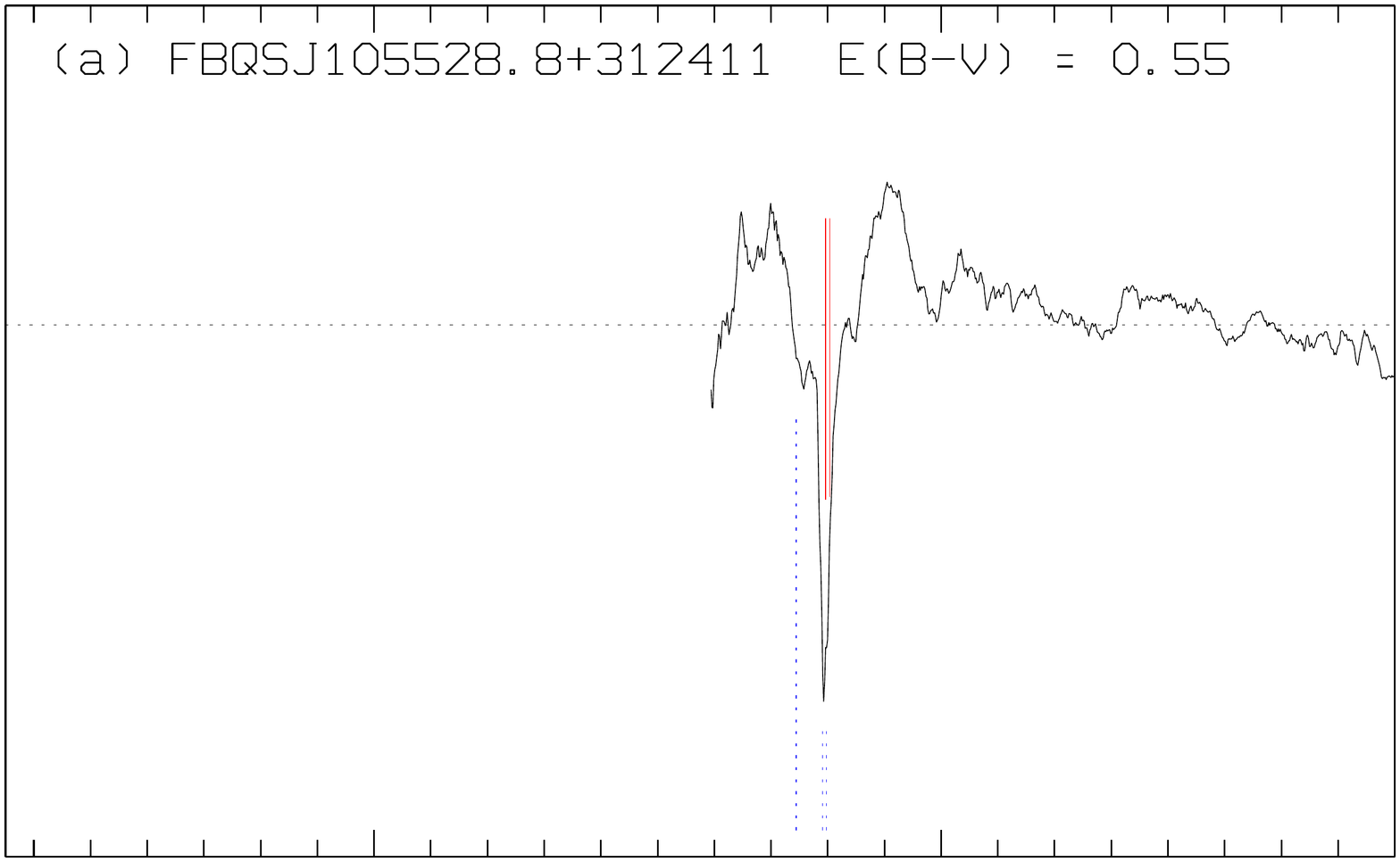}\hfill \=
\includegraphics[bb=53 20 500 770,scale=0.20,angle=270,clip]{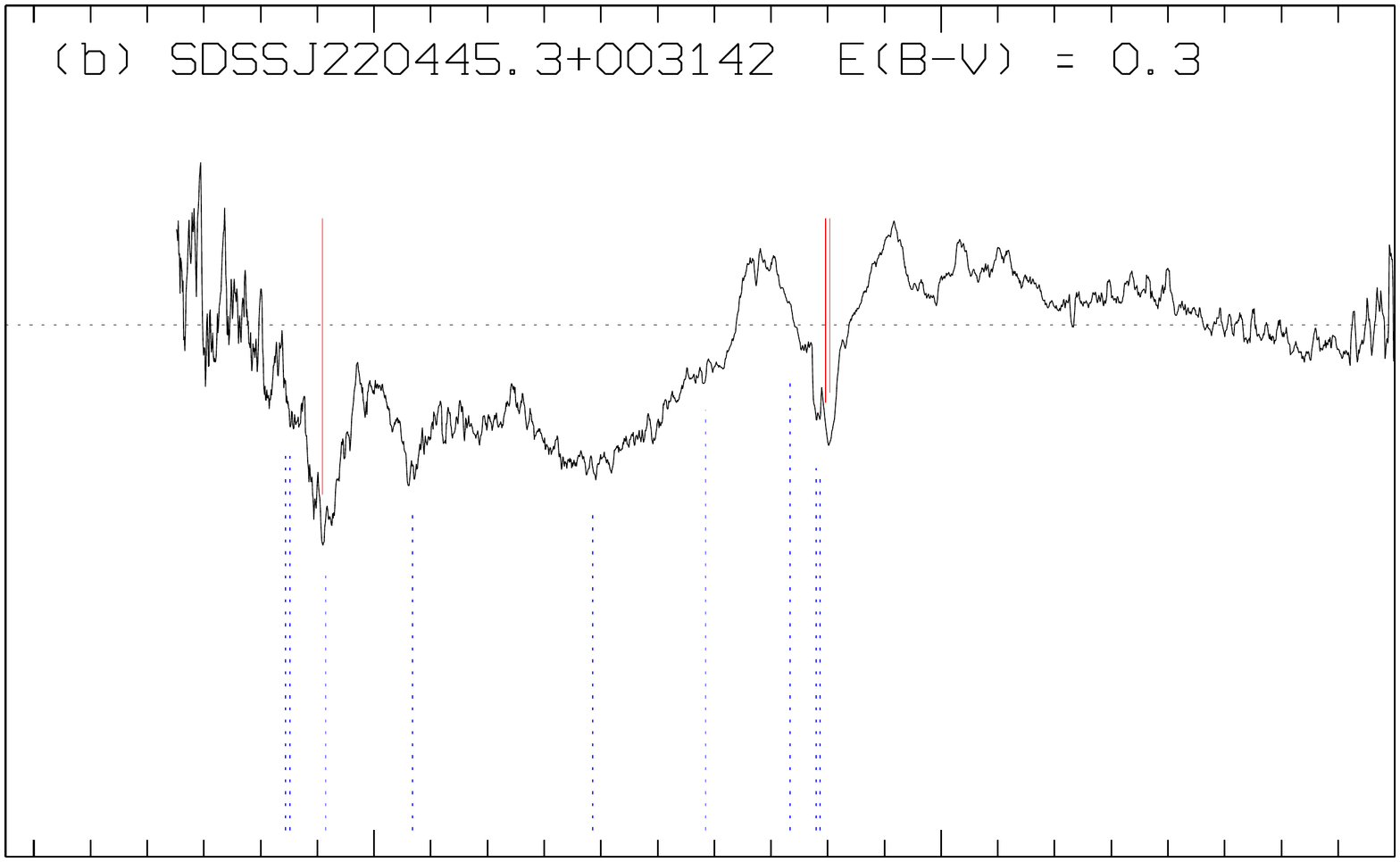}\hfill \=
\includegraphics[bb=53 20 500 770,scale=0.20,angle=270,clip]{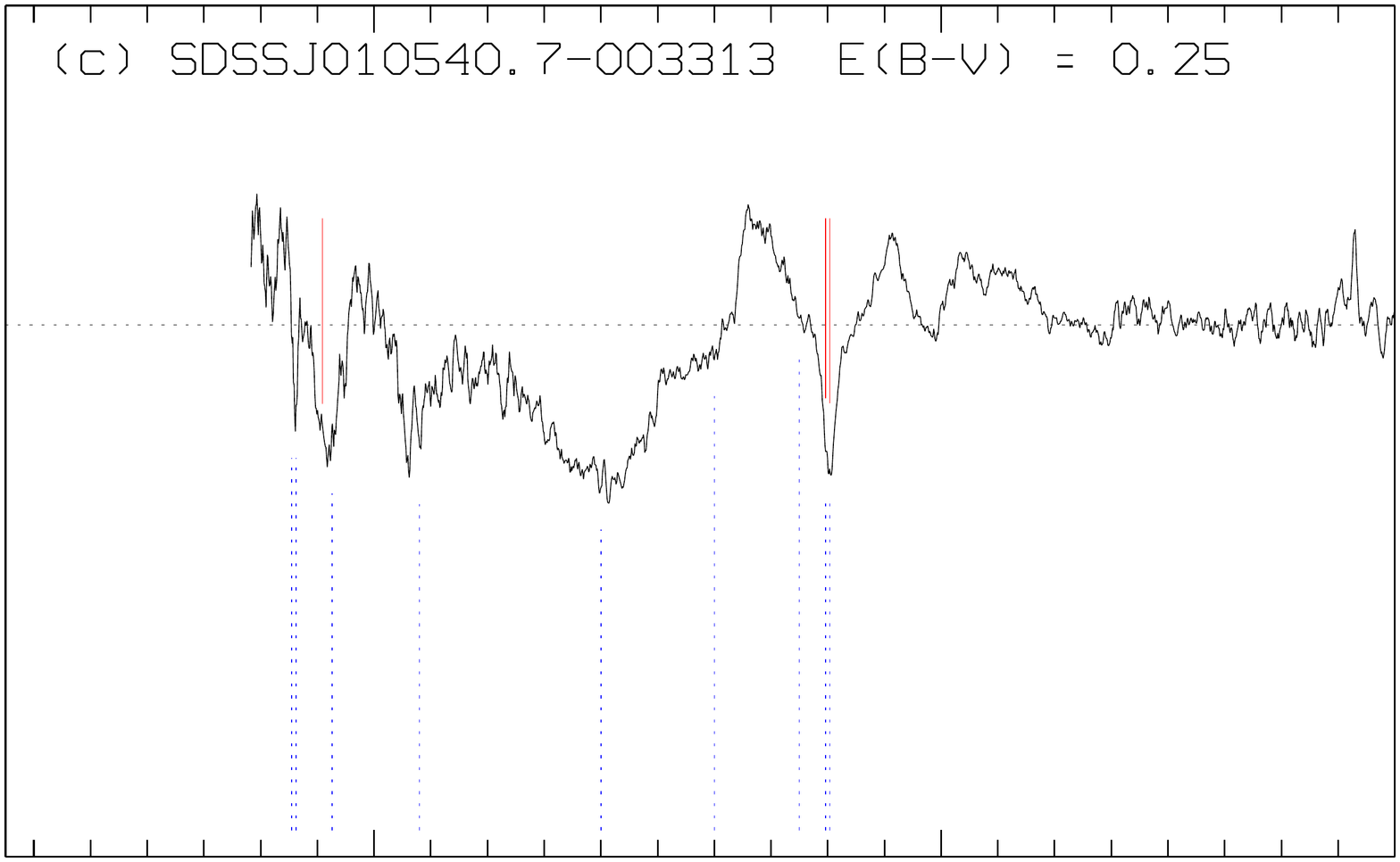}\hfill \\
\includegraphics[bb=53 00 500 770,scale=0.20,angle=270,clip]{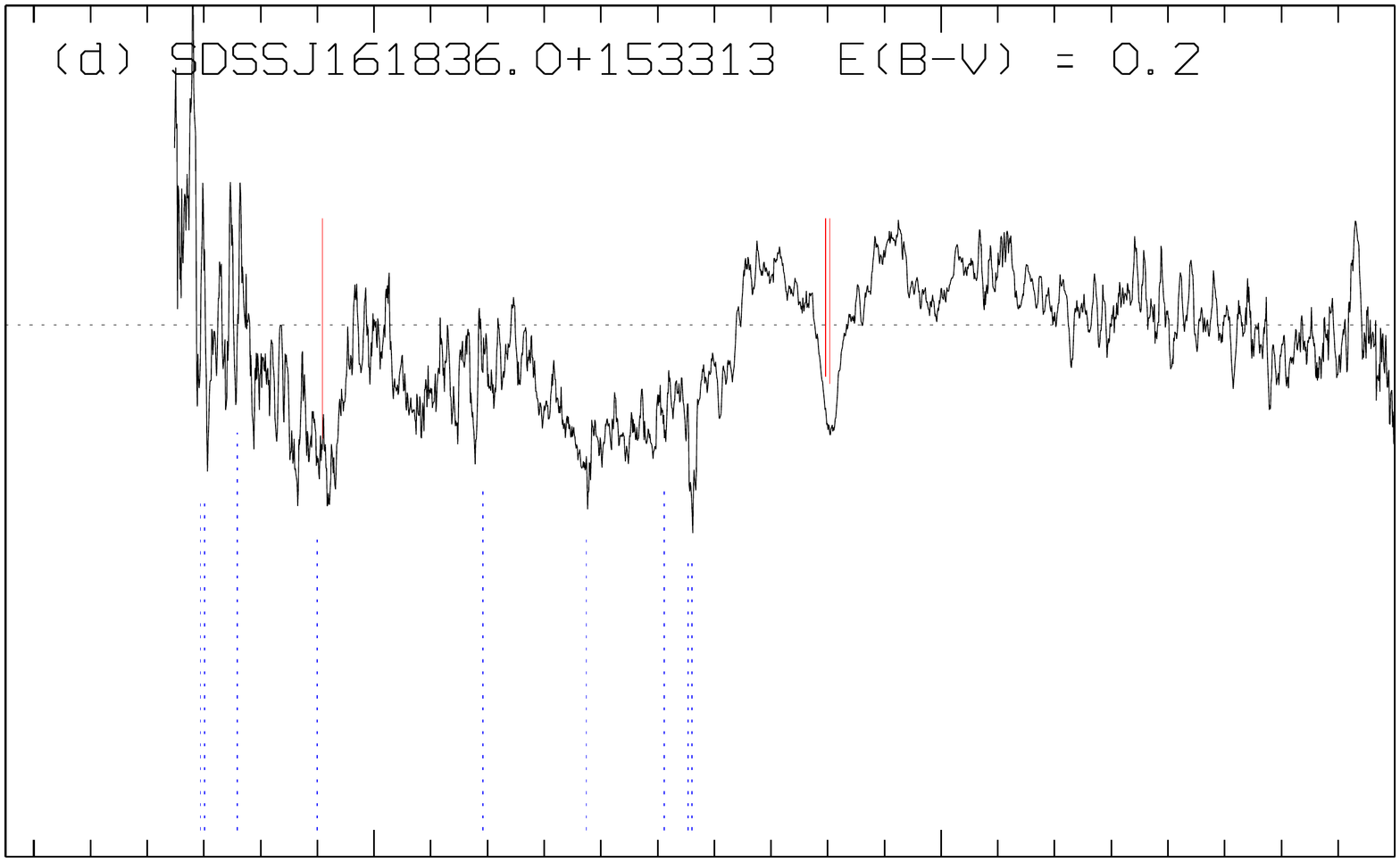}\hfill \=
\includegraphics[bb=53 20 500 770,scale=0.20,angle=270,clip]{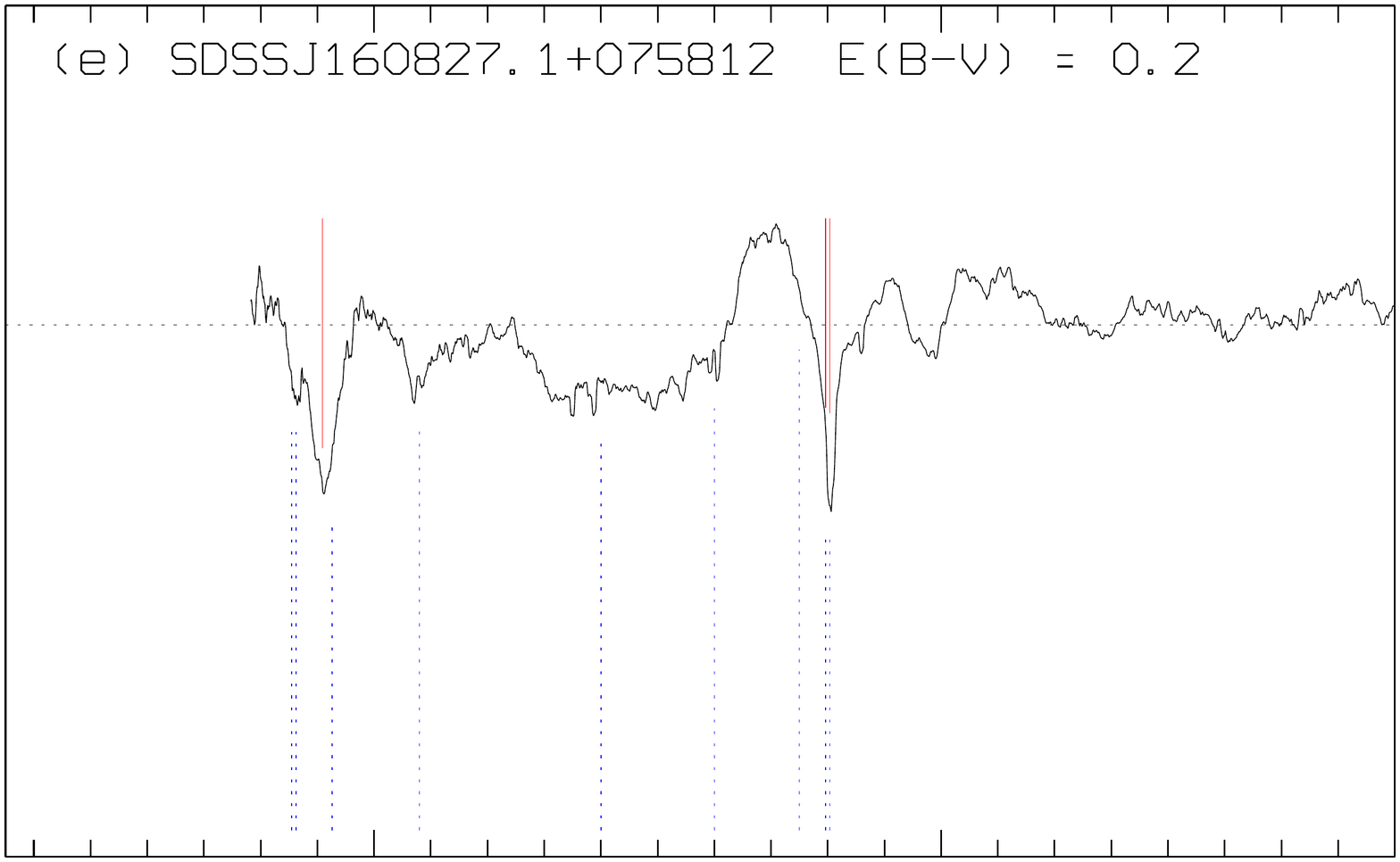}\hfill \=
\includegraphics[bb=53 20 500 770,scale=0.20,angle=270,clip]{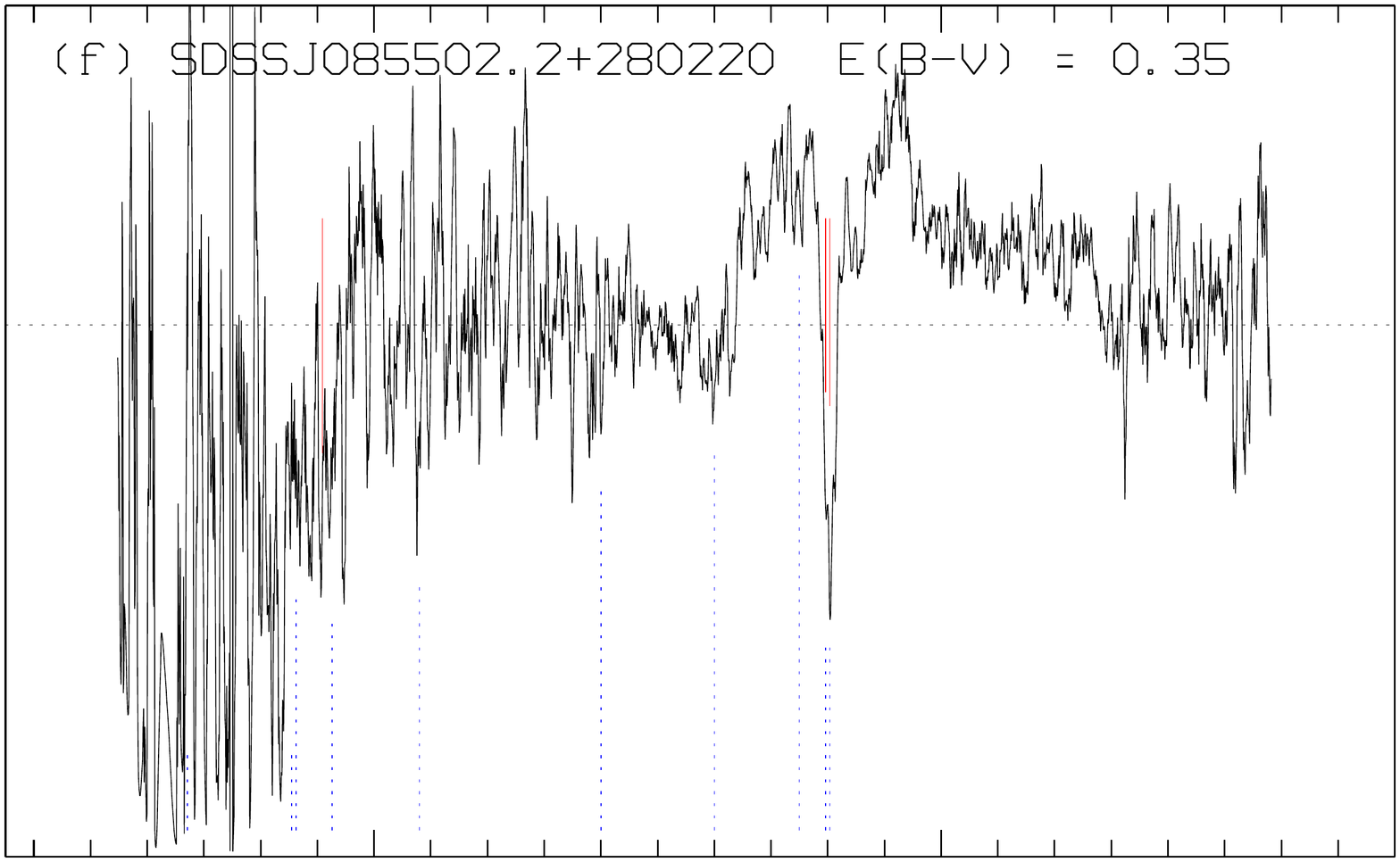}\hfill \\
\includegraphics[bb=53 00 500 770,scale=0.20,angle=270,clip]{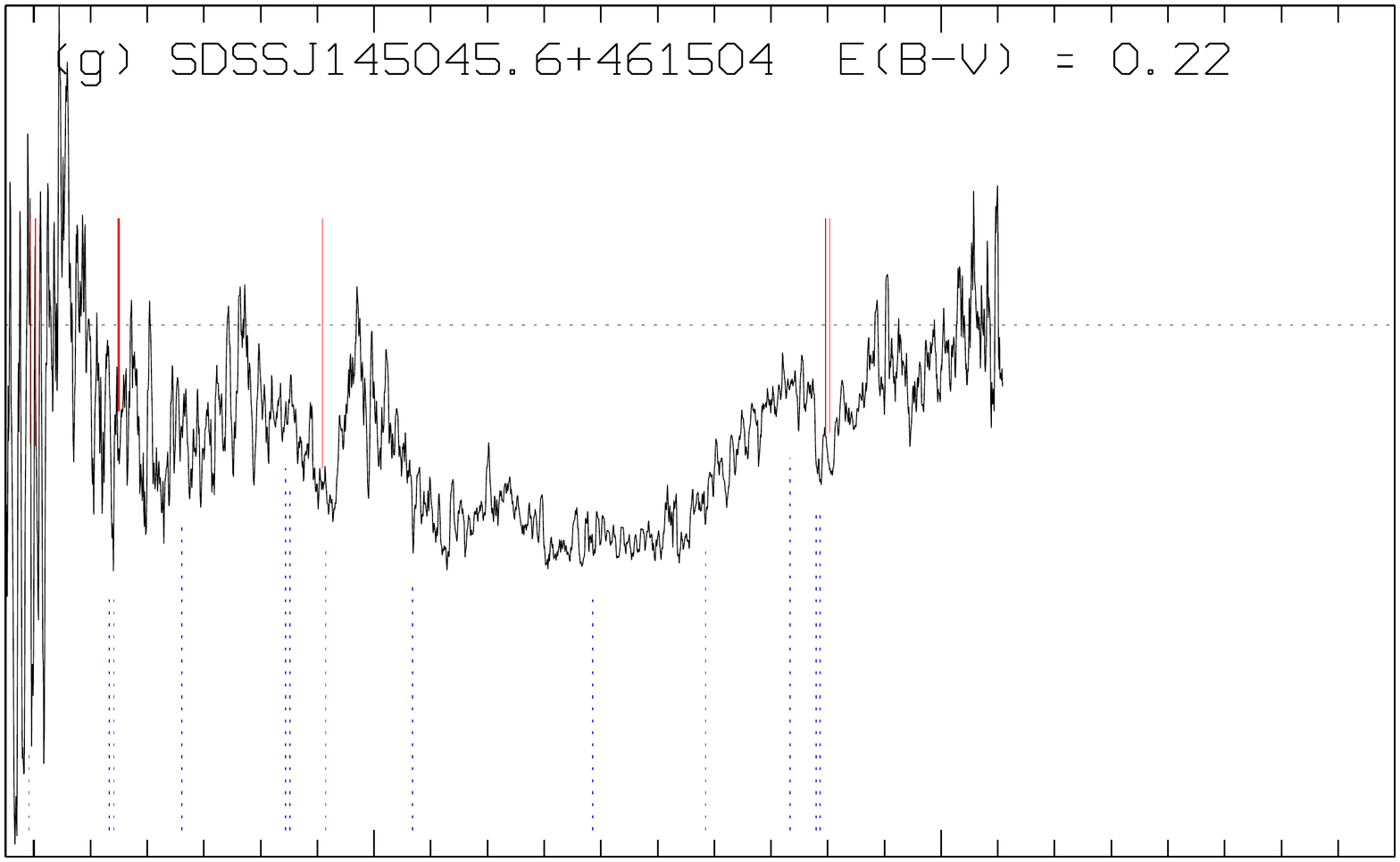}\hfill \=
\includegraphics[bb=53 20 500 770,scale=0.20,angle=270,clip]{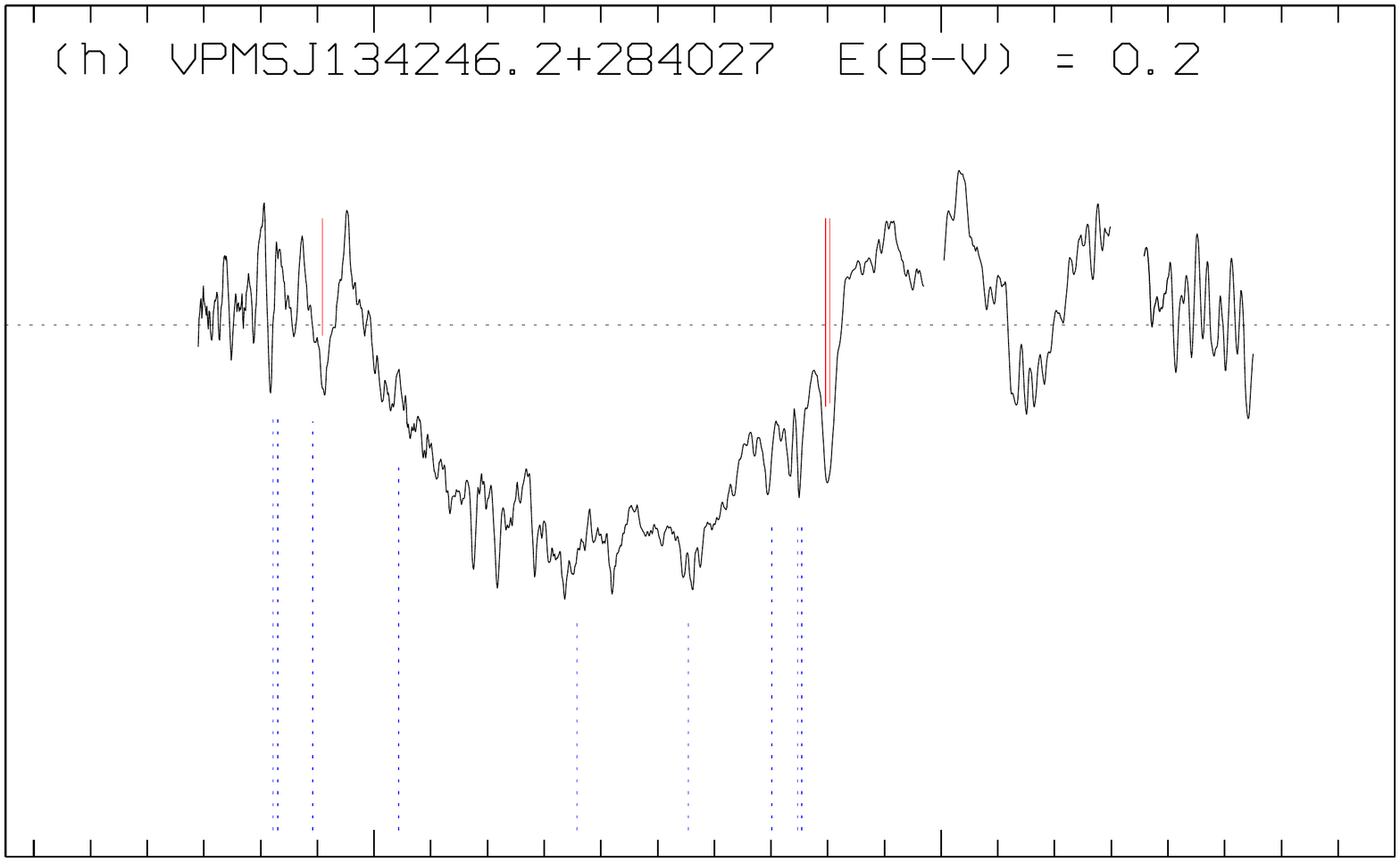}\hfill \=
\includegraphics[bb=53 20 500 770,scale=0.20,angle=270,clip]{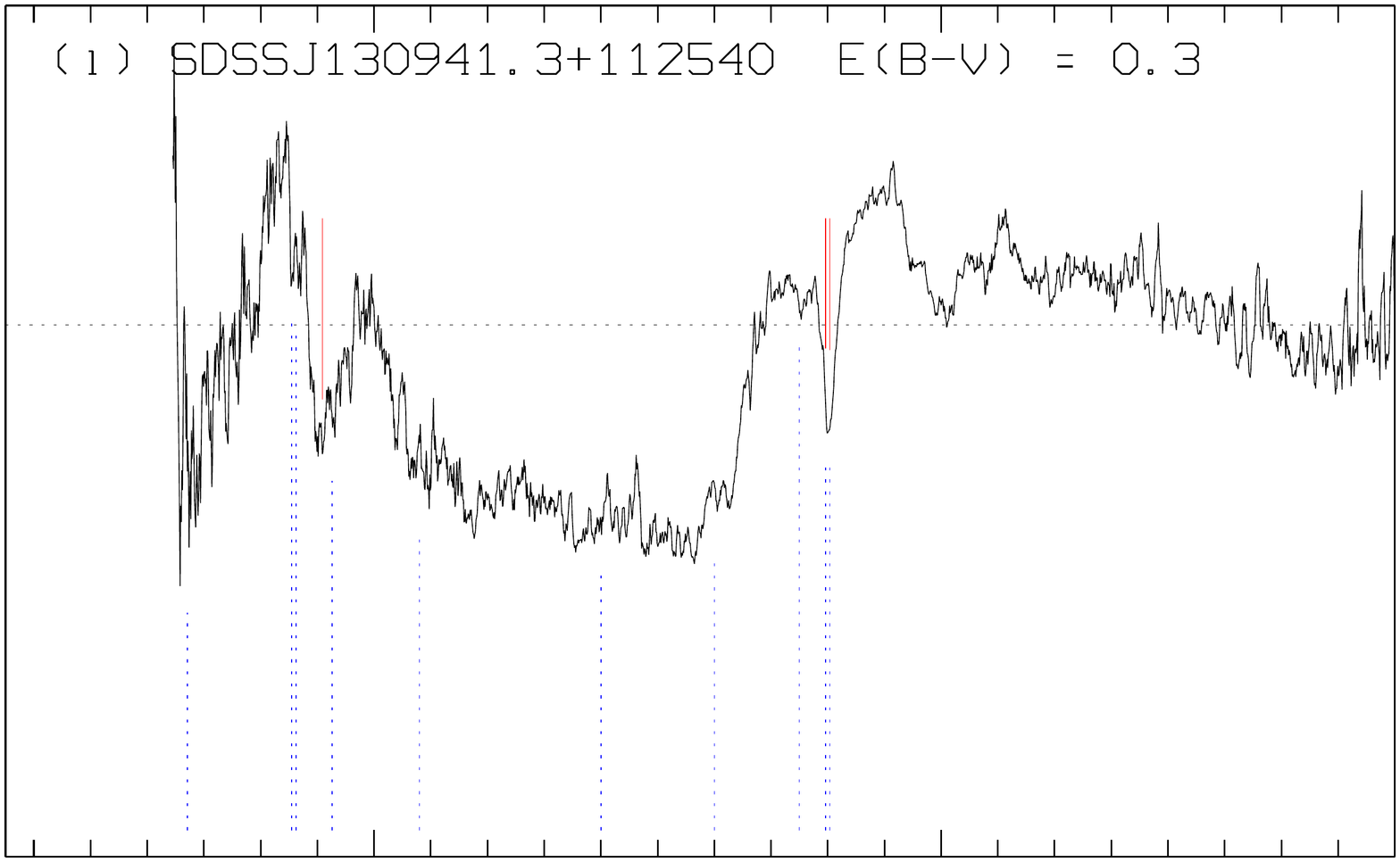}\hfill \\
\includegraphics[bb=53 00 500 770,scale=0.20,angle=270,clip]{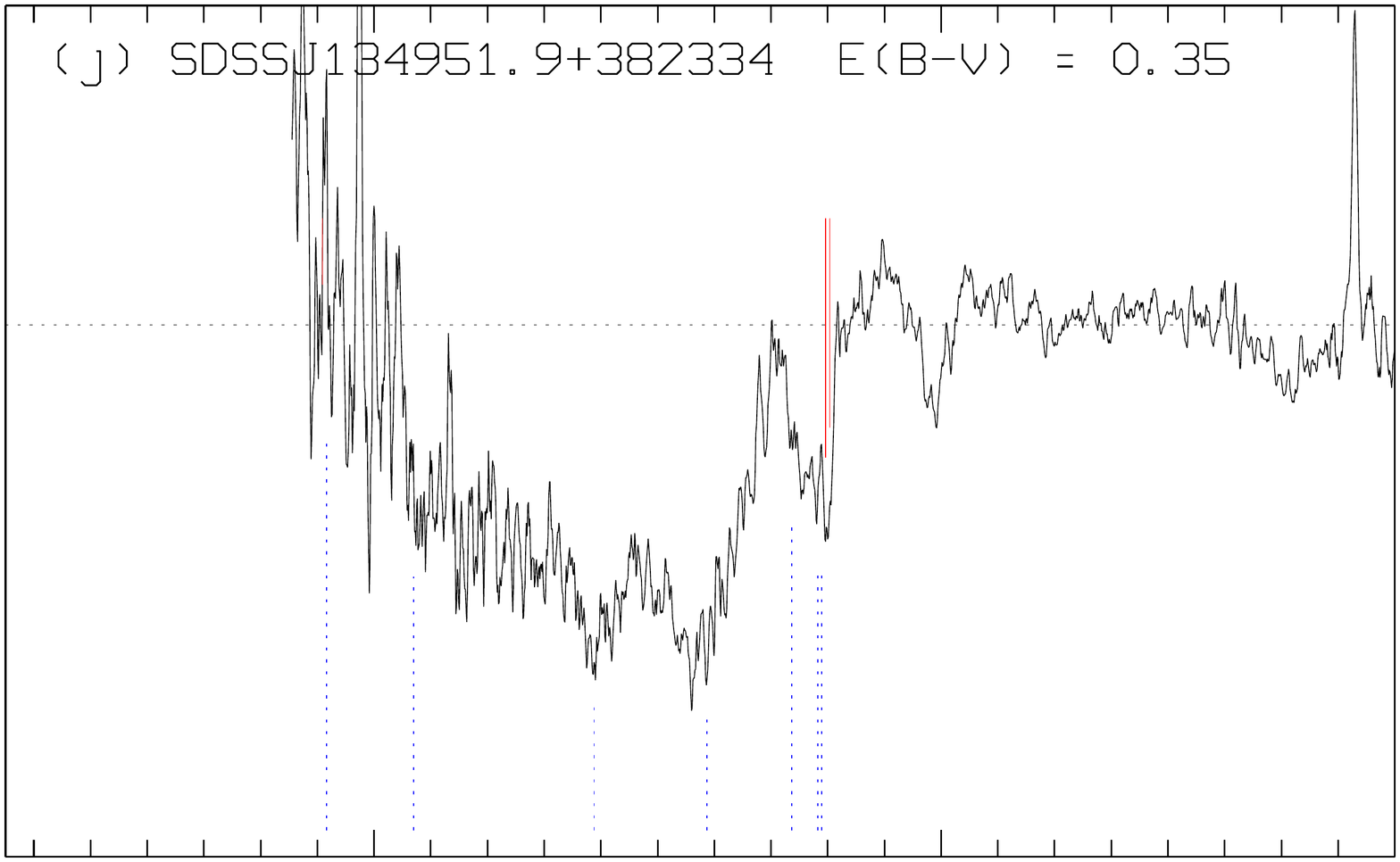}\hfill \=
\includegraphics[bb=53 20 500 770,scale=0.20,angle=270,clip]{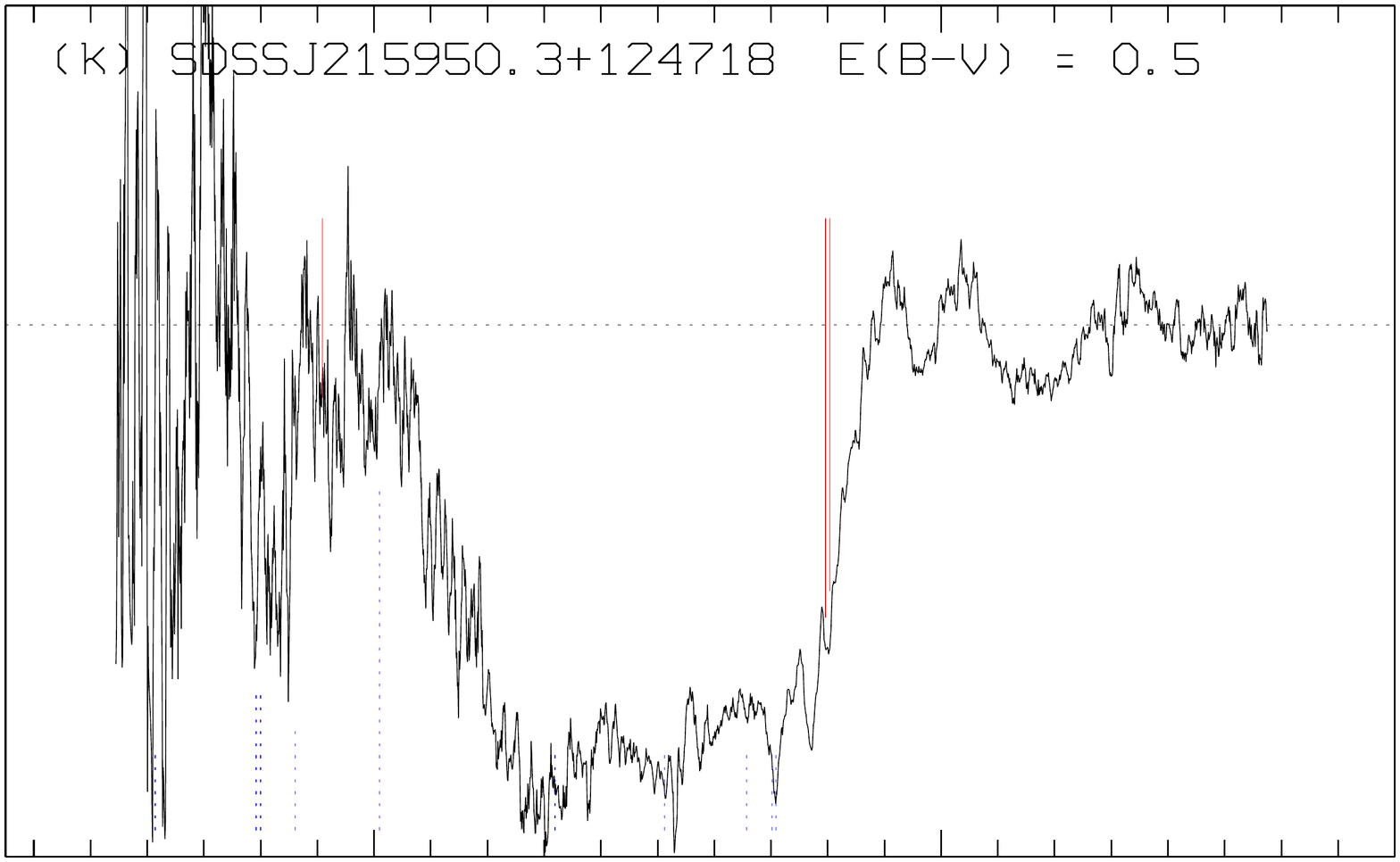}\hfill \=
\includegraphics[bb=53 20 500 770,scale=0.20,angle=270,clip]{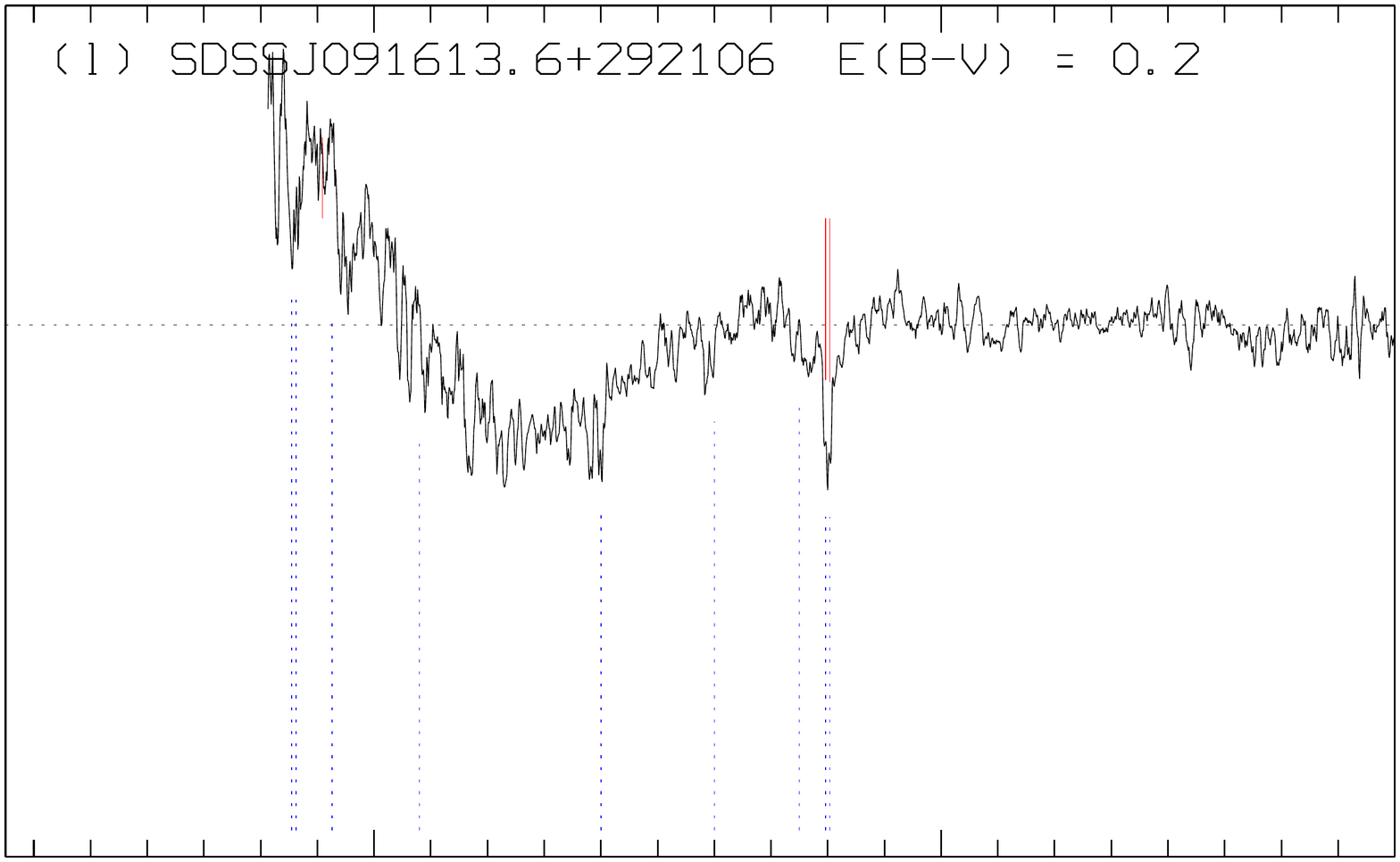}\hfill \\
\includegraphics[bb=53 00 500 770,scale=0.20,angle=270,clip]{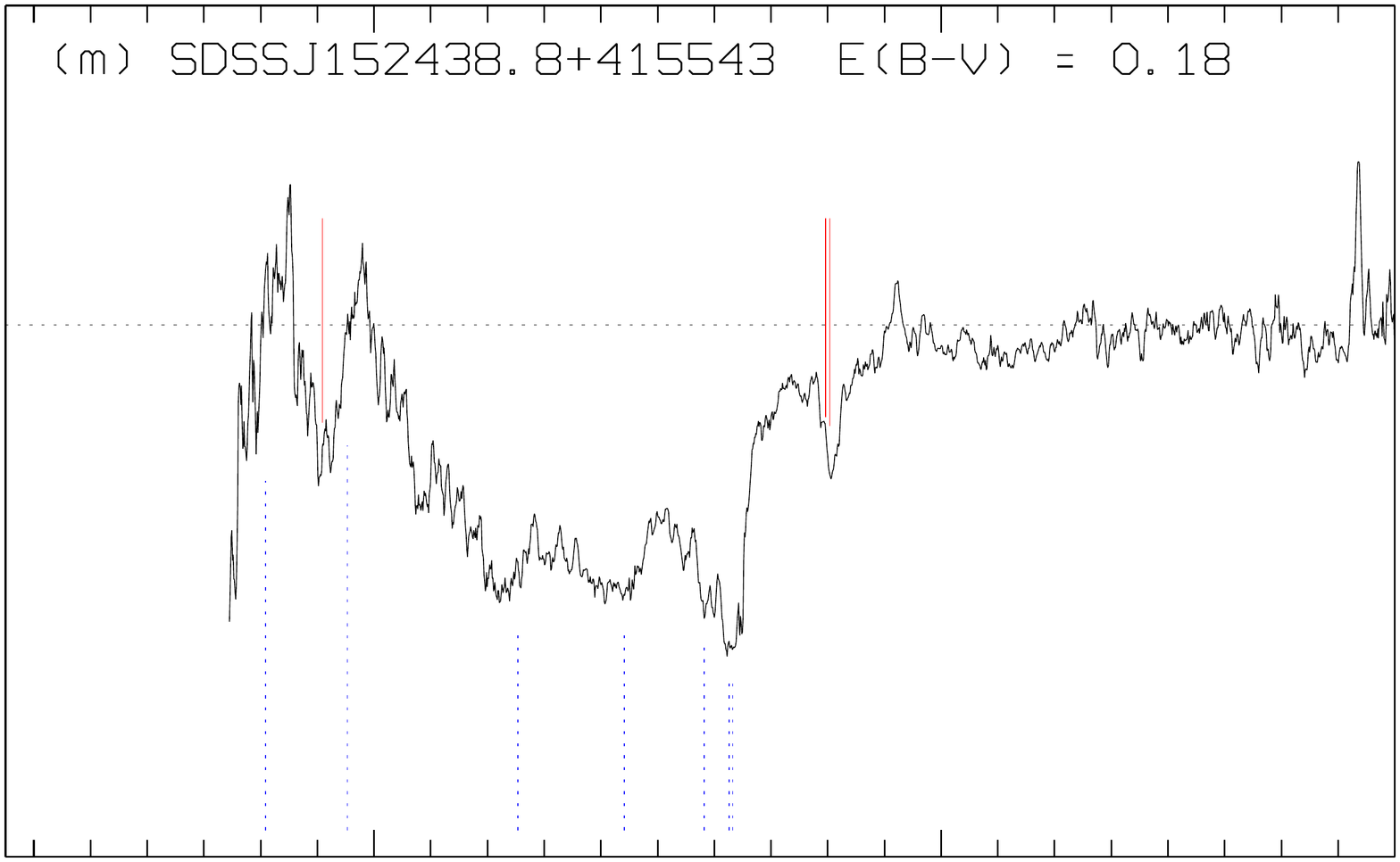}\hfill \=
\includegraphics[bb=53 20 500 770,scale=0.20,angle=270,clip]{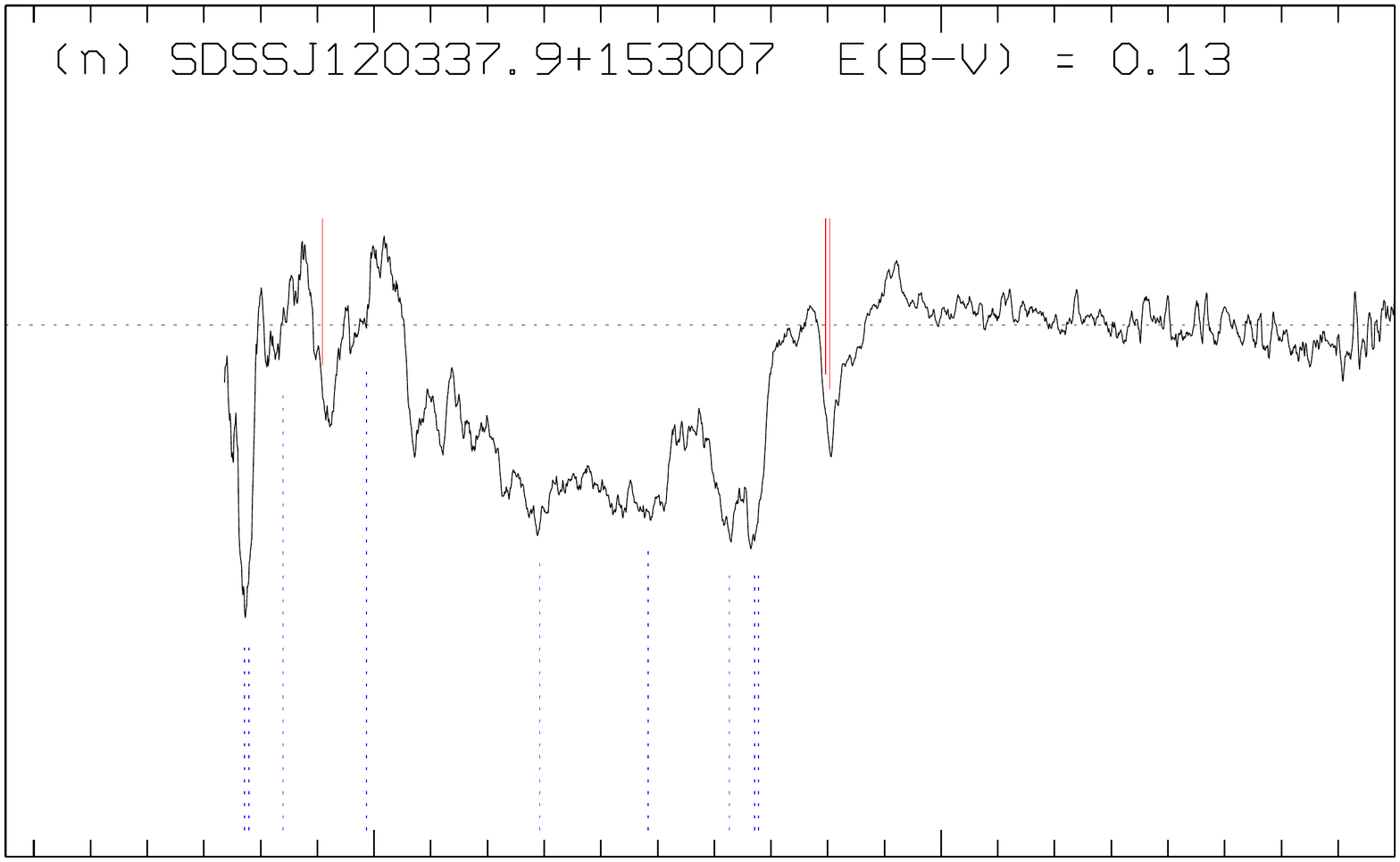}\hfill \=
\includegraphics[bb=53 20 500 770,scale=0.20,angle=270,clip]{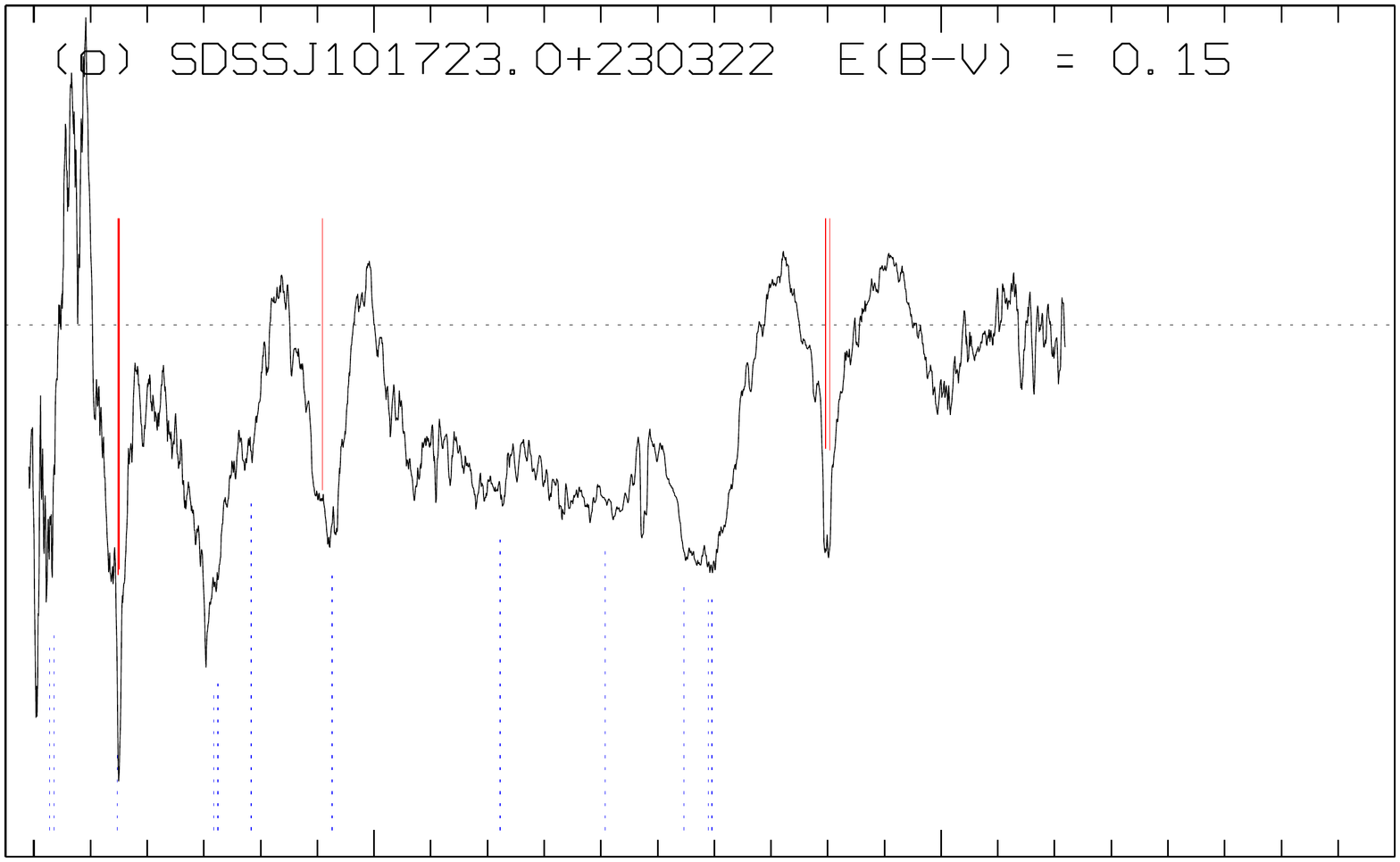}\hfill \\
\includegraphics[bb=53 00 500 770,scale=0.20,angle=270,clip]{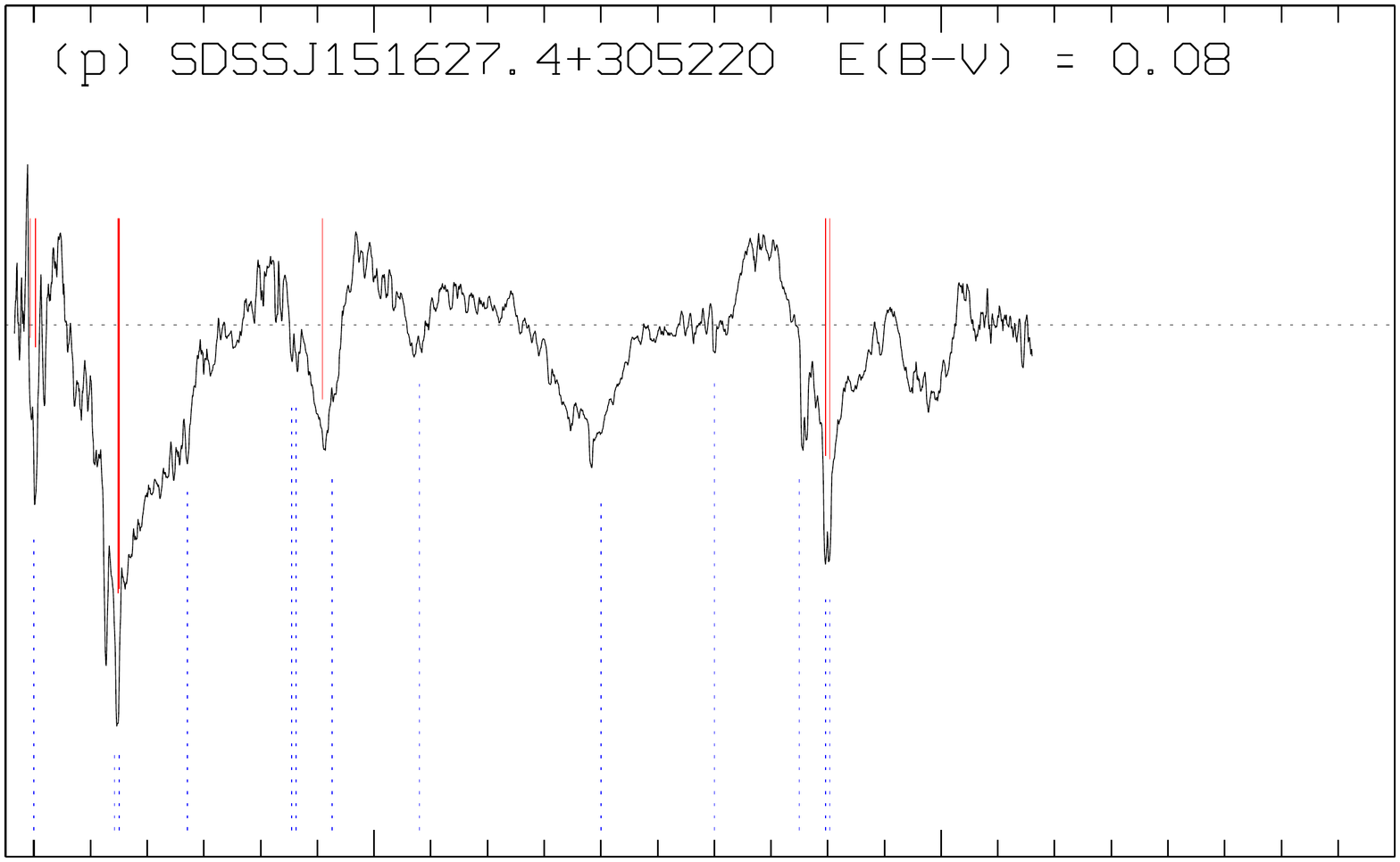}\hfill \=
\includegraphics[bb=53 20 500 770,scale=0.20,angle=270,clip]{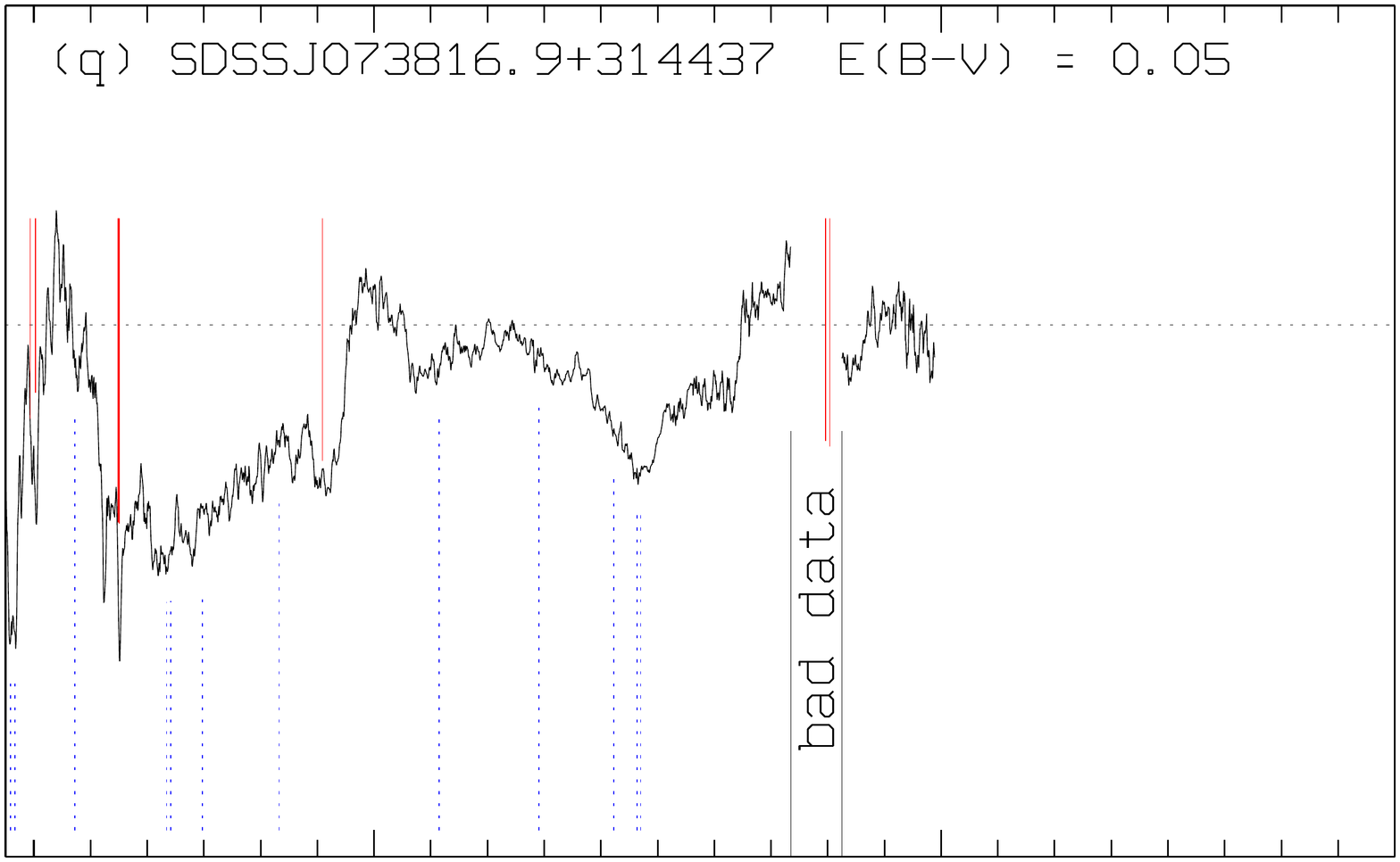}\hfill \=
\includegraphics[bb=53 20 500 770,scale=0.20,angle=270,clip]{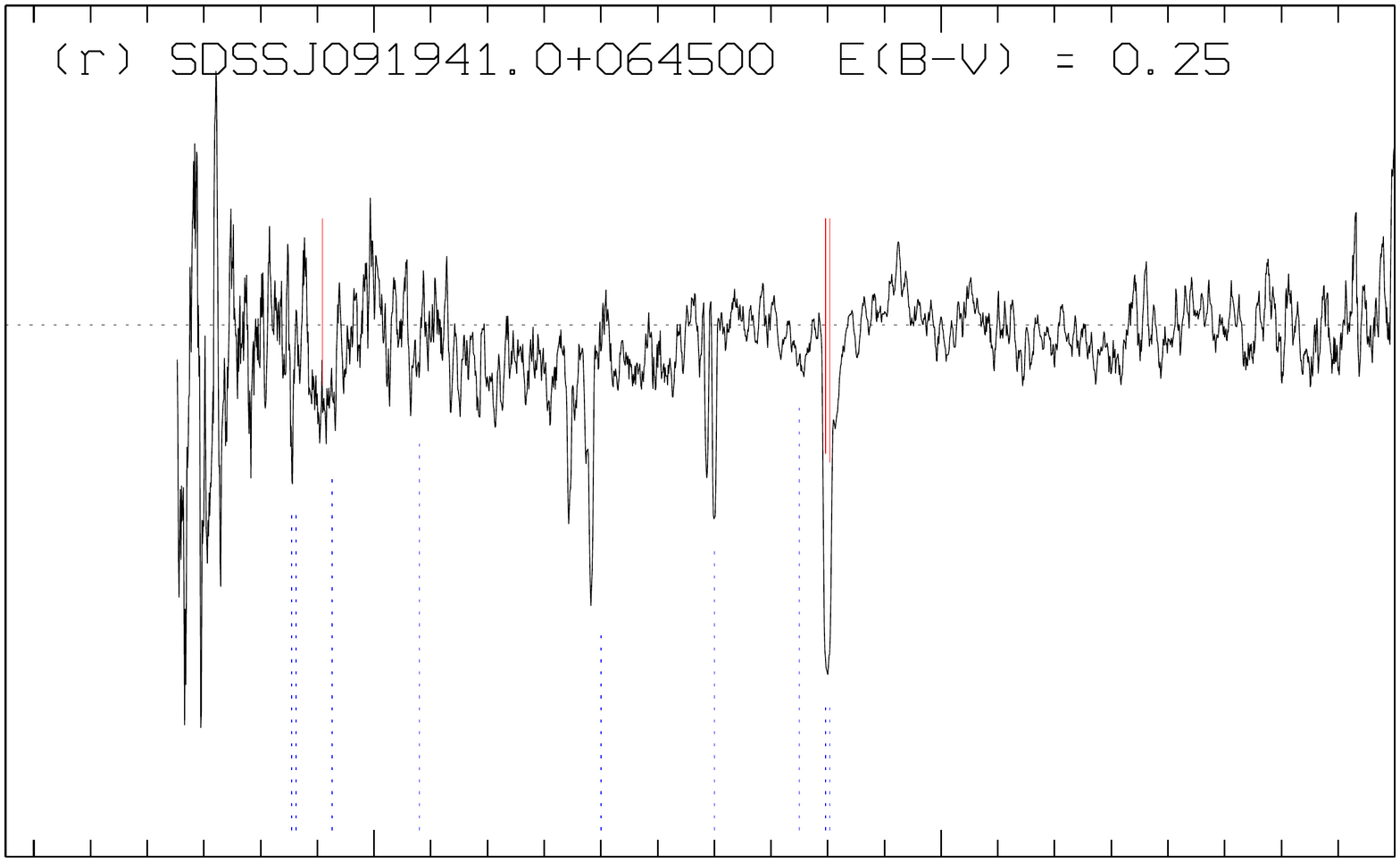}\hfill \\
\includegraphics[bb=53 00 570 770,scale=0.20,angle=270,clip]{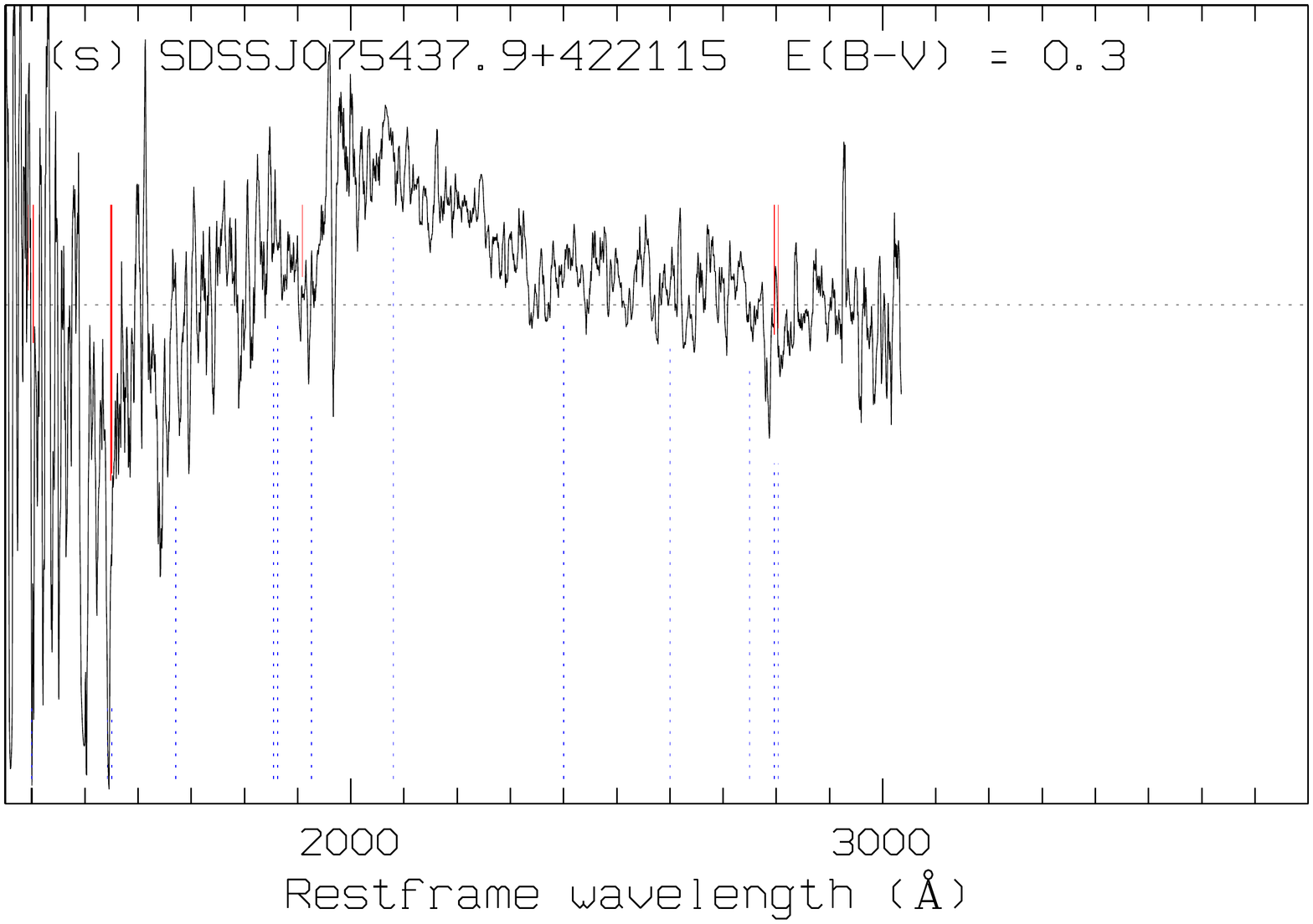}\hfill \=
\includegraphics[bb=53 20 570 770,scale=0.20,angle=270,clip]{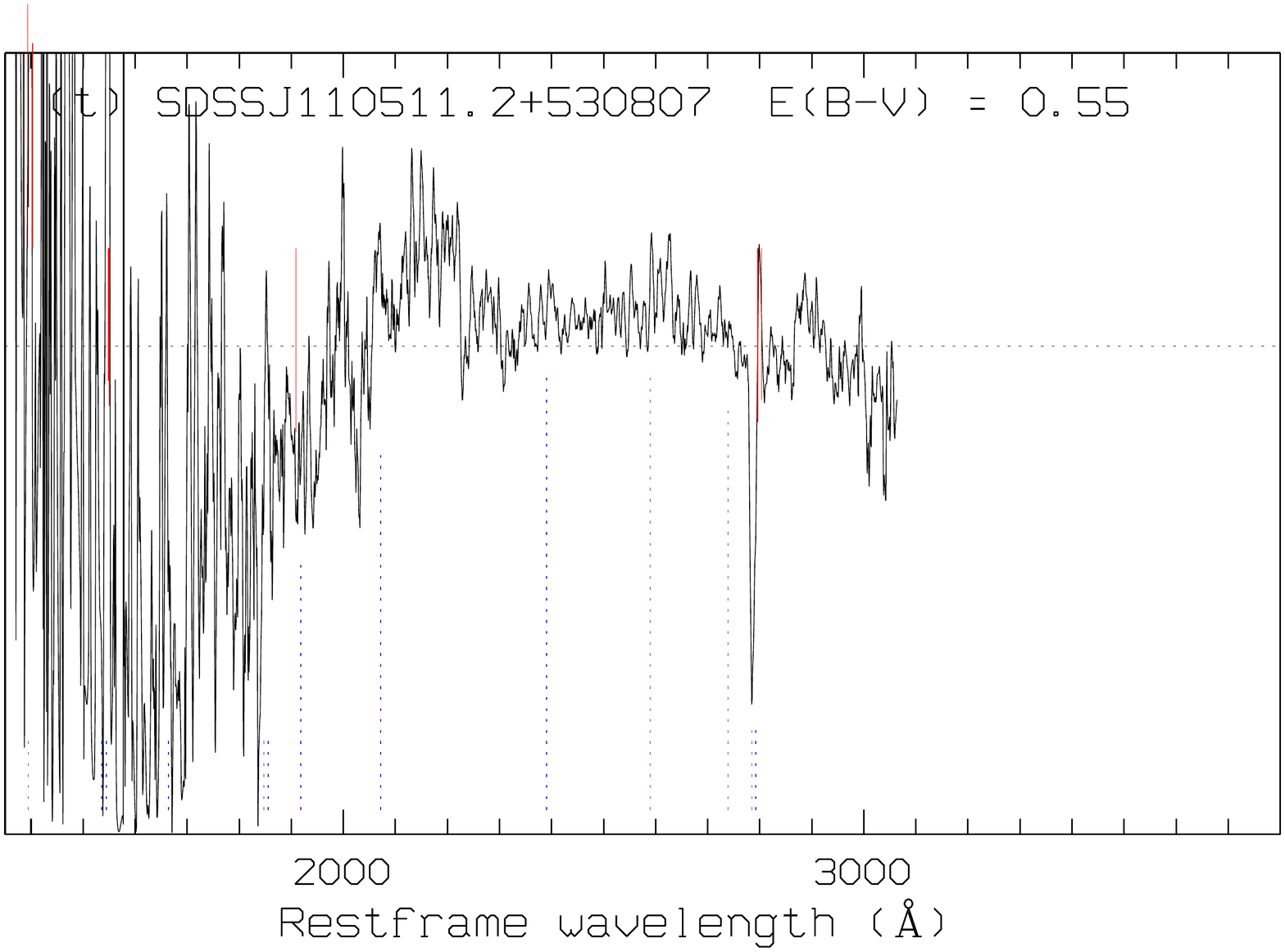}\hfill \=
\includegraphics[bb=53 20 570 770,scale=0.20,angle=270,clip]{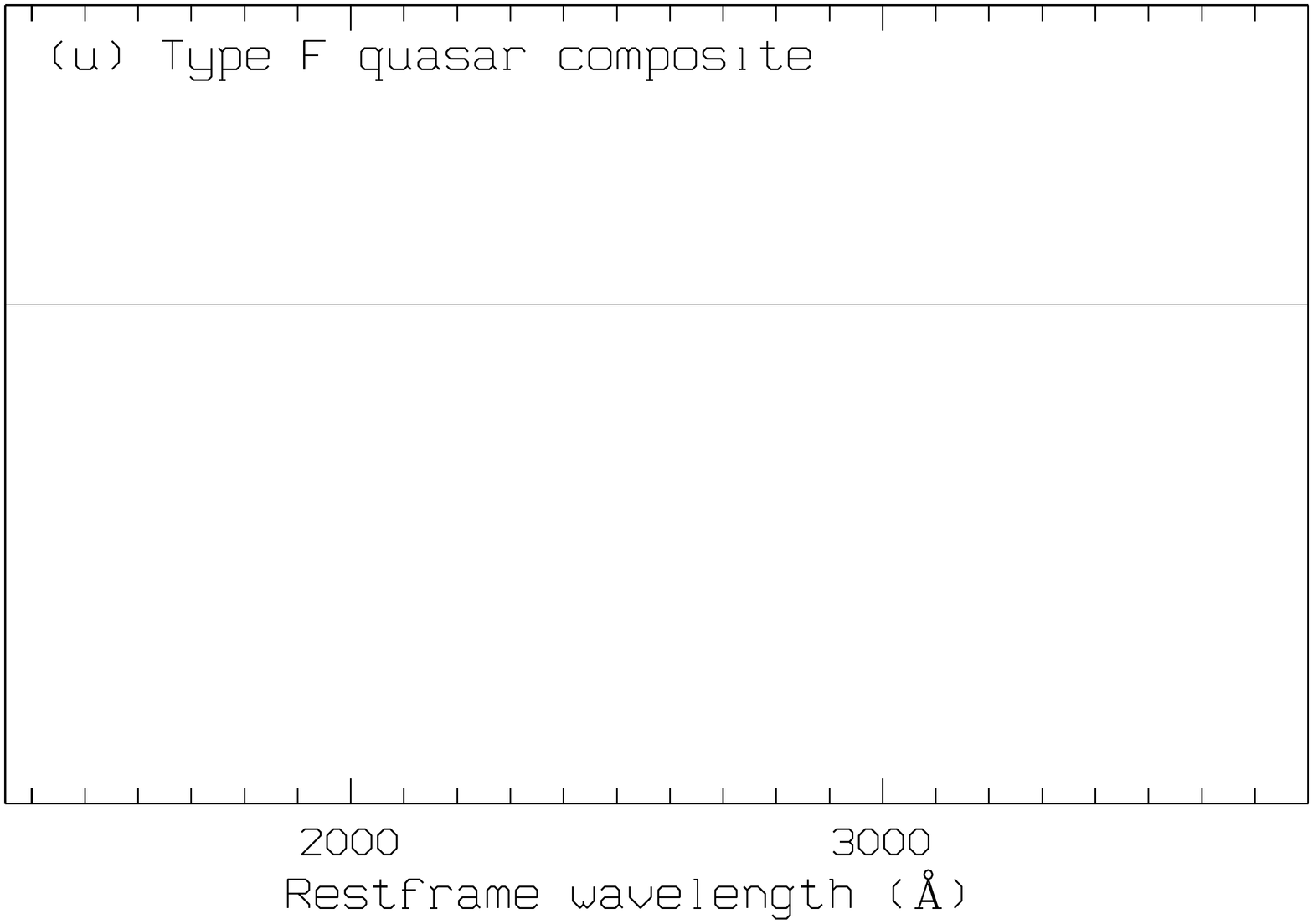}\hfill \\
\end{tabbing}
\caption{As Fig.\,\ref{fig:mysts} but for the ratio of the dereddened
spectra to the type F composite.}
\label{fig:ratio_type_F_mysts}
\end{figure*}

Table\,\ref{tab:mysts} lists 9 mysterious objects and another 11 objects
that may represent links between the mysterious and other unconventional
quasar types.\footnote{Since the selection criteria are not
sharply defined, a few more quasars may be found with more or less 
similar peculiar spectra, such as 
\object{SDSS J110711.40+082331.2},
\object{SDSS\,J140800.43+345124.7}, or
\object{SDSS\,J100353.68+515457.0}.}
The following five objects from this list are not included in Tab.\,\ref{tab:catalogue}:
\object{VPMS J134246.24+284027.5} was not targeted by SDSS,
\object{SDSS J$010540.75-003313.9$} and \object{SDSS J073816.91+314437.0} were
not classified as quasars in the SDSS DR7, and
for both \object{SDSS J130941.35+112540.1} and \object{SDSS J161836.09+153313.5}
the (wrong) redshifts given in the DR7 ($z$ = 4.39 and 4.376, respectively)  
exceed our upper selection limit of $z\le4.3$ (Sect.\,\ref{subsec:selection}).
Tab.\,\ref{tab:mysts} does not include  \object{FBQS 1503+2330} because the 
SDSS spectrum does not cover much of the wavelength range shortwards of
\ion{Mg}{ii}. The objects that are not part of our unusual quasar sample are 
flagged in the first column of Tab.\,\ref{tab:mysts}. For clarity, the 
second column gives the panel in Figs.\,\ref{fig:mysts} to \ref{fig:ratio_type_F_mysts} 
in which we plot the spectrum.

For 11 quasars, the redshift could be obtained from emission lines. 
In 9 cases (marked in Tab.\,\ref{tab:mysts}), 
the systemic redshifts were 
computed from the [\ion{O}{ii}] $\lambda$3728 line, for the other
two spectra from \ion{Mg}{ii} and \ion{C}{iii}]. 
The redshifts of the remaining 9 objects
were identified with those of the highest-$z$ absorption features.
For a few objects, e.g., VPMS J134246.24\-+284027.5,
the redshift is fairly uncertain.

All objects with dubious spectra were checked for proper motions (pm) as described in 
Sect.\,\ref{subsec:contaminants}. The pm data for \object{VPMS J134246.24+284027.5} were
taken from Scholz et al. (\cite{Scholz97}). A significant non-zero pm was found
for the object \object{SDSS J020731.81+004941.6} (Fig.\,\ref{fig:stellar-cont}),
which was rejected thereafter from the quasar list. The determined pm components
$\mu_\alpha \cos\delta$ and $\mu_\delta$ are listed in Tab.\,\ref{tab:mysts},
along with the formal errors and the number of epochs $N_{\rm e}$. 
The column ``pm?'' contains our conclusion, which is not based 
on the formal error alone (see Sect.\,\ref{subsec:contaminants}).
For 13 objects (76\%), the results are clearly consistent with the conclusion that
no significant pm is detected. For three objects, non-zero pm in 
one direction cannot be definitely excluded.  A formally significant pm was 
measured for \object{SDSS J075437.85+422115.3}. This result is doubtful, however,
because the measurements are likely affected by a faint source close to the sightline.
For the three objects without pm data in Tab.\,\ref{tab:mysts}, the extragalactic origin
is clearly indicated by redshifted emission lines.

The partial spectra in the rest-frame wavelength interval 
1350\,\AA\ - 3800\,\AA\ are shown in Fig.\,\ref{fig:mysts}.
The spectra were arbitrarily normalised and slightly smoothed by a
three-pixel boxcar, with the exception of the noisy spectrum of
\object{SDSS J085502.20+280219.6} where a seven-pixel boxcar
was applied. For comparison, the SDSS quasar composite spectrum is shown
in the last panel (bottom right).
The vertical bars above the spectra indicate the typical 
strong quasar emission lines
\ion{Si}{iv}  $\lambda\lambda$1394,1403,
\ion{C}{iv}   $\lambda\lambda$1548,1551, 
\ion{C}{iii}] $\lambda$1909,
and 
\ion{Mg}{ii}  $\lambda\lambda$2796,2804.
The dashed lines below the spectra mark the same absorption lines as in 
Fig.\,\ref{fig:so}. If the redshift was obtained from absorption features, 
the lines above and below the spectrum refer to the same $z = z_{\rm abs}$. 
Exceptions are \object{SDSS J101723.04+230322.1} and \object{SDSS J073816.91+314437.0},
where the solid lines mark the highest-$z$ and the dashed ones the 
strongest absorption troughs. 
 
The three new mysterious objects (Fig.\,\ref{fig:mysts} {\bf d}-{\bf f}) are 
\object{SDSS J160827.08+075811.5} ($z$ = 1.18),
\object{SDSS J161836.09+153313.5} ($z$ = 1.358), and
\object{SDSS J085502.20+280219.6} ($z$ = 1.511).
Both \object{SDSS J160827.08+075811.5} and \object{SDSS J161836.09+153313.5} 
are similar to the Hall et al. object \object{SDSS J$010540.75-003313.9$}. 
A remarkable difference between these two spectra and the SDSS J$010540.75-003313.9$
spectrum is the more abrupt transition between the blue part
and the red part of the continuum. At first glance,
it is tempting to ascribe the pronounced peak to the \ion{Mg}{ii} line. 
However, if we adopt the redshift from the [\ion{O}{ii}] line as the systemic 
redshift, the peak is clearly placed shortwards of $2800$\AA\ in either case.
For \object{SDSS J160827.08+075811.5}, the redshift of the [\ion{O}{ii}] line 
agrees very well with that of an (associated?) absorption line system seen
in \ion{Mg}{ii}, \ion{Mg}{i}, and \ion{Fe}{ii}. 
Another new mysterious object is \object{SDSS J085502.20+280219.6}, which
also resembles \object{SDSS J$010540.75-003313.9$} but has
a stronger depression at shorter wavelengths.

Similar to the mysterious objects in Fig.\,\ref{fig:mysts}\,{\bf a}-{\bf i},  
the spectra of \object{SDSS J134951.93+382334.1} and \object{SDSS J215950.30+124718.4} 
in panels {\bf (j)} and {\bf (k)}, respectively, show a break in their continuum around the
position of the \ion{Mg}{ii} line. The redshift of $z = 1.516$ for 
\object{SDSS J215950.30+124718.4} was derived from a clearly identified narrow 
\ion{Mg}{ii} absorption doublet at the position of the break.
Both \object{SDSS J134951.93+382334.1} and \object{SDSS J215950.30+124718.4} 
are overlapping trough LoBAL quasars (see bottom row of Fig.\,\ref{fig:so})
and possibly mark a link between overlapping-trough objects and mysterious
objects. (The dropoff at \ion{Mg}{ii} in \object{SDSS J215950.30+124718.4}
is more abrupt than in any of the latter objects.) From this point of view, 
the spectra in panels {\bf (h)} to {\bf (k)} seem to represent a sequence of  
overlapping-trough absorption from a low covering factor in 
\object{VPMS J134246.24+284027.5} to the highest covering factor in 
\object{SDSS J215950.30+124718.4}.  

The spectra in Fig.\,\ref{fig:mysts}\,{\bf m}-{\bf o} show narrower 
BAL troughs with low covering factors.
\object{SDSS J151627.40+305219.7} (panel {\bf p}) is very similar
to \object{SDSS J073816.91+314437.0} (panel {\bf q}) from Hall et al. 
(\cite{Hall04}). The high-velocity BAL in \object{SDSS J101723.04+230322.1} 
(panel {\bf o}) makes it plausible that high-velocity BALs contribute 
to the former two spectra.
  
Since intrinsic reddening seems to be present in all spectra in Fig.\,\ref{fig:mysts}, 
we attempted to correct for it and compare the dereddened spectra with the
typical quasar spectrum, following Hall et al. (\cite{Hall02}). As in 
Sect.\,\ref{subsec:composites}, we adopted the SMC extinction curve at the
quasar redshift with $E^{\rm (intr)}_{\rm B-V}$ as a free parameter.   
For $z \la 1.5$, the object spectrum was fitted to the SDSS composite spectrum at
wavelengths $\lambda \ga 3000$\AA. For the higher-$z$ quasars, we simply 
tried to fit at the longest wavelengths what almost certainly amounts to an
underestimated amount of reddening. We note that
the method of deriving $E^{\rm (intr)}_{\rm B-V}$ is only approximate and
the results are expected to be rather uncertain, at least for some objects. 
The mean value $\langle E^{\rm (intr)}_{\rm B-V} \rangle = 0.27$ mag 
of the individually estimated reddening parameters of the 8 type D quasars 
from Tab.\,\ref{tab:mysts} is somewhat larger than the ensemble average of 
0.20 mag derived from the geometric mean composite spectrum 
of the type D quasars (Tab.\,\ref{tab:groups} and Sect.\,\ref{subsec:composites}). 
For the five type A quasars in Tab.\,\ref{tab:mysts},
we have $\langle E^{\rm (intr)}_{\rm B-V} \rangle = 0.22$ mag compared to 0.18 mag
for the whole sample.

The ratios of the thus dereddened spectra to the SDSS composite are shown 
in Fig.\,\ref{fig:ratio_mysts} in the same style as in Fig.\,\ref{fig:mysts}. 
The individual intrinsic reddening parameters are listed in Tab.\,\ref{tab:mysts}. 
A perfect fit would result in a horizontal line at the level of unity.
All dereddened spectra, perhaps with the exception of 
\object{SDSS J091940.97+064459.9}  in panel {\bf (r)}, show a depression 
of the flux relative to the composite at short wavelengths. 
In most but not all cases, this sharp decline sets in at wavelengths around the
\ion{Mg}{ii} line. This behaviour may be caused by very wide iron absorption
troughs that overlap at substantially different velocities 
(Hall et al. \cite{Hall02}). This depression is also seen in the spectra in
rows 4 and 5, where the presence of absorption troughs 
from \ion{Mg}{ii} and \ion{Fe}{ii}  UV1 and UV2  (plus \ion{Si}{ii} and 
\ion{Fe}{iii} UV34, UV48 in the case of \object{SDSS J120337.91+153006.6},
panel {\bf (n)})\footnote{\object{SDSS J152438.79+415543.0}
and \object{SDSS J120337.91+153006.6} are another two sdvdpc-candidates with 
deeper \ion{Fe}{ii} UV2 than UV1 (Sect.\,\ref{so})} is more obvious.
The combination of reddening and partial-covering overlapping-troughs could 
perhaps explain such objects as \object{SDSS J145045.56+461504.2} (panel {\bf g}) and
\object{SDSS J$010540.75-003313.9$} (panel {\bf c}), while narrower troughs
could perhaps explain the spectra in panels {\bf (m)} to {\bf (o)}.
As a remarkable exception, the unusual spectrum of \object{SDSS J091940.97+064459.9} 
can be explained by the 2175\AA\ absorption trough at the quasar redshift
(at which narrow associated \ion{Mg}{ii} absorption is seen).

Hall et al. (\cite{Hall02}) discussed, among others, the possibility that
the strange continua could represent reddened versions
of quasars with weak \ion{Mg}{ii} but strong broad \ion{Fe}{ii} emission. To check this
idea for the quasars in Fig.\,\ref{fig:mysts}, we repeated the dereddening 
procedure with two modifications. First, the composite spectrum of strong 
iron emitters (type F) was used as reference instead of the SDSS quasar composite.
Second, the individual
quasar spectrum was fitted to the reference spectrum at either end.
Fig.\,\ref{fig:ratio_type_F_mysts} shows the ratios of the dereddened spectra
to the reference spectrum. A good fit was achieved for \object{SDSS J091940.97+064459.9}.
However, for most of the spectra the depression in the observed spectrum 
shortwards of \ion{Mg}{ii} was obviously not eliminated. Moreover,
strong dereddening produces an unexplained bump at 2700\AA.  
As argued by Hall et al. (\cite{Hall02}), \ion{Fe}{ii} emission may be present
longwards of \ion{Mg}{ii} (near 3200\AA) but is not expected immediately shortwards of
\ion{Mg}{ii}. In agreement with Hall et al., we conclude that the majority 
of the spectra cannot be explained by reddened strong iron emission alone.   

For several objects in Figs.\,\ref{fig:mysts} to \ref{fig:ratio_type_F_mysts}, the
shape of the spectrum around \ion{Mg}{ii} resembles a broad double-line profile. 
The most interesting case is \object{SDSS J101723.04+230322.1}. 
Very broad double-shouldered  \ion{Mg}{ii} emission, in 
combination with reddening and broad, shallow absorption troughs, 
is another possible scenario mentioned by Hall et al. (\cite{Hall02}) to explain their 
two mysterious objects. As can be seen in Figs.\,\ref{fig:ratio_mysts} and 
\ref{fig:ratio_type_F_mysts}, the spectrum of \object{SDSS J101723.04+230322.1} 
cannot be understood as being caused by these double-shouldered line profiles alone or
in combination with dust reddening, strong iron emission and associated 
narrow-line absorption. In addition, overlapping broad absorption troughs 
from various elements, in particular \ion{Fe}{ii}, are definitely needed to 
explain the complex shape of this spectrum. Remarkably, the apparent 
double-line profile in \object{SDSS J101723.04+230322.1}
is seen not only in the \ion{Mg}{ii} line but also for \ion{C}{iii}] and \ion{C}{iv}.  
Only a few quasars have been reported so far to definitely 
display this structure in other lines than H$\alpha$ and H$\beta$
(Halpern et al. \cite{Halpern96}; 
Strateva et al. \cite{Strateva03}; 
Luo et al. \cite{Luo09}; 
Chornock et al. \cite{Chornock10}).  
The selection of double-peaked AGNs has been primarily based on 
Balmer lines (Halpern et al. \cite{Halpern96}; Eracleous et al. \cite{Eracleous04}).
Ultraviolet observations have shown that their high-ionisation lines (e.g., \ion{C}{iv})
frequently lack double-peaked profiles. However, double-peaked AGNs 
appear to be rather inhomogeneous as a class, thus we feel that
this difference does not conclusively rule out the double-shouldered hypothesis for 
\object{SDSS J101723.04+230322.1}. It would be interesting to study the
line profiles of the Balmer lines via infrared spectroscopy.

Most of the quasars in Tab.\,\ref{tab:mysts} are FIRST sources, but
radio-weak, i.e. close to the threshold between radio-quiet and radio-loud quasars. 
The mean radio loudness of the seven radio-detected 
mysterious objects is  $\langle R_{\rm i} \rangle = 0.97 \pm 0.28$.
Among the three new mysterious objects, only \object{SDSS J160827.08+075811.5} 
was detected by FIRST.  \object{SDSS J085502.20+280219.6} is 2.6 mag 
fainter in $i$ than \object{SDSS J160827.08+075811.5} 
(and a FIRST flux as small as 0.2 mJy thus still corresponds to 
$R_{\rm i} = 0.4$). 
The third quasar, \object{SDSS J215950.30+124718.4}, is not in the FIRST area.
The mean radio loudness parameter of the 15 FIRST sources in 
Tab.\,\ref{tab:mysts} is $\langle R_{\rm i} \rangle = 1.02\ (\pm 0.41)$. 
The strongest radio source, \object{SDSS J091940.97+064459.9}, was classified 
by Plotkin et al. (\cite{Plotkin08}) as a lower-confidence BL Lac object.
None of the objects from Tab.\,\ref{tab:mysts} could be identified 
in the {\it ROSAT} All-Sky Survey Faint Source Catalog 
(Voges et al. \cite{Voges00}). Weak X-ray emission is a characteristic 
property of BAL quasars and is usually attributed to strong X-ray absorption, which 
enables and enhances the acceleration of the BAL outflow driven by radiation 
pressure (Murray \& Chiang \cite{Murray98}; Proga et al. \cite{Proga00}; 
Gallagher et al. \cite{Gallagher06}; Gibson et al. \cite{Gibson09};
Streblyanska et al. \cite{Streblyanska10}).

%
\section{Summary and conclusions}\label{sec:summary}
%

We have compiled a catalogue of unusual quasars from the unprecedented 
spectroscopic database of the SDSS DR7. Unusual quasars were selected from $\sim 10^5$ 
spectra with $z=0.6$\ to 4.3 classified as quasars by the SDSS spectroscopic pipeline. 
The selection method is essentially a combination of the power of Kohonen's
(Kohonen \cite{Kohonen82},\cite{Kohonen01}) self-organising maps and
the detailed visual inspection of the selected spectra. 
We paid particular attention to reject contaminants, i.e. spectra from 
rare spectral types of other astrophysical objects (e.g., white dwarfs 
and hybrid spectra of two objects within the fibre aperture). Proper motions
were checked first by cross-correlating our sample with the PPMXL catalogue (R\"oser et al. 
\cite{Roeser10}), which contains $\sim 90$\% of the selected objects. 
In addition, and more importantly, we used all available multi-epoch positions
from different sky surveys to estimate the proper motions of 121 selected sources.
The results of the pm determination led to the 
rejection of seven objects from the quasar sample.  

The final catalogue contains 1005 quasars, which were classified
into 7 different types: 
(A) LoBALs and unusual BALs (21\%),
(B) strong normal BALs, HiBALs (21\%), 
(C) red quasars (12\%), 
(D) quasars which appear red shortwards of $\sim 3000$\AA\ only (15\%),
(E) weak-line quasars (18\%),  
(F) strong iron emitters (11\%), and
(M) miscellaneous. 
Our approach was primarily aimed at striking outliers and the 
catalogue is expected to be largely complete in this respect. However, 
the selection is not based on sharply defined quantitative criteria and the 
selected sample is thus incomplete in a quantifiable sense. Nevertheless, the compilation
is expected to be very useful for studying relations between the various 
types of spectral peculiarities or selecting particularly interesting individual
objects for detailed investigations. 
   
The analysis of this sample yields the following conclusions for the
main types A to F:
\begin{itemize}
\item 
The arithmetic median composite spectra (Sect.\,\ref{subsec:composites}) 
clearly differ from type to type 
and from the SDSS composite spectrum of normal quasars. For type A, 
the huge spectrum-to-spectrum variation reflects a high level of diversity.  
The spectral peculiarity, measured by the mean square deviation 
of the individual spectrum from the SDSS composite, is on average largest 
for types C,D, and A and smallest for type E. Unusual BALs are frequently 
accompanied by red continua and strong iron emission.
\item 
We combined the SDSS spectra with the fluxes from 2MASS to compute  
geometric mean spectra with significantly improved wavelength coverage
(Sect.\,\ref{subsec:composites}). These composite spectra were used to 
estimate the intrinsic reddening (Sect.\,\ref{subsec:reddening}). 
In agreement with previous studies, we found that BAL quasars have, on
average, significantly redder continua than normal quasars with
$E(B-V) \approx 0.1$ mag for type B adopting SMC extinction. The unusual
BAL quasars (type A), which are mostly LoBALs, are even redder with 
$E(B-V) \approx 0.18$ mag. The decline in the dereddened type A composite
in the UV is ascribed to the combined effect of the UV absorption troughs.
The strongest mean reddening is found for the type C with
$E(B-V) \approx 0.38$ mag. On the other side, weak-line quasars
are on average insignificantly redder than normal ($E(B-V) \approx 0$).
\item 
For the quasars with substantial reddening in the UV only (type D), 
the continuum slope of the SDSS quasar composite could be fitted only
with an extinction curve that is steeper in the UV
than for the SMC (Sect.\,\ref{subsec:reddening}). 
\item
After the (statistical) correction for intrinsic extinction, 
the KS test for the absolute magnitudes yielded that all six types
are on average more luminous than comparison samples of normal quasars
(Sects.\,\ref{subsec:redshifts} and \ref{subsec:reddening}).
\item 
The fraction of quasars with radio detections increases with the spectral
peculiarity. This reflects most likely a selection bias: a substantial fraction
of the quasars with very unusual spectra were selected for foloow-up
spectroscopy because they are radio sources. On the other hand, unusual BALs
(as well as  quasars with  strong iron emission)
tend not to have high radio luminosities. A similar, though not statistically
significant trend was found for the red quasars. It is thus tempting to
speculate that there exists a larger population of unusual quasars that
have not yet been discovered simply because they are too faint radio sources
(Sect.\,\ref{subsec:radio_luminosity}).
\item
To estimate the impact of foreground galaxies along the line of sight
on the spectral peculiarities, we have checked both the SDSS images and spectra for
finger prints of such intervening matter (Sect.\,\ref{subsec:foreground}). 
About 1\% of the quasar images show nearby (usually faint) structures 
and for $\sim 8$\% we detected narrow absorption lines that are distinct
from the quasar lines. There are at least two quasars in the sample where the spectral
peculiarity is clearly caused by foreground galaxies. Moreover, two of nine high-$z$
quasars identified on images from the Hubble Space Telescope show extended 
structures very close to the sightline (Appendix A). 
Nevertheless, we did not find any conclusive evidence that the sample properties
are significantly affected by the extragalactic foreground.
\item
Our sample contains several objects with very peculiar spectra, 
among them FeLoBAL quasars with large numbers of narrow absorption troughs or 
extremely wide overlapping troughs. The most spectacular example of the latter type
is \object{SDSS J094317.59+541705.1} at $z = 2.22$ where the continuum is strongly
depressed from \ion{Mg}{ii} at red edge of the spectrum all the way down to
Ly$\alpha$ at the blue edge (Sect.\,\ref{so}).
\item
We constructed a small sample of nine quasars with spectral properties similar to the two
``mysterious'' objects discovered by Hall et al. (\cite{Hall02}) and another 11 
more or less similar quasars (Sect.\,\ref{subsect:myst}).
The majority (75\%) of these objects are radio sources, but radio-weak ($R_{\rm i} \approx 1$).
Both moderate reddening, FeLoBAL features, and strong Fe emission were
frequently indicated. This combination may explain the shape of the continuum with the
characteristic drop-off shortwards of the position of the \ion{Mg}{ii} line
but does not account for the lack of typical quasar emission lines and the shape 
of the spectrum around the position of the \ion{Mg}{ii} line. Broad double-peaked
emission, in addition to other effects, is an attractive idea (Hall et al.
\cite{Hall02}). Double-shouldered \ion{Mg}{ii} emission seems to be indicated in
some spectra. The best case is \object{SDSS J101723.04+230322.1} where similar
profiles are, however, seen as well around \ion{C}{iv} and \ion{C}{iii}], which
is quite unusual for the known double-peaked quasars.  Moreover, we have also presented
a few quasars where the continuum drop-off appears at much shorter wavelengths,
which seems difficult to explain in this way. A satisfactory explanation of these
puzz\-ling spectra remains a challenge.
\end{itemize}


\begin{acknowledgements}

We thank the referee, Patrick B. Hall, for providing 
many constructive comments and suggestions that significantly 
improved this paper.
This research is based on the Sloan Digital Sky Survey (SDSS).
Funding for the SDSS and SDSS-II has been provided by 
the Alfred P. Sloan Foundation, the Participating Institutions 
(see below), the National Science Foundation, the National 
Aeronautics and Space Administration, the U.S. Department 
of Energy, the Japanese Monbukagakusho, the Max Planck 
Society, and the Higher Education Funding Council for 
England. The SDSS Web site is http://www.sdss.org/.
The SDSS is managed by the Astrophysical Research 
Consortium (ARC) for the Participating Institutions. 
The Participating Institutions are: the American 
Museum of Natural History, Astrophysical Institute 
Potsdam, University of Basel, University of Cambridge 
(Cambridge University), Case Western Reserve University, 
the University of Chicago, the Fermi National 
Accelerator Laboratory (Fermilab), the Institute 
for Advanced Study, the Japan Participation Group, 
the Johns Hopkins University, the Joint Institute 
for Nuclear Astrophysics, the Kavli Institute for 
Particle Astrophysics and Cosmology, the Korean 
Scientist Group, the Los Alamos National Laboratory, 
the Max-Planck-Institute for Astronomy (MPIA), 
the Max-Planck-Institute for Astrophysics (MPA), 
the New Mexico State University, the Ohio State 
University, the University of Pittsburgh, University 
of Portsmouth, Princeton University, the United 
States Naval Observatory, and the University of 
Washington. 

This publication makes use of data products from the 
Wide-field Infrared Survey Explorer, which is a joint 
project of the University of California, Los Angeles, and 
the Jet Propulsion Laboratory/California Institute of Technology, 
funded by the National Aeronautics and Space Administration.
This research is based also on observations with the NASA/ESA Hubble 
Space Telescope, and obtained from the Hubble Legacy Archive, which 
is a collaboration between the Space Telescope Science Institute 
(STScI/NASA), the Space Telescope European Coordinating Facility 
(ST-ECF/ESA) and the Canadian Astronomy Data Centre (CADC/NRC/CSA).
Finally, this research has made use of the SIMBAD database,
operated at CDS, Strasbourg, France.

\end{acknowledgements}

\appendix
\section{Images of unusual quasars in the SDSS and in the Hubble Legacy Archive}

Unusual spectral properties can be the result of a positional coincidence
with a foreground object in combination with gravitational 
lensing (Irwin et al. \cite{Irwin98}; Chartas \cite{Chartas00}).
We used the SDSS Explorer and found that for $\sim 20$\% of the quasars, 
with only slight fluctuations from type to type,
additional objects are seen within 12\arcsec$\times$ 12\arcsec\ fields
centred on the quasar. A substantial fraction is expected to be faint 
stars from our galaxy. For 1.3\% of the quasars,
we found another object closer than $\sim 2$\arcsec\ or extended structures 
that likely overlap the line of sight towards the quasar.
In this case, the fraction is largest for type C (3.4\%), but the samples are much
too small for a statistical analysis. These 13 quasars are listed 
in Tab.\,\ref{tab:fg_sdss}. 
Among them are the two known lensed quasars
\object{SDSS J081959.80+535624.2} (Inada et al. \cite{Inada10}) and 
\object{SDSS J090334.94+502819.3} (Johnston et al. \cite{Johnston03}).
The SDSS image of \object{SDSS J081959.80+535624.2} shows a double structure 
with a separation $\la 4$\arcsec. The spectrum indicates that there is an
absorption line system at $z_{\rm abs} = 0.294$.
\object{SDSS J090334.94+502819.3} was classified as an unusual quasar 
of type\,C because of its red continuum between 
Ly\,$\alpha$ and \ion{C}{iv}. In addition, the \ion{C}{iv} emission line is 
absorbed at the red wing. However, as demonstrated by 
Johnston et al. (\cite{Johnston03}), decomposing the spectrum into a luminous 
red galaxy component at $z=0.388$ and the background  quasar component 
yields a normal quasar spectrum.
Another interesting case is the quasar \object{SDSS J$024230.65-000029.7$}, which 
is seen through an outer spiral arm  of NGC1068.

\begin{table}[hhh]
\caption{
Unusual quasars on {\it HST} images. 
}
\begin{flushleft}
\begin{tabular}{ccclc}
\hline\hline
SDSS J               & $z$   & $T$ & Structure          & Ref.\\
\hline
$005006.34-005319.3$ & 4.347 & B   &  ...                  &\\
$090334.94+502819.3$ & 3.579 & C   & gal. group            & 1\\
$114756.00-025023.4$ & 2.560 & C   & faint nearby          & 2\\
$122622.03+662018.0$ & 3.874 & B   &  ...                  &\\
$134026.43+634433.1$ & 2.784 & A   & faint 5\arcsec        &\\
$150424.98+102939.1$ & 1.835 & E   & gal $\sim 4$\arcsec   &\\
$155633.78+351757.3$ & 1.499 & A   &  ...                  &\\
$173049.10+585059.5$ & 2.033 & A   &  ...                  &\\
$220557.02+121239.7$ & 1.382 & E   &  ...                  &\\
\hline
\end{tabular}
\end{flushleft}
\tiny{
{\bf References.} (1) Johnston et al. (\cite{Johnston03});
(2) Bentz et al. (\cite{Bentz08})
}
\label{tab:fg_hst}
\end{table}

\begin{figure}[hbtp]
\begin{tabbing}
\fbox{\includegraphics[bb=110 320 480 689,scale=0.34,clip]{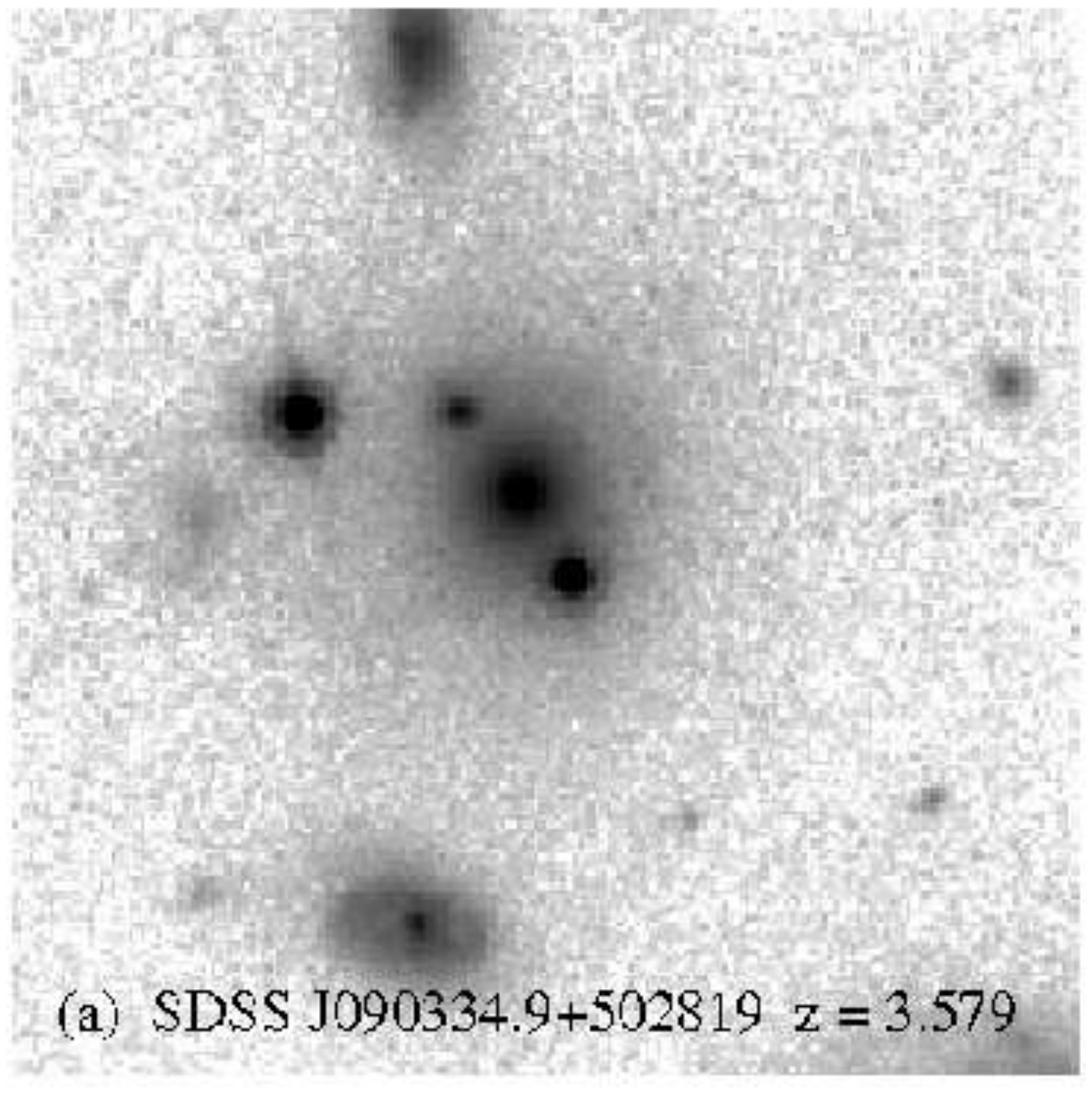}} \=
\fbox{\includegraphics[bb=110 320 480 689,scale=0.34,clip]{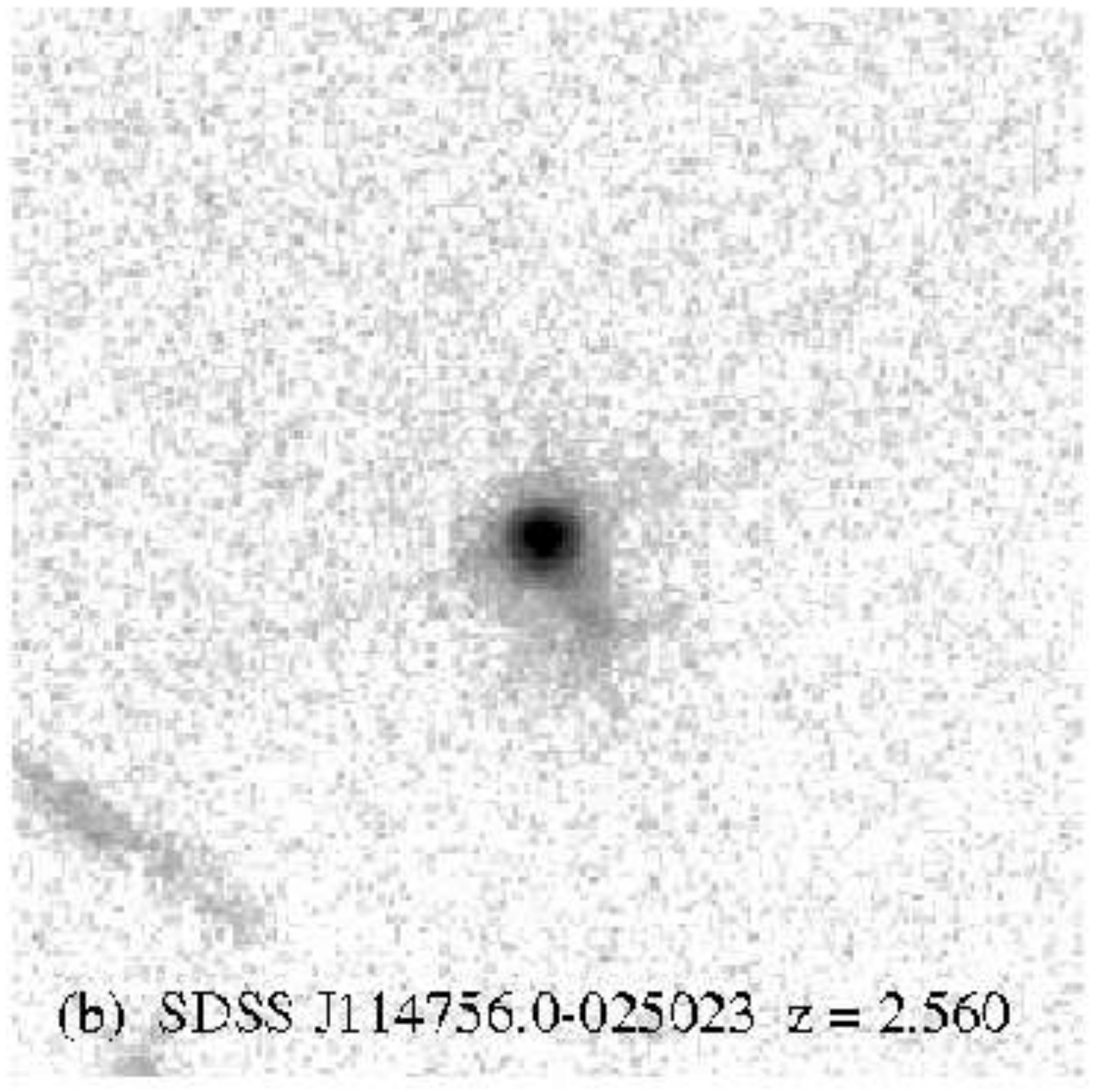}} \= \\
\fbox{\includegraphics[bb=110 320 480 689,scale=0.34,clip]{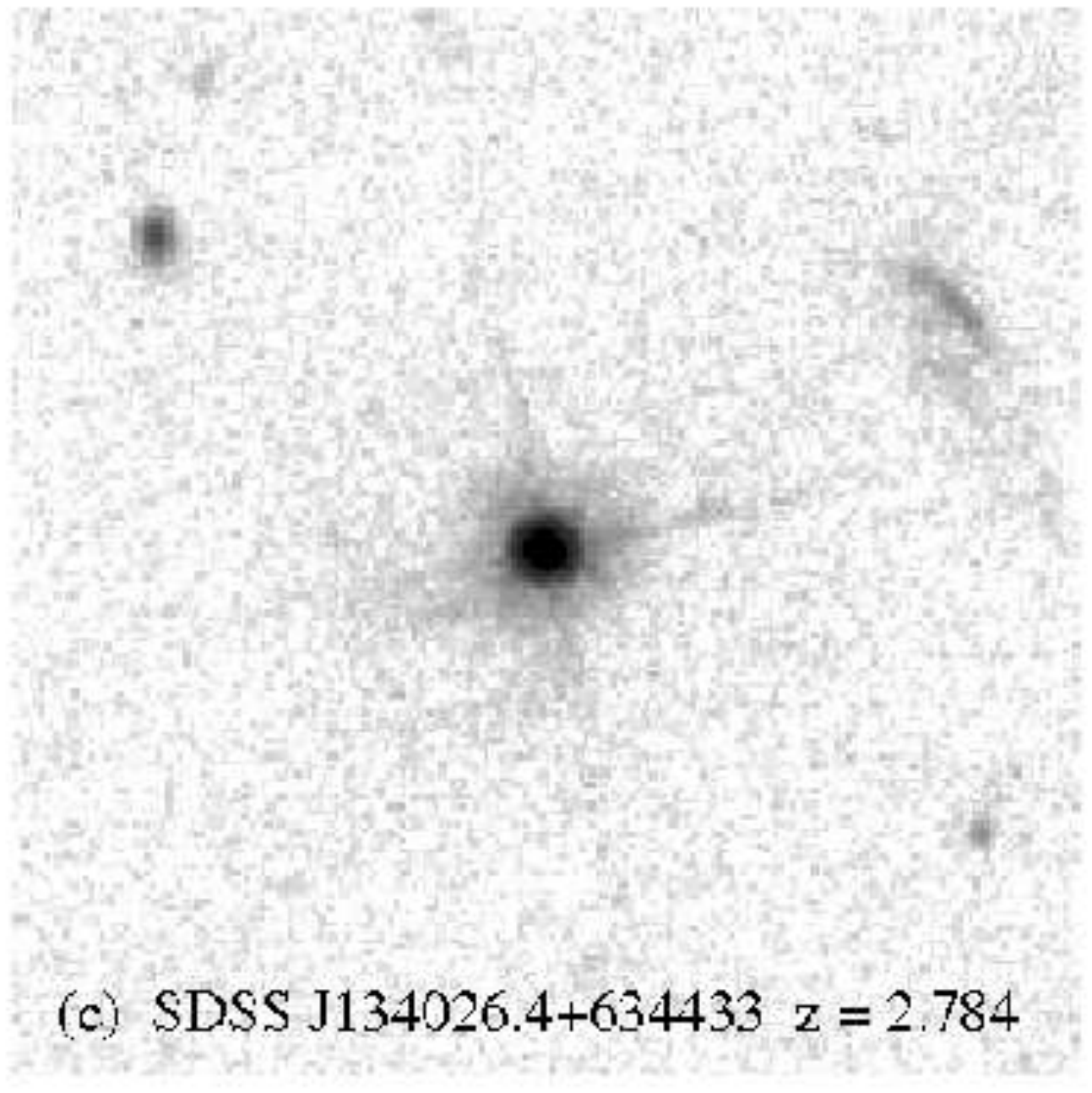}} \>
\fbox{\includegraphics[bb=110 320 480 689,scale=0.34,clip]{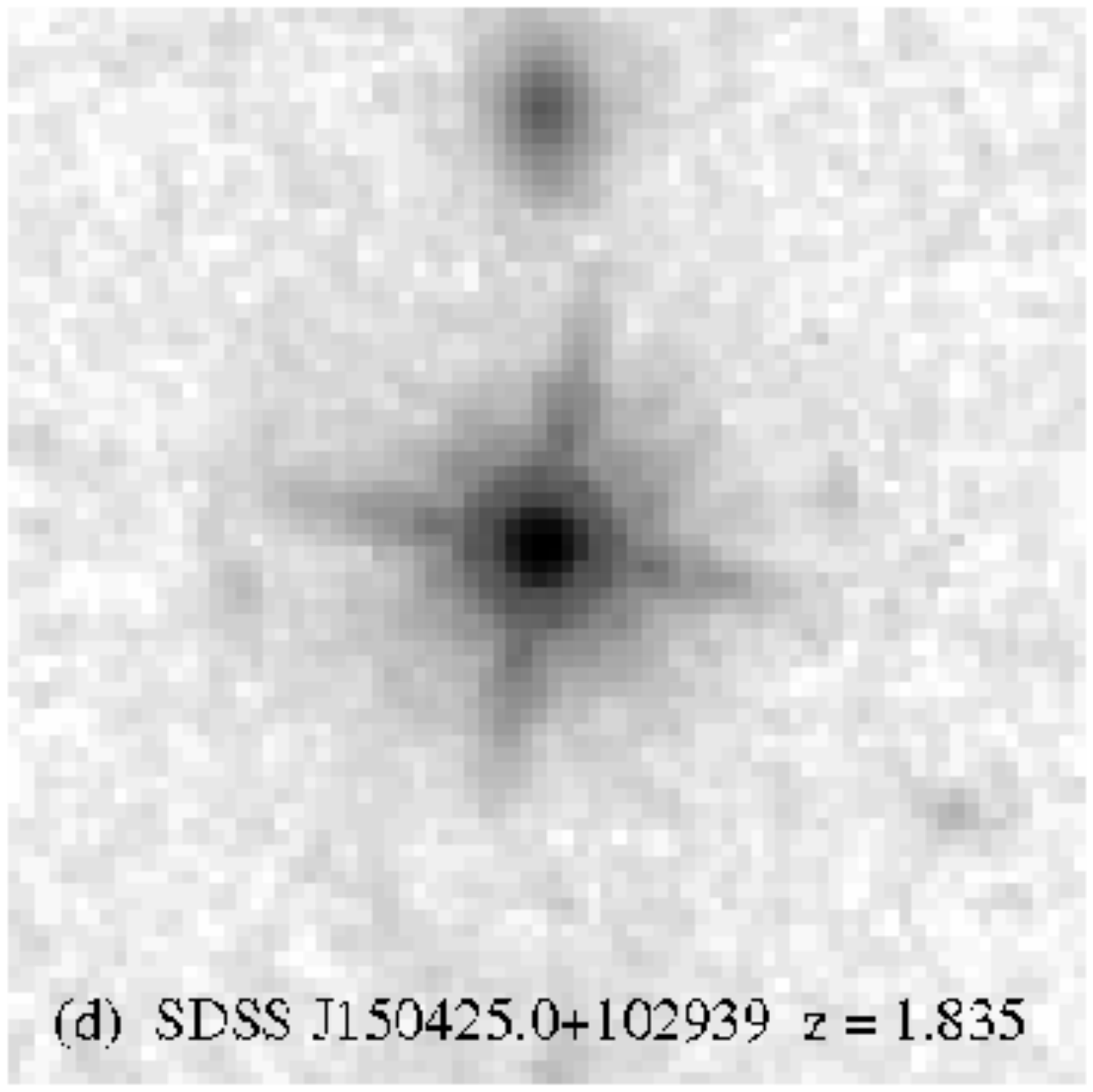}} \> \\
\end{tabbing}
\caption{
$10\arcsec \times 10\arcsec$ cutouts from the {\it Hubble Space Telescope} 
images of four quasars (unresolved objects close to the centre) 
showing extended structures within 5\arcsec\ from the line of sight.
((a) to (c): ACS, (d): WFC3).
}
\label{fig:HST}
\end{figure}

One of us (B.K.) has performed a systematic search of the Hubble Legacy Archive
for images from the {\it Hubble Space Telescope} covering quasars from our 
catalogue. Nine detections were found (Tab.\,\ref{tab:fg_hst}).
Two quasars clearly show extended structures close to the sightline
(Fig.\,\ref{fig:HST}). The gravitationally lensed quasar
\object{SDSS J090334.9+502819} mentioned above is seen through 
the inner region of a luminous featureless galaxy.  
The other object, \object{SDSS J$114756.0-025023$} at $z = 2.560$, is connected with a faint
structure, which may be at the redshift of the quasar (Bentz et al. \cite{Bentz08}).
Both spectra show red continua of type C and line absorption 
at $z_{\rm abs} \approx z_{\rm em}$. Extended structures are also seen
around \object{SDSS J134026.4+634433} and \object{SDSS J150425.0+102939}, 
although at somewhat larger distances of $\sim 4$\arcsec\ from the line of sight.
There are no clear cases of the
signatures of foreground objects in the spectra of these quasars. 

\vspace{0.5cm}
\noindent
{\it Note added in proof.}
After this paper has been accepted for publication, we became aware of the paper by Plotkin et al.
(\cite{Plotkin10}) on optically selected BL Lac candidates from the SDSS. As a side-product, these authors discuss
a small sample of serendipitously recovered/discovered higher-redshift objects that show extreme drop-offs
in their continua bluewards of restframe 2800 \AA. We realised that eight out of their nine objects are
part of our sample of mysterious and related objects. Among them are four objects (SDSS J160827.08+075811.5,
SDSS J161836.09+153313.5, SDSS J134951.93+382334.1, and SDSS J215950.30+124718.4) which were supposed
to be newly discovered by the present study.


\end{document}